\documentclass[11pt,a4paper,english,nofootinbib]{revtex4}
\usepackage{lmodern}
\usepackage{lmodern}

\usepackage[T1]{fontenc}
\usepackage[latin9]{inputenc}
\usepackage{color}
\usepackage{babel}
\usepackage{verbatim}
\usepackage{amsmath}
\usepackage{amssymb}
\usepackage{stmaryrd}
\usepackage{graphicx}
\usepackage{esint}
\usepackage{subscript}
\usepackage[unicode=true,pdfusetitle,
 bookmarks=true,bookmarksnumbered=false,bookmarksopen=false,
 breaklinks=false,pdfborder={0 0 1},backref=false,colorlinks=false]
 {hyperref}

\makeatletter

\pdfpageheight\paperheight
\pdfpagewidth\paperwidth

\@ifundefined{textcolor}{}
{%
 \definecolor{BLACK}{gray}{0}
 \definecolor{WHITE}{gray}{1}
 \definecolor{RED}{rgb}{1,0,0}
 \definecolor{GREEN}{rgb}{0,1,0}
 \definecolor{BLUE}{rgb}{0,0,1}
 \definecolor{CYAN}{cmyk}{1,0,0,0}
 \definecolor{MAGENTA}{cmyk}{0,1,0,0}
 \definecolor{YELLOW}{cmyk}{0,0,1,0}
}

\usepackage{babel}
\usepackage{babel}
\usepackage{babel}
\usepackage{babel}
\usepackage{babel}
\usepackage{latexsym}\usepackage{bm}
\usepackage{xcolor}

\makeatother

\begin{document}
\title{Conserved Charges in Extended Theories of Gravity}
\author{Hamed Adami}
\email{hamed.adami@yahoo.com}

\affiliation{Department of Science,\\
 University of Kurdistan, Sanandaj, Iran\\
 {Research Institute for Astronomy and Astrophysics of Maragha (RIAAM),
P.O. Box 55134-441, Maragha, Iran }}
\author{Mohammad Reza Setare}
\email{rezakord@ipm.ir}

\affiliation{{Department of Science,\\
 Campus of Bijar, University of Kurdistan, Bijar, Iran\\
 Research Institute for Astronomy and Astrophysics of Maragha (RIAAM),
P.O. Box 55134-441, Maragha, Iran }}
\author{Tahsin Ça\u{g}r\i{} \c{S}i\c{s}man}
\email{tahsin.c.sisman@gmail.com}

\affiliation{Department of Astronautical Engineering,\\
 University of Turkish Aeronautical Association, 06790 Ankara, Turkey}
\author{Bayram Tekin}
\email{btekin@metu.edu.tr}

\affiliation{Department of Physics,\\
 Middle East Technical University, 06800 Ankara, Turkey}
\date{\today}
\begin{abstract}
We give a detailed review of construction of conserved quantities
in extended theories of gravity for asymptotically maximally symmetric
spacetimes and carry out explicit computations for various solutions.
Our construction is based on the Killing charge method, and a proper
discussion of the conserved charges of extended gravity theories with
this method requires studying the corresponding charges in Einstein's
theory with or without a cosmological constant. Hence we study the
ADM charges (in the asymptotically flat case but in generic viable
coordinates), the AD charges (in generic Einstein spaces, including
the anti-de Sitter spacetimes) and the ADT charges in anti-de Sitter
spacetimes. We also discuss the conformal properties and the behavior
of these charges under large gauge transformations as well as the
linearization instability issue which explains the vanishing charge
problem for some particular extended theories. We devote a long discussion
to the quasi-local and off-shell generalization of conserved charges
in the 2+1 dimensional Chern-Simons like theories and suggest their
possible relevance to the entropy of black holes. 
\[
\]
\tableofcontents{} 
\[
\]
\end{abstract}
\maketitle

\section{Reading Guide and Conventions}

This review is naturally divided into two parts: In Part I, we discuss
global conserved charges for generic modified or extended gravity
theories. Global here refers to the fact that the integrals defining
the charges are on a spacelike surface on the boundary of the space.
In Part II, we discuss quasi-local and off-shell conserved charges;
particularly, for 2+1 dimensional gravity theories with Chern-Simons
like actions. In what follows, the meaning of these concepts will
be elaborated. Here, let us briefly denote our conventions of the
signature of the metric and the Riemann curvature: We use the mostly
plus signature with $g=\text{diag}(-,+,...,+)$ and take the Riemann
tensor to be defined as 
\begin{equation}
\left[\nabla_{\mu},\nabla_{\nu}\right]V^{\rho}=R_{\mu\nu\phantom{\rho}\sigma}^{\phantom{\mu\nu}\rho}V^{\sigma},
\end{equation}
and the Ricci tensor is defined as 
\begin{equation}
R_{\mu\nu}=R_{\phantom{\rho}\mu\rho\nu}^{\rho},
\end{equation}
while the scalar curvature is 
\begin{equation}
R=R_{\phantom{\mu}\mu}^{\mu}.
\end{equation}

\part{GLOBAL CONSERVED CHARGES}

\section{Introduction}

The first ``casualty of gravity'' is Noether's theorem \cite{Noether},
well at least for the case of rigid spacetime symmetries and their
corresponding conserved charges. The ``Equivalence Principle'' in
any form is both a blessing and a curse: it makes gravity locally
trivial (essentially a coordinate effect in the extreme limit of going
to a point-like lab) but then it also, for the same reason, makes
it hard to find local observables in gravity. In fact all the observables
must be global. To heuristically understand this, as an example, consider
the energy momentum\emph{ density} of the gravitational field: Being
a bosonic (spin-2) field, its kinetic energy density is expected to
be of the form $K\sim\partial g\partial g$ which makes no covariant
sense as it can be set to zero in a locally inertial frame (or in
Riemann normal coordinates). This state of affairs of course affects
both the classical gravity and the would-be quantum gravity. The first
thing one wants to know and compute in any theory are the ``observables''
and it turns out generically, a spacetime has no observables and it
would not make sense to construct classical and quantum theories for
those spacetimes. For example, in a spacetime such as the de Sitter
spacetime which does not have a global timelike Killing vector or
an asymptotic spatial infinity, defining a global conserved positive
energy is not possible \cite{Witten_dS}.

In the absence of a tensorial quantity that can represent local gravitational
energy-mass, momentum \emph{etc}, one necessarily resorts to new ideas;
two of which are \emph{quasilocal} expressions that try to capture
gravitational quantities contained in a \emph{finite} region of space
or the \emph{global} expressions that are assigned to the totality
of spacetime. Both of these involve integrations over some regions
of space, and they are not tensorial. But, that does not mean that
they are physically irrelevant. On the contrary, they are designed
to answer physical questions such as what is the mass of a black hole
or how much energy is radiated if two black holes merge as in the
recent observations by the LIGO detectors, of course with the assumption
that these systems can be isolated from the rest of the Universe?
Especially, global expressions that we shall discuss, the Arnowitt-Deser-Misner
(ADM) \cite{adm} and the Abbott-Deser (AD) conserved charges \cite{Abbott},
and their generalizations to higher derivative models the Abbott-Deser-Tekin
(ADT) conserved charges \cite{Deser_Tekin-PRL,Deser_Tekin-PRD} are
well-defined for asymptotically flat and anti-de Sitter spaces, given
the proper decay conditions on the metric and the extrinsic curvature
of the Cauchy surface.

The celebrated ADM mass assigned to an asymptotically flat manifold
exactly coincides with our expectation of an isolated gravitational
system's mass, such as the mass of a black hole. It also turns out
that the ADM mass is a geometric invariant of an asymptotically flat
Riemannian manifold as long as the proper decay conditions are satisfied
\cite{Bartnik}. It is interesting to note that the ADM mass of an
asymptotically flat manifold also became an important part of differential
geometry and it took a long time by differential geometers to prove
the positiveness of this quantity (under certain assumptions) as was
expected to be positive from the physical arguments such as the stability
of the Minkowski spacetime, or supergravity arguments which require
the Hamiltonian to be positive definite. An account of these discussions
as well as a review of the conserved charges in Einstein's theory
can be found in the relevant chapters of the book \cite{Ashtekar}.

After the series of ADM papers, whose results were summarized in \cite{adm}
a great deal of effort was devoted to better understand the Hamiltonian
structure of gravity theories. The literature is too great to do any
justice here; but several works stand out which we now briefly note:
Regge and Teitelboim \cite{Regge:1974zd} realized that the Hamiltonian
of general relativity on a space with a boundary does not have well-defined
functional derivatives; therefore, Hamilton's equations for the phase
space fields are not equal to the Euler-Lagrange equations. Their
detailed analysis shows that the remedy is to add a boundary term,
and that boundary term is exactly the ADM mass of the asymptotically
flat spacetime. This approach is rather revolutionary for two reasons:
first naively, due to the constraints, the bulk Hamiltonian vanishes
exactly in general relativity; and therefore, if one does not add
a boundary term the spacetime is devoid of any energy. On the other
hand, once a boundary term is added, the full Hamiltonian when evaluated
on-shell reduces to the boundary term. Thus, the numerical value of
the total Hamiltonian is the ADM energy. Secondly, the ADM energy,
even though it looks linear in the metric (deviation) at infinity,
captures all the nonlinear energy in the bulk. This is evident from
the Regge-Teitelboim approach.

The next generalization was that of the AD \cite{Abbott} mass for
asymptotically de Sitter or anti-de Sitter spacetimes. In fact, Abbott
and Deser were interested in the stability of the de Sitter spacetime;
however, the AD expression is also well-defined for anti-de Sitter
(AdS) spacetimes which have a spatial infinity at which the surface
integrals can be defined in a coordinate invariant manner. Following
the Regge-Teitelboim \cite{Regge:1974zd}, the extension of the Hamiltonian
approach to the anti-de Sitter spacetime was given bu Henneaux and
Teitelboim in \cite{hen84}, and more recently in \cite{jamsin} which
also includes relevant references.

The next development, which is the main focus of this review, was
the construction of conserved charges in generic higher curvature
theories in various dimensions using the AD methods. These higher
derivative theories picked up interest due to their potential to behave
better at high and low energies where Einstein's gravity suffer; hence,
the construction of conserved quantities in these theories became
important. The Hamiltonian treatment following \cite{Regge:1974zd,hen84}
in the extended gravity theories is highly complicated due to their
higher derivative nature. On the other hand, the AD approach is relatively
straightforward. Deser and Tekin \cite{Deser_Tekin-PRL,Deser_Tekin-PRD}
extended the AD Killing charge construction for gravity theories that
have quadratic or more powers in spacetimes which are asymptotically
AdS. It is important to understand that the cosmological constant
plays a crucial role in all these conserved quantities. For example,
generically for a nonzero cosmological constant $\Lambda$, a term
in the action of the form $\alpha\left({\rm Riem}\right)^{n}$ contributes
to the conserved charges as much as $\alpha\Lambda^{n-1}$ times the
Einstein's theories contribution. Therefore, a given spacetime might
have quite different conserved quantities in different theories.

We would like make one disclaimer before we lay out the computations
leading to the conserved quantities in generic gravity theories: our
basic tool is the Killing charge construction as employed most clearly
in \cite{Abbott,Deser_Tekin-PRD}. As our main focus is generic higher
derivative theories, the review basically evolves around the AD work
and its extension the ADT work. Of course there are other approaches
to the same problem which we have not followed here such as the works
of \cite{Magnon,HawkingHorowitz,97,Aros,40}; elsewhere comparisons
of various methods are discussed in detail; see for example \cite{Hollands}
by Hollands, Ishibashi and Marolf. For a recent review on conserved
charge and related topics in general relativity, please see \cite{Tekin_et_al},
\cite{Banados_rev}, and \cite{Compere}.

\section{ADT Energy}

Let us consider a generic gravity theory defined by the field equations
\begin{equation}
\mathcal{E}_{\mu\nu}\left(g,R,\nabla R,R^{2},...\right)=\kappa\tau_{\mu\nu},\label{eq:EoM}
\end{equation}
where $\mathcal{E}_{\mu\nu}$ is a divergence-free, symmetric two-tensor
possibly coming from an action. Here, $R$ denotes the Riemann tensor
or its contractions, $\kappa$ is the $n$-dimensional Newton's constant
and $\tau_{\mu\nu}$ represents the nongravitational matter source.
It is clear that $\nabla_{\mu}\mathcal{E}^{\mu\nu}=0$ does not yield
a globally conserved charge without further structure. A moment of
reflection shows that given an exact Killing vector $\xi^{\mu}$,
one can construct a covariantly-conserved current $J^{\mu}:=\xi_{\nu}\mathcal{E}^{\mu\nu}$
yielding a partially-conserved current density $\mathcal{J}^{\mu}:=\sqrt{-g}\xi_{\nu}\mathcal{E}^{\mu\nu}$
since $\partial_{\mu}\left(\sqrt{-g}\xi_{\nu}\mathcal{E}^{\mu\nu}\right)=\sqrt{-g}\nabla_{\mu}\left(\xi_{\nu}\mathcal{E}^{\mu\nu}\right)=0$.
But, as this current identically vanishes outside the sources, it
does not lead to a physically meaningful nontrivial conserved quantity.
The compromise is to follow the ADM construction for asymptotically
flat geometries, and AD and ADT constructions for asymptotically (anti-)de
Sitter {[}(A)dS{]} geometries, and split the metric into a background
$\bar{g}_{\mu\nu}$ and a fluctuation $h_{\mu\nu}$ as 
\begin{equation}
g_{\mu\nu}=\bar{g}_{\mu\nu}+\kappa h_{\mu\nu}.\label{eq:Metric_decomp}
\end{equation}
Here and in what follows, without worrying about the units of $\kappa$,
we shall use it as a perturbation order counting parameter. Clearly,
the decomposition of the metric $g$ as (\ref{eq:Metric_decomp})
is coordinate dependent, and hence a different set of coordinates
would lead to different $\bar{g}_{\mu\nu}$ and $h_{\mu\nu}$ tensors.
But, it is a simple exercise to show that infinitesimal change of
coordinates on the manifold can be seen as a gauge transformation
of the form $h_{\mu\nu}^{\prime}\left(x\right)=h_{\mu\nu}\left(x\right)+\bar{\nabla}_{\mu}\zeta_{\nu}\left(x\right)+\bar{\nabla}_{\nu}\zeta_{\mu}\left(x\right)$
where $\bar{\nabla}_{\mu}$ and all the barred quantities refer to
the background metric $\bar{g}_{\mu\nu}$. Therefore, $h_{\mu\nu}$
in this setting is a background two-tensor. At this stage, even though
$\bar{g}_{\mu\nu}$ is defined by the equation 
\begin{equation}
\mathcal{E}_{\mu\nu}\left(\bar{g},\bar{R},\bar{\nabla}\bar{R},\bar{R}^{2},...\right)=0,\label{eq:Background_EoM}
\end{equation}
clearly there is a still a large and perhaps infinite degeneracy in
the choice of the background metric. In principle any solution of
the above equation can be taken as a background and we will consider
any Einstein metric as a background in Einstein's theory; however,
in a generic gravity theory obtaining nontrivial solutions is itself
an outstanding problem; therefore, in what follows we will choose
the background to be a maximally symmetric spacetime for which the
complicated field equations of a generic gravity theory highly simplify
and yield a solution. Thus, we will make this choice, and often refer
to $\bar{g}$ to be the (classical) vacuum with the following properties
\begin{equation}
\bar{R}_{\mu\alpha\nu\beta}=\frac{2}{\left(n-2\right)\left(n-1\right)}\Lambda\left(\bar{g}_{\mu\nu}\bar{g}_{\alpha\beta}-\bar{g}_{\mu\beta}\bar{g}_{\alpha\nu}\right),\qquad\bar{R}_{\mu\nu}=\frac{2}{n-2}\Lambda\bar{g}_{\mu\nu},\qquad\bar{R}=\frac{2n\Lambda}{n-2}.\label{eq:Max_sym_background}
\end{equation}
Inserting these into (\ref{eq:Background_EoM}), one arrives at an
equation $f\left(\Lambda\right)\bar{g}_{\mu\nu}=0$ which determines
the effective cosmological constant $\Lambda$. Of course, depending
on the parameters of the theory, there could be no real-valued solution
or many solutions. The case of no maximally symmetric solution is
an interesting one which we shall not discuss here. If there are many
real-valued solutions for $\Lambda$, at this stage there is no compelling
reason to choose one over the other. Therefore, any such vacuum would
be viable. Hence, as far as the charge construction is concerned,
we shall simply consider a generic $\Lambda$ which is a real-valued
solution to the vacuum equation (\ref{eq:Background_EoM}).\footnote{Of course, some of these vacua can be eliminated on the basis of some
other criteria such as their instability against small fluctuations.} Once such a $\Lambda$ exists, there are $n\left(n+1\right)/2$ background
Killing symmetries which we shall collectively denote as $\bar{\xi}_{a}^{\mu}$.
But, not to clutter the notation we will simply omit the index $a$.

The splitting (\ref{eq:Metric_decomp}) as applied to the field equations
(\ref{eq:EoM}) yields 
\begin{equation}
\mathcal{E}_{\mu\nu}\left(\bar{g}\right)+\kappa\mathcal{E}_{\mu\nu}^{\left(1\right)}\left(h\right)+\kappa^{2}\mathcal{E}_{\mu\nu}^{\left(2\right)}\left(h\right)+O\left(\kappa^{3}\right)=\kappa\tau_{\mu\nu}.
\end{equation}
We can move all the nonlinear terms to the right-hand side and recast
the equation as a linear operator in $h$ with a nontrivial source
term that includes not only the matter source, but also the gravitational
energy and momentum, \emph{etc} sourced by the gravitational self-interaction:
\begin{equation}
\mathcal{E}_{\mu\nu}^{\left(1\right)}\left(h\right):=T_{\mu\nu},
\end{equation}
where $T_{\mu\nu}=\tau_{\mu\nu}+\kappa\mathcal{E}_{\mu\nu}^{\left(2\right)}\left(h\right)+O\left(\kappa^{2}\right)$.
We must also split the Bianchi identity $\nabla_{\mu}\mathcal{E}^{\mu\nu}=0$
as 
\begin{align}
\nabla_{\mu}\mathcal{E}^{\mu\nu} & =\bar{\nabla}_{\mu}\bar{\mathcal{E}}^{\mu\nu}+\kappa\bar{\nabla}_{\mu}\mathcal{E}^{\left(1\right)\mu\nu}+\kappa\left(\nabla_{\mu}\right)_{\left(1\right)}\bar{\mathcal{E}}^{\mu\nu}+O\left(\kappa^{2}\right)=0,\label{eq:Bianchi_id_exp}
\end{align}
where $\left(\nabla_{\mu}\right)_{\left(1\right)}\bar{\mathcal{E}}^{\mu\nu}:=\left(\Gamma_{\mu\lambda}^{\mu}\right)_{\left(1\right)}\bar{\mathcal{E}}^{\lambda\nu}+\left(\Gamma_{\mu\lambda}^{\nu}\right)_{\left(1\right)}\bar{\mathcal{E}}^{\mu\lambda}$
and the linearized Christoffel connection is 
\begin{align}
\left(\Gamma_{\mu\nu}^{\rho}\right)_{\left(1\right)} & =\frac{1}{2}\left(\bar{\nabla}_{\mu}h_{\nu}^{\rho}+\bar{\nabla}_{\nu}h_{\mu}^{\rho}-\bar{\nabla}^{\rho}h_{\mu\nu}\right).
\end{align}
From (\ref{eq:Bianchi_id_exp}) and making use of the vacuum field
equation, one arrives at the linearized Bianchi identity 
\begin{equation}
\bar{\nabla}_{\mu}\mathcal{E}^{\left(1\right)\mu\nu}=0.
\end{equation}
Now, we can define a nontrivial partially-conserved current as 
\begin{equation}
\mathcal{J}^{\mu}:=\sqrt{-\bar{g}}\bar{\xi}_{\nu}\mathcal{E}^{\left(1\right)\mu\nu},
\end{equation}
satisfying desired equation 
\begin{equation}
\partial_{\mu}\left(\sqrt{-\bar{g}}\bar{\xi}_{\nu}\mathcal{E}^{\left(1\right)\mu\nu}\right)=\sqrt{-\bar{g}}\bar{\nabla}_{\mu}\left(\bar{\xi}_{\nu}\mathcal{E}^{\left(1\right)\mu\nu}\right)=0.
\end{equation}
Observe that outside the matter source, one has 
\begin{equation}
T_{\mu\nu}\rightarrow\sum_{i=1}^{\infty}\kappa^{i}\mathcal{E}_{\mu\nu}^{\left(i+1\right)}\left(h\right),
\end{equation}
and therefore, 
\begin{equation}
\mathcal{J}^{\mu}\rightarrow\sqrt{-\bar{g}}\bar{\xi}_{\nu}\sum_{i=1}^{\infty}\kappa^{i}\mathcal{E}^{\left(i+1\right)\mu\nu}\left(h\right),
\end{equation}
is generically nonzero. Now, we can use the Stokes' theorem 
\begin{equation}
0=\int_{\bar{\mathcal{M}}}d^{n}x\,\partial_{\mu}\mathcal{J}^{\mu}=\int_{\mathcal{\partial\bar{M}}}d^{n-1}y\,\bar{n}_{\mu}\mathcal{J}^{\mu}=\int_{\mathcal{\partial\bar{M}}}d^{n-1}y\,\sqrt{\left|\bar{\gamma}\right|}\,\bar{n}_{\mu}\,\bar{\xi}_{\nu}\,\mathcal{E}^{\left(1\right)\mu\nu},\label{eq:Stokes_thm_first_level}
\end{equation}
where $\mathcal{\partial\bar{M}}$ represents the $\left(n-1\right)$-dimensional
boundary of the background manifold $\bar{\mathcal{M}}$, $\bar{n}^{\mu}$
is the unit normal vector to the boundary which we assume to be non-null,
and $\bar{\gamma}$ is the induced metric on $\mathcal{\partial\bar{M}}$.
Given a spacelike hypersurface $\bar{\Sigma}$, one can define global
charges up to an overall normalization as 
\begin{equation}
Q\left(\bar{\xi}\right):=\int_{\bar{\Sigma}}d^{n-1}y\,\sqrt{\bar{\gamma}}\bar{n}_{\mu}\bar{\xi}_{\nu}\mathcal{E}^{\left(1\right)\mu\nu},
\end{equation}
under the assumption of $\mathcal{J}^{\mu}$ vanishing at spacelike
infinity. 
\begin{figure}
\begin{centering}
\includegraphics[scale=0.3]{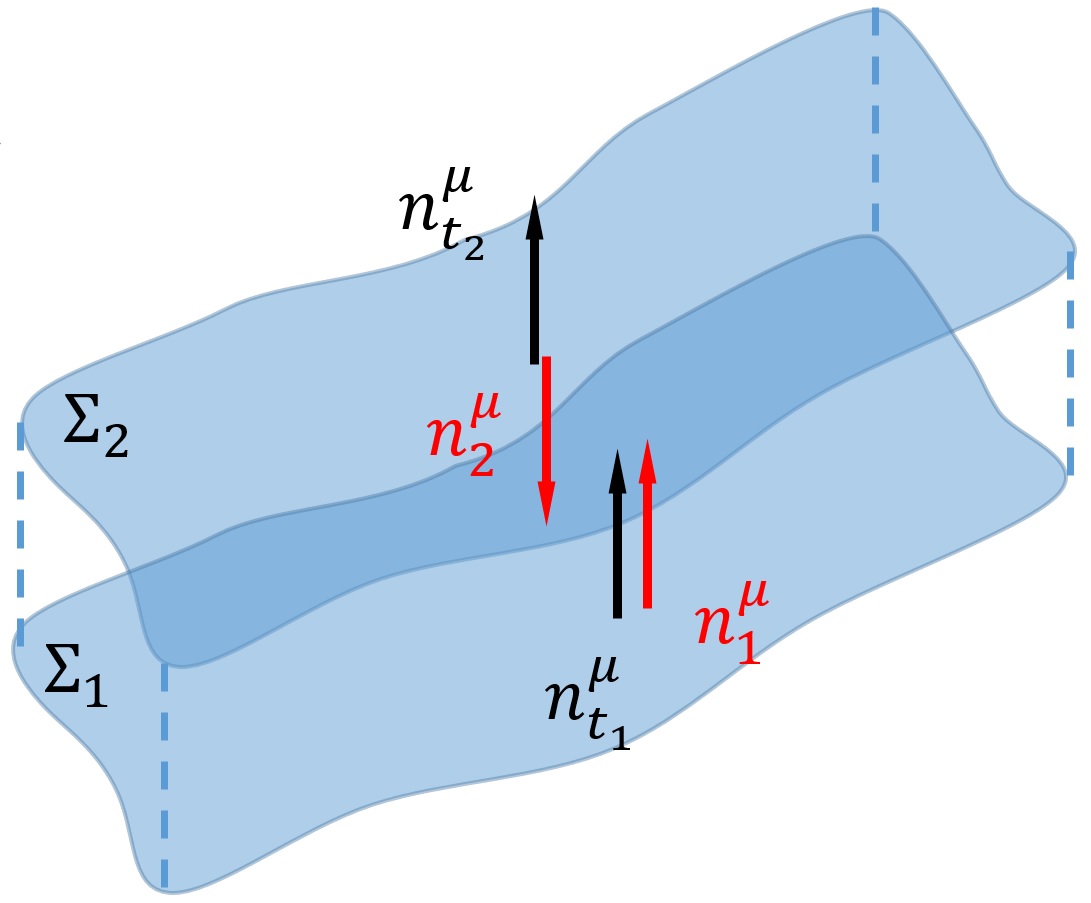} 
\par\end{centering}
\caption{Disjoint hypersurfaces $\Sigma_{1}$ and $\Sigma_{2}$ over which
the integration is taken. Considering these hypersurfaces are obtained
by timelike slicing, then they are spacelike and their normal vectors
$n_{t_{1}}^{\mu}$ and $n_{t_{2}}^{\mu}$, respectively, are timelike.
For spacelike hypersurfaces, as a convention the surface normals are
taken into the bulk of spacetime in the application of the Stokes'
theorem, and these surface normals are represented as $n_{1}^{\mu}$
and $n_{2}^{\mu}$ in the Figure.}
\end{figure}

A crucial point here is, as can be seen in Figure 1, $\bar{\Sigma}$
is generically not equal to $\partial\mathcal{\bar{M}}$; therefore,
it can have a boundary of its own. In fact, with this definition,
(\ref{eq:Stokes_thm_first_level}) becomes 
\begin{equation}
Q_{\bar{\Sigma}_{2}}\left(\bar{\xi}\right)=Q_{\bar{\Sigma}_{1}}\left(\bar{\xi}\right),
\end{equation}
which is a statement of charge conservation. In that case, as $\bar{\xi}_{\nu}\mathcal{E}^{\left(1\right)\mu\nu}\equiv\bar{\nabla}_{\nu}{\cal F}^{\mu\nu}\left(\bar{\xi}\right)$,
where ${\cal F}^{\mu\nu}$ is an antisymmetric two-tensor, one can
use the Stokes' theorem one more time to arrive at 
\begin{equation}
Q\left(\bar{\xi}\right)=\int_{\partial\bar{\Sigma}}d^{n-2}z\,\sqrt{\bar{\gamma}^{\left(\partial\bar{\Sigma}\right)}}\,\bar{n}_{\mu}\,\bar{\sigma}_{\nu}\,{\cal F}^{\mu\nu}\left(\bar{\xi}\right),
\end{equation}
where $\bar{\sigma}_{\nu}$ is the outward unit vector on the $\left(n-2\right)$-dimensional
spacelike surface $\partial\bar{\Sigma}$. Of course, the explicit
form of $\mathcal{F}^{\mu\nu}\left(\bar{\xi}\right)$ can only be
found from the field equations once the theory is given. For a more
compact notation, one can define an anti-symmetric binormal vector
\begin{equation}
\bar{\epsilon}_{\mu\nu}:=\frac{1}{2}\left(\bar{n}_{\mu}\bar{\sigma}_{\nu}-\bar{n}_{\nu}\bar{\sigma}_{\mu}\right),\label{eq:Anti-sym_binormal_vector}
\end{equation}
and write the charge as 
\begin{equation}
Q\left(\bar{\xi}\right)=\int_{\partial\bar{\Sigma}}d^{n-2}z\,\sqrt{\bar{\gamma}^{\left(\partial\bar{\Sigma}\right)}}\bar{\epsilon}_{\mu\nu}{\cal F}^{\mu\nu}\left(\bar{\xi}\right).
\end{equation}
It is apt now to give the explicit expressions of ${\cal F}^{\mu\nu}\left(\bar{\xi}\right)$
in the cosmological Einstein's gravity, and in the vanishing cosmological
constant case, obtain the usual expressions for the asymptotically
flat spacetimes.

\subsection{AD and ADM Conserved Charges\label{subsec:AD-and-ADM}}

Cosmological Einstein's gravity is defined with the action 
\begin{equation}
I=\frac{1}{4\Omega_{n-2}G_{n}}\int d^{n}x\,\sqrt{-g}\left(R-2\Lambda\right),\label{eq:Cosmological_Einstein_action}
\end{equation}
where $G_{n}$ is $n$-dimensional Newton's constant and $\Omega_{n-2}$
is the solid angle. The matter coupled field equations following from
this action are 
\begin{equation}
G_{\mu\nu}:=R_{\mu\nu}-\frac{1}{2}g_{\mu\nu}R+\Lambda g_{\mu\nu}=2\Omega_{n-2}G_{n}\tau_{\mu\nu}.
\end{equation}
The linearization of $G_{\mu\nu}$ about its unique maximally symmetric
vacuum (\ref{eq:Max_sym_background}) gives 
\begin{equation}
{\cal G}_{\mu\nu}^{\left(1\right)}:=R_{\mu\nu}^{\left(1\right)}-\frac{1}{2}\bar{g}_{\mu\nu}R^{\left(1\right)}-\frac{2}{n-2}\Lambda h_{\mu\nu},\label{eq:Lin_Eins_tensor}
\end{equation}
where the linearized Ricci tensor and the scalar curvature are 
\begin{align}
R_{\mu\nu}^{\left(1\right)} & =\frac{1}{2}\left(\bar{\nabla}_{\rho}\bar{\nabla}_{\mu}h_{\nu}^{\rho}+\bar{\nabla}_{\rho}\bar{\nabla}_{\nu}h_{\mu}^{\rho}-\bar{\Box}h_{\mu\nu}-\bar{\nabla}_{\nu}\bar{\nabla}_{\mu}h\right),\label{eq:Lin_Ricci}\\
R^{\left(1\right)} & =\bar{g}^{\mu\nu}R_{\mu\nu}^{\left(1\right)}-h^{\mu\nu}\bar{R}_{\mu\nu}=\bar{g}^{\mu\nu}R_{\mu\nu}^{\left(1\right)}-\frac{2}{n-2}\Lambda h.\label{eq:Lin_R}
\end{align}
In \cite{Abbott}, the conserved current $\bar{\xi}_{\nu}\mathcal{G}^{\left(1\right)\mu\nu}$
was written as 
\begin{equation}
\bar{\xi}_{\nu}{\cal G}^{\left(1\right)\mu\nu}=\bar{\nabla}_{\alpha}\left(\bar{\xi}_{\nu}\bar{\nabla}_{\beta}K^{\mu\alpha\nu\beta}-K^{\mu\beta\nu\alpha}\bar{\nabla}_{\beta}\bar{\xi}_{\nu}\right),
\end{equation}
where the superpotential $K^{\mu\alpha\nu\beta}$, having the symmetries
of the Riemann tensor, is 
\begin{equation}
K^{\mu\alpha\nu\beta}:=\frac{1}{2}\left(\bar{g}^{\alpha\nu}\tilde{h}^{\mu\beta}+\bar{g}^{\mu\beta}\tilde{h}^{\alpha\nu}-\bar{g}^{\alpha\beta}\tilde{h}^{\mu\nu}-\bar{g}^{\mu\nu}\tilde{h}^{\alpha\beta}\right),\qquad\tilde{h}^{\mu\nu}:=h^{\mu\nu}-\frac{1}{2}\bar{g}^{\mu\nu}h.
\end{equation}
Note that the superpotential can be compactly written by use of the
Kulkarni-Nomizu product \cite{Besse} of $\bar{g}$ and $\tilde{h}$
as 
\begin{equation}
K_{\mu\alpha\nu\beta}=-\left(\bar{g}\owedge\tilde{h}\right)_{\mu\alpha\nu\beta}.
\end{equation}
Therefore, one arrives at the conserved charges in cosmological Einstein's
gravity in the following compact form 
\begin{equation}
\boxed{\phantom{\frac{\frac{\xi}{\xi}}{\frac{\xi}{\xi}}}Q\left(\bar{\xi}\right)=\frac{1}{2\Omega_{n-2}G_{n}}\int_{\partial\bar{\Sigma}}d^{n-2}z\,\sqrt{\bar{\gamma}^{\left(\partial\bar{\Sigma}\right)}}\bar{\epsilon}_{\mu\nu}\left(\bar{\xi}_{\alpha}\bar{\nabla}_{\beta}K^{\mu\nu\alpha\beta}-K^{\mu\beta\alpha\nu}\bar{\nabla}_{\beta}\bar{\xi}_{\alpha}\right),\phantom{\frac{\frac{\xi}{\xi}}{\frac{\xi}{\xi}}}}\label{eq:Conserved_charges}
\end{equation}
which is called the Abbott-Deser charge. Here, $\bar{\epsilon}_{\mu\nu}$
is defined in (\ref{eq:Anti-sym_binormal_vector}) and other aspects
of the integration are defined around equation (\ref{eq:Anti-sym_binormal_vector}).
In \cite{Deser_Tekin-PRL,Deser_Tekin-PRD}, another commonly used
form of (\ref{eq:Conserved_charges}) which suits for an extension
to generic $f\left({\rm Riemann}\right)$ type theories was given
as 
\begin{align}
Q\left(\bar{\xi}\right)=\frac{1}{4\Omega_{n-2}G_{n}}\int_{\partial\bar{\Sigma}}dS_{i}\biggl( & \bar{\xi}_{\nu}\bar{\nabla}^{0}h^{i\nu}-\bar{\xi}_{\nu}\bar{\nabla}^{i}h^{0\nu}+\bar{\xi}^{0}\bar{\nabla}^{i}h-\bar{\xi}^{i}\bar{\nabla}^{0}h+h^{0\nu}\bar{\nabla}^{i}\bar{\xi}_{\nu}\nonumber \\
 & -h^{i\nu}\bar{\nabla}^{0}\bar{\xi}_{\nu}+\bar{\xi}^{i}\bar{\nabla}_{\nu}h^{0\nu}-\bar{\xi}^{0}\bar{\nabla}_{\nu}h^{i\nu}+h\bar{\nabla}^{0}\bar{\xi}^{i}\biggr).\label{eq:ADT_charge}
\end{align}
Let us note a few observations before we compute the charges of some
interesting spacetimes:
\begin{itemize}
\item There are $n\left(n+1\right)/2$ independent nontrivial conserved
charges corresponding to each Killing vector of a maximally symmetric
spacetime: $n$ of these result from spacetime ``translations''
and the $\left(n-1\right)\left(n-2\right)/2$ of these are from ``rotations''
summing up to $\left(n^{2}-n+2\right)/2$ conserved charges; the rest
are $\left(n-1\right)$ ``boosts'' whose details were studied in
\cite{Henneaux85}.
\item The maximally symmetric spacetime which we denoted as $\bar{g}$ is
assigned to have zero charges.\footnote{Note that in the holographic definition of the conserved charges consistency
might require that the background has a nonzero charge; for example,
as in the case of $\text{AdS}_{5}$ \cite{97}.}
\item $Q\left(\bar{\xi}\right)$ is invariant under the infinitesimal gauge
transformations $\delta_{\zeta}h_{\mu\nu}\left(x\right)=\bar{\nabla}_{\mu}\zeta_{\nu}\left(x\right)+\bar{\nabla}_{\nu}\zeta_{\mu}\left(x\right)$
by construction since $\delta_{\zeta}\mathcal{G}_{\mu\nu}^{\left(1\right)}=0$
even though $K^{\mu\alpha\nu\beta}$ is not gauge invariant.
\item We have derived (\ref{eq:Conserved_charges}) for $\Lambda\ne0$,
but the expression is valid for the $\Lambda=0$ case. Therefore,
$Q\left(\bar{\xi}\right)$ represents the ADM charges for asymptotically
flat spacetimes in generic coordinates. If one chooses the Cartesian
coordinates and takes the timelike Killing vector $\xi^{\mu}=\left(-1,0,\dots,0\right)$,
then (\ref{eq:Conserved_charges}) reduces to the celebrated ADM mass
\begin{eqnarray}
M_{{\rm ADM}}=\frac{1}{2\Omega_{n-2}G_{n}}\int_{S^{n-2}} & dS_{i} & \left(\partial_{j}h^{ij}-\partial^{i}h_{jj}\right),\label{eq:ADM_mass}
\end{eqnarray}
where the integral is to be evaluated on a sphere at spatial infinity.
Similarly, angular momentum (or momenta) has a similar expression
\begin{eqnarray}
J=\frac{1}{2\Omega_{n-2}G_{n}}\int_{S^{n-2}} & dS_{i} & \left(\bar{\xi}^{i}\partial_{j}h^{0j}-\bar{\xi}_{j}\partial^{i}h^{0j}\right),\label{eq:ADM_angular_mom}
\end{eqnarray}
where $\bar{\xi}^{i}$ is the corresponding Killing vector. In practice,
to compute the conserved charges of a given solution, using the coordinate
independent expression (\ref{eq:Conserved_charges}) is much more
efficient, especially if the metric is complicated.
\item Even if one takes the background spacetime to be nonmaximally symmetric,
but with at least one Killing vector field, one arrives at the same
charge expression (\ref{eq:Conserved_charges}). This is a highly
nontrivial point and not immediately clear, but one can run the same
computation as above without the assumption of maximal symmetry and
observe that one has additional parts 
\begin{align}
\bar{\xi}_{\nu}{\cal G}_{L}^{\mu\nu}= & \bar{\nabla}_{\alpha}\left(\bar{\xi}_{\nu}\bar{\nabla}_{\beta}K^{\mu\alpha\nu\beta}-K^{\mu\beta\nu\alpha}\bar{\nabla}_{\beta}\bar{\xi}_{\nu}\right)\nonumber \\
 & +\bar{\xi}_{\nu}h^{\mu\alpha}\bar{G}_{\alpha}^{\phantom{\alpha}\nu}+\frac{1}{2}\bar{\xi}^{\mu}h^{\rho\sigma}\bar{G}_{\rho\sigma}-\frac{1}{2}\bar{\xi}_{\rho}\bar{G}^{\mu\rho}h,\label{eq:Arbitrary_background_Fmn}
\end{align}
but once the background field equations are used, the additional parts
drop out. Hence, the earlier result derived for the maximally symmetric
vacuum is valid for all Einstein spaces with a Killing vector field
\cite{Bouchareb,Nazaroglu,Devecioglu}. To see this result explicitly,
let us lay out the details of the computation since it will be relevant
for theories that extend Einstein's theory and have nonmaximally symmetric
vacua (for example, in three dimensions topologically massive gravity
is an example). For a nonmaximally symmetric background $\bar{g}_{\mu\nu}$,
the linearized Einstein tensor have the form 
\begin{equation}
{\cal G}_{\mu\nu}^{\left(1\right)}=R_{\mu\nu}^{\left(1\right)}-\frac{1}{2}\bar{g}_{\mu\nu}R_{\left(1\right)}+\Lambda h_{\mu\nu}-\frac{1}{2}h_{\mu\nu}\bar{R},
\end{equation}
with the help of (\ref{eq:Lin_Ricci}) and the first equality of (\ref{eq:Lin_R}),
${\cal G}_{\mu\nu}^{\left(1\right)}$ reduces to 
\begin{align}
{\cal G}_{\mu\nu}^{\left(1\right)}= & \frac{1}{2}\left(\bar{\nabla}_{\rho}\bar{\nabla}_{\mu}h_{\nu}^{\rho}+\bar{\nabla}_{\rho}\bar{\nabla}_{\nu}h_{\mu}^{\rho}-\bar{\Box}h_{\mu\nu}-\bar{\nabla}_{\nu}\bar{\nabla}_{\mu}h\right)\nonumber \\
 & -\frac{1}{2}\bar{g}_{\mu\nu}\left(\bar{\nabla}_{\rho}\bar{\nabla}_{\sigma}h^{\rho\sigma}-\bar{\Box}h-h^{\rho\sigma}\bar{R}_{\rho\sigma}\right)+\Lambda h_{\mu\nu}-\frac{1}{2}h_{\mu\nu}\bar{R}.
\end{align}
We assume that the background has at least one Killing vector field.
In trying to find a form as $\bar{\xi}_{\nu}{\cal G}^{\left(1\right)\mu\nu}=\bar{\nabla}_{\nu}\mathcal{F}^{\mu\nu}$
where $\mathcal{F}^{\mu\nu}$ is antisymmetric, following \cite{Abbott}
it is better to write $\bar{\xi}_{\nu}{\cal G}^{\left(1\right)\mu\nu}$
as $\bar{\xi}_{\nu}{\cal G}^{\left(1\right)\mu\nu}\sim\bar{\nabla}_{\alpha}\bar{\nabla}_{\beta}K^{\alpha\beta\mu\nu}+X^{\mu\nu}$
since one just needs to handle an interchange of order in derivatives
for the term $\bar{\xi}_{\nu}{\cal G}^{\left(1\right)\mu\nu}\sim\bar{\xi}_{\nu}\bar{\nabla}_{\alpha}\bar{\nabla}_{\beta}K^{\alpha\beta\mu\nu}+\bar{\xi}_{\nu}X^{\mu\nu}$
as 
\begin{equation}
\bar{\xi}_{\nu}{\cal G}^{\left(1\right)\mu\nu}\sim\bar{\xi}_{\nu}\bar{\nabla}_{\alpha}\bar{\nabla}_{\beta}K^{\alpha\beta\mu\nu}+\bar{\xi}_{\nu}X^{\mu\nu}=\bar{\nabla}_{\alpha}\left(\bar{\xi}_{\nu}\bar{\nabla}_{\beta}K^{\alpha\beta\mu\nu}\right)-\left(\bar{\nabla}_{\alpha}\bar{\xi}_{\nu}\right)\bar{\nabla}_{\beta}K^{\alpha\beta\mu\nu}+\bar{\xi}_{\nu}X^{\mu\nu}.
\end{equation}
To work out what $K^{\alpha\beta\mu\nu}$ and $X^{\mu\nu}$ are for
the Einstein's theory, ${\cal G}_{L}^{\mu\nu}$ can be rewritten as
the derivative terms plus nonderivative terms: 
\begin{align}
{\cal G}_{L}^{\mu\nu}= & \frac{1}{2}\bar{\nabla}_{\alpha}\bar{\nabla}_{\beta}\left(\bar{g}^{\nu\beta}h^{\mu\alpha}+\bar{g}^{\mu\beta}h^{\alpha\nu}-\bar{g}^{\alpha\beta}h^{\mu\nu}-\bar{g}^{\mu\nu}h^{\alpha\beta}+\bar{g}^{\mu\nu}\bar{g}^{\alpha\beta}h-\bar{g}^{\mu\alpha}\bar{g}^{\nu\beta}h\right)\nonumber \\
 & +\frac{1}{2}\bar{g}^{\mu\nu}h^{\rho\sigma}\bar{R}_{\rho\sigma}+\Lambda h_{\mu\nu}-\frac{1}{2}h_{\mu\nu}\bar{R}.
\end{align}
From the first line, we want to define the $K$ tensor such that it
has the same symmetries as the Riemann tensor. Comparing with the
maximally symmetric Riemann tensor $\bar{R}_{\mu\alpha\nu\beta}\sim\Lambda\left(\bar{g}_{\mu\nu}\bar{g}_{\alpha\beta}-\bar{g}_{\mu\beta}\bar{g}_{\alpha\nu}\right),$
one can see that except the first and the last terms, the others combine
to define the desired $K^{\mu\alpha\nu\beta}$. Order change for the
last term is automatic, and once we have the order change in the first
term, ${\cal G}_{L}^{\mu\nu}$ becomes 
\[
{\cal G}_{L}^{\mu\nu}=\bar{\nabla}_{\alpha}\bar{\nabla}_{\beta}K^{\mu\alpha\nu\beta}+X^{\mu\nu},
\]
where 
\begin{equation}
K^{\mu\alpha\nu\beta}\equiv\frac{1}{2}\left(\bar{g}^{\alpha\nu}\tilde{h}^{\mu\beta}+\bar{g}^{\mu\beta}\tilde{h}^{\alpha\nu}-\bar{g}^{\alpha\beta}\tilde{h}^{\mu\nu}-\bar{g}^{\mu\nu}\tilde{h}^{\alpha\beta}\right),\qquad\tilde{h}^{\mu\nu}\equiv h^{\mu\nu}-\frac{1}{2}\bar{g}^{\mu\nu}h,
\end{equation}
and the ``fudge'' tensor reads 
\begin{align*}
X^{\mu\nu} & \equiv\frac{1}{2}\bar{g}^{\nu\beta}\left[\bar{\nabla}_{\alpha},\bar{\nabla}_{\beta}\right]h^{\mu\alpha}+\frac{1}{2}\bar{g}^{\mu\nu}h^{\rho\sigma}\bar{R}_{\rho\sigma}+\Lambda h_{\mu\nu}-\frac{1}{2}h_{\mu\nu}\bar{R}\\
 & =\frac{1}{2}\left(h^{\mu\alpha}\bar{R}_{\alpha}^{\phantom{\alpha}\nu}-\bar{R}^{\mu\alpha\nu\beta}h_{\alpha\beta}\right)+\frac{1}{2}\bar{g}^{\mu\nu}h^{\rho\sigma}\bar{R}_{\rho\sigma}+\Lambda h^{\mu\nu}-\frac{1}{2}h^{\mu\nu}\bar{R}.
\end{align*}
Note that both $\bar{\nabla}_{\alpha}\bar{\nabla}_{\beta}K^{\mu\alpha\nu\beta}$
and $X^{\mu\nu}$ are not symmetric in $\mu$ and $\nu$. Now, let
us find the antisymmetric tensor $\mathcal{F}^{\mu\nu}$ by rearranging
$\bar{\xi}_{\nu}{\cal G}_{L}^{\mu\nu}$ with the help of the identity
$\bar{\nabla}_{\beta}\bar{\nabla}_{\alpha}\bar{\xi}_{\nu}=\bar{R}_{\phantom{\rho}\beta\alpha\nu}^{\rho}\bar{\xi}_{\rho}$
as 
\begin{equation}
\bar{\xi}_{\nu}{\cal G}_{L}^{\mu\nu}=\bar{\nabla}_{\alpha}\left(\bar{\xi}_{\nu}\bar{\nabla}_{\beta}K^{\mu\alpha\nu\beta}-K^{\mu\beta\nu\alpha}\bar{\nabla}_{\beta}\bar{\xi}_{\nu}\right)+K^{\mu\alpha\nu\beta}\bar{R}_{\phantom{\rho}\beta\alpha\nu}^{\rho}\bar{\xi}_{\rho}+X^{\mu\nu}\bar{\xi}_{\nu},
\end{equation}
where the first term is in the desired form. Note that $K^{\mu\alpha\nu\beta}\bar{R}_{\phantom{\rho}\beta\alpha\nu}^{\rho}$
can also be written as $K^{\mu\alpha\nu\beta}\bar{R}_{\phantom{\rho}\alpha\beta\nu}^{\rho}/2$.
Using the background cosmological Einstein tensor 
\begin{equation}
\bar{G}_{\mu\nu}=\bar{R}_{\mu\nu}-\frac{1}{2}\bar{g}_{\mu\nu}\bar{R}+\Lambda\bar{g}_{\mu\nu},
\end{equation}
we can recast $\bar{\xi}_{\nu}{\cal G}_{L}^{\mu\nu}$ to (\ref{eq:Arbitrary_background_Fmn}).
As a summary, all of this says that the AD conserved charges (\ref{eq:Conserved_charges})
or (\ref{eq:ADT_charge}) are valid for all Einstein spaces with at
least one Killing vector field. 
\end{itemize}

\subsection{Large diffeomorphisms (gauge transformations)}

In the above construction of the conserved charges, we noted that
the results are invariant under small diffeomorphisms that act on
$h_{\mu\nu}$ as $\delta_{\zeta}h_{\mu\nu}\left(x\right)=\bar{\nabla}_{\mu}\zeta_{\nu}\left(x\right)+\bar{\nabla}_{\nu}\zeta_{\mu}\left(x\right)$
which is valid as long as $\left|\bar{\nabla}_{\mu}\zeta_{\nu}\left(x\right)\right|\ll\left|h_{\mu\nu}\left(x\right)\right|$
is satisfied in the given coordinate system for the given chart. Here,
we briefly address the case of large diffeomorphisms and show how
they split the charges assigned to a given geometry into various sectors.\footnote{Here, what we mean by various sectors is that the large diffeomorphisms
should be considered as two classes: first, the ones which are not
allowed by the boundary conditions; and second, the ones that are
compatible with the boundary conditions and the associated finite
charges as in the case of \cite{39}. The following computation basically
is a way to explore allowed diffeomorphisms consistent with the charge
definition in the ADM approach.} For the sake of simplicity, we discuss the Minkowski spacetime and
follow the arguments in \cite{ST} which was based on \cite{DenisovSolovev,BrayChrusciel}.
In spherical coordinates, the four-dimensional flat Minkowski spacetime
has the metric 
\begin{equation}
{\rm d}s^{2}=-{\rm d}t^{2}+{\rm d}r^{2}+r^{2}{\rm d}\Omega,\qquad{\rm d}\Omega:={\rm d}\theta^{2}+\sin^{2}\theta{\rm d}\phi^{2}\label{eq:Min_spherical}
\end{equation}
Needless to say, any coordinate change will keep the Riemann tensor
of this spacetime to be zero. Consider the following specific coordinate
transformations 
\[
t=f\left(\tau,\rho\right),\qquad r=k\left(\tau,\rho\right),
\]
and $\theta$, $\phi$ are kept intact. Then, the metric in these
new coordinates $\left(\tau,\rho,\theta,\phi\right)$ reads 
\begin{equation}
{\rm d}s^{2}=\left(\dot{k}^{2}-\dot{f}^{2}\right)\,{\rm d}\tau^{2}+2\left(\dot{k}k^{\prime}-\dot{f}f^{\prime}\right)\,{\rm d}\tau{\rm d}\rho+\left(k^{\prime}{}^{2}-f^{\prime}{}^{2}\right)\,{\rm d}\rho^{2}+k^{2}{\rm d}\Omega,\label{eq:Minkowski_transfromed}
\end{equation}
where we have suppressed the arguments of the functions, and defined
$\dot{f}:=\partial f/\partial\tau$ and $f^{\prime}:=\partial f/\partial\rho$.
Taking (\ref{eq:Min_spherical}) as the background geometry with zero
mass, we can calculate the mass of (\ref{eq:Minkowski_transfromed})
with respect to the background. More properly, we can identify the
background metric from (\ref{eq:Minkowski_transfromed}) as $f=\tau$
and $k=\rho$. Then, the deviation from the background reads 
\begin{equation}
h_{\mu\nu}dx^{\mu}dx^{\nu}=\left(1+\dot{k}^{2}-\dot{f}^{2}\right)\,{\rm d}\tau^{2}+2\left(\dot{k}k^{\prime}-\dot{f}f^{\prime}\right)\,{\rm d}\tau{\rm d}\rho+\left(-1+k^{\prime}{}^{2}-f^{\prime}{}^{2}\right)\,{\rm d}\rho^{2}+\left(-\rho^{2}+k^{2}\right){\rm d}\Omega.
\end{equation}
Using (\ref{eq:ADM_mass}) and taking the background Killing vector
as $\bar{\xi}^{\mu}=\left(-1,0,0,0\right)$, the ADM mass-energy of
(\ref{eq:Minkowski_transfromed}) becomes 
\begin{equation}
E=\left.\frac{1}{2G\rho}\left(-\rho^{2}\left(f^{\prime}\right)^{2}+\left(k-\rho k^{\prime}\right)^{2}\right)\right|_{\rho\rightarrow\infty}.
\end{equation}
It is clear that $M$ can take any value depending on the choices
of the functions $f$ and $k$. Let us consider a specific example
with 
\begin{equation}
f\left(\tau,\rho\right)=\tau,\qquad\qquad k\left(\tau,\rho\right)=\rho y\left(\tau\right)+2\sqrt{2Gm\rho},
\end{equation}
where $y$ is an arbitrary function of $\tau$ and $m$ is a positive
constant. For this choice $M=m$. More generally, for 
\begin{equation}
k\left(\tau,\rho\right)=\rho y\left(\tau\right)+\left(Gm\rho\right)^{1-s},
\end{equation}
$E$ diverges for $s<1/2$ and vanishes for $s>1/2$. Even negative
values for the total mass is possible. For example, choosing 
\begin{equation}
k\left(\tau,\rho\right)=\rho,\qquad\qquad f\left(\tau,\rho\right)=y\left(\tau\right)+2\sqrt{2Gm\rho},
\end{equation}
yields a negative mass of $E=-m$.

Without much ado, let us give a similar example for the case of angular
momentum: Let us consider the following coordinate transformations
\begin{equation}
r=k\left(t,\rho\right)=\rho\sqrt{mt}+\sqrt{Gm\rho},\qquad\phi=p\left(\rho,\psi\right)=\psi+Gm^{3}a\left(m\rho\right)^{s},
\end{equation}
where $a$ is a parameter with dimension of length. Note that we kept
$t$, $\theta$ intact. Then, the metric in these new coordinates
read 
\begin{align}
{\rm d}s^{2}= & -\left(1-\frac{m\rho^{2}}{4t}\right)\text{d\ensuremath{t}}^{2}+\left(1+\frac{1}{2}\sqrt{\frac{G}{t\rho}}\right)m\rho\text{d\ensuremath{\rho}}\text{d\ensuremath{t}}\nonumber \\
 & +\left(m^{2}a^{2}s^{2}\sin^{2}\theta\left(1+\sqrt{\frac{G}{t\rho}}\right)^{2}\left(m\rho\right)^{2s}+\left(1+\frac{1}{2}\sqrt{\frac{G}{t\rho}}\right)^{2}\right)m\,t\text{d\ensuremath{\rho}}^{2}\nonumber \\
 & +2m\,a\,s\sin^{2}\theta\left(1+\sqrt{\frac{G}{t\rho}}\right)^{2}\left(m\rho\right)^{s+1}t\text{d\ensuremath{\rho}}\text{d\ensuremath{\psi}}+\left(1+\sqrt{\frac{G}{t\rho}}\right)^{2}m\rho^{2}t{\rm d}\Omega,
\end{align}
which is diffeomorphic to the Minkowski spacetime and of course it
is Riemann flat. Using the conserved charge expression (\ref{eq:Conserved_charges}),
for the Killing vector fields $\bar{\xi}^{\mu}=\left(-1,0,0,0\right)$
and $\bar{\xi}^{\mu}=\left(0,0,0,1\right)$, one finds 
\begin{equation}
E=\frac{m}{4}+\frac{2}{3}a^{2}m^{6}s^{2}\left(m\rho\right)^{2s+1}t,\qquad J=-\frac{m\,a\,s\left(m\rho\right)^{s+3}}{3},
\end{equation}
which must be evaluated in the limit $\rho\rightarrow\infty$. Both
of the terms become finite for $s=-3$ which yields 
\begin{equation}
E=\frac{m}{4},\qquad J=m\,a.
\end{equation}

All these arguments show that the flat Minkowski spacetime has infinitely
many diffeomorphic copies that all have vanishing Riemann tensor but
with different ADM energies and angular momenta. In the light of this
discussion, the conventional positive energy theorem that assigns
zero energy, and linear and angular momenta to the flat Minkowski
spacetime \cite{Schoen,Witten,Deser,Grisaru,Parker,Nester} also assumes
that the metric components in the example discussed above decay sufficiently
fast at infinity, that is the $s>1/2$ case with vanishing energy.

\subsection{Schwarzschild-(A)dS Black Holes in $n$-dimensions}

Let us apply the AD construction to the immediate natural setting
of the static spherically symmetric black hole in (A)dS. The metric
in static coordinates reads 
\begin{equation}
{\rm d}s^{2}=-\left(1-\left(\frac{r_{0}}{r}\right)^{n-3}+\frac{r^{2}}{\ell^{2}}\right){\rm d}t^{2}+\left(1-\left(\frac{r_{0}}{r}\right)^{n-3}+\frac{r^{2}}{\ell^{2}}\right)^{-1}{\rm d}r^{2}+r^{2}{\rm d}\Omega_{n-2}^{2},\label{eq:SAdS}
\end{equation}
where $\ell^{2}:=-\frac{(n-2)(n-1)}{2\Lambda}$ and $n>3$. The background
metric corresponds to $r_{0}=0$ with the timelike Killing vector
$\xi^{\mu}=\left(-1,\vec{0}\right)$ satisfying $\bar{g}_{\mu\nu}\bar{\xi}^{\mu}\bar{\xi}^{\nu}=-\left(1+\frac{r^{2}}{\ell^{2}}\right)$.
{[}Formally, we can carry out the same construction for the de Sitter
case with $\ell\rightarrow i\ell$, but clearly due to the cosmological
horizon at a finite $r$ coordinate, $\bar{\xi}^{\mu}$ fails to be
timelike everywhere, and hence a global time does not exist. Therefore,
for black holes in dS only small black hole limit can be approximately
considered, for more on this see \cite{Deser_Tekin-PRD}.{]} For the
case of AdS background, the result of the energy computation based
on (\ref{eq:Conserved_charges}) turns out to be as expected 
\begin{equation}
E=\frac{n-2}{4G_{n}}r_{0}^{n-3},\label{eq:E_of_SAdS}
\end{equation}
and in four dimensions $r_{0}=2Gm$, and $E=M$ as expected.

\subsection{Kerr-AdS Black Holes in $n$-dimensions}

Let us now consider a more complicated example by adding rotations
to the $n$-dimensional black hole. The metric was given in \cite{Gibbons}
in the Kerr-Schild form \cite{KerrSchild,GursesGursey} as 
\begin{equation}
{\rm d}s^{2}={\rm d}\bar{s}^{2}+\frac{2mG_{n}}{U}\left(k_{\mu}{\rm d}x^{\mu}\right)^{2},\label{eq:KS}
\end{equation}
where $m$ is a real parameter, $k_{\mu}{\rm d}x^{\mu}$ is a one-form
defined below, $U$ is a function of the coordinate $r$ and the direction
cosines $\mu_{i}$s as 
\begin{equation}
U:=r^{\epsilon}\sum_{i=1}^{N+\epsilon}\frac{\mu_{i}^{2}}{r^{2}+a_{i}^{2}}\prod_{j=1}^{N}\left(r^{2}+a_{j}^{2}\right),\qquad\sum_{i=1}^{N+\epsilon}\mu_{i}^{2}=1,
\end{equation}
with $\epsilon=0/1$ for odd/even dimensions and $n=2N+1+\epsilon$.
Here, $N$ refers to the number of rotation parameters $a_{i}$ associated
to the azimuthal angles $\phi_{i}$s. For even $n$, one should note
that $a_{N+1}=0$ since $\phi_{N+1}$ does not exist. The background
metric reads 
\begin{align}
{\rm d}\bar{s}^{2}= & -W\left(1+\frac{r^{2}}{\ell^{2}}\right){\rm d}t^{2}+F{\rm d}r^{2}+\sum_{i=1}^{N+\epsilon}\frac{r^{2}+a_{i}^{2}}{1-\frac{a_{i}^{2}}{\ell^{2}}}{\rm d}\mu_{i}^{2}+\sum_{i=1}^{N}\frac{r^{2}+a_{i}^{2}}{1-\frac{a_{i}^{2}}{\ell^{2}}}\mu_{i}^{2}{\rm d}\phi_{i}^{2}\nonumber \\
 & -\frac{1}{W\ell^{2}\left(1+\frac{r^{2}}{\ell^{2}}\right)}\left(\sum_{i=1}^{N+\epsilon}\frac{\left(r^{2}+a_{i}^{2}\right)\mu_{i}{\rm d}\mu_{i}}{1-\frac{a_{i}^{2}}{\ell^{2}}}\right)^{2},\label{eq:dS}
\end{align}
where again $W$ and $F$ are functions of the coordinates $r$ and
$\mu_{i}$s as 
\begin{equation}
W:=\sum_{i=1}^{N+\epsilon}\frac{\mu_{i}^{2}}{1-\frac{a_{i}^{2}}{\ell^{2}}},\qquad\qquad\qquad F:=\frac{1}{1+\frac{r^{2}}{\ell^{2}}}\sum_{i=1}^{N+\epsilon}\frac{r^{2}\mu_{i}^{2}}{r^{2}+a_{i}^{2}}.
\end{equation}
The one-form $k_{\mu}{\rm d}x^{\mu}$ is given as 
\begin{equation}
k_{\mu}{\rm d}x^{\mu}=F{\rm d}r+W{\rm d}t-\sum_{i=1}^{N}\frac{a_{i}\mu_{i}^{2}}{1-\frac{a_{i}^{2}}{\ell^{2}}}{\rm d}\phi_{i},
\end{equation}
which is null and geodesic for both the background and the full metrics
which conveys the spirit of the Kerr-Schild construction. The $\ell\rightarrow\infty$
limit of (\ref{eq:KS}) yields the Myers-Perry black hole \cite{MyersPerry}
and $a_{i}\rightarrow0$ yields the Schwarzschild-Tangherlini black
holes. Defining the perturbation as 
\begin{equation}
h_{\mu\nu}=\frac{2mG_{n}}{U}k_{\mu}k_{\nu},
\end{equation}
which yields $h=0$. Note that the Kerr-Schild form is quite convenient
for perturbative calculations as the perturbation $h_{\mu\nu}$ is
exact.

We can now run the machinery and calculate the energy and angular
momenta of these solutions (\ref{eq:KS}). For the energy, we shall
take $\bar{\xi}^{\mu}=(-1,\vec{0})$, and then the energy integral
becomes 
\[
E=\frac{1}{4\Omega_{n-2}G_{n}}\int_{\partial\bar{\Sigma}}dS_{r}\left(\bar{g}_{00}\bar{\nabla}^{0}h^{r0}+\bar{g}_{00}\bar{\nabla}^{r}h^{00}+h^{0\nu}\bar{\nabla}^{r}\bar{\xi}_{\nu}-h^{r\nu}\bar{\nabla}^{0}\bar{\xi}_{\nu}+\bar{\nabla}_{\nu}h^{r\nu}\right),
\]
and expanding the covariant derivatives, one arrives at 
\begin{align}
E=\frac{1}{4\Omega_{n-2}G_{n}}\int_{\partial\bar{\Sigma}}d\Omega_{n-2}\sqrt{\left|\bar{g}\right|}\Biggl( & \bar{g}_{00}\bar{g}^{rr}\partial_{r}h^{00}+\frac{1}{2}h^{00}\bar{g}^{rr}\partial_{r}\bar{g}_{00}-\frac{m}{U}\bar{g}^{00}\partial_{r}\bar{g}_{00}+2m\partial_{r}U^{-1}\nonumber \\
 & +\frac{2m}{U}\bar{g}^{rr}\partial_{r}\bar{g}_{rr}-\frac{m}{U}g^{rr}k^{i}k^{j}\partial_{r}\bar{g}_{ij}+\frac{m}{U}\bar{g}^{ij}\partial_{r}\bar{g}_{ij}\Biggr).\label{eq:ed}
\end{align}
The integral is to be computed on the boundary $\partial\bar{\Sigma}$
which is located at $r\rightarrow\infty$. The integration, whose
details can be found in \cite{Kanik,Tekin_et_al}, yields the energy
of the $n$-dimensional rotating black hole as 
\begin{equation}
E_{n}=\frac{m}{\Xi}\sum_{i=1}^{N}\left(\frac{1}{\Xi_{i}}-\frac{1}{2}\left(1-\epsilon\right)\right).\label{eq:finalen-1}
\end{equation}
where we have defined 
\begin{equation}
\Xi\equiv\prod_{i=1}^{N}\left(1-\frac{a_{i}^{2}}{\ell^{2}}\right),\qquad\Xi_{i}\equiv1-\frac{a_{i}^{2}}{\ell^{2}}.
\end{equation}
In four dimensions, one gets a modified energy expression due to the
rotation 
\begin{equation}
E=\frac{m}{\left(1-\frac{a^{2}}{\ell^{2}}\right)^{2}}.
\end{equation}
which in the $\ell\rightarrow\infty$ limit or $a\rightarrow0$ limit
yields the expected result. Note that $a\ne\ell$ for the metric to
be nonsingular. In addition, this result agrees (up to a constant
factor) with those of \cite{Pope,Deruelle}.

Similarly, taking the Killing vector $\xi_{(i)}^{\mu}=\left(0,...,0,1_{i},0,..\right)$
corresponding to the $i^{\mbox{th}}$ azimuthal angle $\phi_{i}$
out of $N$ possibilities, one finds the corresponding Killing charge
to be 
\begin{align}
Q_{i}= & \frac{1}{4\Omega_{n-2}G_{n}}\int_{\partial\bar{\Sigma}}dS_{r}\left(\bar{g}_{\phi_{i}\phi_{i}}\bar{\nabla}^{0}h^{r\phi_{i}}-\bar{g}_{\phi_{i}\phi_{i}}\bar{\nabla}^{r}h^{0\phi_{i}}+h^{0\nu}\bar{\nabla}^{r}\bar{\xi}_{\nu}-h^{r\nu}\bar{\nabla}^{0}\bar{\xi}_{\nu}\right)\nonumber \\
= & \frac{1}{4\Omega_{n-2}G_{n}}\int_{\partial\bar{\Sigma}}d\Omega_{n-2}\sqrt{\left|\bar{g}\right|}\left(-\bar{g}_{\phi_{i}\phi_{i}}\bar{g}^{rr}\bar{g}^{00}\partial_{r}h_{0}^{\;\phi_{i}}\right).
\end{align}
which yields \cite{Kanik,Tekin_et_al} 
\begin{equation}
J_{i}=\frac{ma_{i}}{\Xi\Xi_{i}},\label{eq:ang}
\end{equation}
in agreement with \cite{Pope,Deruelle}. We can relate the total energy
and angular momenta as 
\begin{equation}
E=\sum_{i=1}^{N}\frac{J_{i}}{a_{i}},
\end{equation}
for even dimensions and for the odd case, one has 
\begin{equation}
E=\sum_{i=1}^{N}\frac{J_{i}}{a_{i}}-\frac{Nm}{2\Xi},
\end{equation}
where there is an additional piece independent of the angular momenta,
but for any $J_{i}\rightarrow0$, one must take the limit $J_{i}/a_{i}\rightarrow m$.

\subsection{Computation of the charges for the solitons}

We can use the AD formalism to compute the conserved charges of solitonic
(nonsingular) solutions of gravity theories. These are quite interesting
objects for various reasons; one of them being their possible negative
but bounded mass. Here, we follow \cite{Cebeci}.

\subsubsection{The $AdS$ Soliton}

Horowitz and Myers \cite{Horowitz} analytically continued the near
extremal $p$-brane solutions to obtain the following metric the so-called
``AdS Soliton\textquotedbl ; 
\begin{eqnarray}
ds^{2}=\frac{r^{2}}{\ell^{2}}\left[\left(1-\frac{r_{0}^{p+1}}{r^{p+1}}\right){\rm d}\tau^{2}+\sum_{i=1}^{p-1}\left({\rm d}x^{i}\right)^{2}-{\rm d}t^{2}\right]+\left(1-\frac{r_{0}^{p+1}}{r^{p+1}}\right)^{-1}\,\frac{\ell^{2}}{r^{2}}\,{\rm d}r^{2}, & \qquad & \left(r\ge r_{0}\right)\label{adssoliton}
\end{eqnarray}
The coordinates on the $p$-brane are $t$ and $x^{i}$ ($i=1,...,p-1$).
The conical singularity at $r=r_{0}$ is removed if $\tau$ has a
period $\beta=4\pi\ell^{2}/(r_{0}(p+1))$. Using the Killing vector
$\bar{\xi}^{\mu}=(-1,0,...,0)$ in (\ref{eq:Conserved_charges}) and
taking the background to be $r_{0}=0$, one arrives at the energy
density of the solitonic brane $E/V_{n-3}$ as 
\begin{equation}
\frac{E}{V_{n-3}}=-\frac{\pi}{(n-1)\,\Omega_{n-2}\,G_{n}}\,\frac{r_{0}^{n-2}}{\ell^{n-2}}\,,
\end{equation}
which matches the result of \cite{Horowitz} computed with the energy
definition of Hawking-Horowitz \cite{HawkingHorowitz} up to trivial
normalization factors.

\subsubsection{Eguchi-Hanson Solitons}

In odd dimensional cosmological spacetimes, Clarkson and Mann \cite{clarkson}
found solitonic solutions that are called Eguchi-Hanson solitons since
they are analogs to the even dimensional Eguchi-Hanson metrics \cite{eguchi}
which approach asymptotically to $AdS/Z_{p}$, where $p\ge3$, instead
of the global AdS spacetime. These solitons have lower energy than
the global AdS spacetime as was demonstrated by Clarkson and Mann
with the boundary counter-term method of \cite{henningson,97} for
the case of five dimensions. This method requires each dimension to
be worked out separately as the counter-terms vary when the dimension
changes. But, the AD method can be applied for a generic dimensional
solution. The Eguchi-Hanson soliton metric is 
\begin{eqnarray}
ds^{2} & = & -g\,{\rm d}t^{2}+\left(\frac{2r}{n-1}\right)^{2}\,f\,\left[{\rm d}\psi+\sum_{i=1}^{(n-3)/2}\cos\theta_{i}\,{\rm d}\phi_{i}\right]^{2}\nonumber \\
 &  & +\frac{dr^{2}}{g\,f}+\frac{r^{2}}{n-1}\,\sum_{i=1}^{(n-3)/2}\,\left({\rm d}\theta_{i}^{2}+\sin^{2}\theta_{i}\,{\rm d}\phi_{i}^{2}\right)\,,\label{EH d-dim}
\end{eqnarray}
where 
\begin{equation}
g(r)=1\mp\frac{r^{2}}{\ell^{2}}\;\;,\quad f(r)=1-\left(\frac{r_{0}}{r}\right)^{n-1}\,.\label{eh}
\end{equation}
The solution exists both in dS and AdS; but, let us concentrate in
the AdS case and to remove the string-like singularity at $r=r_{0}$,
$\psi$ must have a period $4\pi/p$ and the constant $r_{0}$ is
given as 
\begin{equation}
r_{0}^{2}=\ell^{2}\,(\frac{p^{2}}{4}-1)\,.
\end{equation}
Taking the background to be $r_{0}=0$, the energy for $\bar{\xi}^{\mu}=(-1,0,...,0)$
is 
\begin{equation}
E=-\frac{(4\pi)^{(n-1)/2}\,r_{0}^{n-1}}{p\,\ell^{2}\,(n-1)^{(n-1)/2}\,\Omega_{n-2}\,G_{n}}\,,
\end{equation}
which reduces to 
\begin{equation}
E=-\frac{r_{0}^{4}}{4\,p\,\ell^{2}\,G_{5}}\,,
\end{equation}
for $n=5$. In the counter-term method, the global AdS has a finite
energy whereas in our method it has zero energy by definition. Once
that is taken into account, both methods give the same result up to
a trivial normalization factor.

\subsection{Energy of the Taub-NUT-Reissner-{Nordstr$\ddot{\text{o}}$m} metric}

As some other nontrivial applications of the AD formalism for spacetimes
with nontrivial topology, let us consider the four and six dimensional
Taub-NUT-Reissner-{Nordstr$\ddot{\text{o}}$m} solutions \cite{stelea1}.
For $n=4$, the metric is 
\begin{equation}
ds^{2}=-f\,\Bigl({\rm d}t-2\,N\,\cos\theta\,{\rm d}\phi\Bigr)^{2}+\frac{{\rm d}r^{2}}{f}+\left(r^{2}+N^{2}\right)\,\left({\rm d}\theta^{2}+\sin^{2}\theta\,{\rm d}\phi^{2}\right)\,,\label{eq:Taub-NUT-etc_metric}
\end{equation}
where $N$ is the nut charge and 
\begin{equation}
f(r)=\frac{r^{4}+(\ell^{2}+6N^{2})\,r^{2}-2\,m\,\ell^{2}\,r-3N^{4}+\ell^{2}\,(q^{2}-N^{2})}{\ell^{2}\,(r^{2}+N^{2})}\,.
\end{equation}
The naive identification of the background as $m=q=N=0$ yields a
divergent energy for any nonzero $N$ and $m$ configuration. This
is because $N=0$ and $N\ne0$ spacetimes are not in the same topological
class; hence, for a meaningful use of the formalism one should work
within a fixed $N$ sector and choose the background to be $m=q=0$;
but, keep $N$ finite. This yields for (\ref{eq:Taub-NUT-etc_metric})
\begin{equation}
E=\frac{m}{G_{4}}\,,
\end{equation}
again for the timelike Killing vector $\bar{\xi}=(-1,0,0,0)$. In
$n=6$, the metric reads 
\begin{eqnarray}
ds^{2} & = & -f\,\Bigl({\rm d}t-2\,N\,\cos\theta_{1}\,{\rm d}\phi_{1}-2\,N\,\cos\theta_{2}\,{\rm d}\phi_{2}\Bigr)^{2}+\frac{{\rm d}r^{2}}{f}\nonumber \\
 &  & +\left(r^{2}+N^{2}\right)\,\left(d\theta_{1}^{2}+\sin^{2}\theta_{1}\,{\rm d}\phi_{1}^{2}+{\rm d}\theta_{2}^{2}+\sin^{2}\theta_{2}\,{\rm d}\phi_{2}^{2}\right)\,,
\end{eqnarray}
where the metric function $f$ is 
\begin{eqnarray*}
f(r) & = & \frac{q^{2}\,\left(3r^{2}+N^{2}\right)}{\left(r^{2}+N^{2}\right)^{4}}\\
 &  & +\frac{1}{3\ell^{2}\left(r^{2}+N^{2}\right)^{2}}\left(\ell^{2}\left(-3N^{4}-6mr+6N^{2}r^{2}+r^{4}\right)-15N^{6}+45N^{4}r^{2}+15N^{2}r^{4}+3r^{6}\right).
\end{eqnarray*}
Similarly, choosing the background to be $m=q=0$ but $N\ne0$, and
the energy can be computed as 
\begin{equation}
E=12\,\frac{m}{G_{6}}\,.
\end{equation}
Next we turn our attention to generic gravity theories with more powers
of curvature.

\section{Charges of Quadratic Curvature Gravity}

We will work out the construction of conserved charges for generic
gravity theories with an action of the form 
\begin{equation}
I=\int d^{D}x\,\sqrt{-\left|g\right|}\,f\left(R_{\rho\sigma}^{\mu\nu}\right),\label{eq:F(Riem)}
\end{equation}
where $F$ is a smooth function of the Riemann tensor and its contractions
given in the form $R_{\rho\sigma}^{\mu\nu}:=R_{\phantom{\mu\nu}\rho\sigma}^{\mu\nu}$,
and hence contractions do not require the metric. For example, $R_{\phantom{\nu}\sigma}^{\nu}=R_{\mu\sigma}^{\mu\nu}$
and $R=R_{\mu\nu}^{\mu\nu}$. Here, we start with the quadratic theory
\begin{equation}
I=\int d^{n}x\sqrt{-g}\left(\frac{1}{\kappa}\left(R-2\Lambda_{0}\right)+\alpha R^{2}+\beta R_{\mu\nu}R^{\mu\nu}+\gamma\left(R_{\mu\nu\rho\sigma}R^{\mu\nu\rho\sigma}-4R_{\mu\nu}R^{\mu\nu}+R^{2}\right)\right),\label{eq:Quad_act}
\end{equation}
whose results can be easily extended to the above more general action
as we shall lay out in more detail. We have organized the last part
into the Gauss-Bonnet combination which does not contribute to the
field equations for $n\le4$ (in fact, it vanishes identically in
$n\le3$ and becomes a surface term for $n=4$). Arguments from string
theory require $\alpha$ and $\beta$ to vanish \cite{Zwiebach},
or in general using field redefinitions in a generic theory which
is not truncated at the quadratic order as (\ref{eq:Quad_act}), one
can get rid off the $\alpha$ and $\beta$ terms. But, here our goal
is to consider (\ref{eq:Quad_act}) to be exact (with all its deficiencies
such as having a ghost for nonzero $\beta$) and construct the conserved
charges. For this purpose, we need the field equations which read
\begin{align}
\frac{1}{\kappa}\left(R_{\mu\nu}-\frac{1}{2}g_{\mu\nu}R+g_{\mu\nu}\Lambda_{0}\right)+2\alpha R\left(R_{\mu\nu}-\frac{1}{4}g_{\mu\nu}R\right)+\left(2\alpha+\beta\right)\left(g_{\mu\nu}\Box-\nabla_{\mu}\nabla_{\nu}\right)R\nonumber \\
+2\gamma\left(RR_{\mu\nu}-2R_{\mu\sigma\nu\rho}R^{\sigma\rho}+R_{\mu\sigma\rho\tau}R_{\nu}^{\phantom{\nu}\sigma\rho\tau}-2R_{\mu\sigma}R_{\phantom{\sigma}\nu}^{\sigma}-\frac{1}{4}g_{\mu\nu}\left(R_{\alpha\beta\rho\sigma}R^{\alpha\beta\rho\sigma}-4R_{\alpha\beta}R^{\alpha\beta}+R^{2}\right)\right)\nonumber \\
+\beta\Box\left(R_{\mu\nu}-\frac{1}{2}g_{\mu\nu}R\right)+2\beta\left(R_{\mu\sigma\nu\rho}-\frac{1}{4}g_{\mu\nu}R_{\sigma\rho}\right)R^{\sigma\rho} & =\tau_{\mu\nu}.\label{eq:EoM_quad}
\end{align}
In general, there are two maximally symmetric vacua whose effective
cosmological constant $\Lambda$ is determined by the quadratic equation
\begin{equation}
\frac{\Lambda-\Lambda_{0}}{2\kappa}+k\Lambda^{2}=0,\qquad k:=\left(n\alpha+\beta\right)\frac{\left(n-4\right)}{\left(n-2\right)^{2}}+\gamma\frac{\left(n-3\right)\left(n-4\right)}{\left(n-1\right)\left(n-2\right)}.\label{eq:Eff_Lambda}
\end{equation}
Linearizing (\ref{eq:EoM_quad}) about one of these vacua and collecting
all the higher order terms to the right, one arrives at 
\begin{equation}
c\,\mathcal{G}_{\mu\nu}^{\left(1\right)}+\left(2\alpha+\beta\right)\left(\bar{g}_{\mu\nu}\bar{\square}-\bar{\nabla}_{\mu}\bar{\nabla}_{\nu}+\frac{2\Lambda}{n-2}\bar{g}_{\mu\nu}\right)R_{\left(1\right)}+\beta\left(\bar{\square}\mathcal{G}_{\mu\nu}^{\left(1\right)}-\frac{2\Lambda}{n-1}\bar{g}_{\mu\nu}R_{\left(1\right)}\right)=:T_{\mu\nu},\label{Linearized_eom}
\end{equation}
where the constant $c$ in-front of the linearized Einstein tensor
reads 
\begin{equation}
c:=\frac{1}{\kappa}+\frac{4\Lambda n}{n-2}\alpha+\frac{4\Lambda}{n-1}\beta+\frac{4\Lambda\left(n-3\right)\left(n-4\right)}{\left(n-1\right)\left(n-2\right)}\gamma.\label{eq:c}
\end{equation}
Observe that one has $\bar{\nabla}^{\mu}T_{\mu\nu}=0$ which follows
from the linearized Bianchi identity that we worked out before in
the previous part 
\[
\bar{\nabla}_{\mu}{\cal G}^{\left(1\right)\mu\nu}=0,
\]
and the following identities which can be easily derived by commuting
the derivatives with the help of the Ricci identity 
\begin{align*}
\bar{\nabla}^{\mu}\left(\bar{g}_{\mu\nu}\bar{\Box}-\bar{\nabla}_{\mu}\bar{\nabla}_{\nu}+\frac{2\Lambda}{n-2}\bar{g}_{\mu\nu}\right)R_{\left(1\right)} & =0\\
\bar{\nabla}^{\mu}\left(\bar{\Box}{\cal G}_{\mu\nu}^{\left(1\right)}-\frac{2\Lambda}{n-1}\bar{g}_{\mu\nu}\right)R_{\left(1\right)} & =0,
\end{align*}
where of course $\bar{g}^{\mu\nu}{\cal G}_{\mu\nu}^{\left(1\right)}=\frac{2-n}{2}R_{\left(1\right)}$.
To be able to write $\bar{\xi}_{\mu}T^{\mu\nu}$ as a boundary term,
one needs the following identities 
\begin{align}
\bar{\xi}_{\nu}\bar{\Box}{\cal G}^{\left(1\right)\mu\nu}= & \bar{\nabla}_{\alpha}\left(\bar{\xi}_{\nu}\bar{\nabla}^{\alpha}{\cal G}_{\left(1\right)}^{\mu\nu}-\bar{\xi}_{\nu}\bar{\nabla}^{\mu}{\cal G}_{\left(1\right)}^{\alpha\nu}-{\cal G}_{\left(1\right)}^{\mu\nu}\bar{\nabla}^{\alpha}\bar{\xi}_{\nu}+{\cal G}_{\left(1\right)}^{\alpha\nu}\bar{\nabla}^{\mu}\bar{\xi}_{\nu}\right)\nonumber \\
 & +{\cal G}_{\left(1\right)}^{\mu\nu}\Box\bar{\xi}_{\nu}+\bar{\xi}_{\nu}\bar{\nabla}_{\alpha}\bar{\nabla}^{\mu}{\cal G}_{\left(1\right)}^{\alpha\nu}-{\cal G}_{\left(1\right)}^{\alpha\nu}\bar{\nabla}_{\alpha}\bar{\nabla}^{\mu}\bar{\xi}_{\nu},
\end{align}
\begin{equation}
\bar{\nabla}_{\alpha}\bar{\nabla}_{\beta}\bar{\xi}_{\nu}=\bar{R}_{\phantom{\mu}\nu\beta\alpha}^{\mu}\bar{\xi}_{\mu}=\frac{2\Lambda}{\left(n-2\right)\left(n-1\right)}\left(\bar{g}_{\nu\alpha}\bar{\xi}_{\beta}-\bar{g}_{\alpha\beta}\bar{\xi}_{\nu}\right),\qquad\qquad\bar{\Box}\bar{\xi}_{\mu}=-\frac{2\Lambda}{n-2}\bar{\xi}_{\mu},
\end{equation}
\begin{equation}
\bar{\xi}_{\nu}\bar{\nabla}_{\alpha}\bar{\nabla}^{\mu}{\cal G}_{\left(1\right)}^{\alpha\nu}=\frac{2\Lambda n}{\left(n-2\right)\left(n-1\right)}\bar{\xi}_{\nu}{\cal G}_{\left(1\right)}^{\mu\nu}+\frac{\Lambda}{n-1}\bar{\xi}^{\mu}R_{\left(1\right)}.
\end{equation}
Collecting all the pieces together, the conserved charges of quadratic
gravity for asymptotically (A)dS spacetimes read 
\begin{align}
Q_{{\rm QG}}\left(\bar{\xi}\right)= & \left(c+\frac{4\Lambda\beta}{\left(n-1\right)\left(n-2\right)}\right)\int d^{n-1}x\sqrt{-\bar{g}}\bar{\xi}_{\nu}{\cal G}_{\left(1\right)}^{0\nu}\nonumber \\
 & +\left(2\alpha+\beta\right)\int_{\partial\bar{\Sigma}}dS_{i}\left(\bar{\xi}^{0}\bar{\nabla}^{i}R_{\left(1\right)}+R_{\left(1\right)}\bar{\nabla}^{0}\bar{\xi}^{i}-\bar{\xi}^{i}\bar{\nabla}^{0}R_{\left(1\right)}\right)\nonumber \\
 & +\beta\int_{\partial\bar{\Sigma}}dS_{i}\left(\bar{\xi}_{\nu}\bar{\nabla}^{i}{\cal G}_{\left(1\right)}^{0\nu}-\bar{\xi}_{\nu}\bar{\nabla}^{0}{\cal G}_{\left(1\right)}^{i\nu}-{\cal G}_{\left(1\right)}^{0\nu}\bar{\nabla}^{i}\bar{\xi}_{\nu}+{\cal G}_{\left(1\right)}^{i\nu}\bar{\nabla}^{0}\bar{\xi}_{\nu}\right),\label{fullcharge}
\end{align}
where the first line was given before in (\ref{eq:ADT_charge}); hence,
we have not depicted it as a surface integral here. For asymptotically
(A)dS spacetimes, the last two lines in (\ref{fullcharge}) vanish
due to the fact that ${\cal G}_{\mu\nu}^{\left(1\right)}\rightarrow O\left(r^{-n+1}\right)$.
Therefore, the final result is quite nice: defining an effective Newton's
constant as 
\begin{equation}
\frac{1}{\kappa_{\text{eff}}}:=\frac{1}{\kappa}+\frac{4\Lambda\left(n\alpha+\beta\right)}{n-2}+\frac{4\Lambda\left(n-3\right)\left(n-4\right)}{\left(n-1\right)\left(n-2\right)}\gamma=c+\frac{4\Lambda\beta}{\left(n-1\right)\left(n-2\right)},\label{eq:kappa_eff_for_conserved_charges}
\end{equation}
in quadratic gravity, the effect of quadratic terms is encoded in
the $\kappa_{\text{eff}}$ and the final result can be succinctly
written as\footnote{It is important to realize that the $\kappa_{{\rm eff}}$ is not the
inverse of $c$ that appears in front of the linearized Einstein tensor
in equation (\ref{Linearized_eom}) as one would naively expect from
the weak field limit: the cosmological constant brings nontrivial
contributions from the higher curvature terms as can be seen (\ref{eq:kappa_eff_for_conserved_charges}).} 
\begin{equation}
\boxed{\phantom{\frac{\frac{\xi}{\xi}}{\frac{\xi}{\xi}}}Q_{{\rm QG}}\left(\bar{\xi}\right)=\frac{\kappa}{\kappa_{{\rm eff}}}Q_{\text{Einstein}}\left(\bar{\xi}\right).\phantom{\frac{\frac{\xi}{\xi}}{\frac{\xi}{\xi}}}}\label{eq:Q_quad}
\end{equation}
For asymptotically flat backgrounds, $\kappa_{{\rm eff}}=\kappa$
and the higher order terms do not contribute to the charges. Let us
apply this construction to the Boulware-Deser black hole solution
\cite{BoulwareDeser} of the Einstein--Gauss-Bonnet theory for which
we take $\alpha=\beta=\Lambda_{0}=0$ and the solution reads 
\begin{equation}
{\rm d}s^{2}=g_{00}{\rm d}t^{2}+g_{rr}{\rm d}r^{2}+r^{2}{\rm d}\Omega_{n-2},
\end{equation}
with 
\begin{equation}
-g_{00}=g_{rr}^{-1}=1+\frac{r^{2}}{4\kappa\gamma\left(n-3\right)\left(n-4\right)}\left(1\pm\left(1+8\gamma\left(n-3\right)\left(n-4\right)\frac{r_{0}^{n-3}}{r^{n-1}}\right)^{\frac{1}{2}}\right).\label{eq:BD_BH}
\end{equation}
Note that the $\pm$ branches are quite different: the minus branch
referring to the asymptotically flat Schwarzschild black hole and
the plus branch referring to the asymptotically Schwarzschild-AdS
black hole, more explicitly, asymptotically one has 
\begin{equation}
-g_{00}\rightarrow1-\left(\frac{r_{0}}{r}\right)^{n-3},\qquad\qquad-g_{00}\rightarrow1+\left(\frac{r_{0}}{r}\right)^{n-3}+\frac{r^{2}}{\kappa\gamma\left(n-3\right)\left(n-4\right)},\label{eq:asymptotic}
\end{equation}
where we have set $\Lambda=-\frac{\left(n-1\right)\left(n-2\right)}{2\kappa\gamma\left(n-3\right)\left(n-4\right)}$
coming from the vacuum equation. For the asymptotically flat Schwarzschild
black hole, we have computed the mass before (\ref{eq:E_of_SAdS}).
For the (A)dS case, naively one gets a negative energy, but we should
use (\ref{eq:Q_quad}) with $\kappa_{{\rm eff}}=-\kappa$; therefore,
\begin{equation}
E=\frac{\left(n-2\right)}{4G_{n}}r_{0}^{n-3},
\end{equation}
and energy is positive for both branches.

The above charge construction for the quadratic curvature gravity
is widely used in the literature; for example, see \cite{Bouchareb,17,Dias,Critical4D,Critical,Ghodsi1,Ghodsi2,Lu,Ghodsi3,Kim,Goya,Ghodsi4,Gim,GiribetVasquez,Ayon-Beato,Bravo-Gaete,Vasquez,Guajardo,Cisterna,Ghodsi5}.
Now, let us generalize the charge construction for the quadratic curvature
gravity to the higher curvature gravity with the Lagrangian density
constructed from the Riemann tensor and its contractions.

\section{Charges of $f\left(R_{\rho\sigma}^{\mu\nu}\right)$ Gravity}

In principle, using the above construction for quadratic gravity,
we can find the conserved charges for asymptotically (A)dS spacetimes
of the more generic theory defined by the action 
\begin{equation}
I=\int d^{n}x\,\sqrt{-g}\,f\left(R_{\rho\sigma}^{\mu\nu}\right),\label{eq:F(Riem)_action}
\end{equation}
whose field equation can be found from the following variation 
\begin{equation}
\delta_{g}I=\int d^{n}x\,\left(\delta\sqrt{-g}f\left(R_{\alpha\beta}^{\mu\nu}\right)+\sqrt{-g}\frac{\partial f}{\partial R_{\rho\sigma}^{\mu\nu}}\delta R_{\rho\sigma}^{\mu\nu}\right).\label{eq:Var_action}
\end{equation}
For generic variations of the metric, the variation of the Riemann
tensor reads 
\begin{align}
\delta R_{\rho\sigma}^{\mu\nu}= & \frac{1}{2}\left(g_{\alpha\rho}\nabla_{\sigma}\nabla^{\nu}-g_{\alpha\sigma}\nabla_{\rho}\nabla^{\nu}\right)\delta g^{\mu\alpha}-\frac{1}{2}\left(g_{\alpha\rho}\nabla_{\sigma}\nabla^{\mu}-g_{\alpha\sigma}\nabla_{\rho}\nabla^{\mu}\right)\delta g^{\alpha\nu}\nonumber \\
 & -\frac{1}{2}R_{\rho\sigma\phantom{\nu}\alpha}^{\phantom{\rho\sigma}\nu}\delta g^{\mu\alpha}+\frac{1}{2}R_{\rho\sigma\phantom{\mu}\alpha}^{\phantom{\rho\sigma}\mu}\delta g^{\alpha\nu},\label{eq:Var_Riem}
\end{align}
which follows from usual easy to obtain-expression (but beware of
the location of the indices!) 
\begin{equation}
\delta R_{\phantom{\mu}\nu\rho\sigma}^{\mu}=\nabla_{\rho}\delta\Gamma_{\nu\sigma}^{\mu}-\nabla_{\sigma}\delta\Gamma_{\nu\rho}^{\mu}.
\end{equation}
After inserting (\ref{eq:Var_Riem}) into (\ref{eq:Var_action}) and
making repeated use of integration by parts, one arrives at the full
non-linear equations 
\begin{align}
\mathcal{E}_{\mu\nu}:=\frac{1}{2}\left(g_{\nu\rho}\nabla^{\lambda}\nabla_{\sigma}-g_{\nu\sigma}\nabla^{\lambda}\nabla_{\rho}\right)\frac{\partial f}{\partial R_{\rho\sigma}^{\mu\lambda}}-\frac{1}{2}\left(g_{\mu\rho}\nabla^{\lambda}\nabla_{\sigma}-g_{\mu\sigma}\nabla^{\lambda}\nabla_{\rho}\right)\frac{\partial f}{\partial R_{\rho\sigma}^{\lambda\nu}}\nonumber \\
-\frac{1}{2}\left(\frac{\partial f}{\partial R_{\rho\sigma}^{\mu\lambda}}R_{\rho\sigma\phantom{\lambda}\nu}^{\phantom{\rho\sigma}\lambda}-\frac{\partial f}{\partial R_{\rho\sigma}^{\lambda\nu}}R_{\rho\sigma\phantom{\lambda}\mu}^{\phantom{\rho\sigma}\lambda}\right)-\frac{1}{2}g_{\mu\nu}f\left(R_{\rho\sigma}^{\alpha\beta}\right) & =\tau_{\mu\nu},\label{eq:EoM_of_fRiem}
\end{align}
where we have also added a source term. For the construction of the
conserved charges, we need the following information: $\mathcal{E}_{\mu\nu}\left(\bar{g}\right)=0$
and $\mathcal{E}_{\mu\nu}^{\left(1\right)}\left(h\right)=:T_{\mu\nu}\left(\tau,h\right)$
where $T_{\mu\nu}$ starts at $O\left(h^{2}\right)$. To find the
effective cosmological constant of the maximally symmetric solution,
we set $\mathcal{E}_{\mu\nu}\left(\bar{g}\right)=0$ and note that
the first line of (\ref{eq:EoM_of_fRiem}) vanishes and the equation
reduces to 
\begin{equation}
\left[\frac{\partial f}{\partial R_{\rho\sigma}^{\mu\lambda}}\right]_{\bar{g}}\bar{R}_{\rho\sigma\phantom{\lambda}\nu}^{\phantom{\rho\sigma}\lambda}-\left[\frac{\partial f}{\partial R_{\rho\sigma}^{\lambda\nu}}\right]_{\bar{g}}\bar{R}_{\rho\sigma\phantom{\lambda}\mu}^{\phantom{\rho\sigma}\lambda}+\bar{g}_{\mu\nu}f\left(\bar{R}_{\rho\sigma}^{\alpha\beta}\right)=0,\label{eq:Vacuum_eq}
\end{equation}
where the barred quantities refer to the maximally symmetric metric
$\bar{g}_{\mu\nu}$ with the Riemann tensor 
\begin{equation}
\bar{R}_{\rho\sigma}^{\mu\nu}=\frac{2\Lambda}{\left(n-1\right)\left(n-2\right)}\left(\delta_{\rho}^{\mu}\delta_{\sigma}^{\nu}-\delta_{\sigma}^{\mu}\delta_{\rho}^{\nu}\right).\label{AdS_background}
\end{equation}
Contemplating on the vacuum equation (\ref{eq:Vacuum_eq}), one realizes
that the information regarding the theory enters through only the
following two background evaluated quantities 
\begin{equation}
\bar{f}:=f\left(\bar{R}_{\rho\sigma}^{\alpha\beta}\right),\qquad\qquad\left[\frac{\partial f}{\partial R_{\rho\sigma}^{\mu\nu}}\right]_{\bar{g}},\label{eq:fbar_defn}
\end{equation}
which are the zeroth order and the first order terms in the Taylor
series expansion of $f$ in the particular form of the Riemann tensor
$R_{\rho\sigma}^{\alpha\beta}$ around the maximally symmetric background
(\ref{AdS_background}). This simple observation is quite useful since
it tells us that if these quantities are the same for two given theories
then they have the same maximally symmetric vacua. Therefore, to obtain
the vacuum of the theory, one can simply consider the following ``Einstein-Hilbert''
action with a specific Newton's constant and a bare cosmological constant
as 
\begin{equation}
I=\int d^{n}x\,\sqrt{-g}\left(\bar{f}+\left[\frac{\partial f}{\partial R_{\rho\sigma}^{\mu\nu}}\right]_{\bar{g}}\left(R_{\rho\sigma}^{\mu\nu}-\bar{R}_{\rho\sigma}^{\mu\nu}\right)\right).\label{eq:ELA_Taylor_expanded}
\end{equation}
This can be seen as follows: Here, the term $\left[\partial f/\partial R_{\rho\sigma}^{\mu\nu}\right]_{\bar{g}}$
is proportional to the Kronecker-deltas such as $\delta_{\mu}^{\rho}\delta_{\nu}^{\sigma}$,
and considering the symmetries of the Riemann tensor, one should antisymmetrize
accordingly as $\delta_{\mu}^{[\rho}\delta_{\nu}^{\sigma]}$ and set
\begin{equation}
\left[\frac{\partial f}{\partial R_{\rho\sigma}^{\mu\nu}}\right]_{\bar{g}}=:\frac{1}{\kappa_{l}}\delta_{\mu}^{[\rho}\delta_{\nu}^{\sigma]},\label{eq:kappa_l_defn}
\end{equation}
which defines $\kappa_{l}$. Observe that one then has 
\begin{equation}
\left[\frac{\partial f}{\partial R_{\rho\sigma}^{\mu\nu}}\right]_{\bar{g}}R_{\rho\sigma}^{\mu\nu}=\frac{1}{\kappa_{l}}R,\label{eq:Taylor_exp_1st_ord}
\end{equation}
and so (\ref{eq:ELA_Taylor_expanded}) can be recast as 
\begin{equation}
I=\frac{1}{\kappa_{l}}\int d^{n}x\,\sqrt{-g}\left(R-\bar{R}+\kappa_{l}\bar{f}\right),\label{eq:ELA}
\end{equation}
from which one can find the effective cosmological constant as $\Lambda=\frac{1}{2}\left(\bar{R}-\kappa_{l}\bar{f}\right)$
yielding 
\begin{equation}
\boxed{\phantom{\frac{\frac{\xi}{\xi}}{\frac{\xi}{\xi}}}\Lambda=\frac{n-2}{4}\bar{f}\kappa_{l}\phantom{\frac{\frac{\xi}{\xi}}{\frac{\xi}{\xi}}}},\label{eq:Vacua_of_f(Riem)}
\end{equation}
which is naturally the solution of (\ref{eq:Vacuum_eq}) when (\ref{eq:kappa_l_defn})
used. Note that (\ref{eq:Vacua_of_f(Riem)}) is generically a polynomial
equation in $\Lambda$ if the action is a higher curvature theory
in powers of the Riemann tensor. To summarize, given a Lagrangian
density $f\left(R_{\rho\sigma}^{\mu\nu}\right)$, the effective Newton's
constant (as it appears in the Einstein-Hilbert action) in any of
its maximally symmetric vacua can be computed from the following formula
\begin{equation}
\boxed{\phantom{\frac{\bar{R}_{\rho}^{\mu}}{\bar{R}_{\rho}^{\mu}}}\frac{1}{\kappa_{l}}=\frac{\bar{R}_{\rho\sigma}^{\mu\nu}}{\bar{R}}\left[\frac{\partial f}{\partial R_{\rho\sigma}^{\mu\nu}}\right]_{\bar{g}}\phantom{\frac{\bar{R}_{\rho}^{\mu}}{\bar{R}_{\rho}^{\mu}}}},
\end{equation}
and then, the vacua of the theory can be calculated from (\ref{eq:Vacua_of_f(Riem)}).
While we have carried out the above construction for theories whose
action do not depend on the derivatives of the Riemann tensor and
its contractions, the results are valid even for the more generic
case when such terms are present in the action, as they do not contribute
to two parameter $\Lambda$ and $\kappa_{l}$ for maximally symmetric
backgrounds. (Of course, higher derivative terms in the curvature
will contribute to the spectrum of the theory as well as its fully
non-linear solutions besides the vacua.)

Now, let us move on to the linearization of the field equations, that
is $\mathcal{E}_{\mu\nu}^{\left(1\right)}\left(h\right)=T_{\mu\nu}\left(\tau,h\right)$,
for the construction of conserved charges. Although this would be
a straightforward calculation, it is rather cumbersome which motivates
us to follow a shortcut. The shortcut boils down to finding the equivalent
quadratic curvature action (EQCA) for the given higher curvature theory
\cite{Hindawi,Gullu-UniBI,Gullu-AllUni3D,Sisman-AllUni,Senturk,UniBI4D,UniBIanyD}.\footnote{EQCA constructions similar to \cite{Sisman-AllUni} and \cite{Senturk,UniBI4D}
were given in \cite{Bueno} and \cite{Cano}, respectively. } The EQCA by construction has the same vacua and the same linearized
field equations as the given $f\left(R_{\rho\sigma}^{\mu\nu}\right)$
theory. Then, using the conserved charge expressions for the quadratic
curvature theory, one can immediately obtain the conserved charges
of the $f\left(R_{\rho\sigma}^{\mu\nu}\right)$ theory from its EQCA.
Now, let us show how the EQCA captures the linearized field equations,
$\mathcal{E}_{\mu\nu}^{\left(1\right)}\left(h\right)=T_{\mu\nu}\left(\tau,h\right)$.
In linearizing the field equations, one needs the following somewhat
complicated linearized terms 
\begin{align}
\left[g_{\mu\nu}f\left(R_{\alpha\beta}^{\mu\nu}\right)\right]^{\left(1\right)}= & h_{\mu\nu}\bar{f}+\bar{g}_{\mu\nu}\left[\frac{\partial f}{\partial R_{\rho\sigma}^{\alpha\beta}}\right]_{\bar{g}}\left(R_{\rho\sigma}^{\alpha\beta}\right)^{\left(1\right)},
\end{align}
and 
\begin{align}
\left(\frac{\partial f}{\partial R_{\rho\sigma}^{\mu\lambda}}R_{\rho\sigma\phantom{\lambda}\nu}^{\phantom{\rho\sigma}\lambda}\right)^{\left(1\right)}= & \left[\frac{\partial^{2}f}{\partial R_{\alpha\tau}^{\eta\theta}\partial R_{\rho\sigma}^{\mu\lambda}}\right]_{\bar{g}}\left(R_{\alpha\tau}^{\eta\theta}\right)^{\left(1\right)}\bar{R}_{\rho\sigma\phantom{\lambda}\nu}^{\phantom{\rho\sigma}\lambda}+\left[\frac{\partial f}{\partial R_{\rho\sigma}^{\mu\lambda}}\right]_{\bar{g}}\left(R_{\rho\sigma\phantom{\lambda}\nu}^{\phantom{\rho\sigma}\lambda}\right)^{\left(1\right)},
\end{align}
and 
\begin{align}
\left(g_{\nu\rho}\nabla^{\lambda}\nabla_{\sigma}\frac{\partial f}{\partial R_{\rho\sigma}^{\mu\lambda}}\right)^{\left(1\right)}= & \bar{g}_{\nu\rho}\left[\frac{\partial^{2}f}{\partial R_{\alpha\tau}^{\eta\theta}\partial R_{\rho\sigma}^{\mu\lambda}}\right]_{\bar{g}}\bar{\nabla}^{\lambda}\bar{\nabla}_{\sigma}\left(R_{\alpha\tau}^{\eta\theta}\right)^{\left(1\right)}+\bar{g}_{\nu\rho}\left[\frac{\partial f}{\partial R_{\rho\alpha}^{\mu\lambda}}\right]_{\bar{g}}\bar{\nabla}^{\lambda}\left(\Gamma_{\sigma\alpha}^{\sigma}\right)^{\left(1\right)}\nonumber \\
 & -\bar{g}_{\nu\rho}\left[\frac{\partial f}{\partial R_{\rho\sigma}^{\alpha\lambda}}\right]_{\bar{g}}\bar{\nabla}^{\lambda}\left(\Gamma_{\sigma\mu}^{\alpha}\right)^{\left(1\right)}-\bar{g}_{\nu\rho}\left[\frac{\partial f}{\partial R_{\rho\sigma}^{\mu\alpha}}\right]_{\bar{g}}\bar{\nabla}^{\lambda}\left(\Gamma_{\sigma\lambda}^{\alpha}\right)^{\left(1\right)},
\end{align}
where the superscript $\left(1\right)$ represents the linearization
up to and including $O\left(h_{\mu\nu}\right)$ as before. Just like
in the case of the equivalent linear action (ELA) construction, the
crucial observation here is that linearization process of a given
theory can be carried out if the following quantities 
\begin{equation}
\bar{f},\qquad\left[\frac{\partial f}{\partial R_{\rho\sigma}^{\mu\lambda}}\right]_{\bar{g}},\qquad\left[\frac{\partial^{2}f}{\partial R_{\alpha\tau}^{\eta\theta}\partial R_{\rho\sigma}^{\mu\lambda}}\right]_{\bar{g}},
\end{equation}
can be computed. Hence, the shortcut logic that we expounded upon
above applies here verbatim: if these three quantities are the same
for any two gravity theories, then those two theories have the same
linearized field equation around the same vacua, and any quantity
such as conserved charges and spectra computed from the linearized
field equations. {[}For the spectrum calculation and the masses of
all the perturbative excitations in the theory see \cite{Tekin_rap}.{]}
Therefore, one resorts to the following second order Taylor series
expansion of $f\left(R_{\alpha\beta}^{\mu\nu}\right)$ around its
maximally symmetric background as 
\begin{align}
I=\int d^{n}x\,\sqrt{-g} & \left(\bar{f}+\left[\frac{\partial f}{\partial R_{\rho\sigma}^{\lambda\nu}}\right]_{\bar{g}}\left(R_{\rho\sigma}^{\lambda\nu}-\bar{R}_{\rho\sigma}^{\lambda\nu}\right)\right.\nonumber \\
 & \left.+\frac{1}{2}\left[\frac{\partial^{2}f}{\partial R_{\alpha\tau}^{\eta\theta}\partial R_{\rho\sigma}^{\mu\lambda}}\right]_{\bar{g}}\left(R_{\alpha\tau}^{\eta\theta}-\bar{R}_{\alpha\tau}^{\eta\theta}\right)\left(R_{\rho\sigma}^{\mu\lambda}-\bar{R}_{\rho\sigma}^{\mu\lambda}\right)\right),\label{eq:EQCA_as_Taylor_expansion}
\end{align}
which is the equivalent quadratic curvature action (EQCA). Let us
look at the individual terms in the second line of (\ref{eq:EQCA_as_Taylor_expansion})
to reduce it to the usual quadratic gravity theory following the similar
lines we used above: First, $\alpha$, $\beta$, and $\gamma$ parameters
of the quadratic gravity can be defined as 
\begin{equation}
\frac{1}{2}\left[\frac{\partial^{2}f}{\partial R_{\alpha\tau}^{\eta\theta}\partial R_{\rho\sigma}^{\mu\lambda}}\right]_{\bar{g}}R_{\alpha\tau}^{\eta\theta}R_{\rho\sigma}^{\mu\lambda}=:\alpha R^{2}+\beta R_{\sigma}^{\lambda}R_{\lambda}^{\sigma}+\gamma\left(R_{\rho\sigma}^{\eta\lambda}R_{\eta\lambda}^{\rho\sigma}-4R_{\sigma}^{\lambda}R_{\lambda}^{\sigma}+R^{2}\right).\label{eq:a_b_g_definitions}
\end{equation}
Similar to (\ref{eq:kappa_l_defn}), the second order derivative of
$f\left(R_{\alpha\beta}^{\mu\nu}\right)$ evaluated in the background
becomes 
\begin{equation}
\left[\frac{\partial^{2}f}{\partial R_{\alpha\tau}^{\eta\theta}\partial R_{\rho\sigma}^{\mu\lambda}}\right]_{\bar{g}}=2\alpha\delta_{\eta}^{[\alpha}\delta_{\theta}^{\tau]}\delta_{\mu}^{[\rho}\delta_{\lambda}^{\sigma]}+\beta\left(\delta_{[\eta}^{\alpha}\delta_{\theta]}^{[\rho}\delta_{[\mu}^{\left|\tau\right|}\delta_{\lambda]}^{\sigma]}-\delta_{[\eta}^{\tau}\delta_{\theta]}^{[\rho}\delta_{[\mu}^{\left|\alpha\right|}\delta_{\lambda]}^{\sigma]}\right)+12\gamma\delta_{\eta}^{[\alpha}\delta_{\theta}^{\tau}\delta_{\mu}^{\rho}\delta_{\lambda}^{\sigma]},\label{eq:d2f/dRiem2_gbar}
\end{equation}
where in writing the last term we have made use of the fact that $\delta_{\nu_{1}\nu_{2}\nu_{3}\nu_{4}}^{\mu_{1}\mu_{2}\mu_{3}\mu_{4}}R_{\mu_{1}\mu_{2}}^{\nu_{1}\nu_{2}}R_{\mu_{3}\mu_{4}}^{\nu_{3}\nu_{4}}=4\chi_{\text{GB}}$
where $\chi_{\text{GB}}$ is the Gauss-Bonnet combination. The generalized
Kronecker-delta satisfies 
\begin{equation}
\delta_{\nu_{1}\nu_{2}\nu_{3}\nu_{4}}^{\mu_{1}\mu_{2}\mu_{3}\mu_{4}}=\epsilon_{abcd}\delta_{\nu_{a}}^{\mu_{1}}\delta_{\nu_{b}}^{\mu_{2}}\delta_{\nu_{c}}^{\mu_{3}}\delta_{\nu_{d}}^{\mu_{4}}=4!\delta_{\nu_{[a}}^{\mu_{1}}\delta_{\nu_{b}}^{\mu_{2}}\delta_{\nu_{c}}^{\mu_{3}}\delta_{\nu_{d]}}^{\mu_{4}}.
\end{equation}
Using (\ref{eq:kappa_l_defn}) and (\ref{eq:a_b_g_definitions}),
one arrives at \cite{Senturk,UniBI4D} 
\begin{equation}
I_{\text{EQCA}}=\int d^{n}x\,\sqrt{-g}\left(\frac{1}{\tilde{\kappa}}\left(R-2\tilde{\Lambda}_{0}\right)+\alpha R^{2}+\beta R_{\sigma}^{\lambda}R_{\lambda}^{\sigma}+\gamma\chi_{\text{GB}}\right),\label{eq:EQCA}
\end{equation}
where the parameters of the quadratic theory are

\begin{equation}
\frac{1}{\tilde{\kappa}}:=\frac{1}{\kappa_{l}}-\frac{4\Lambda}{n-2}\left(n\alpha+\beta+\gamma\frac{\left(n-2\right)\left(n-3\right)}{\left(n-1\right)}\right),\label{eq:kappa_tilde}
\end{equation}
and 
\begin{equation}
\frac{\tilde{\Lambda}_{0}}{\tilde{\kappa}}:=-\frac{1}{2}\bar{f}+\left(\frac{n\Lambda}{n-2}\right)\frac{1}{\kappa_{l}}-\frac{2\Lambda^{2}n}{\left(n-2\right)^{2}}\left(n\alpha+\beta+\gamma\frac{\left(n-2\right)\left(n-3\right)}{\left(n-1\right)}\right).\label{eq:Lambda_0_tilde}
\end{equation}
The vacua of (\ref{eq:EQCA}) satisfies \cite{Deser_Tekin-PRL,Deser_Tekin-PRD}
\begin{equation}
\frac{\Lambda-\tilde{\Lambda}_{0}}{2\tilde{\kappa}}+\left(\left(n\alpha+\beta\right)\frac{\left(n-4\right)}{\left(n-2\right)^{2}}+\gamma\frac{\left(n-3\right)\left(n-4\right)}{\left(n-1\right)\left(n-2\right)}\right)\Lambda^{2}=0,\label{eq:Vacua_of_EQCA}
\end{equation}
which is the same vacuum equation as that of the $f\left(R_{\alpha\beta}^{\mu\nu}\right)$
theory and its equivalent linearized version (\ref{eq:ELA}). Note
that this is a highly nontrivial result: One should replace (\ref{eq:kappa_tilde})
and (\ref{eq:Lambda_0_tilde}) in (\ref{eq:Vacua_of_EQCA}) to get
(\ref{eq:Vacua_of_f(Riem)}). In the Appendix, we give a direct demonstration
of the equivalence between the linearized field equations of the $f\left(R_{\alpha\beta}^{\mu\nu}\right)$
theory and the EQCA.

In summary, to find the conserved charges of the $f\left(R_{\alpha\beta}^{\mu\nu}\right)$
theory around any one of its maximally symmetric vacua given by (\ref{eq:Vacua_of_f(Riem)}),
one can find the equivalent quadratic curvature action (\ref{eq:EQCA})
whose parameters follow from the expressions (\ref{eq:kappa_l_defn}),
(\ref{eq:a_b_g_definitions}), (\ref{eq:kappa_tilde}), and (\ref{eq:Lambda_0_tilde}),
then find the effective Newton's constant of the theory using (\ref{eq:kappa_eff_for_conserved_charges})
to plug in to the final conserved charge expression (\ref{eq:Q_quad}).
A generalization involving the derivatives of the Riemann tensor in
the action was given in \cite{Azeyanagi}. It would be interesting
to rederive the results of \cite{Azeyanagi} using the methods of
this section.

\subsection{Example 1: Charges of Born-Infeld Extension of New Massive Gravity
(BINMG)}

The following theory was introduced in \cite{Gullu-BINMG} as an extension
of the new massive gravity (NMG) \cite{NMG}; 
\begin{equation}
I_{\text{BINMG}}:=\int d^{3}x\,\sqrt{-g}f\left(R_{\nu}^{\mu}\right)=-4m^{2}\int d^{3}x\,\left[\sqrt{-\det\left(g_{\mu\nu}+\frac{\sigma}{m^{2}}G_{\mu\nu}\right)}-\left(1-\frac{\lambda_{0}}{2}\right)\sqrt{-g}\right],\label{eq:det_BINMG_action}
\end{equation}
where $G_{\mu\nu}:=R_{\mu\nu}-\frac{1}{2}g_{\mu\nu}R$, $\lambda_{0}:=\Lambda_{0}/m^{2}$,
and $\sigma=\pm1$. Just like NMG which is a quadratic theory, this
Born-Infeld type gravity describes a massive spin-2 graviton in the
vacuum of the theory. Surprisingly, unlike NMG which has generically
two vacua, this theory has a single vacuum which we shall show below
using the preceding construction. The theory also has all the rotating
and non-rotating types of the BTZ black hole \cite{BTZ}. Let us apply
our prescription that we laid above to construct the conserved mass
and angular momentum of the BTZ black hole. Since in three dimensions
the Riemann and Ricci tensors are double duals of each other $\left(\frac{1}{4}\epsilon_{\rho\mu\nu}\epsilon_{\sigma\alpha\beta}R^{\mu\nu\alpha\beta}=G_{\rho\sigma}\right)$
they carry the same curvature information; so it is better to use
the Ricci tensor and hence, the relevant quantities for the construction
of charges and the vacuum are the following 
\begin{align}
\bar{f} & =4m^{2}\left[\left(1-\frac{\lambda_{0}}{2}\right)-\left(1-\sigma\lambda\right)^{3/2}\right],\label{eq:f_bar_BINMG}\\
\left[\frac{\partial f}{\partial R_{\beta}^{\alpha}}\right]_{\bar{g}}R_{\beta}^{\alpha} & =\sigma\sqrt{1-\sigma\lambda}R,\label{eq:df/dRic_bar_BINMG}\\
\frac{1}{2}\left[\frac{\partial^{2}f}{\partial R_{\sigma}^{\rho}\partial R_{\beta}^{\alpha}}\right]_{\bar{g}}R_{\sigma}^{\rho}R_{\beta}^{\alpha} & =\frac{1}{m^{2}\sqrt{1-\sigma\lambda}}\left(R_{\nu}^{\mu}R_{\mu}^{\nu}-\frac{3}{8}R^{2}\right),
\end{align}
where $\lambda:=\Lambda/m^{2}$ and $\sigma\lambda>1$ should be satisfied.
Then, the parameters of the equivalent quadratic curvature action
read as 
\begin{align}
\frac{1}{\tilde{\kappa}} & =\frac{\sigma-\frac{\lambda}{2}}{\sqrt{1-\sigma\lambda}},\\
\frac{\tilde{\Lambda}_{0}}{\kappa} & =m^{2}\left[\lambda_{0}-2+\frac{1}{\sqrt{1-\sigma\lambda}}\left(2-\sigma\lambda-\frac{\lambda^{2}}{4}\right)\right],\\
\beta=-\frac{8}{3}\alpha & =\frac{1}{m^{2}\sqrt{1-\sigma\lambda}},
\end{align}
where $\lambda$ is uniquely fixed as\cite{Nam-Extended,Gullu-cfunc}
\begin{equation}
\lambda=\sigma\lambda_{0}\left(1-\frac{\lambda_{0}}{4}\right),\qquad\lambda_{0}<2,\label{eq:BTZ_BINMG}
\end{equation}
which was obtained after using (\ref{eq:f_bar_BINMG}) and (\ref{eq:df/dRic_bar_BINMG})
in (\ref{eq:Vacua_of_f(Riem)}). Under the condition (\ref{eq:BTZ_BINMG})
and $m^{2}\lambda=-1/\ell^{2}<0$, the rotating BTZ black hole is
given with the metric 
\begin{equation}
{\rm d}s^{2}=-N^{2}{\rm d}t^{2}+N^{-2}{\rm d}r^{2}+r^{2}\left(N^{\phi}{\rm d}t+{\rm d}\phi\right)^{2},\label{eq:BTZ_metric}
\end{equation}
where 
\begin{equation}
N^{2}\left(r\right)=-M+\frac{r^{2}}{\ell^{2}}+\frac{j^{2}}{4r^{2}},\qquad N^{\phi}\left(r\right)=-\frac{j}{2r^{2}}.
\end{equation}
In cosmological Einstein's theory, the $M=0$ and $j=0$ solution
corresponds to the vacuum, and the $M=-1$ and $j=0$ solution corresponds
to the global AdS which is a bound state. For BINMG, as we argued
above these Einsteinian charges will be rescaled by the $\kappa_{{\rm eff}}$
(\ref{eq:kappa_eff_for_conserved_charges}) which reads 
\begin{equation}
\frac{1}{\kappa_{\text{eff}}}=\sigma\sqrt{1-\sigma\lambda}.
\end{equation}
Then, from (\ref{eq:Q_quad}), the mass and the angular momentum of
the BTZ black hole in BINMG can be found as 
\begin{equation}
E=\sigma\sqrt{1-\sigma\lambda}M,\qquad J=\sigma\sqrt{1-\sigma\lambda}j.\label{eq:E-L}
\end{equation}
Note that this result matches with the ones computed via the first
law of black hole thermodynamics in \cite{Nam-Extended} and note
also that $Ej=MJ$ as in Einstein's theory in 3D.

The above formalism also gives us the central charge as defined by
Wald \cite{Wald} 
\begin{equation}
c=\frac{\ell}{2G_{3}}\delta_{\nu}^{\mu}\left[\frac{\partial f}{\partial R_{\nu}^{\mu}}\right]_{\bar{g}}=\frac{3\ell}{2G_{3}\kappa_{l}},
\end{equation}
where we will set $G_{3}=1$ as above. For Einstein's theory $c=3\ell/2$
and for BINMG with $1/\kappa_{l}=\sigma\left(1-\sigma\lambda\right)^{1/2}$,
one has 
\begin{equation}
c=\frac{3\ell}{2}\sigma\left(1-\sigma\lambda\right)^{1/2},
\end{equation}
and using (\ref{eq:BTZ_BINMG}), $c$ further reduces to 
\begin{equation}
c=\frac{3\sigma}{4\ell}\left(2-\lambda\right).
\end{equation}

\section{Conserved Charges of Topologically Massive Gravity}

In the above examples, we studied explicitly diffeomorphism invariant
theories, in this section we will study the celebrated topologically
massive gravity (TMG) \cite{DJT-PRL,DJT} in three dimensions whose
action is diffeomorphism invariant only up to a boundary term. The
conserved charges of this theory for flat backgrounds was constructed
in \cite{DJT} and its extension for AdS backgrounds was given in
\cite{DT-TMG} following the AD construction laid out above. The theory
also admits non-Einsteinian nonsingular solutions and the conserved
charges for those cases was given in \cite{Bouchareb} by the same
construction and in \cite{Nazaroglu} using the covariant canonical
formalism which uses the symplectic structure. Here, for the sake
of diversity, we will follow \cite{Nazaroglu} and construct the symplectic
two form (which has potential applications in the quantization of
the theory) to find the conserved charges. At the end, we calculate
the conserved charges of some solutions of the theory which are not
asymptotically AdS. The approach we followed here was used in \cite{Alkac}
for the NMG case.

It is well-known that with the help of a symplectic two-form on the
phase space, one can give a covariant description of the phase space
without defining the canonical coordinates. Here, we follow \cite{Witten-Symp}.

The action for TMG is given by \cite{DJT-PRL} 
\begin{equation}
I=\int_{\mathcal{M}}d^{3}x\left(\sqrt{-g}\left(R-2\Lambda\right)+\frac{1}{2\mu}\epsilon^{\alpha\beta\gamma}\Gamma_{\alpha\nu}^{\mu}\left(\partial_{\beta}\Gamma_{\gamma\mu}^{\nu}+\frac{2}{3}\Gamma_{\beta\rho}^{\nu}\Gamma_{\gamma\mu}^{\rho}\right)\right),\label{lg}
\end{equation}
where $\epsilon^{\alpha\beta\gamma}$ is the totally antisymmetric
tensor density of weight as $\sqrt{-g}$. Variation of the action
with respect to an arbitrary deformation of the metric yields 
\begin{equation}
\delta I_{\mathcal{M}}=\int_{\mathcal{M}}d^{3}x\sqrt{-g}\delta g^{\mu\nu}\left(G_{\mu\nu}+\frac{1}{\mu}C_{\mu\nu}\right),
\end{equation}
where the cosmological Einstein tensor, the Cotton tensor, and the
Schouten tensor, respectively, are 
\begin{align}
G_{\mu\nu} & =R_{\mu\nu}-\frac{1}{2}g_{\mu\nu}R+\Lambda g_{\mu\nu},\label{eq:Cosm_Eins_ten}\\
C^{\mu\nu} & :=\frac{\epsilon^{\mu\beta\gamma}}{\sqrt{-g}}\nabla_{\beta}S_{\ \gamma}^{\nu},\qquad\qquad S_{\mu\nu}:=R_{\mu\nu}-\frac{1}{4}g_{\mu\nu}R.
\end{align}
The variation of the action produces the following boundary term:
\begin{equation}
\delta I_{\partial\mathcal{M}}=\int_{\mathcal{M}}d^{3}x\partial_{\alpha}\left(\sqrt{-g}\left(g^{\mu\nu}\delta\Gamma_{\phantom{\alpha}\mu\nu}^{\alpha}-g^{\alpha\mu}\delta\Gamma_{\phantom{\nu}\mu\nu}^{\nu}\right)-\frac{1}{\mu}\epsilon^{\alpha\nu\sigma}\left(S_{\phantom{\rho}\sigma}^{\rho}\delta g_{\nu\rho}+\frac{1}{2}\Gamma_{\nu\beta}^{\rho}\delta\Gamma_{\phantom{\beta}\sigma\rho}^{\beta}\right)\right).\label{eq:Boundary_term}
\end{equation}
Therefore, the source-free field equations are 
\begin{equation}
G_{\mu\nu}+\frac{1}{\mu}C_{\mu\nu}=0,
\end{equation}
which necessarily yield $R=6\Lambda$. It pays to define the two boundary
terms (as one-form densities on the phase space) coming from the Einstein
and the Chern-Simons parts as 
\begin{eqnarray}
\Lambda_{{\rm EH}}^{\alpha} & := & \sqrt{-g}\left(g^{\mu\nu}\delta\Gamma_{\phantom{\alpha}\mu\nu}^{\alpha}-g^{\alpha\mu}\delta\Gamma_{\phantom{\nu}\mu\nu}^{\nu}\right),\\
\Lambda_{{\rm CS}}^{\alpha} & := & -\frac{1}{\mu}\epsilon^{\alpha\nu\sigma}\left(S_{\phantom{\rho}\sigma}^{\rho}\delta g_{\nu\rho}+\frac{1}{2}\Gamma_{\nu\beta}^{\rho}\delta\Gamma_{\phantom{\beta}\sigma\rho}^{\beta}\right).
\end{eqnarray}
From these two pieces, the \emph{symplectic current} follows as 
\begin{equation}
J^{\alpha}:=-\frac{1}{\sqrt{-g}}\delta\left(\Lambda_{EH}^{\alpha}+\Lambda_{CS}^{\alpha}\right),
\end{equation}
separately, one has the following local symplectic parts 
\begin{equation}
J_{{\rm EH}}^{\alpha}=-\frac{\delta\Lambda_{EH}^{\alpha}}{\sqrt{-g}}=\delta\Gamma_{\phantom{\alpha}\mu\nu}^{\alpha}\wedge\left(\delta g^{\mu\nu}+\frac{1}{2}g^{\mu\nu}\delta\ln g\right)-\delta\Gamma_{\phantom{\nu}\mu\nu}^{\nu}\wedge\left(\delta g^{\alpha\mu}+\frac{1}{2}g^{\alpha\mu}\delta\ln g\right)
\end{equation}
and 
\begin{equation}
J_{{\rm CS}}^{\alpha}=-\frac{\delta\Lambda_{CS}^{\alpha}}{\sqrt{g}}=\frac{1}{\mu}\frac{\epsilon^{\alpha\nu\sigma}}{\sqrt{-g}}\left(\delta S_{\phantom{\rho}\sigma}^{\rho}\wedge\delta g_{\nu\rho}+\frac{1}{2}\delta\Gamma_{\phantom{\rho}\nu\beta}^{\rho}\wedge\delta\Gamma_{\phantom{\beta}\sigma\rho}^{\beta}\right).
\end{equation}
Finally, the sought-after symplectic two-form on the phase space of
TMG is given as an integral over a two-dimensional hypersurface $\Sigma$
as $\omega:=\int_{\Sigma}d\Sigma_{\alpha}J^{\alpha}$ which explicitly
reads 
\begin{equation}
\begin{aligned}\omega=\int_{\Sigma}d\Sigma_{\alpha}\Bigg[ & \delta\Gamma_{\phantom{\alpha}\mu\nu}^{\alpha}\wedge\left(\delta g^{\mu\nu}+\frac{1}{2}g^{\mu\nu}\delta\ln g\right)-\delta\Gamma_{\phantom{\nu}\mu\nu}^{\nu}\wedge\left(\delta g^{\alpha\mu}+\frac{1}{2}g^{\alpha\mu}\delta\ln g\right)\\
 & +\frac{1}{\mu}\frac{\epsilon^{\alpha\nu\sigma}}{\sqrt{-g}}\left(\delta S_{\phantom{\rho}\sigma}^{\rho}\wedge\delta g_{\nu\rho}+\frac{1}{2}\delta\Gamma_{\phantom{\rho}\nu\beta}^{\rho}\wedge\delta\Gamma_{\phantom{\beta}\sigma\rho}^{\beta}\right)\Bigg].
\end{aligned}
\label{eq:Symplectic_two}
\end{equation}
Of course, this ``formal'' symplectic two-form has to satisfy the
following requirements for it to be a viable symplectic structure 
\begin{enumerate}
\item It must be a closed two-form, that is, $\delta\omega=0$, without
the use of field equations or their variations thereof. 
\item $J^{\alpha}$ must be covariantly conserved, that is, $\nabla_{\alpha}J^{\alpha}=0$,
modulo field equations and their variations, $\delta G_{\mu\nu}+\frac{1}{\mu}\delta C_{\mu\nu}=0$. 
\item $\omega$ must be diffeomorphism invariant both in the full solution
space and in the more relevant quotient space of solutions modulo
the diffeomorphism group. 
\end{enumerate}
Some of these requirements are easily seen to be satisfied by $\omega$
given in (\ref{eq:Symplectic_two}), but some require rather lengthy
computations. Let us go over them briefly. It is easy to see that
the $\delta\omega=0$ for any smooth metric and its deformation. Hence,
item 1 is clearly satisfied. Let us now compute the onshell conservation
of $J^{\alpha}$ as follows: Let us define the covariant divergence
of the current as 
\begin{equation}
\nabla_{\alpha}J^{\alpha}:=I_{1}+I_{2}+\frac{1}{\mu}I_{3},
\end{equation}
where 
\begin{align}
I_{1} & :=\frac{1}{2}\nabla_{\alpha}\left(g^{\mu\nu}\delta\Gamma_{\phantom{\alpha}\mu\nu}^{\alpha}\wedge\delta\ln g-g^{\alpha\mu}\delta\Gamma_{\phantom{\alpha}\mu\nu}^{\nu}\wedge\delta\ln g\right),\\
I_{2} & :=\nabla_{\alpha}\left(\delta\Gamma_{\phantom{\alpha}\mu\nu}^{\alpha}\wedge\delta g^{\mu\nu}-\delta\Gamma_{\phantom{\nu}\mu\nu}^{\nu}\wedge\delta g^{\alpha\mu}\right),\\
I_{3} & :=\nabla_{\alpha}\left(\frac{\epsilon^{\alpha\nu\sigma}}{\sqrt{-g}}\left(\delta S_{\phantom{\rho}\sigma}^{\rho}\wedge\delta g_{\nu\rho}+\frac{1}{2}\delta\Gamma_{\phantom{\rho}\nu\beta}^{\rho}\wedge\delta\Gamma_{\phantom{\beta}\sigma\rho}^{\beta}\right)\right).
\end{align}
To be able to use the variation of the field equations, we can recast
$I_{1}$ and $I_{2}$ as follows 
\begin{eqnarray}
I_{1} & = & \frac{1}{2}g^{\mu\nu}\delta R_{\mu\nu}\wedge\delta\ln g+g^{\mu\nu}\delta\Gamma_{\phantom{\alpha}\mu\nu}^{\alpha}\wedge\delta\Gamma_{\phantom{\lambda}\alpha\lambda}^{\lambda},\\
I_{2} & = & \delta R_{\mu\nu}\wedge\delta g^{\mu\nu}-g^{\mu\nu}\delta\Gamma_{\phantom{\alpha}\mu\nu}^{\alpha}\wedge\delta\Gamma_{\phantom{\lambda}\alpha\lambda}^{\lambda},
\end{eqnarray}
where we made use of the Palatini identity, $\delta R_{\mu\nu}=\nabla_{\alpha}\delta\Gamma_{\phantom{\alpha}\mu\nu}^{\alpha}-\nabla_{\mu}\delta\Gamma_{\phantom{\alpha}\nu\alpha}^{\alpha}$
and the explicit form of $\delta\Gamma_{\phantom{\alpha}\mu\nu}^{\alpha}$.
Furthermore, using the variation of the field equation, $I_{1}$ and
$I_{2}$ combine to yield 
\begin{equation}
I_{1}+I_{2}=\frac{\epsilon^{\mu\beta\gamma}}{\mu\sqrt{-g}}\left(S_{\phantom{\sigma}\gamma}^{\sigma}\delta\Gamma_{\phantom{\nu}\beta\sigma}^{\nu}+\nabla_{\beta}\delta S_{\phantom{\nu}\gamma}^{\nu}\right)\wedge\delta g_{\mu\nu},
\end{equation}
where we used the variation of the Cotton tensor; 
\begin{equation}
\delta C^{\mu\nu}=\frac{\epsilon^{\mu\beta\gamma}}{\sqrt{-g}}\left(-\frac{1}{2}\nabla_{\beta}S_{\phantom{\nu}\gamma}^{\nu}\delta\ln g+\nabla_{\beta}\delta S_{\phantom{\nu}\gamma}^{\nu}+S_{\phantom{\sigma}\gamma}^{\sigma}\delta\Gamma_{\phantom{\nu}\beta\sigma}^{\nu}\right).
\end{equation}
Working on $I_{3}$, one has 
\begin{equation}
I_{3}=-\frac{\epsilon^{\mu\beta\gamma}}{\sqrt{-g}}\left(\nabla_{\beta}\delta S_{\phantom{\nu}\gamma}^{\nu}\wedge\delta g_{\mu\nu}+g_{\lambda\mu}\delta S_{\phantom{\nu}\gamma}^{\nu}\wedge\delta\Gamma_{\phantom{\lambda}\beta\nu}^{\lambda}+\delta\Gamma_{\phantom{\nu}\mu\sigma}^{\nu}\wedge\nabla_{\beta}\delta\Gamma_{\phantom{\sigma}\gamma\nu}^{\sigma}\right).
\end{equation}
Finally, combining all these, one arrives at 
\begin{equation}
\nabla_{\alpha}J^{\alpha}=\frac{\epsilon^{\mu\beta\gamma}}{\mu\sqrt{-g}}\delta\Gamma_{\phantom{\nu}\beta\sigma}^{\nu}\wedge\left(\delta\left(g_{\mu\nu}S_{\phantom{\sigma}\gamma}^{\sigma}\right)+\nabla_{\mu}\delta\Gamma_{\phantom{\sigma}\gamma\nu}^{\sigma}\right).\label{conserved}
\end{equation}
To show that the right-hand side vanishes as desired, we still need
to use further three-dimensional identity on the Riemann tensor and
its variation as 
\begin{eqnarray}
\epsilon^{\mu\beta\gamma}R_{\phantom{\sigma}\mu\gamma\nu}^{\sigma} & = & \epsilon^{\mu\beta\gamma}\left(\delta_{\gamma}^{\sigma}S_{\mu\nu}+S_{\phantom{\sigma}\gamma}^{\sigma}g_{\mu\nu}\right),\\
\epsilon^{\mu\beta\gamma}\delta R_{\phantom{\sigma}\mu\gamma\nu}^{\sigma} & = & \epsilon^{\mu\beta\gamma}\delta_{\gamma}^{\sigma}\delta S_{\mu\nu}+\epsilon^{\mu\beta\gamma}\delta\left(S_{\phantom{\sigma}\gamma}^{\sigma}g_{\mu\nu}\right).
\end{eqnarray}
With the help of these, one finally arrives at $\nabla_{\alpha}J^{\alpha}=0$
on shell, fulfilling the second requirement.

The first part of the third item is already satisfied because $\omega$
is constructed from tensors. But, for the second part, namely showing
that $\omega$ is diffeomorphism invariant in the quotient space of
the solutions modulo diffeomorphisms, we need to do further work.
The crux of the computation is to show that for pure gauge directions
$\omega$ vanishes. For this purpose, let us decompose the variation
of the metric into nongauge and pure gauge parts as 
\begin{equation}
\delta g_{\mu\nu}^{\prime}:=\delta g_{\mu\nu}+\nabla_{\mu}\xi_{\nu}+\nabla_{\nu}\xi_{\mu}\label{diffeo}
\end{equation}
where $\xi$ is a one-form on the cotangent space of the phase space.
This decomposition acts as the Lie derivative on the associated tensors
as $\mathcal{L}_{\xi}T$ with respect to the vector $\xi$, and hence
one has 
\begin{equation}
\delta\Gamma_{\phantom{\prime\lambda}\mu\nu}^{\prime\lambda}=\delta\Gamma_{\phantom{\lambda}\mu\nu}^{\lambda}+\nabla_{\mu}\nabla_{\nu}\xi^{\lambda}+R_{\nu\phantom{\lambda}\mu\beta}^{\phantom{\nu}\lambda}\xi^{\beta},
\end{equation}
\begin{equation}
\delta S_{\phantom{\prime\mu}\nu}^{\prime\mu}=\delta S_{\phantom{\mu}\nu}^{\mu}+\xi^{\beta}\nabla_{\beta}S_{\phantom{\mu}\nu}^{\mu}+S_{\phantom{\mu}\beta}^{\mu}\nabla_{\nu}\xi^{\beta}-S_{\nu\beta}\nabla^{\beta}\xi^{\mu}.
\end{equation}
The change in the symplectic current of the Einstein-Hilbert part,
that is $\Delta J_{{\rm EH}}^{\alpha}:=J_{{\rm EH}}^{\prime\alpha}-J_{{\rm EH}}^{\alpha}$,
can be computed as \cite{Witten-Symp} 
\begin{equation}
\Delta J_{{\rm EH}}^{\alpha}=\nabla_{\mu}\mathcal{F}_{{\rm EH}}^{\mu\alpha}+G^{\mu\alpha}\left(\xi_{\mu}\wedge\delta\ln g+2\xi^{\nu}\wedge\delta g_{\mu\nu}\right)-G^{\mu\nu}\xi^{\alpha}\wedge\delta g_{\mu\nu}+2\xi_{\mu}\wedge\delta G^{\alpha\mu},\label{gauge_einstein_in_G}
\end{equation}
where the $\mathcal{F}_{{\rm EH}}^{\mu\alpha}$ is an antisymmetric
tensor defined as: 
\begin{equation}
\mathcal{F}_{{\rm EH}}^{\mu\alpha}:=\nabla^{\mu}\delta g^{\nu\alpha}\wedge\xi_{\nu}+\delta g^{\nu\alpha}\wedge\nabla_{\nu}\xi^{\mu}+\frac{1}{2}\delta\ln g\wedge\nabla^{\alpha}\xi^{\mu}+\nabla_{\nu}\delta g^{\mu\nu}\wedge\xi^{\alpha}+\nabla^{\mu}\delta\ln g\wedge\xi^{\alpha}-\left(\alpha\leftrightarrow\mu\right).
\end{equation}
At this stage, if we consider the pure Einstein-Hilbert theory alone,
then the last four {terms of (\ref{gauge_einstein_in_G})} vanish
on shell $G_{\mu\nu}=0$ and $\delta G_{\mu\nu}=0$. On the other
hand, $\nabla_{\mu}\mathcal{F}_{{\rm EH}}^{\mu\alpha}$ is a boundary
term and vanishes for sufficiently decaying metric variations yielding
$\Delta J_{{\rm EH}}^{\alpha}=0$ and corresponding symplectic two-form
$\omega_{{\rm EH}}$ is diffeomorphism invariant on the quotient space
of classical solutions (Einstein spaces) to the diffeomorphism group.

For the full Chern-Simons theory, the computation is slightly longer:
One finds the change in the Chern-Simons part of the symplectic current
under the splitting (\ref{diffeo}) as 
\begin{equation}
\begin{aligned}\mu\Delta J_{{\rm CS}}^{\alpha}=\frac{\epsilon^{\alpha\nu\sigma}}{\sqrt{-g}}\Bigg[ & \left(-S_{\beta\sigma}\nabla^{\beta}\xi^{\rho}+S_{\phantom{\beta}\beta}^{\rho}\nabla_{\sigma}\xi^{\beta}+\nabla_{\beta}S_{\phantom{\rho}\sigma}^{\rho}\xi^{\beta}\right)\wedge\delta g_{\nu\rho}\\
 & +\delta S_{\phantom{\rho}\sigma}^{\rho}\wedge\left(\nabla_{\rho}\xi_{\nu}+\nabla_{\nu}\xi_{\rho}\right)+\left(\nabla_{\nu}\nabla_{\beta}\xi^{\rho}+R_{\beta\phantom{\rho\ }\nu\gamma}^{\phantom{\beta}\rho}\xi^{\gamma}\right)\wedge\delta\Gamma_{\phantom{\beta}\sigma\rho}^{\beta}\Bigg].
\end{aligned}
\label{eq:Delta_J_CS}
\end{equation}
Using 
\begin{equation}
\nabla_{\beta}\delta S_{\phantom{\beta}\sigma}^{\beta}=\frac{1}{4}\nabla_{\sigma}\delta R+\delta\Gamma_{\phantom{\lambda}\beta\sigma}^{\lambda}S_{\phantom{\beta}\lambda}^{\beta}-\delta\Gamma_{\phantom{\lambda}\beta\lambda}^{\lambda}S_{\phantom{\beta}\sigma}^{\beta},
\end{equation}
and the three-dimensional relation 
\begin{equation}
\epsilon^{\mu\alpha\beta}\xi^{\nu}=g^{\mu\nu}\epsilon^{\rho\alpha\beta}\xi_{\rho}+g^{\alpha\nu}\epsilon^{\mu\rho\beta}\xi_{\rho}+g^{\beta\nu}\epsilon^{\mu\alpha\rho}\xi_{\rho},
\end{equation}
one can recast $\Delta J_{{\rm CS}}^{\alpha}$ as 
\begin{equation}
\mu\Delta J_{CS}^{\alpha}=\nabla_{\mu}\mathcal{F}_{{\rm CS}}^{\mu\alpha}+C^{\mu\alpha}\left(\xi_{\mu}\wedge\delta\ln g+2\xi^{\nu}\wedge\delta g_{\mu\nu}\right)-C^{\mu\nu}\xi^{\alpha}\wedge\delta g_{\mu\nu}+2\xi_{\mu}\wedge\delta C^{\alpha\mu},\label{gauge_chern}
\end{equation}
where the antisymmetric tensor $\mathcal{F}_{CS}^{\mu\alpha}$ is
defined as 
\begin{equation}
\mathcal{F}_{CS}^{\mu\alpha}:=\frac{\epsilon^{\alpha\mu\sigma}}{\sqrt{-g}}\left(-\delta\Gamma_{\phantom{\beta}\sigma\rho}^{\beta}\wedge\nabla_{\beta}\xi^{\rho}+2\delta S_{\phantom{\nu}\sigma}^{\nu}\wedge\xi_{\nu}+S_{\phantom{\rho}\gamma}^{\rho}\delta g_{\sigma\rho}\wedge\xi^{\gamma}+S_{\phantom{\beta}\sigma}^{\beta}\delta g_{\beta\rho}\wedge\xi^{\rho}\right).\label{boundary_chern}
\end{equation}
It is now clear that combining the Chern-Simons part with the Einstein's
part and using the field equations of TMG and their variations, one
arrives at the result that $\Delta J^{\alpha}=0$ for sufficiently
fast decaying metric variations. This says that $\omega$ has no components
in the pure gauge directions for sufficiently fast decaying metric
variations.

Let us now use the above construction to find the conserved charges
of the theory corresponding to the Killing symmetries of a given background.
For this purpose, we choose the following diffeomorphisms which corresponds
to the isometries of the background 
\begin{equation}
\nabla_{\mu}\xi_{\nu}+\nabla_{\nu}\xi_{\mu}=0.
\end{equation}
In the language we used so far, we are setting the nongauge invariant
part of the $\delta g_{\mu\nu}=0$, and hence, by definition $\Delta J^{\alpha}=0$
yielding 
\begin{equation}
\nabla_{\mu}\left(\mathcal{F}_{{\rm EH}}^{\mu\alpha}+\frac{1}{\mu}\mathcal{F}_{{\rm CS}}^{\mu\alpha}\right)=0,
\end{equation}
on shell yielding the local conservation 
\begin{equation}
\partial_{\mu}\left[\sqrt{-g}\left(\mathcal{F}_{{\rm EH}}^{\mu\alpha}+\frac{1}{\mu}\mathcal{F}_{{\rm CS}}^{\mu\alpha}\right)\right]=0,
\end{equation}
from which we can define a globally conserved total charge for each
$\xi^{\mu}$. Identifying $\delta g_{\mu\nu}\rightarrow h_{\mu\nu}$,
where $h_{\mu\nu}$ is a perturbation around a given background $\bar{g}$
with Killing symmetries, and keeping the $\xi^{\mu}$ terms on the
same side of the wedge products before dropping them yields the final
result as 
\begin{equation}
Q\left(\bar{\xi}\right)=\frac{1}{2\pi G_{3}}\int_{\partial\Sigma}dS_{\alpha}\left(\mathcal{F}_{{\rm EH}}^{\alpha0}+\frac{1}{\mu}\mathcal{F}_{{\rm CS}}^{\alpha0}\right),\label{eq:Charge}
\end{equation}
which more explicitly reads 
\begin{equation}
\begin{aligned}Q\left(\bar{\xi}\right)= & \frac{1}{2\pi G_{3}}\int_{\Sigma}d^{2}x\sqrt{-\bar{g}}\bar{\xi}_{\nu}{\cal G}_{\left(1\right)}^{0\nu}\\
 & +\frac{1}{2\pi G_{3}\mu}\int_{\partial\Sigma}dS_{\alpha}\epsilon^{0\alpha\sigma}\left(-\delta\Gamma_{\phantom{\beta}\sigma\rho}^{\beta}\bar{\nabla}_{\beta}\bar{\xi}^{\rho}+2\delta S_{\phantom{\nu}\sigma}^{\nu}\bar{\xi}_{\nu}+S_{\phantom{\rho}\gamma}^{\rho}h_{\sigma\rho}\bar{\xi}^{\gamma}+S_{\phantom{\beta}\sigma}^{\beta}h_{\beta\rho}\bar{\xi}^{\rho}\right)\Bigg],
\end{aligned}
\label{eq:Charge_explicit}
\end{equation}
where we have left the first line as an integral over the hypersurface
$\Sigma$, but as we discussed in Sec.~\ref{subsec:AD-and-ADM} ,
it can be written as a surface term as (\ref{eq:Conserved_charges})
or (\ref{eq:ADT_charge}). It is important to note that for a generic
background, the Einsteinian part and the Chern-Simons part are not
separately gauge-invariant as we have seen above. But, for AdS backgrounds,
since the linearized cosmological Einstein tensor ${\cal G}_{\mu\nu}^{\left(1\right)}$
and the linearized Cotton tensor are separately gauge-invariant, we
can recast (\ref{eq:Charge_explicit}) in an explicitly gauge-invariant
form as 
\begin{equation}
\begin{aligned}Q\left(\bar{\xi}\right)=Q_{\text{Einstein}}\left(\bar{\Xi}\right) & +\frac{1}{\mu}\int_{\partial\Sigma}dS_{i}\left(\epsilon^{0i\beta}\mathcal{G}_{\nu\beta}^{\left(1\right)}\bar{\xi}^{\nu}+\epsilon^{\nu i\beta}\mathcal{G}_{\phantom{\left(1\right)0}\beta}^{\left(1\right)0}\bar{\xi}_{\nu}+\epsilon^{0\nu\beta}\mathcal{G}_{\phantom{\left(1\right)i}\beta}^{\left(1\right)i}\bar{\xi}_{\nu}\right)\end{aligned}
,\label{eq:Charge_DT_TMG}
\end{equation}
where $\bar{\Xi}^{\mu}:=\bar{\xi}^{\mu}+\frac{1}{\mu}\frac{\epsilon^{\mu\alpha\beta}}{\sqrt{-\bar{g}}}\bar{\nabla}_{\alpha}\bar{\xi}_{\beta}$
is also a Killing vector field. Note that for asymptotically AdS spaces,
the second term on the right-hand side vanishes; therefore, the effect
of the Chern-Simons part is represented solely in the twist part of
the $\bar{\Xi}^{\mu}$ vector as far as conserved charges are concerned.
For generic spacetimes which are not asymptotically AdS, $\bar{\Xi}^{\mu}$
fails to be a Killing vector field.

An immediate application of the above construction is to the BTZ black
hole which is a solution to TMG for any value of $\mu$ since it is
a locally AdS\textsubscript{3} spacetime. Recall that the rotating
BTZ black hole metric is given as 
\begin{equation}
ds^{2}=-\left(-G_{3}M+\frac{r^{2}}{\ell^{2}}+\frac{j^{2}G_{3}^{2}}{4r^{2}}\right){\rm d}t^{2}+\left(-G_{3}M+\frac{r^{2}}{\ell^{2}}+\frac{j^{2}G_{3}^{2}}{4r^{2}}\right)^{-1}{\rm d}r^{2}+r^{2}\left(-\frac{jG_{3}}{2r^{2}}{\rm d}t+{\rm d}\phi\right)^{2}.\label{eq:BTZ_bh}
\end{equation}
Taking the background to be as before with $M=0$ and $j=0$, one
finds by using (\ref{eq:Charge_explicit}) or equivalently (\ref{eq:Charge_DT_TMG})
that the BTZ metric (\ref{eq:BTZ_bh}) receives nontrivial corrections
to its conserved energy and angular momentum from the Cotton part
\cite{Kanik,Olmez}: 
\begin{equation}
E=M-\frac{j}{\mu\ell^{2}},\qquad\qquad J=j-\frac{M}{\mu}.
\end{equation}
For $M=j\mu$ and $\mu\ell=1$, $E=0$ and $J=0$, namely the black
hole is degenerate with the vacuum. This particular limit was studied
in \cite{ChiralGrav} where it was shown that there is a single boundary
conformal field theory with a right moving Virasoro algebra with the
central charge $c_{R}=3\ell/G_{3}$ and the energy of all bulk excitations
vanish. This theory is called the chiral gravity.

Let us give some further examples which are solutions to full TMG
equations but do not solve the cosmological Einstein theory.

\subsubsection{Logarithmic solution of TMG at the chiral point}

At the chiral point $\mu\ell=1$, the metric 
\begin{equation}
ds^{2}=-N{\rm d}t^{2}+\frac{{\rm d}r^{2}}{N}+r^{2}\left(N_{\theta}{\rm d}t-{\rm d}\theta\right)^{2}+N_{k}\left({\rm d}t-\ell{\rm d}\theta\right)^{2},
\end{equation}
with 
\begin{equation}
N=-G_{3}m+\frac{r^{2}}{\ell^{2}}+\frac{m^{2}\ell^{2}G_{3}^{2}}{4r^{2}},\qquad N_{\theta}=\frac{m\ell G_{3}}{2r^{2}},\qquad N_{k}=kG_{3}\ln\left(\frac{2r^{2}-m\ell^{2}G_{3}}{2r_{0}^{2}}\right),
\end{equation}
was shown to solve TMG \cite{Giribet}. $m=k=0$ defines the background
metric. For the Killing vector $\xi^{\mu}=\left(-1,0,0\right)$, one
obtains the energy 
\begin{equation}
E=4k,
\end{equation}
and for the Killing vector $\xi^{\mu}=\left(0,0,1\right)$, the angular
momentum becomes 
\begin{equation}
J=4k\ell.
\end{equation}
These charges are found in \cite{Giribet} with the counter-term approach,
in \cite{MiskovicOlea} with the first order formalism, and in \cite{Cvetkovic}
with Nester's definition of conserved charges \cite{Nester-Charge}.

\subsubsection{Null warped AdS$_{3}$}

At the tuned value of $\mu\ell=-3$, the metric (called the null warped
AdS\textsubscript{3}) 
\begin{equation}
\frac{ds^{2}}{\ell^{2}}=-2r{\rm d}t{\rm d}\theta+\frac{d{\rm r^{2}}}{4r^{2}}+\left(r^{2}+r+k\right){\rm d}\theta^{2},
\end{equation}
solves TMG whose detailed description can be found in \cite{Anninos}.
We can compute its conserved charges by taking $k=0$ as the background
metric. The result yields 
\begin{equation}
E=0,\qquad J=-\frac{8k\ell}{3},
\end{equation}
which are the same as the ones given in \cite{Bouchareb,Anninos}.

\subsubsection{Spacelike stretched black holes}

By the works of Nutku \cite{Nutku} and Gurses \cite{Gurses}, we
know that for arbitrary $\mu$ the following metric solves TMG 
\begin{equation}
ds^{2}=-N{\rm d}t^{2}+\ell^{2}R\left({\rm d}\theta+N^{\theta}{\rm d}t\right)^{2}+\frac{\ell^{4}{\rm d}r^{2}}{4RN},
\end{equation}
where the functions are 
\begin{eqnarray}
R & := & \frac{r}{4}\left(3\left(\nu^{2}-1\right)r+\left(\nu^{2}+3\right)\left(r_{+}+r_{-}\right)-4\nu\sqrt{r_{+}r_{-}\left(\nu^{2}+3\right)}\right),\\
N & := & \frac{\ell^{2}\left(\nu^{2}+3\right)\left(r-r_{+}\right)\left(r-r_{-}\right)}{4R},\qquad N^{\theta}:=\frac{2\nu r-\sqrt{r_{+}r_{-}\left(\nu^{2}+3\right)}}{2R},
\end{eqnarray}
with $\nu=-\frac{\mu\ell}{3}$. The solution describes a spacelike
stretched black hole for $\nu^{2}>1$ with $r_{\pm}$ as outer and
inner horizons. The details of this metric such as conserved charges
were studied in \cite{Bouchareb,MiskovicOlea,Cvetkovic,Anninos}.
Defining the background with $r_{\pm}=0$ and using (\ref{eq:Charge_explicit}),
for the Killing vector $\xi^{\mu}=\left(-1,0,0\right)$, the energy
can be calculated as 
\begin{equation}
E=\frac{\left(3+\nu^{2}\right)\ell}{3\nu}\left(\nu\left(r_{+}+r_{-}\right)-\sqrt{\left(3+\nu^{2}\right)r_{+}r_{-}}\right),
\end{equation}
which is the same as the result given in \cite{Bouchareb,Cvetkovic,Anninos}.
For the Killing vector $\xi^{\mu}=\left(0,0,1\right)$, the angular
momentum can be computed to be

\begin{multline}
J=\frac{\ell}{24\nu}\Biggl(2\left(10\nu^{4}-15\nu^{2}+9\right)\left(r_{+}^{2}+r_{-}^{2}\right)+18\left(\nu^{2}-1\right)\left(\nu^{2}-2\right)r_{+}r_{-}\\
+\nu\left(5\nu^{2}-9\right)\left(r_{+}+r_{-}\right)\sqrt{\left(3+\nu^{2}\right)r_{+}r_{-}}\Biggr),
\end{multline}
which differs from the one, $\mathcal{J}$, given in \cite{Bouchareb,Cvetkovic,Anninos},
but they are related as $\mathcal{J}=c_{1}\left(\nu\right)J+c_{2}\left(\nu\right)E$
where $c_{1}$ and $c_{2}$ were given in \cite{Nazaroglu}. The crucial
point here is that both $J$ and $E$ are finite in full TMG even
though they are divergent separately in Einstein's theory and pure
Cotton theory.

\section{Charges in Scalar-Tensor Gravities }

Let us consider a generic scalar-tensor modification of Einstein's
theory given by the action 
\begin{equation}
I=\frac{1}{2\kappa}\int d^{n}x\sqrt{-g}\,U\left(\phi\right)\left(R\left(g\right)-2\Lambda_{0}-W\left(\phi\right)\partial_{\mu}\phi\partial^{\mu}\phi-V\left(\phi\right)+H\left(\phi\right){\cal L}_{\mbox{matter}}\left(\psi\right)\right),\label{stringframe}
\end{equation}
where ${\cal L}_{\mbox{matter}}\left(\psi\right)$ represents all
the matter besides the scalar. One can add higher order curvature
terms, but for the sake of simplicity we will consider the above theory
and study the conformal properties of the charges we defined above.
With the following conformal transformation to the so-called Einstein
frame 
\begin{equation}
g_{\mu\nu}^{E}:=U\left(\phi\right)^{\frac{2}{n}}g_{\mu\nu},\label{scaling}
\end{equation}
the action can be recast as 
\begin{equation}
I=\frac{1}{2\kappa}\int d^{n}x\sqrt{-g^{E}}\left(R\left(g^{E}\right)-2\Lambda_{0}\right)+I_{M},\label{einsteinframe}
\end{equation}
where $I_{M}$ now has the form 
\begin{equation}
I_{M}=\frac{1}{2\kappa}\int d^{n}x\sqrt{-g^{E}}\left(A\left(\phi\right)\partial_{\mu}\phi\partial^{\mu}\phi+X\left(\phi\right)+Z\left(\phi\right){\cal L}_{\mbox{matter}}\left(\psi\right)\right).
\end{equation}
Now, the conserved charge of (\ref{einsteinframe}) is represented
by (\ref{eq:Conserved_charges}) or equivalently (\ref{eq:ADT_charge}).
Under the conformal transformation (\ref{scaling}), the antisymmetric
tensor $\mathcal{F}^{\nu\mu}$ defining the conserved charge (\ref{eq:ADT_charge})
transforms as 
\begin{align}
\sqrt{-\bar{g}}\mathcal{F}^{\nu\mu}\left(\bar{\xi}\right)=U^{-\frac{2}{n}}\sqrt{-\bar{g}^{E}}\Biggl( & \mathcal{F}^{\nu\mu}\left(\bar{\xi}^{E}\right)-\frac{3}{n}\bar{\xi}_{\alpha}^{E}h_{E}^{\nu\alpha}\partial^{\mu}\log U+\frac{3}{n}\xi_{\alpha}^{E}h_{E}^{\mu\alpha}\partial^{\nu}\log U\nonumber \\
 & -\frac{n-1}{n}\xi_{E}^{\nu}h_{E}^{\mu\alpha}\partial_{\alpha}\log U+\frac{n-1}{n}\xi_{E}^{\mu}h_{E}^{\nu\alpha}\partial_{\alpha}\log U\Biggr),\label{confchargetransform}
\end{align}
where we have assumed that the conformal transformation did not eliminate
the Killing vectors. Therefore, the conserved charges are conformally
invariant if $U\left(\infty\right)=1$; namely, $g_{\mu\nu}$ and
$g_{\mu\nu}^{E}$ have the same charges. If on the other hand $U\left(\infty\right)$
is some arbitrary constant, then the charges of these two metrics
differ by a multiplicative constant which in any case can be attributed
to the normalization of the Killing vector. More details can be found
in \cite{DeserTekin-Conformal} where $F\left(R\right)$ type theories
and quadratic theories were also discussed. Note that the effect of
conformal transformations on the surface gravity and the temperature
of stationary black holes were discussed before in \cite{JacobsonKang}
with a similar conclusion that they are invariant given that the conformal
transformation approaches to unity at infinity. For more generic matter
fields, see an extended discussion of conserved charges in \cite{Murata}.

\section{Conserved Charges in the First Order Formulation}

It is well-known that when fermions are introduced to gravity, the
metric formulation is not sufficient and one has to introduce the
vierbein and spin connection. As this is the case in the supergravity
theories, we will introduce the construction of conserved charges
in the first order formalism in this section. This will also be a
useful background material for the rest of this review where we shall
define the three-dimensional Chern-Simons like theories in this formalism.
The following discussion was given in \cite{Cebeci} which we closely
follow.

What we shall describe can be generalized to any geometric theory
of gravity (such as the higher order ones) but for the sake of simplicity
let us consider the cosmological Einstein's theory and reproduce the
first order form of the Abbott-Deser charges in the form given by
Deser and Tekin. Of course, the major difference between the first
order and the send order (metric formulation) of gravity theories
arises for generic theories while they yield the same result in Einstein's
gravity. Nevertheless it pays to study the most relevant case here.
We shall introduce the notation as we go along. The $n$-dimensional
field equations read

\begin{equation}
G_{a}+\Lambda\,\star\,e_{a}=\kappa\,T_{a}\,,\label{eineq}
\end{equation}
where $G_{a}$ is $\left(n-1\right)$-form Einstein tensor, and $\star$
is the Hodge star operator, $e_{a}$ is a 1-form. It is clear that
equation (\ref{eineq}) comes from the variation of an action $I=\int\mathcal{L}$
where $\mathcal{L}$ is an $n$-form as expected. The metric tensor
can be written as 
\[
g=\eta_{ab}\,e^{a}\otimes e^{b},
\]
where the 1-forms $e^{a}$ are the orthonormal coframe fields and
in general they do not exist globally. Let $\bar{e}^{a}$ the background
orthonormal coframe which satisfies (\ref{eineq}) for $T_{a}=0$.
Then, the full coframe 1-form can be expanded as 
\begin{equation}
e^{a}:=\bar{e}^{a}+\varphi^{a}\,_{b}\,\bar{e}^{b}\,=\left(\delta_{b}^{a}+\varphi^{a}\,_{b}\right)\,\bar{e}^{b},\label{cofram}
\end{equation}
where $\varphi^{a}\,_{b}$ are the 0-forms with proper decay condition.
It is clear that we can always write (\ref{cofram}) as it is; but,
let us show that this is possible. We have in local coordinates 
\[
e^{a}=\bar{e}^{a}+\psi^{a}\,_{\mu}\,dx^{\mu},
\]
but since $dx^{\mu}=\bar{E}^{\mu}\,_{b}\,\bar{e}^{b}$, we have $\varphi^{a}\,_{b}=\psi^{a}\,_{\mu}\,\bar{E}^{\mu}\,_{b}$.
The splitting (\ref{cofram}) when inserted in (\ref{eineq}) leads
to 
\[
G_{a}^{\left(1\right)}(\varphi^{b}\,_{c})=\kappa\,\tau_{a}\,,
\]
where again $G_{a}^{\left(1\right)}(\varphi^{b}\,_{c})$ is a $(n-1)$-form
differential operator which is linear in the deviation parts $\varphi^{b}\,_{c}$
and $\tau_{a}$ includes all the higher order terms in $\varphi^{b}\,_{c}$
as well as the compactly supported matter part $T_{a}$. To get a
conserved charge expression we need extra structure such as symmetries
and local conserved currents. Let $\bar{D}$ denote the covariant
derivative with respect to the Levi-Civita 1-forms $\bar{\omega}^{a}\,_{b}$
which satisfy the Cartan structure equation 
\[
\bar{D}\,\bar{e}^{a}=0=d\bar{e}^{a}+\bar{\omega}^{a}\,_{b}\wedge\bar{e}^{b}.
\]
Following the same reasoning as before, we assume that the background
spacetime has some symmetries denoted by the Killing vectors $\bar{\xi}_{a}$
ant the Killing equation in this language reads 
\begin{equation}
\bar{D}_{a}\,\bar{\xi}_{b}\,^{(I)}+\bar{D}_{b}\,\bar{\xi}_{a}\,^{(I)}=0\,,\label{kill}
\end{equation}
where we used the definition $\bar{D}_{a}:=\bar{\iota}_{a}\,\bar{D}$
and $\bar{\iota}_{a}$ is the interior-product with respect to the
background frame vector: For example, $\bar{\iota}_{b}\,\bar{e}^{a}=\delta_{b}\,^{a}$.
From the full Bianchi identity one obtains 
\[
\bar{D}G_{\left(1\right)}^{a}=0,
\]
and to convert it to a partial current conservation we employ the
Killing vectors to get 
\[
\bar{D}\,(\bar{\xi}_{a}\,G_{\left(1\right)}^{a})=d(\bar{\xi}_{a}\,G_{\left(1\right)}^{a})=0.
\]
Furthermore, we can define $G_{\left(1\right)}^{a}:=G^{ab}\,\bar{\star}\,\bar{e}_{b}$
where $G^{ab}$ are 0-forms. Leaving the details to \cite{Cebeci},
let us write the final expression for the conserved charges in terms
of $\varphi^{a}\,_{b}$ for the cosmological Einstein's theory 
\begin{eqnarray}
Q(\bar{\xi}) & = & \frac{1}{4\,\Omega_{n-2}\,G_{n}}\,\int_{\partial\bar{\Sigma}}\,dS_{i}\,\left(-\bar{\xi}^{0}\,\bar{D}^{b}\,\varphi^{i}\,_{b}+\varphi^{bi}\,\bar{D}_{b}\,\bar{\xi}^{0}-\varphi^{b}\,_{b}\,\bar{D}^{i}\,\bar{\xi}^{0}+\bar{\xi}^{0}\,\bar{D}^{i}\,\varphi^{b}\,_{b}\right.\nonumber \\
 &  & \left.\qquad\qquad\qquad-\bar{\xi}^{i}\,\bar{D}^{0}\,\varphi^{b}\,_{b}+\bar{\xi}^{i}\,\bar{D}^{b}\,\varphi^{0}\,_{b}-\bar{\xi}_{b}\,\bar{D}^{i}\,\varphi^{0b}+\varphi^{0b}\,\bar{D}^{i}\,\bar{\xi}_{b}+\bar{\xi}_{b}\,\bar{D}^{0}\,\varphi^{ib}\right)\,.\label{char}
\end{eqnarray}
This is of the same form as the Abbott-Deser or Deser-Tekin expression
for cosmological Einstein's gravity given in the metric formulation.
But, one should bear in mind that the equality of the first order
and the second order (metric) formulation is valid only for a small
class of theories, such as the Einstein's gravity. For generic gravity
theories, first and second order formulations yield completely different
theories. If fermions are to be coupled to gravity, it is clear that
the first order formulation must be used. In that case, the above
procedure is more apt for the construction of charges.

It is also known that in the first order formulation, if the vierbein
is allowed to be non-invertible, some gravity theories can be mapped
to gauge theories and quantized exactly. As two examples see \cite{Witten-3DCS}
where the 3 dimensional Einstein's gravity (with or without a cosmological
constant) was mapped to a non-compact Chern-Simons theory and \cite{Horne-Witten}
where 3 dimensional conformal gravity was mapped to a non-compact
Chern-Simons theory.

\section{Vanishing Conserved Charges and Linearization Instability}

The astute reader might have realized that in the above construction
of global conserved charges for Einstein's theory or for generic gravity
theories, two crucial ingredients are the Stokes' theorem and the
existence of asymptotic rigid symmetries (or Killing vectors). Once
Stokes' theorem is invoked, one necessarily resorts to perturbative
methods: namely, a background spacetime $\left(\mathcal{M},\bar{g}\right)$
is assigned zero charges and subsequently the conserved charges of
a perturbed spacetime $\left(\mathcal{M},g\right)$ that has the same
topology as the background and with the metric $g=\bar{g}+h$ are
measured with respect to the background charges. Clearly, if $\mathcal{M}$
does not have a spatial boundary; namely, if its topology is of the
form $\mathbb{R}\times\Sigma$ where $\Sigma$ is a closed Riemannian
manifold, one is forced to conclude that it must have zero charges
for all metrics (solutions of the theory). This is simple to understand
as there is no boundary to integrate over the charge densities. This
leads to an interesting conundrum: bulk and ``boundary'' expressions
of the charges may fail to give the same results. But, of course,
this is not possible and the resolution of the paradox comes from
an apparently unexpected analysis which was worked out in 1970s \cite{Brill_Deser,Marsden_Fischer,Moncrief,Marsden_Arms,Marsden,Marsden_lectures}
and came to be well-understood by the beginning of 1980s for Einstein's
gravity with compact Cauchy surfaces without a boundary. Here we shall
explain this issue without going into too much detail as the subject
requires another review of its own.

The idea is the following: in certain nonlinear theories, such as
Einstein's gravity, perturbation theory can fail for certain backgrounds.
More concretely, if the background spacetime has a compact Cauchy
surface with at least one Killing vector field, then the linearized
equations of the nonlinear theory have more (spurious) solutions than
the ones that can actually be obtained from the linearization of exact
solutions. Namely, $h$, the perturbed metric obtained as a solution
to linearized Einstein equations cannot be obtained from the linearization
of an exact metric $g$. Such a phenomenon is called \textit{linearization
instability } and it is completely different from a \textit{dynamical
instability } in the sense that the former refers to the failure of
the perturbative techniques about a special exact solution while the
latter refers to an actual instability of a given solution. Linearization
instability issue is a rather long and beautiful subject which deserves
a separate discussion. See the recent work by Altas and Tekin \cite{Altas}
for references and the situation for generic gravity theories in which
novel forms of this phenomenon arise even for non-compact initial
data surface.

Here, we would just like to allude to the subject and briefly explain
the important issue of nonvacuum solutions of a theory having exactly
vanishing charges. Apparently, something like that would mean that
the vacuum is infinitely degenerate but this is a red-herring. Let
us give a simple example: Consider the Maxwell electrodynamics with
charges and currents in not $\mathbb{R}^{1,3}$ but on $\mathbb{R}\times S^{3}$
where we have compactified the space. Since there is no spatial boundary
all fluxes vanish and the total electric charge must be zero: but
there can be dipoles etc. This is the global picture, on the other
hand, locally Maxwell equations admit solutions with apparently $E\simeq q/r^{2}$
type electric fields which have non-zero charges. So, clearly such
local solutions do not satisfy the global integral constraint on the
vanishing of the total electric charge. A similar, but due to the
nonlinearities of the theory, more complicated example was found in
Einstein's gravity by Brill and Deser \cite{Brill_Deser} who showed
that on $\mathbb{R}\times T^{3}$, there are quadratic constraints
on the perturbations $h$ that solve the linearized Einstein equation.
The meaning of the constraints when carefully studied is that $\mathbb{R}\times T^{3}$
is an \emph{isolated} solution of Einstein's equations which does
not admit any perturbation whatsoever. A perhaps better understanding
of the linearization instability issue was achieved with the help
of the following vantage point mainly put forward by Fischer, Marsden,
Moncrief, Arms \cite{Marsden_Fischer,Moncrief,Marsden_Arms,Marsden,Marsden_lectures}
and some others.

Let $\mathcal{E}$ be the \emph{set }of solutions of Einstein's equations,
when does this set form a \emph{manifold}? It turns out this set has
some conical structure, but in general, save these conical singularities,
it is an infinite dimensional manifold. The conical singularity arises
exactly at metrics $g$ that have compact Cauchy surfaces and Killing
fields. This is a necessary and sufficient condition. Since the set
of metrics that have symmetries on a given manifold is set of measure
zero, linearization stability is a generic property of Einstein's
equations \cite{SaraykarRai}.

To understand this issue more rigorously from a well-defined mathematical
point of view, one can split Einstein's equations into the 3+1 form
and study the constraints on a Cauchy surface $\Sigma$ and evolution
equations off the surface. Such a splitting reduces the problem into
an analysis of the solution set of the constraints (as opposed to
the full Einstein equations). The problem then reduces to a problem
in elliptic operator theory and can be stated as follows: given a
solution $\bar{\gamma}$ and $\bar{K}$ to the constraints (where
$\bar{\gamma}$ is the induced metric of the Cauchy surface and $\bar{K}$
is the extrinsic curvature), is this solution isolated or is there
an open subset of solutions around this solution? Then, the question
reduces further to the linearization of the constraint equations around
$\bar{\gamma}$ and $\bar{K}$, and eventually boils down to checking
the surjectivity of the linearized constraint operators. Surjectivity
is required to show that the tangent space around the given solution
point has the correct dimensionality which means the solution set
being a manifold around that point. The problem is somewhat complicated
due to the gauge issues, but we now have a complete understanding
of how and if perturbation theory can fail in Einstein's gravity.
For the case of Einstein's gravity without matter, we refer the reader
to \cite{Choquet-Bruhat} and with matter to \cite{Girbau}.

The origin of the linearization instability in Einstein's gravity
is compactness of the Cauchy surface and the Killing symmetries of
it as noted above. For example, Minkowski spacetime with its noncompact
Cauchy surface is linearization stable \cite{Deser_Choquet-Bruhat}.
On the other hand, for generic gravity theories linearization instability
can take place even for spacetimes that have noncompact Cauchy surfaces.
This is due to the fact that, as we have seen in the charge construction
each rank two tensor added to the Einstein tensor in the field equations
bring a contribution to the conserved charges in an additive manner;
and hence, at certain parameter values of the theory, all the charges
at the boundary of the spatial hypersurface vanish while their bulk
version do not for nonvacuum solutions. This is a subtle point and
requires a little bit of computation which we reproduce here following
\cite{Altas}. Let $\mathcal{E}_{\mu\nu}\left[g\right]=0$ is a our
generic covariant field equations which has the property $\nabla^{\mu}\mathcal{E}_{\mu\nu}=0$.
Let $\bar{g}$ solve this equation and constitute the background about
which we shall carry out perturbation theory. Defining 
\begin{equation}
g_{\mu\nu}=\bar{g}_{\mu\nu}+\lambda h_{\mu\nu}+\frac{\lambda^{2}}{2}k_{\mu\nu},
\end{equation}
where $\lambda$ is a small dimensionless parameter. Expanding the
field equation to second order in $\lambda$, one arrives at 
\begin{equation}
\bar{\mathcal{E}}_{\mu\nu}\left[\bar{g}\right]+\lambda\mathcal{E}_{\mu\nu}^{\left(1\right)}\left[h\right]+\lambda^{2}\left(\mathcal{E}_{\mu\nu}^{\left(2\right)}\left[h\right]+\mathcal{E}_{\mu\nu}^{\left(1\right)}\left[k\right]\right)+O\left(\lambda^{3}\right)=0.\label{eq:lambda2_expansion_of_EoM}
\end{equation}
By definition, the first term is zero while the second one is set
to zero to determine $h$. At order $O\left(\lambda^{2}\right)$,
given $h$, if one can find a $k$, then the expansion is consistent.
On the other hand, if such a $k$ does not exist, then one has an
inconsistency. In this formulation, it is difficult to show that there
is or there is no $k$ for every solution of the linearized field
equations. Therefore, we can find a global constraint on $h$ without
referring to $k$ as follows. Given a Killing vector field $\bar{\xi}^{\mu}$
of the background $\bar{g}$, we can contract $O\left(\lambda^{2}\right)$
of (\ref{eq:lambda2_expansion_of_EoM}) to get 
\begin{equation}
\bar{\xi}^{\nu}\mathcal{E}_{\mu\nu}^{\left(2\right)}\left[h\right]+\bar{\xi}^{\nu}\mathcal{E}_{\mu\nu}^{\left(1\right)}\left[k\right]=0,
\end{equation}
and integrate over $\bar{\Sigma}$ to obtain 
\begin{equation}
\int_{\bar{\Sigma}}d^{n-1}y\,\sqrt{\bar{\gamma}^{\bar{\Sigma}}}\bar{n}_{\mu}\bar{\xi}_{\nu}\left(\mathcal{E}_{\left(2\right)}^{\mu\nu}\left[h\right]+\mathcal{E}_{\left(1\right)}^{\mu\nu}\left[k\right]\right)=0.\label{eq:Integrated_lambda2_EoM}
\end{equation}
The first term of this equation represents the so called \textit{Taub
charge } defined as 
\begin{equation}
Q_{{\rm Taub}}\left[\bar{\text{\ensuremath{\xi}}},h\right]:=\int_{\bar{\Sigma}}d^{n-1}y\,\sqrt{\bar{\gamma}^{\bar{\Sigma}}}\bar{n}_{\mu}\bar{\xi}_{\nu}\mathcal{E}_{\left(2\right)}^{\mu\nu}\left[h\right].
\end{equation}
The second term $\bar{\xi}^{\nu}\mathcal{E}_{\mu\nu}^{\left(1\right)}\left[k\right]$
can be written as 
\begin{equation}
\bar{\text{\ensuremath{\xi}}}^{\mu}\mathcal{E}_{\mu\nu}^{\left(1\right)}\left[k\right]=c\left(\alpha_{i},\Lambda\right)\bar{\nabla}^{\mu}\mathcal{F}_{\mu\nu}^{E}\left[\bar{\text{\ensuremath{\xi}}},k\right]+\bar{\nabla}^{\mu}\mathcal{F}_{\mu\nu}^{{\rm Mod}}\left[\bar{\text{\ensuremath{\xi}}},k\right],
\end{equation}
where $c\left(\alpha_{i},\Lambda\right)$ is a constant that is function
of the theory parameters $\alpha_{i}$ and the effective cosmological
constant $\Lambda$, the $\mathcal{F}_{\mu\nu}^{E}$ tensor is an
antisymmetric tensor having the Einsteinian form 
\begin{equation}
\mathcal{F}_{\mu\nu}^{E}\left[\bar{\text{\ensuremath{\xi}}},k\right]:=\bar{\xi}_{\nu}\bar{\nabla}_{\beta}K^{\mu\alpha\nu\beta}-K^{\mu\beta\nu\alpha}\bar{\nabla}_{\beta}\bar{\xi}_{\nu},
\end{equation}
and $\mathcal{F}_{\mu\nu}^{{\rm Mod}}$ is also an antisymmetric tensor
determined by the higher derivative terms. For asymptotically AdS
spacetimes, $\mathcal{F}_{\mu\nu}^{{\rm Mod}}$ vanishes identically
at the boundary, so $\mathcal{F}_{\mu\nu}^{E}$ is the only surviving
piece for the second term in (\ref{eq:Integrated_lambda2_EoM}) for
not so fast decaying $k$. Then, (\ref{eq:Integrated_lambda2_EoM})
takes the form 
\begin{equation}
Q_{{\rm Taub}}\left[\bar{\text{\ensuremath{\xi}}},h\right]+c\left(\alpha_{i},\Lambda\right)\int_{\partial\bar{\Sigma}}d^{n-2}z\,\sqrt{\bar{\gamma}^{\left(\partial\bar{\Sigma}\right)}}\bar{\epsilon}_{\mu\nu}\left(\bar{\xi}_{\alpha}\bar{\nabla}_{\beta}K^{\mu\nu\alpha\beta}-K^{\mu\beta\alpha\nu}\bar{\nabla}_{\beta}\bar{\xi}_{\alpha}\right)=0.\label{eq:Integrated_lambda2_EoM_final_form}
\end{equation}
In general, this equation determines the second order perturbation
$k$ from the predetermined first order perturbation $h$ which is
supposed to be found from the first order equation $\mathcal{E}_{\mu\nu}^{\left(1\right)}\left[h\right]=0$.
However, there are cosmological higher derivative theories for which
the prefactor $c\left(\alpha_{i},\Lambda\right)$ vanishes identically
for certain choices of the parameters. This means that these theories
have vanishing conserved charges for all of their solutions. For these
theories, which are commonly dubbed as critical theories, (\ref{eq:Integrated_lambda2_EoM_final_form})
reduces to 
\begin{equation}
Q_{{\rm Taub}}\left[\bar{\text{\ensuremath{\xi}}},h\right]=0,\label{eq:Constraint_on_h}
\end{equation}
which is nothing but a second order constraint on the already determined
$h$. For the generic case, it is very hard to satisfy this second
order constraint for all $h$, and this would be an indication of
the failure of the perturbative scheme about the given AdS background,
that is the theory has linearization instability about its AdS solution.
It is important to emphasize that we have not assumed the compactness
of the Cauchy surfaces; hence, for the case of critical theories,
the linearization instability may arise even for noncompact Cauchy
surfaces.

Two explicit examples of this type of linearization instability for
quadratic gravity in generic dimensions and chiral gravity in three
dimensions are given in \cite{Altas}.

\part{CONSERVED CHARGES FOR EXTENDED 3D GRAVITIES: QUASI-LOCAL APPROACH}

\section{Introduction}

As we have noted already in Part I, to get fully gauge covariant or
coordinate independent expressions for conserved quantities, one has
to carry out integrations over the boundary of spacetime. Even this
was a subtle task, as we have shown that large gauge transformations
can change the value of the integrals. Given this fact, one still
would like to find some meaningful integrals that are not taken at
infinity but at a finite distance from, say, a black hole. Such an
approach would be aptly called quasi-local and it produced very interesting
results. For example, Wald showed that diffeomorphism invariance of
the theory leads to the entropy of bifurcate horizons \cite{Wald},
and this result matches with that of the Bekenstein-Hawking area formula
\cite{99} for Einstein's gravity and some other theories. Therefore,
even though at the moment, no satisfactory quasilocal formulation
of mass of a spacetime exists, one should not underestimate the usefulness
of the quasi-local approach especially in the context of black hole
thermodynamics. We can also mention solution phase space method introduced
in \cite{Hajian}, as a method which relaxes Wald integrations from
the asymptotic or horizon, in parallel with the quasi-local method
in the ADT context. In what follows, we shall review the quasi-local
and off-shell extension of the ADT method and mostly apply these to
the 2+1 dimensional gravity theories that have received a great deal
of interest in the recent literature.

A nice detailed account of quasi-local approaches were given in the
review \cite{95}. For example, one such approach is that of Brown
and York \cite{96} and this formalism is extended by introducing
counter terms for the case of asymptotically AdS spacetimes \cite{97}.
Also, higher derivative gravity examples of such a method can be found
in \cite{Nojiri,Cvetic}. An earlier quasi-local approach is that
of {Komar \cite{101a}}. Covariant quasi-local energy as Noether
charge associated with some time-like Killing vector in diffeomorphism-invariant
actions has been developed by Iyer and Wald \cite{9}. The key idea
is that one can always construct a Noether current which is conserved
when the field equations hold. Based on Barnich-Brandt-Henneaux uniqueness
\cite{Barnich:1995ap}, one can deduce that this mass is unique and
can be defined on any codimension 2 surface that encloses the sources
\cite{Barnich:2003xg}. In the covariant phase space method which
is based on the symplectic structure of the underlying gravity theory,
the conserved charges are calculated using the Noether potential \cite{9,33,34,Wald}.
As noted above, via Wald's formulation, black hole entropy is a conserved
charge of diffeomorphism invariance given that the Lagrangian is diffeomorphism
invariant \emph{exactly} not up to a boundary term.

When the Chern-Simons term is present in the action, the theory becomes
diffeomorphism invariant only up to a boundary term, and hence, Wald's
formulation needs to be modified. This was done by Tachikawa \cite{14}
and the result is that the black hole entropy is still a conserved
charge with modifications coming from the Chern-Simons term. There
is an interesting connection between the on-shell ADT density and
the linearized Noether potential. Indeed, it is observed that, at
the asymptotic boundary, when the linearized potential around the
background (which is a solution for the equations of motion) is combined
with the surface term, the result is the ADT charge density \cite{40,Barnich:2003xg,102,103}.
Although this connection is very interesting, it has been indirect
and has been shown to exist only in Einstein's gravity. Recently,
a non-trivial generalization of the mentioned relation was presented
in covariant and non-covariant theories of gravity \cite{Kim,32,36}.
This generalization was achieved by promoting the on-shell ADT charge
density to the off-shell level. By expressing the linearized Noether
potential and combining it with the surface term, the relation can
be directly understood. By integrating from the ADT charge density
along a single-parameter path in the solution space, one obtains an
expression for the quasi-local conserved charge which is identical
with the result coming from the covariant phase space method. This
result confirms that the extended off-shell ADT formalism is equivalent
to the covariant phase space method at the on-shell level.

As our applications will be mostly in the 2+1 dimensional gravity
theories with the Chern-Simons term, it turns out the first order
formulation in terms of the dreibein and the spin connection instead
of the metric is more convenient. Therefore, we first start introducing
the basic elements of this formalism which can also be found in many
gravity textbooks.

\section{Chern-Simons-like theories of gravity}

\label{S1.0} There is a class of gravitational theories in (2+1)-dimensions
that are naturally expressed in terms of first order formalism. Some
examples of such theories are topological massive gravity (TMG) \cite{DJT},
new massive gravity (NMG) \cite{2}, minimal massive gravity (MMG)
\cite{3}, zwei-dreibein gravity (ZDG) \cite{4}, generalized minimal
massive gravity (GMMG) \cite{5}, etc. We shall refer to all such
theories as Chern-Simons-like theories of gravity \cite{11}.

\subsection{First order formalism of gravity theories}

\label{S1.1} Given a manifold $\mathcal{M}$ with a metric $g$.
One can work with the coordinate adapted basis for the tangent space
\cite{6,7}. $\left\{ \partial/\partial x^{\mu}\right\} $ and the
cotangent space $\left\{ dx^{\mu}\right\} $ for which one has 
\begin{equation}
g_{\mu\nu}:=g\left(\frac{\partial}{\partial x^{\mu}},\frac{\partial}{\partial x^{\nu}}\right).
\end{equation}
But, instead of that, one can also work in a non-coordinate basis
$e_{a}$ for the tangent space and $e^{a}$ for the cotangent space
in which the metric components are constant and one has 
\begin{equation}
\eta_{ab}=g\left(e_{a},e_{b}\right).
\end{equation}
The relation between these two basis vectors at each point of spacetime
reads 
\begin{equation}
e_{a}=e_{a}^{\phantom{a}\mu}\frac{\partial}{\partial x^{\mu}},\label{eq:Inverse_vielbein}
\end{equation}
where $e_{a}^{\phantom{a}\mu}$ is an $n\times n$ matrix which is
called the inverse vielbein. The inverse of (\ref{eq:Inverse_vielbein})
defines the vielbein 
\begin{equation}
\frac{\partial}{\partial x^{\mu}}=e_{\phantom{a}\mu}^{a}e_{a}.
\end{equation}
Then, one can write in a local patch 
\begin{equation}
g_{\mu\nu}=e_{\phantom{a}\mu}^{a}e_{\phantom{b}\nu}^{b}\eta_{ab}.
\end{equation}
In this formulation, it is clear that besides the usual coordinate
transformations, one has the local Lorentz rotations which act on
the so-called flat indices which we denoted by the Latin letters.
Determinant of metric $g=\det(g_{\mu\nu})$ is related to vielbein
as 
\begin{equation}
\sqrt{-g}\varepsilon_{\mu_{1}\cdots\mu_{n}}=\varepsilon_{a_{1}\cdots a_{n}}e_{\hspace{2mm}\mu_{1}}^{a_{1}}\cdots e_{\hspace{2mm}\mu_{n}}^{a_{n}}\label{1.1}
\end{equation}
where $\varepsilon_{\mu_{1}\cdots\mu_{n}}$ and $\varepsilon_{a_{1}\cdots a_{n}}$
denote Levi-Civita symbols. We can consider vielbein $e_{\hspace{1.5mm}\mu}^{a}$
as an $n\times n$ matrix. One can deduce from \eqref{1.1} that $\det\left(e_{\hspace{1.5mm}\mu}^{a}\right)=\sqrt{-g}$
therefore $e_{\hspace{1.5mm}\mu}^{a}$ is invertible when spacetime
metric is non-singular. By assuming spacetime metric is non-singular
then we can find invert of vielbein $e_{\hspace{1.5mm}a}^{\mu}$ so
that 
\begin{equation}
e_{\hspace{1.5mm}\mu}^{a}e_{\hspace{1.5mm}b}^{\mu}=\delta_{\hspace{1.5mm}b}^{a}\hspace{0.5cm}\text{and}\hspace{0.5cm}e_{\hspace{1.5mm}a}^{\mu}e_{\hspace{1.5mm}\nu}^{a}=\delta_{\hspace{1.5mm}\nu}^{\mu}\label{1.2}
\end{equation}
where $\delta_{\hspace{1.5mm}b}^{a}$ and $\delta_{\hspace{1.5mm}\nu}^{\mu}$
denote Kronecker delta. Now consider the following transformation
from one vielbein to another 
\begin{equation}
\tilde{e}_{\hspace{1.5mm}\mu}^{a}=\Lambda_{\hspace{1.5mm}b}^{a}e_{\hspace{1.5mm}\mu}^{b},\label{1.3}
\end{equation}
which is of course a local transformation. The requirement that the
components of the spacetime metric remain intact under such a transformation
imposes the usual orthogonality condition 
\begin{equation}
\Lambda_{\hspace{1.5mm}a}^{c}\eta_{cd}\Lambda_{\hspace{1.5mm}b}^{d}=\eta_{ab},\label{1.4}
\end{equation}
which just says that $\Lambda_{\hspace{1.5mm}b}^{a}$ is an element
of Lorentz group, i.e. $\Lambda\in SO(n-1,1)$. Therefore, clearly
as an \textquotedbl internal transformation\textquotedbl{} among the
dynamical fields of the theory, we can see this a gauge transformation.
Let us now move on to defining the proper derivative operations in
this setting. Let $\nabla_{\mu}$ denote covariant derivative. If
this derivative acts on a mixed tensor that has Lorentz indices, the
resulting object will not transform properly under the Lorentz transformations
which can be remedied by introducing a spin-connection $\omega_{\hspace{1.5mm}b\mu}^{a}$
which plays the role of connection for Lorentz indices. Therefore
one can define the following (total) linear covariant derivative {$\mathcal{T}_{\nu b}^{\mu a}$}
as 
\begin{equation}
{\nabla_{\lambda}^{(T)}\mathcal{T}_{\nu b}^{\mu a}=\nabla_{\lambda}\mathcal{T}_{\nu b}^{\mu a}+\omega_{\hspace{1.5mm}c\mu}^{a}\mathcal{T}_{\nu b}^{\mu c}-\omega_{\hspace{1.5mm}b\mu}^{c}\mathcal{T}_{\nu c}^{\mu a},}\label{1.5}
\end{equation}
The spin-connection is an invariant quantity under coordinate transformations
so \eqref{1.5} is covariant under the Lorentz gauge transformations
provided that the spin-connection obeys the following transformation
for \eqref{1.3} 
\begin{equation}
\tilde{\omega}_{\hspace{1.5mm}b\mu}^{a}=\Lambda_{\hspace{1.5mm}c}^{a}\omega_{\hspace{1.5mm}d\mu}^{c}\Lambda_{b}^{\hspace{1.5mm}d}+\Lambda_{\hspace{1.5mm}c}^{a}\partial_{\mu}\Lambda_{b}^{\hspace{1.5mm}c},\label{1.6}
\end{equation}
where $\Lambda_{b}^{\hspace{1.5mm}a}=\left(\Lambda^{-1}\right)_{\hspace{1.5mm}b}^{a}$.
The metric-connection compatibility condition, $\nabla_{\lambda}^{(T)}g_{\mu\nu}=0$,
ensures that one can use $g_{\mu\nu}$ and its inverse $g^{\mu\nu}$
to lower and raise coordinate indices. Similarly, in order to use
$\eta_{ab}$ and its inverse $\eta^{ab}$ to lower and raise the Lorentz
indices we must impose $\nabla_{\lambda}^{(T)}\eta_{ab}=0$ which
is satisfied for 
\begin{equation}
\omega_{ab\mu}=-\omega_{ba\mu}.\label{1.7}
\end{equation}
The connection $\Gamma_{\mu\nu}^{\alpha}$ can be decomposed into
two parts as 
\begin{equation}
\Gamma_{\hspace{1.5mm}\mu\nu}^{\alpha}=\hat{\Gamma}_{\hspace{1.5mm}\mu\nu}^{\alpha}+C_{\hspace{1.5mm}\mu\nu}^{\alpha}\label{1.8}
\end{equation}
where $\hat{\Gamma}_{\hspace{1.5mm}\mu\nu}^{\alpha}$ is the Levi-Civita
connection and $C_{\hspace{1.5mm}\mu\nu}^{\alpha}$ is the so called
contorsion tensor defined as 
\begin{equation}
C_{\hspace{1.5mm}\mu\nu}^{\alpha}=T_{\hspace{1.5mm}\mu\nu}^{\alpha}+T_{\mu\hspace{2mm}\nu}^{\hspace{1.5mm}\alpha}+T_{\nu\hspace{2mm}\mu}^{\hspace{1.5mm}\alpha}.\label{1.9}
\end{equation}
Here $T_{\hspace{1.5mm}\mu\nu}^{\alpha}=\Gamma_{\hspace{1.5mm}[\mu\nu]}^{\alpha}$
is Cartan's torsion tensor. It immediately follows that 
\begin{equation}
C_{\hspace{1.5mm}[\mu\nu]}^{\alpha}=T_{\hspace{1.5mm}\mu\nu}^{\alpha}.\label{1.10}
\end{equation}
which we shall use. The vielbein spin-connection compatibility condition
reads 
\begin{equation}
\nabla_{\mu}^{(T)}e_{\hspace{1.5mm}\nu}^{a}=\partial_{\mu}e_{\hspace{1.5mm}\nu}^{a}-\Gamma_{\hspace{1.5mm}\mu\nu}^{\alpha}e_{\hspace{1.5mm}\alpha}^{a}+\omega_{\hspace{1.5mm}b\mu}^{a}e_{\hspace{1.5mm}\nu}^{b}=0,\label{1.11}
\end{equation}
which guarantees the orthogonality relation when the vielbein is transported
along a world line by parallel transport. The condition \eqref{1.11}determines
the spin-connection as 
\begin{equation}
\omega_{\hspace{1.5mm}b\mu}^{a}=e_{\hspace{1.5mm}\nu}^{a}\nabla_{\mu}e_{\hspace{1.5mm}b}^{\nu}.\label{1.12}
\end{equation}
One can use \eqref{1.8} to recast \eqref{1.12}: 
\begin{equation}
\omega_{\hspace{1.5mm}b\mu}^{a}=\Omega_{\hspace{1.5mm}b\mu}^{a}+C_{\hspace{1.5mm}\mu b}^{a}\label{1.13}
\end{equation}
with 
\begin{equation}
\Omega_{\hspace{1.5mm}b\mu}^{a}=e_{\hspace{1.5mm}\nu}^{a}\hat{\nabla}_{\mu}e_{\hspace{1.5mm}b}^{\nu}\label{1.14}
\end{equation}
where $\hat{\nabla}$ denotes the covariant derivative with respect
to the torsion-free connection $\hat{\Gamma}$. Therefore, $\Omega_{\hspace{1.5mm}b\mu}^{a}$
is the torsion-free spin-connection. One can formulate Einstein's
gravity using \eqref{1.12} which would amount to the second order
formalism. Instead, if the vielbein and spin-connection are taken
as two independent dynamical fields, then the obtained formalism is
called the first order formulation.

Now let us define the following 1-forms: $e^{a}=e_{\hspace{1.5mm}\mu}^{a}dx^{\mu}$,
$\omega_{\hspace{1.5mm}b}^{a}=\omega_{\hspace{1.5mm}b\mu}^{a}dx^{\mu}$,
$\Omega_{\hspace{1.5mm}b}^{a}=\Omega_{\hspace{1.5mm}b\mu}^{a}dx^{\mu}$
and $C_{\hspace{1.5mm}b}^{a}=C_{\hspace{1.5mm}\mu b}^{a}dx^{\mu}$,
so that we have $\omega_{\hspace{1.5mm}b}^{a}=\Omega_{\hspace{1.5mm}b}^{a}+C_{\hspace{1.5mm}b}^{a}$.
One can define an exterior Lorentz covariant derivative (ELCD). Let
{$\mathcal{A}_{b}^{a}$} be a Lorentz-tensor-valued $p$-form, then
one has 
\begin{equation}
{D(\omega)\mathcal{A}_{b}^{a}=d\mathcal{A}_{b}^{a}+\omega_{\hspace{1.5mm}c}^{a}\wedge\mathcal{A}_{b}^{c}-\omega_{\hspace{1.5mm}b}^{c}\wedge\mathcal{A}_{c}^{a}}\label{1.15}
\end{equation}
where $d$ denotes the exterior derivative. Since the spin-connection
1-form transforms as \eqref{1.6} under gauge transformations, the
ELCD is indeed Lorentz covariant. Curvature 2-form can be defined
as 
\begin{equation}
R_{\hspace{1.5mm}b}^{a}(\omega)=d\omega_{\hspace{1.5mm}b}^{a}+\omega_{\hspace{1.5mm}c}^{a}\wedge\omega_{\hspace{1.5mm}b}^{c}\label{1.16}
\end{equation}
and the torsion 2-form can be defined as 
\begin{equation}
T^{a}=D(\omega)e^{a}\label{1.17}
\end{equation}
which are both Lorentz and coordinate invariant quantities. One can
use \eqref{1.11} to show that \eqref{1.17} can be written as $T^{a}=e_{\hspace{1.5mm}\alpha}^{a}\Gamma_{\hspace{1.5mm}\mu\nu}^{\alpha}dx^{\mu}\wedge dx^{\nu}$.
The Bianchi identities follow as 
\begin{equation}
D(\omega)R_{\hspace{1.5mm}b}^{a}(\omega)=0,\hspace{0.7cm}D(\omega)T^{a}(\omega)=R_{\hspace{1.5mm}b}^{a}(\omega)\wedge e^{b}.\label{1.18}
\end{equation}
which will be play the role of the constraints in the theory.

Finally we can write the action of general relativity with a cosmological
constant in $n$-dimensions as a functional of the vielbein and the
spin-connection 
\begin{equation}
S[e,\omega]=\int L[e,\omega]\label{1.19}
\end{equation}
where $L[e,\omega]$ is the Lagrangian $n$-form which reads 
\begin{equation}
\begin{split}L_{\text{EC-}\Lambda}[e,\omega]=\frac{1}{16\pi G}\biggl( & -\varepsilon_{a_{1}\cdots a_{n}}e^{a_{1}}\wedge\cdots\wedge e^{a_{n-2}}\wedge R^{a_{n-1}a_{n}}\\
 & +\frac{\Lambda_{0}}{n!}\varepsilon_{a_{1}\cdots a_{n}}e^{a_{1}}\wedge\cdots\wedge e^{a_{n}}\biggr).
\end{split}
\label{1.20}
\end{equation}
As usual, $G$ and $\Lambda_{0}$ are the Newton's constant and the
cosmological constant. This action is explicitly invariant under the
Lorentz gauge transformations and the diffeomorphisms as desired.
The relation between the curvature two form $R^{ab}(\Omega)$ and
the Riemann curvature tensor $\mathcal{R}_{\alpha\beta\mu\nu}$ is
also useful to note 
\begin{equation}
R^{ab}(\Omega)=\frac{1}{2}e_{\hspace{1.5mm}\alpha}^{a}e_{\hspace{1.5mm}\beta}^{b}\mathcal{R}_{\hspace{3mm}\mu\nu}^{\alpha\beta}dx^{\mu}\wedge dx^{\nu}\label{1.21}
\end{equation}
which can be obtained by substituting \eqref{1.14} into \eqref{1.16}
with $\omega=\Omega$.

\subsection{Lorentz-Lie derivative and total variation}

\label{S1.2} Let $\pounds_{\xi}$ denote the Lie derivative along
a vector field $\xi$. Lie derivative of a differential form, say
$V$, is given by $\pounds_{\xi}V=di_{\xi}V+i_{\xi}dV$, where $i_{\xi}$
denotes interior product with $\xi$. This Lie derivative of a Lorentz
tensor-valued $p$-form is not covariant under the discussed Lorentz
gauge transformations. Therefore we need to modify it.

Let $\lambda_{\hspace{1.5mm}b}^{a}$ be the generator of Lorentz gauge
transformations $\Lambda_{\hspace{1.5mm}b}^{a}$, i.e. $\Lambda=\exp(\lambda)$.
Then we define the \textit{Lorentz-Lie derivative} (LL-derivative)
of a Lorentz tensor-valued $p$-form {$\mathcal{A}_{b}^{a}$} as
\cite{8} 
\begin{equation}
{\mathfrak{L}_{\xi}\mathcal{A}_{b}^{a}=\pounds_{\xi}\mathcal{A}_{b}^{a}+\lambda_{\xi\hspace{1mm}c}^{\hspace{1mm}a}\mathcal{A}_{b}^{c}-\lambda_{\xi\hspace{1mm}b}^{\hspace{1mm}c}\mathcal{A}_{c}^{a}}.\label{1.22}
\end{equation}
This derivative is under the Lorentz gauge transformations when $\lambda_{\xi\hspace{1mm}b}^{\hspace{1mm}a}$
transforms like a connection: 
\begin{equation}
\tilde{\lambda}_{\xi}=\Lambda\lambda_{\xi}\Lambda^{-1}+\Lambda\pounds_{\xi}\Lambda^{-1}.\label{1.23}
\end{equation}
The generator of the Lorentz gauge transformation $\lambda$ is an
arbitrary function of coordinates. One can see from \eqref{1.23}
that the change of $\lambda$, under infinitesimal Lorentz gauge transformations,
is given by $\delta\lambda=-\pounds_{\xi}\lambda$. Therefore, $\lambda$
is expected to be a function of $\xi$, i.e. $\lambda=\lambda_{\xi}(x)$.
Now, we impose a compatibility condition so that the LL-derivative
of the Minkowski metric $\eta_{ab}$ vanishes. This condition implies
that $\lambda_{\xi}$ must be anti-symmetric, that is $\lambda_{\xi}^{ab}=-\lambda_{\xi}^{ba}$.

Consider two variations $\delta_{\text{D}}$ and $\delta_{\text{L}}$
as variations due to the diffeomorphisms and the infinitesimal Lorentz
gauge transformations, respectively. $\delta_{\text{D}}$ is generated
by a vector field $\xi$ and it is equal to the Lie derivative along
$\xi$, i.e. for a Lorentz tensor-valued $p$-form {$\mathcal{A}_{b}^{a}$
we have $\delta_{\text{D}}\mathcal{A}_{b}^{a}=\pounds_{\xi}\mathcal{A}_{b}^{a}$
\cite{9,10}. Since $\mathcal{A}_{b}^{a}$ transforms as 
\begin{equation}
\tilde{\mathcal{A}}_{b}^{a}=\Lambda_{\hspace{1.5mm}c}^{a}(\Lambda^{-1})_{\hspace{1.5mm}b}^{d}\mathcal{A}_{d}^{c},\label{1.24}
\end{equation}
under the Lorentz gauge transformations, the variation of a Lorentz
tensor-valued $p$-form induced by this generator is 
\begin{equation}
{\delta_{\text{L}}\mathcal{A}_{b}^{a}=\lambda_{\xi\hspace{1mm}c}^{\hspace{1mm}a}\mathcal{A}_{b}^{c}-\lambda_{\xi\hspace{1mm}b}^{\hspace{1mm}c}\mathcal{A}_{c}^{a}}.\label{1.25}
\end{equation}
We can introduce the total variation induced by a vector field $\xi$
as $\delta_{\xi}=\delta_{\text{D}}+\delta_{\text{L}}$. Clearly, the
total variation of a Lorentz tensor-valued $p$-form is equal to its
LL-derivative : 
\begin{equation}
{\delta_{\xi}\mathcal{A}_{b}^{a}=\mathfrak{L}_{\xi}\mathcal{A}_{b}^{a}.}\label{1.26}
\end{equation}
The spin-connection is invariant under the diffeomorphisms, hence
one has $\delta_{\text{D}}\omega_{\hspace{1.5mm}b}^{a}=\mathfrak{L}_{\xi}\omega_{\hspace{1.5mm}b}^{a}$.
But since the spin-connection transforms like \eqref{1.6} under the
Lorentz gauge transformations, therefore 
\begin{equation}
\delta_{\text{L}}\omega_{\hspace{1.5mm}b}^{a}=\lambda_{\xi\hspace{1mm}c}^{\hspace{1mm}a}\omega_{\hspace{1.5mm}b}^{c}-\lambda_{\xi\hspace{1mm}b}^{\hspace{1mm}c}\omega_{\hspace{1.5mm}c}^{a}-d\lambda_{\xi\hspace{1mm}b}^{\hspace{1mm}a}.\label{1.27}
\end{equation}
Therefore its total variation is 
\begin{equation}
\delta_{\xi}\omega_{\hspace{1.5mm}b}^{a}=\mathfrak{L}_{\xi}\omega_{\hspace{1.5mm}b}^{a}-d\lambda_{\xi\hspace{1mm}b}^{\hspace{1mm}a}.\label{1.28}
\end{equation}
The extra term in \eqref{1.28} comes from the transformation of the
spin-connection under the Lorentz gauge transformations. The total
variation of $e^{a}$ and $\omega_{\hspace{1.5mm}b}^{a}$ are covariant
under the Lorentz gauge transformations as well as the diffeomorphisms.

\subsection{Gravity in three dimensions}

\label{S1.3} It is convenient to use a 3 dimensional vector algebra
notation for the Lorentz vectors in which contractions with $\eta_{ab}$
and $\varepsilon_{abc}$ are denoted by dots and crosses, respectively.
From now on, we drop the wedge product for simplicity. One can work
with the dual spin-connection and the dual curvature 2-form which
are defined as 
\begin{equation}
\omega^{a}=\frac{1}{2}\varepsilon_{\hspace{1.5mm}bc}^{a}\omega^{bc},\hspace{0.7cm}R^{a}=\frac{1}{2}\varepsilon_{\hspace{1.5mm}bc}^{a}R^{bc}.\label{1.29}
\end{equation}
In this way, \eqref{1.13} becomes 
\begin{equation}
\omega^{a}=\Omega^{a}+k^{a}\label{1.30}
\end{equation}
where 
\begin{equation}
\Omega^{a}=\frac{1}{2}\varepsilon_{\hspace{1.5mm}bc}^{a}\Omega^{bc}=\frac{1}{2}e_{\hspace{1.5mm}\alpha}^{a}\epsilon_{\hspace{1.5mm}\nu\beta}^{\alpha}e_{\hspace{1.5mm}c}^{\beta}\hat{\nabla}_{\mu}e^{c\nu}dx^{\mu}\label{1.52}
\end{equation}
and 
\begin{equation}
k^{a}=\frac{1}{2}\varepsilon_{\hspace{1.5mm}bc}^{a}C^{bc}=\frac{1}{2}e_{\hspace{1.5mm}\sigma}^{a}\epsilon^{\sigma\alpha\beta}C_{\alpha\mu\beta}dx^{\mu}\label{1.53}
\end{equation}
are the dual torsion-free spin-connection and the dual contorsion
1-forms, respectively. The dual curvature 2-form and torsion 2-form
can be written in terms of the dreibein and dual spin-connection as
\begin{equation}
R(\omega)=d\omega+\frac{1}{2}\omega\times\omega,\label{1.31}
\end{equation}
\begin{equation}
T(\omega)=D(\omega)e=de+\omega\times e.\label{1.32}
\end{equation}
One can use \eqref{1.30} to relate the torsion 2-form with the dual
contorsion 1-form in 3 dimensions: 
\begin{equation}
T(\omega)=k\times e.\label{1.54}
\end{equation}
where we used the fact that $\Omega$ is the torsion-free dual spin-connection,
for which we have $T(\Omega)=0$. Then, as before, the Lagrangian
3-form of cosmological Einstein's gravity reads 
\begin{equation}
L_{\text{EC-}\Lambda}=-e\cdot R(\omega)+\frac{\Lambda_{0}}{6}e\cdot e\times e,\label{1.33}
\end{equation}
Bianchi identities \eqref{1.18} for 3 dimensions become 
\begin{equation}
D(\omega)R(\omega)=0,\hspace{1cm}D(\omega)T(\omega)=R(\omega)\times e.\label{1.51}
\end{equation}

Up to a boundary term, the GR action can be written as a CS theory
both with no dynamical degrees of freedom in the bulk relativity \cite{12,Witten-3DCS}.
The CS action is defined as 
\begin{equation}
S_{\text{CS}}[\textbf{A}]=\frac{k}{4\pi}\int\textbf{tr}\left(\textbf{A}d\textbf{A}+\frac{2}{3}\textbf{A}\textbf{A}\textbf{A}\right)\label{1.34}
\end{equation}
where $k=\frac{1}{4G}$ is the CS-level and $\textbf{A}=e^{a}\textbf{P}_{a}+\omega^{a}\textbf{J}_{a}$
is a Lie algebra valued connection 1-form. $\textbf{P}_{a}$ and $\textbf{J}_{a}$
satisfy the algebra 
\begin{equation}
\left[\textbf{P}_{a},\textbf{P}_{b}\right]=-\Lambda_{0}\varepsilon_{abc}\textbf{J}^{c},\hspace{0.7cm}\left[\textbf{P}_{a},\textbf{J}_{b}\right]=\varepsilon_{abc}\textbf{P}^{c},\hspace{0.7cm}\left[\textbf{J}_{a},\textbf{J}_{b}\right]=\varepsilon_{abc}\textbf{J}^{c},\label{1.35}
\end{equation}
which corresponds to the Lie algebras $so(2,2)$, $so(3,1)$ and $iso(2,1)$
for $\Lambda_{0}<0$, $\Lambda_{0}>0$ and $\Lambda_{0}=0$, respectively.
The trace denotes a non-degenerate bilinear form on the algebra normalized
as 
\begin{equation}
\textbf{tr}(\textbf{P}_{a}\textbf{J}_{b})=\eta_{ab},\hspace{0.7cm}\textbf{tr}(\textbf{P}_{a}\textbf{P}_{b})=\textbf{tr}(\textbf{J}_{a}\textbf{J}_{b})=0.\label{1.36}
\end{equation}
Now we consider the negative cosmological constant case for which
we write $\Lambda_{0}=-\frac{1}{l^{2}}$, where $l$is the radius
of AdS$_{3}$. $so(2,2)$ is isomorphic to $sl(2,\mathbb{R})\times sl(2,\mathbb{R})$.
Therefore, the generators of two $sl(2,\mathbb{R})$ algebras are
related to $\textbf{P}_{a}$ and $\textbf{J}_{a}$ as 
\begin{equation}
\textbf{J}_{a}^{\pm}=\frac{1}{2}\left(\textbf{J}_{a}\pm l\textbf{P}_{a}\right).\label{1.37}
\end{equation}
This identification diagonalizes the algebra as 
\begin{equation}
\left[\textbf{J}_{a}^{\pm},\textbf{J}_{b}^{\pm}\right]=\varepsilon_{abc}\textbf{J}^{\pm c},\hspace{0.7cm}\left[\textbf{J}_{a}^{+},\textbf{J}_{b}^{-}\right]=0,\label{1.38}
\end{equation}
and the traces become 
\begin{equation}
\textbf{tr}(\textbf{J}_{a}^{\pm}\textbf{J}_{b}^{\pm})=\pm\frac{1}{2}\eta_{ab},\hspace{0.7cm}\textbf{tr}(\textbf{J}_{a}^{+}\textbf{J}_{b}^{-})=0.\label{1.39}
\end{equation}
One has (up to a boundary term) 
\begin{equation}
S_{\text{EC-}\Lambda}[e,\omega]=S_{\text{CS}}[\textbf{A}^{+}]-S_{\text{CS}}[\textbf{A}^{-}]\label{1.40}
\end{equation}
where the gauge fields are given as 
\begin{equation}
\textbf{A}^{\pm}=A^{\pm a}\textbf{J}_{a}\label{1.41}
\end{equation}
with the components 
\begin{equation}
A^{\pm a}=\omega^{a}\pm\frac{1}{l}e^{a}.\label{1.42}
\end{equation}
Note that, in \eqref{1.41}, $\textbf{J}_{a}$ is in fact $\textbf{J}_{a}^{+}$
and we dropped the superscript plus for simplicity. Moreover in \eqref{1.40},
the CS-level is $k=\frac{l}{4G}$, which is a dimensionless quantity.
Hence the to dimensionful parameters in Einstein's gravity $(G,l)$
combine to make a dimensionless parameter in the Chern-Simons formulations.

\subsection{{Chern-Simons-like theories of gravity} as the general}

\label{S1.4} As alluded before, one can formulate a class of gravity
theories in terms of $e^{a}$, $\omega^{a}$ and some Lorentz vector
valued 1-form auxiliary fields (for instance, $h^{a}$, $f^{a}$ and
so on). The Lagrangian 3-form of such theories is defined as 
\begin{equation}
L=\frac{1}{2}\tilde{g}_{rs}a^{r}\cdot da^{s}+\frac{1}{6}\tilde{f}_{rst}a^{r}\cdot a^{s}\times a^{t},\label{1.43}
\end{equation}
where $a^{ra}=a_{\hspace{3mm}\mu}^{ra}dx^{\mu}$ are Lorentz vector-valued
one-forms, where $r=1,...,N$ refers to the flavor index. $\tilde{g}_{rs}$
is a symmetric constant metric in flavor space and $\tilde{f}_{rst}$
is the totally symmetric \textquotedbl flavor tensor\textquotedbl .
Note that in this notation $a^{ra}$ denotes a collection of dreibein,
dual spin-connection and the auxiliary fields. The Lagrangian contains
terms like $f\cdot R=f\cdot d\omega+\frac{1}{2}f\cdot\omega\times\omega$,
$f\cdot D(\omega)h=f\cdot dh+\omega\cdot f\times h$, $\omega\cdot d\omega+\frac{1}{2}\omega\cdot\omega\times\omega$.
It can be seen that all of these combinations obey the equation $\tilde{f}_{\omega rs}=\tilde{g}_{rs}$.
From \eqref{1.26} and \eqref{1.28} one can work out the total variation
of $a^{ra}$ which yields 
\begin{equation}
\delta_{\xi}a^{ra}=\mathfrak{L}_{\xi}a^{ra}-\delta_{\omega}^{r}d\chi_{\xi}^{a},\label{1.44}
\end{equation}
where $\delta_{s}^{r}$ is the Kronecker delta and we introduced 
\begin{equation}
\chi_{\xi}^{a}=\frac{1}{2}\varepsilon_{\hspace{1.5mm}bc}^{a}\lambda_{\xi}^{ab}\label{1.45}
\end{equation}
Total variations of the dreibein $e^{a}$, the dual spin connection
$\omega^{a}$ and a generic Lorentz vector-valued 1-form $h^{a}$
follow along the similar lines: 
\begin{equation}
\delta_{\xi}e=D(\omega)i_{\xi}e+i_{\xi}T(\omega)+(\chi_{\xi}-i_{\xi}\omega)\times e,\label{1.55}
\end{equation}
\begin{equation}
\delta_{\xi}\omega=i_{\xi}R(\omega)+D(\omega)(i_{\xi}\omega-\chi_{\xi}),\label{1.56}
\end{equation}
\begin{equation}
\delta_{\xi}h=D(\omega)i_{\xi}h+i_{\xi}D(\omega)h+(\chi_{\xi}-i_{\xi}\omega)\times h,\label{1.57}
\end{equation}
where we have made use of \eqref{1.44} and \eqref{1.22}. Plugging
\eqref{1.30} and \eqref{1.54} into \eqref{1.55}-\eqref{1.57},
one arrives at somewhat reduced forms of the these total variations
\begin{equation}
\delta_{\xi}e=D(\Omega)i_{\xi}e+(\chi_{\xi}-i_{\xi}\Omega)\times e,\label{1.58}
\end{equation}
\begin{equation}
\delta_{\xi}\Omega=i_{\xi}R(\Omega)+D(\Omega)(i_{\xi}\Omega-\chi_{\xi}),\label{1.59}
\end{equation}
\begin{equation}
\delta_{\xi}h=D(\Omega)i_{\xi}h+i_{\xi}D(\Omega)h+(\chi_{\xi}-i_{\xi}\Omega)\times h,\label{1.60}
\end{equation}
On the other hand, the total variation of $a^{ra}$ is covariant under
the Lorentz gauge transformations and diffeomorphisms. Finally, the
Lagrangian 3-form \eqref{1.42} varies as a total derivative part
and field equation part: 
\begin{equation}
\delta L=\delta a^{r}\cdot E_{r}+d\Theta(a,\delta a),\label{1.46}
\end{equation}
where the field equation part reads 
\begin{equation}
E_{r}^{\hspace{1.5mm}a}=\tilde{g}_{rs}da^{sa}+\frac{1}{2}\tilde{f}_{rst}(a^{s}\times a^{t})^{a},\label{1.47}
\end{equation}
and the surface part reads 
\begin{equation}
\Theta(a,\delta a)=\frac{1}{2}\tilde{g}_{rs}\delta a^{r}\cdot a^{s}.\label{1.48}
\end{equation}
Setting $\delta$ in \eqref{1.46} to be the total variation $\delta_{\xi}$,
one arrives at 
\begin{equation}
\delta_{\xi}L=\mathfrak{L}_{\xi}a^{r}\cdot E_{r}+d\Theta(a,\mathfrak{L}_{\xi}a)-d\chi_{\xi}\cdot E_{\omega}-d\left(\frac{1}{2}\tilde{g}_{\omega r}d\chi_{\xi}\cdot a^{r}\right),\label{E.1}
\end{equation}
where we made use of \eqref{1.44}. On the other hand, the LL-derivative
of the Lagrangian 3-form can be computed to be 
\begin{equation}
\mathfrak{L}_{\xi}L=\mathfrak{L}_{\xi}a^{r}\cdot E_{r}+d\Theta(a,\mathfrak{L}_{\xi}a)+\frac{1}{2}\tilde{g}_{rs}a^{r}\cdot\left[\mathfrak{L}_{\xi},d\right]a^{s},\label{E.2}
\end{equation}
where $\left[\mathfrak{L}_{\xi},d\right]=\mathfrak{L}_{\xi}d-d\mathfrak{L}_{\xi}$.
The exterior derivative and the LL-derivative do not commute, in fact
the commutator yields 
\begin{equation}
\left[\mathfrak{L}_{\xi},d\right]a^{r}=-d\chi_{\xi}\times a^{r}.\label{E.3}
\end{equation}
Using \eqref{E.1},\eqref{E.2} and \eqref{E.3}, one finds 
\begin{equation}
\delta_{\xi}L-\mathfrak{L}_{\xi}L=d\left(\frac{1}{2}\tilde{g}_{\omega r}d\chi_{\xi}\cdot a^{r}\right)+\frac{1}{2}\left(\tilde{g}_{rs}-\tilde{f}_{\omega rs}\right)d\chi_{\xi}\cdot a^{r}\times a^{s}.\label{E.4}
\end{equation}
All the interesting theories of this type has $\tilde{g}_{rs}=\tilde{f}_{\omega rs}$.
Using this, the total variation of the Lagrangian induced by the diffeomorphism
generator $\xi$ can be written as a total derivative term 
\begin{equation}
\delta_{\xi}L=\mathfrak{L}_{\xi}L+d\psi_{\xi}=d\left(i_{\xi}L+\psi_{\xi}\right),\label{1.49}
\end{equation}
where 
\begin{equation}
\psi_{\xi}(a)=\frac{1}{2}\tilde{g}_{\omega r}d\chi_{\xi}\cdot a^{r}.\label{1.50}
\end{equation}
The Lagrangian is not invariant under general coordinates transformations
and/or general Lorentz gauge transformations, but since the total
variation is written as a surface term \eqref{1.49}, $\xi$ is an
infinitesimal symmetry of the theory: as the field equations that
follow from these variations are generally covariant. This was a general
discussion, it turns out for some models $\psi_{\xi}$ vanishes and
we refer to them as the Lorentz-diffeomorphism covariant theories
because they are globally covariant under the Lorenz gauge transformations
as well as diffeomorphisms.

\section{Some examples of Chern-Simons-like theories of gravity}

\label{S2.0} Here, we will discuss briefly some examples of the above
construction. Further examples can be found in \cite{15}.

\subsection{Example.1: {Einstein's gravity with a negative cosmological constant}}

\label{S2.1} In this case \eqref{1.32} one has $\Lambda_{0}=-l^{-2}$,
$a^{r}=\{e,\omega\}$ and the nontrivial components of flavor metric
and the tensor are 
\begin{equation}
\tilde{g}_{e\omega}=-1,\hspace{0.7cm}\tilde{f}_{eee}=-\frac{1}{l^{2}}.\label{2.1}
\end{equation}
The field equations \eqref{1.46} reduce to 
\begin{equation}
R(\omega)+\frac{1}{2l^{2}}e\times e=0,\label{2.2}
\end{equation}
\begin{equation}
T(\omega)=0.\label{2.3}
\end{equation}
The last equation is the torsion-free condition and hence $\omega=\Omega$
and \eqref{2.2} becomes 
\begin{equation}
R(\Omega)+\frac{1}{2l^{2}}e\times e=0.\label{2.4}
\end{equation}
Therefore solutions of Einstein gravity with negative cosmological
constant must satisfy \eqref{2.4}.

\subsubsection{{Banados-Teitelboim-Zanelli black hole}}

\label{S.2.1.1} We discussed the BTZ black hole before, but here
we rewrite the metric in a slightly different form as 
\begin{equation}
\begin{split}ds^{2}= & -\frac{\left(r^{2}-r_{+}^{2}\right)\left(r^{2}-r_{-}^{2}\right)}{l^{2}r^{2}}dt^{2}+\frac{l^{2}r^{2}}{\left(r^{2}-r_{+}^{2}\right)\left(r^{2}-r_{-}^{2}\right)}dr^{2}\\
 & +r^{2}\left(d\phi-\frac{r_{+}r_{-}}{l^{2}r^{2}}\right)^{2},
\end{split}
\label{2.4.1}
\end{equation}
where $r_{\pm}$ are outer/inner horizon radii, respectively. The
dreibeins can be chosen as 
\begin{equation}
\begin{split} & e^{0}=\left(\frac{(r^{2}-r_{+}^{2})(r^{2}-r_{-}^{2})}{l^{2}r^{2}}\right)^{\frac{1}{2}}dt\\
 & e^{1}=r\left(d\phi-\frac{r_{+}r_{-}}{lr^{2}}dt\right)\\
 & e^{2}=\left(\frac{l^{2}r^{2}}{(r^{2}-r_{+}^{2})(r^{2}-r_{-}^{2})}\right)^{\frac{1}{2}}dr.
\end{split}
\label{2.4.2}
\end{equation}
By substituting \eqref{2.4.2} into \eqref{2.4}, one can check that
the dreibeins \eqref{2.4.2} solve the field equations.

\subsection{Example.2: {Minimal Massive Gravity}}

\label{S.2.2} For minimal massive gravity (MMG) \cite{3}, we need
three flavors of one-forms $a^{r}=\{e,\omega,h\}$, where $h$ is
an auxiliary 1-form field. Also, the non-zero components of the flavor
metric and the flavor tensor are 
\begin{equation}
\begin{split} & \tilde{g}_{e\omega}=-\sigma,\hspace{1cm}\tilde{g}_{eh}=1,\hspace{1cm}\tilde{g}_{\omega\omega}=\frac{1}{\mu},\hspace{1cm}\tilde{f}_{eee}=\Lambda_{0}\\
 & \tilde{f}_{e\omega\omega}=-\sigma,\hspace{1cm}\tilde{f}_{eh\omega}=1,\hspace{1cm}\tilde{f}_{\omega\omega\omega}=\frac{1}{\mu},\hspace{1cm}\tilde{f}_{ehh}=\alpha,
\end{split}
\label{2.5}
\end{equation}
where $\sigma$, is $\pm$ and $\Lambda_{0}$, $\mu$ and $\alpha$
denote the cosmological parameter with dimension of mass squared,
the mass parameter of the Lorentz Chern-Simons term and a dimensionless
parameter, respectively. The field equations \eqref{1.46} become
\begin{equation}
-\sigma R(\omega)+\frac{\Lambda_{0}}{2}e\times e+D(\omega)h+\frac{\alpha}{2}h\times h=0,\label{2.6}
\end{equation}
\begin{equation}
-\sigma T(\omega)+\frac{1}{\mu}R(\omega)+e\times h=0,\label{2.7}
\end{equation}
\begin{equation}
T(\omega)+\alpha e\times h=0.\label{2.8}
\end{equation}
The last equation says that this is not a torsion-free theory but
we can introduce the following connection 
\begin{equation}
\omega=\Omega-\alpha h,\label{2.9}
\end{equation}
to obtain $T(\Omega)=0$ hence $\Omega$ is the torsion-free spin-connection.
It is easy to show that \eqref{2.6} and \eqref{2.7} are equivalent
to the following equations 
\begin{equation}
R(\Omega)+\frac{\alpha\Lambda_{0}}{2}e\times e+\mu(1+\sigma\alpha)^{2}e\times h=0,\label{2.10}
\end{equation}
\begin{equation}
D(\Omega)h-\frac{\alpha}{2}h\times h+\sigma\mu(1+\sigma\alpha)e\times h+\frac{\Lambda_{0}}{2}e\times e=0,\label{2.11}
\end{equation}
Assuming that $(1+\sigma\alpha)\neq0$, \eqref{2.10} and \eqref{2.11}
can be solved (see \cite{3}). In components, \eqref{2.10} yields
\begin{equation}
h_{\hspace{1.5mm}\mu}^{a}=-\frac{1}{\mu(1+\sigma\alpha)^{2}}\left[S_{\hspace{1.5mm}\mu}^{a}+\frac{\alpha\Lambda_{0}}{2}e_{\hspace{1.5mm}\mu}^{a}\right],\label{2.12}
\end{equation}
where $S_{\mu\nu}$ is 3D Schouten tensor 
\begin{equation}
S_{\mu\nu}=\mathcal{R}_{\mu\nu}-\frac{1}{4}g_{\mu\nu}\mathcal{R}.\label{2.13}
\end{equation}
By plugging \eqref{2.13} into \eqref{2.11}, one recovers the field
equations of MMG in terms of the metric tensor $g_{\mu\nu}$. The
BTZ black hole, as a locally AdS$_{3}$ metric, is a solution of MMG\footnote{In \cite{Yekta1}, MMG was studied using the Hamiltonian formalism.}.
Note that MMG reduces to TMG for $\alpha=0$.

\subsection{Example.3: {New Massive Gravity}}

\label{S.2.3} New massive gravity (NMG) was introduced in \cite{2}
and it needs four flavors of one-forms $a^{r}=\{e,\omega,h,f\}$ to
be described in the first-order formalism.\footnote{In \cite{Yekta2}, by using the Hamiltonian formalism, the canonical
structure of NMG was discussed in warped AdS\textsubscript{3} sector,
and the charges of warped black holes were constructed.} Here, the nonzero components of the flavor metric and the flavor
tensor can be found as 
\begin{equation}
\begin{split} & \tilde{g}_{e\omega}=-\sigma,\hspace{1cm}\tilde{g}_{\omega f}=-\frac{1}{m^{2}},\hspace{1cm}\tilde{g}_{eh}=1,\\
 & \tilde{f}_{e\omega\omega}=-\sigma,\hspace{1cm}\tilde{f}_{f\omega\omega}=-\frac{1}{m^{2}},\hspace{1cm}\tilde{f}_{eh\omega}=1,\\
 & \tilde{f}_{eee}=\Lambda_{0},\hspace{1cm}\tilde{f}_{eff}=-\frac{1}{m^{2}},
\end{split}
\label{2.14}
\end{equation}
Plugging these into \eqref{1.46}, one arrives at the field equations
\begin{equation}
-\sigma R(\omega)+\frac{\Lambda_{0}}{2}e\times e+D(\omega)h-\frac{1}{2m^{2}}f\times f=0,\label{2.15}
\end{equation}
\begin{equation}
-\sigma T(\omega)-\frac{1}{m^{2}}D(\omega)f+e\times h=0,\label{2.16}
\end{equation}
\begin{equation}
-\frac{1}{m^{2}}\left(R(\omega)+e\times f\right)=0,\label{2.17}
\end{equation}
\begin{equation}
T(\omega)=0,\label{2.18}
\end{equation}
where the last equation \eqref{2.18} yields $\omega=\Omega$, and
hence the connection is torsion-free. The auxiliary fields $f$ and
$h$ can be found from \eqref{2.17} and \eqref{2.16} as 
\begin{equation}
f^{a}=-S^{a},\hspace{1cm}h^{a}=-\frac{1}{m^{2}}C^{a}.\label{2.19}
\end{equation}
Here, $C_{\mu\nu}$ is the Cotton tensor.

\subsubsection{{Rotating Oliva-Tempo-Troncoso black hole as a solution of new massive
gravity}}

\label{S.2.3.1} The NMG admits two maximally symmetric spacetimes
which degenerates into one if the following tuning is made \cite{2}
\begin{equation}
\sigma=1,\hspace{0.7cm}m^{2}=\frac{1}{2l^{2}},\hspace{0.7cm}\Lambda_{0}=-\frac{1}{2l^{2}}.\label{2.21}
\end{equation}
In this limit, there arises a new black hole solution the so called
rotating Oliva-Tempo-Troncoso (OTT) spacetime \cite{17,18,19} with
the metric 
\begin{equation}
ds^{2}=-N(r)^{2}F(r)^{2}dt^{2}+F(r)^{-2}dr^{2}+r^{2}\left(d\phi+N^{\phi}(r)dt\right)^{2},\label{2.22}
\end{equation}
where the metric functions are given as 
\begin{equation}
\begin{split} & F(r)=\frac{H(r)}{r}\sqrt{\frac{H(r)^{2}}{l^{2}}+\frac{b}{2}H(r)(1+\eta)+\frac{b^{2}l^{2}}{16}(1-\eta)^{2}-\mu\eta},\\
 & N(r)=1+\frac{bl^{2}}{4H(r)}(1-\eta),\\
 & N^{\phi}(r)=\frac{l}{2r^{2}}\sqrt{1-\eta^{2}}\left(\mu-bH(r)\right),\\
 & H(r)=\sqrt{r^{2}-\frac{\mu l^{2}}{2}(1-\eta)-\frac{b^{2}l^{4}}{16}(1-\eta)^{2}}.
\end{split}
\label{2.23}
\end{equation}
Here, $\mu$, $b$, and $\eta$ are free parameters, and for the particular
case of $\eta=1$, the solution is static. For the case of $b=0$,
the solution reduces to the rotating BTZ black hole \cite{BTZ}. The
dreibeins can be chosen as 
\begin{equation}
e^{0}=N(r)F(r)dt,\hspace{0.7cm}e^{1}=F(r)^{-1}dr,\hspace{0.7cm}e^{2}=r\left(d\phi+N^{\phi}(r)dt\right).\label{2.24}
\end{equation}
Note that this is a locally conformally flat solution; and hence,
the Cotton tensor vanishes identically.

\subsection{Example.4: {Generalized Massive Gravity}}

\label{S.2.4} Generalized massive gravity was introduced in \cite{NMG},
and elaborated more in \cite{21}. It also needs four flavor of one-forms,
$a^{r}=\{e,\omega,h,f\}$, and the nonvanishing components of the
flavor metric and the flavor tensor are 
\begin{equation}
\begin{split} & \tilde{g}_{e\omega}=-\sigma,\hspace{0.5cm}\tilde{g}_{eh}=1,\hspace{0.5cm}\tilde{g}_{f\omega}=-\frac{1}{m^{2}},\hspace{0.5cm}\tilde{g}_{\omega\omega}=\frac{1}{\mu},\\
 & \tilde{f}_{e\omega\omega}=-\sigma,\hspace{0.5cm}\tilde{f}_{eh\omega}=1,\hspace{0.5cm}\tilde{f}_{f\omega\omega}=-\frac{1}{m^{2}},\hspace{0.5cm}\tilde{f}_{\omega\omega\omega}=\frac{1}{\mu},\\
 & \tilde{f}_{eee}=\Lambda_{0},\hspace{0.5cm}\tilde{f}_{eff}=-\frac{1}{m^{2}}.
\end{split}
\label{2.25}
\end{equation}
Again, the theory is torsion-free and the auxiliary fields $h$ and
$f$ can be solved as 
\begin{equation}
h^{a}=-\frac{1}{\mu}S^{a}-\frac{1}{m^{2}}C^{a},\hspace{1cm}f^{a}=-S^{a}.\label{2.26}
\end{equation}

\subsection{Example.5: {Generalized Minimal Massive Gravity}}

\label{S2.5} Generalized minimal massive gravity was introduced in
\cite{5} as a model which is free of the negative-energy bulk modes,
and the Hamiltonian analysis showed that it does not have a Boulware-Deser
ghost at the nonlinear level; hence, the theory propagates two healthy
physical degrees of freedom. Just like the above two cases, it is
described by four flavor of one-forms, and the nonzero components
of the flavor metric and the flavor tensor are 
\begin{equation}
\begin{split} & \tilde{g}_{e\omega}=-\sigma,\hspace{0.5cm}\tilde{g}_{eh}=1,\hspace{0.5cm}\tilde{g}_{f\omega}=-\frac{1}{m^{2}},\hspace{0.5cm}\tilde{g}_{\omega\omega}=\frac{1}{\mu},\\
 & \tilde{f}_{e\omega\omega}=-\sigma,\hspace{0.5cm}\tilde{f}_{eh\omega}=1,\hspace{0.5cm}\tilde{f}_{f\omega\omega}=-\frac{1}{m^{2}},\hspace{0.5cm}\tilde{f}_{\omega\omega\omega}=\frac{1}{\mu},\\
 & \tilde{f}_{eee}=\Lambda_{0},\hspace{0.5cm}\tilde{f}_{ehh}=\alpha,\hspace{0.5cm}\tilde{f}_{eff}=-\frac{1}{m^{2}}.
\end{split}
\label{2.27}
\end{equation}
For $\alpha=0$, the theory reduces to the general massive gravity
case. Plugging \eqref{2.27} into \eqref{1.43}, one arrives at the
Lagrangian of the theory 
\begin{equation}
\begin{split}L= & -\sigma e\cdot R(\omega)+\frac{\Lambda_{0}}{6}e\cdot e\times e+\frac{1}{2\mu}\left(\omega\cdot d\omega+\frac{1}{3}\omega\cdot\omega\times\omega\right)\\
 & -\frac{1}{m^{2}}\left(f\cdot R+\frac{1}{2}e\cdot f\times f\right)+h\cdot T(\omega)+\frac{\alpha}{2}e\cdot h\times h,
\end{split}
\label{2.28}
\end{equation}
up to a total derivative. The field equations becomes 
\begin{equation}
E_{e}=-\sigma R(\omega)+\frac{\Lambda_{0}}{2}e\times e+D(\omega)h-\frac{1}{2m^{2}}f\times f+\frac{\alpha}{2}h\times h=0,\label{2.29}
\end{equation}
\begin{equation}
E_{\omega}=-\sigma T(\omega)+\frac{1}{\mu}R(\omega)-\frac{1}{m^{2}}D(\omega)f+e\times h=0,\label{2.30}
\end{equation}
\begin{equation}
E_{f}=-\frac{1}{m^{2}}\left(R(\omega)+e\times f\right)=0,\label{2.31}
\end{equation}
\begin{equation}
E_{h}=T(\omega)+\alpha e\times h=0.\label{2.32}
\end{equation}
The last equation is similar to \eqref{2.8}; therefore, we can use
a torsion-free spin connection to recast the field equations as 
\begin{equation}
-\sigma R(\Omega)+(1+\sigma\alpha)D(\Omega)h-\frac{1}{2}\alpha(1+\sigma\alpha)h\times h+\frac{\Lambda_{0}}{2}e\times e-\frac{1}{2m^{2}}f\times f=0,\label{2.33}
\end{equation}
\begin{equation}
-e\times f+\mu(1+\sigma\alpha)e\times h-\frac{\mu}{m^{2}}D(\Omega)f+\frac{\mu\alpha}{m^{2}}h\times f=0,\label{2.34}
\end{equation}
\begin{equation}
R(\Omega)-\alpha D(\Omega)h+\frac{1}{2}\alpha^{2}h\times h+e\times f=0,\label{2.35}
\end{equation}
General solution for the auxiliary fields is nontrivial to find; therefore,
in the next section we shall consider certain specific solutions.

\subsubsection{{Solutions of Einstein gravity with negative cosmological constant}}

\label{S2.5.1} Here, we will show that all Einstein metrics solve
the field equations of generalized minimal massive gravity. For this
purpose, we consider the following ansatz for auxiliary fields 
\begin{equation}
f^{a}=Fe^{a},\hspace{1cm}h^{a}=He^{a},\label{2.36}
\end{equation}
which reduce the equations \eqref{2.33}-\eqref{2.35} to 
\begin{equation}
\frac{\sigma}{l^{2}}-\alpha(1+\sigma\alpha)H^{2}+\Lambda_{0}-\frac{F^{2}}{m^{2}}=0,\label{2.37}
\end{equation}
\begin{equation}
-\frac{1}{\mu l^{2}}+2(1+\sigma\alpha)H+\frac{2\alpha}{m^{2}}FH+\frac{\alpha^{2}}{\mu}H^{2}=0,\label{2.38}
\end{equation}
\begin{equation}
-F+\mu(1+\sigma\alpha)H+\frac{\mu\alpha}{m^{2}}FH=0.\label{2.39}
\end{equation}
This set of equations can be solved to arrive at 
\begin{equation}
l_{\pm}^{2}=\left[\alpha^{2}H^{2}+2\sigma m^{2}\pm2\sqrt{m^{2}\left(\sigma^{2}m^{2}-\alpha H^{2}+\Lambda_{0}\right)}\right]^{-1},\label{2.40}
\end{equation}
\begin{equation}
F_{\pm}=\sigma m^{2}\pm\sqrt{m^{2}\left(\sigma^{2}m^{2}-\alpha H^{2}+\Lambda_{0}\right)},\label{2.41}
\end{equation}
where $H$ satisfies an algebraic equation as 
\begin{equation}
\begin{split} & -\Lambda_{0}m^{4}+2\mu m^{2}\left[\alpha\Lambda_{0}-\sigma m^{2}(1+\alpha\sigma)\right]H\\
 & +\left[\mu^{2}m^{2}(1+\alpha\sigma)(1+3\alpha\sigma)+\alpha\left(m^{4}-\alpha\mu^{2}\Lambda_{0}\right)\right]H^{2}\\
 & -2\mu\alpha^{2}m^{2}H^{3}+\mu\alpha^{3}H^{4}=0.
\end{split}
\label{2.42}
\end{equation}
with the following conditions on the roots 
\begin{equation}
\begin{split} & H^{2}\leq\alpha^{-1}\left(\sigma^{2}m^{2}+\Lambda_{0}\right)\hspace{0.5cm}\text{for}\hspace{0.5cm}\alpha>0,\\
 & H^{2}\geq\alpha^{-1}\left(\sigma^{2}m^{2}+\Lambda_{0}\right)\hspace{0.5cm}\text{for}\hspace{0.5cm}\alpha<0,
\end{split}
\label{2.43}
\end{equation}
Here, $l^{2}$, $H$ and $F$ are real since they appear in physical
quantities.

\subsubsection{Warped black holes}

\label{S2.5.2} The spacelike warped AdS$_{3}$ black hole is described
by the metric \cite{22} 
\begin{equation}
\frac{ds^{2}}{l^{2}}=-N(r)^{2}dt^{2}+\frac{dr^{2}}{4N(r)^{2}R(r)^{2}}+R(r)^{2}\left(d\phi+N^{\phi}(r)dt\right)^{2},\label{2.44}
\end{equation}
with the metric functions 
\begin{equation}
\begin{split}R(r)^{2}= & \frac{1}{4}\zeta^{2}r\left[\left(1-\nu^{2}\right)r+\nu^{2}\left(r_{+}+r_{-}\right)+2\nu\sqrt{r_{+}r_{-}}\right],\\
N(r)^{2}= & \zeta^{2}\nu^{2}\frac{\left(r-r_{+}\right)\left(r-r_{-}\right)}{4R(r)^{2}},\qquad\qquad N^{\phi}(r)=\left|\zeta\right|\frac{r+\nu\sqrt{r_{+}r_{-}}}{2R(r)^{2}},
\end{split}
\label{2.45}
\end{equation}
where $r_{\pm}$ are outer/inner horizon radii, respectively. To keep
the same convention as \cite{23,Bouchareb,25,26}\footnote{For further discussion, see \cite{22}.},
we used the parameters $\zeta$ and $\nu$. The isometry group of
this metric is $SL(2,\mathbb{R})\times U(1)$. One can introduce a
symmetric-two tensor $\mathcal{T}_{\mu\nu}$ as \cite{27} 
\begin{equation}
\mathcal{T}_{\mu\nu}=a_{1}g_{\mu\nu}+a_{2}J_{\mu}J_{\nu},\label{2.46}
\end{equation}
with $J=J^{\mu}\partial_{\mu}=\partial_{t}$ and $J_{\mu}J^{\mu}=l^{2}$.
One also has 
\begin{equation}
\hat{\nabla}_{\mu}J_{\nu}=\frac{\left|\zeta\right|}{2l}\epsilon_{\mu\nu\lambda}J^{\lambda}.\label{2.48}
\end{equation}
One can compute the Ricci tensor and scalar curvature as 
\begin{equation}
\mathcal{R}_{\mu\nu}=\frac{\zeta^{2}}{2l^{2}}\left(1-2\nu^{2}\right)g_{\mu\nu}-\frac{\zeta^{2}}{l^{4}}\left(1-\nu^{2}\right)J_{\mu}J_{\nu},\label{2.49}
\end{equation}
\begin{equation}
\mathcal{R}=\frac{\zeta^{2}}{2l^{2}}\left(1-4\nu^{2}\right).\label{2.50}
\end{equation}
The dreibeins can be chosen as 
\begin{equation}
\begin{split}e^{0}= & lN(r)dt,\\
e^{1}= & \frac{l}{2R(r)N(r)}dr,\\
e^{2}= & lR(r)\left(d\phi+N^{\phi}dt\right).
\end{split}
\label{2.51}
\end{equation}
Now, one can consider the following ansatz for auxiliary fields 
\begin{equation}
\begin{split}h_{\hspace{1.5mm\mu}}^{a}= & H_{1}e_{\hspace{1.5mm\mu}}^{a}+H_{2}J^{a}J_{\mu},\\
f_{\hspace{1.5mm\mu}}^{a}= & F_{1}e_{\hspace{1.5mm\mu}}^{a}+F_{2}J^{a}J_{\mu},
\end{split}
\label{2.52}
\end{equation}
where $H_{1}$, $H_{2}$, $F_{1}$, $F_{2}$ are constants, and $J^{a}=e_{\hspace{1.5mm\mu}}^{a}J^{\mu}$.
Using the equations \eqref{2.48}-\eqref{2.52}, the field equations
of the generalized minimal massive gravity reduce to 
\begin{equation}
\frac{\zeta^{2}}{4l^{2}}-\frac{1}{2}\alpha l\left|\zeta\right|H_{2}-\alpha^{2}H_{1}\left(H_{1}+l^{2}H_{2}\right)-\left(2F_{1}+l^{2}F_{2}\right)=0,\label{2.53}
\end{equation}
\begin{equation}
-\frac{\zeta^{2}}{l^{4}}\left(1-\nu^{2}\right)+\frac{3\alpha}{2l}\left|\zeta\right|H_{2}+\alpha^{2}H_{1}H_{2}+F_{2}=0,\label{2.54}
\end{equation}
\begin{equation}
\begin{split} & \frac{1}{\mu}\left(2F_{1}+l^{2}F_{2}\right)-\left(1+\alpha\sigma\right)\left(2H_{1}+l^{2}H_{2}\right)-\frac{l}{2m^{2}}\left|\zeta\right|F_{2}\\
 & -\frac{\alpha}{m^{2}}\left[2H_{1}F_{1}+l^{2}\left(H_{1}F_{2}+H_{2}F_{1}\right)\right]=0,
\end{split}
\label{2.55}
\end{equation}
\begin{equation}
-\frac{1}{\mu}F_{2}+\left(1+\alpha\sigma\right)H_{2}+\frac{3}{2lm^{2}}\left|\zeta\right|F_{2}+\frac{\alpha}{m^{2}}\left(H_{1}F_{2}+H_{2}F_{1}\right)=0,\label{2.56}
\end{equation}
\begin{equation}
\begin{split} & -\frac{\zeta^{2}}{4l^{2}}\sigma+\frac{1}{2}\left(1+\alpha\sigma\right)l\left|\zeta\right|H_{2}+\alpha\left(1+\alpha\sigma\right)H_{1}\left(H_{1}+l^{2}H_{2}\right)\\
 & -\Lambda_{0}+\frac{1}{m^{2}}F_{1}\left(F_{1}+l^{2}F_{2}\right)=0,
\end{split}
\label{2.57}
\end{equation}
\begin{equation}
\frac{\zeta^{2}}{l^{4}}\left(1-\nu^{2}\right)\sigma-\frac{3}{2l}\left(1+\alpha\sigma\right)\left|\zeta\right|H_{2}-\alpha\left(1+\alpha\sigma\right)H_{1}H_{2}-\frac{1}{m^{2}}F_{1}F_{2}=0.\label{2.58}
\end{equation}
As a result, the metric \eqref{2.44} provides a solution to generalized
minimal massive gravity when the parameters satisfy equations \eqref{2.53}-\eqref{2.58}.
More details are given in Appendix \ref{A}.

\section{Extended off-shell ADT current}

\label{S.3.0} Historically, Regge and Teitelboim were the pioneers
who showed that it is possible to write the charges as boundary terms
in gravity theories \cite{Regge:1974zd,Regge:1974otg}. Arnowitt,
Deser, and Misner (ADM)\cite{adm} then introduced a way to associate
a conserved mass for an asymptotically flat spacetime in general relativity.
Their construction was generalized to the asymptotically AdS spacetimes
for the cosmological Einstein's gravity \cite{Abbott,28}. In a further
development Deser and Tekin generalized the ADM's construction to
all extended theories of gravity with many powers of curvature and
we have discussed these developments in the first part of this work
\cite{Deser_Tekin-PRD,Deser_Tekin-PRL,Senturk}. It is worth to mention
that Iyer and Wald \cite{9}, before the work by Deser and Tekin,
found an expression for the conserved quantities in any generally
covariant theory of gravity based on \cite{33}. The original ADM
mass is not explicitly covariant in the sense that the expression
is written in Cartesian coordinates, but it is known to be covariant
(coordinate independent, modulo the proper decay conditions discussed
before). The ADT formalism is explicitly covariant and in this sense,
extends the ADM mass to any viable coordinate system and more over
it also incorporates the conserved angular momenta. The authors of
\cite{Kim} extended ADT formalism to the off-shell level, in any
diffeomorphism-invariant theory of gravity. They have extended this
work to a theory of gravity containing a gravitational Chern-Simons
term in \cite{32} in the metric formulation. In what follows the
off-shell ADT method will be constructed in the first order formalism.

The variation of the Lagrangian \eqref{1.46} can be generated by
a vector field $\xi$ as 
\begin{equation}
\delta_{\xi}L=\delta_{\xi}a^{r}\cdot E_{r}+d\Theta(a,\delta_{\xi}a).\label{3.1}
\end{equation}
We assume that $\xi$ is a function of the dynamical fields $a$ and
$x$, that is $\xi=\xi(a,x)$. Using \eqref{1.44} and \eqref{1.49}
in \eqref{3.1}, after a somewhat lengthy computation whose details
are given in \ref{B}, one arrives 
\begin{equation}
\begin{split}dJ_{\xi}= & \left(i_{\xi}\omega-\chi_{\xi}\right)\cdot\left(D(\omega)E_{\omega}+e\times E_{e}+a^{r^{\prime}}\times E_{r^{\prime}}\right)\\
 & +i_{\xi}a^{r^{\prime}}\cdot D(\omega)E_{r^{\prime}}-i_{\xi}D(\omega)a^{r^{\prime}}\cdot E_{r^{\prime}}\\
 & +i_{\xi}e\cdot D(\omega)E_{e}-i_{\xi}T(\omega)\cdot E_{e}-i_{\xi}R(\omega)\cdot E_{\omega},
\end{split}
\label{3.2}
\end{equation}
with 
\begin{equation}
J_{\xi}=\Theta(a,\delta_{\xi}a)-i_{\xi}L-\psi_{\xi}+i_{\xi}a^{r}\cdot E_{r}-\chi_{\xi}\cdot E_{\omega}.\label{3.3}
\end{equation}
Here, $r^{\prime}$ does not run over the flavors $e$ and $\omega$.
Note that $dJ_{\xi}$ becomes zero since it is simply the linear combination
of Bianchi identities given in \eqref{1.51} \cite{33}.

Let us elaborate on this point. Consider a generic action for a Chern-Simons
like gravity theory $S=\int_{\mathcal{M}}L$ having the variation
\begin{equation}
\delta_{\xi}S=\int_{\mathcal{M}}\delta_{\xi}L=\int_{\mathcal{M}}\left[\delta_{\xi}a^{r}\cdot E_{r}+d\Theta(a,\delta_{\xi}a)\right],\label{3.5}
\end{equation}
which can be put in the form 
\begin{equation}
\begin{split}\int_{\partial\mathcal{M}}J_{\xi}=\int_{\mathcal{M}}\bigl[ & \left(i_{\xi}\omega-\chi_{\xi}\right)\cdot(D(\omega)E_{\omega}+e\times E_{e}+a^{r^{\prime}}\times E_{r^{\prime}})\\
 & +i_{\xi}a^{r^{\prime}}\cdot D(\omega)E_{r^{\prime}}-i_{\xi}D(\omega)a^{r^{\prime}}\cdot E_{r^{\prime}}\\
 & +i_{\xi}e\cdot D(\omega)E_{e}-i_{\xi}T(\omega)\cdot E_{e}-i_{\xi}R(\omega)\cdot E_{\omega}\bigr].
\end{split}
\label{3.6}
\end{equation}
Assuming vanishing fields on the boundary $\partial\mathcal{M}$,
one has 
\begin{equation}
\begin{split}0=\int_{\mathcal{M}}\bigl[ & \left(i_{\xi}\omega-\chi_{\xi}\right)\cdot(D(\omega)E_{\omega}+e\times E_{e}+a^{r^{\prime}}\times E_{r^{\prime}})\\
 & +i_{\xi}a^{r^{\prime}}\cdot D(\omega)E_{r^{\prime}}-i_{\xi}D(\omega)a^{r^{\prime}}\cdot E_{r^{\prime}}\\
 & +i_{\xi}e\cdot D(\omega)E_{e}-i_{\xi}T(\omega)\cdot E_{e}-i_{\xi}R(\omega)\cdot E_{\omega}\bigr].
\end{split}
\label{3.7}
\end{equation}
To have a vanishing integral for arbitrary diffeomorphism generator
$\xi$, the integrand which is proportional to $\xi$ must vanish
as a result of Bianchi identities; and hence, this yields 
\begin{equation}
dJ_{\xi}=0.\label{3.8}
\end{equation}
$J_{\xi}$ is an off-shell Noether conserved current that can be written
in an exact form as 
\begin{equation}
J_{\xi}=dK_{\xi},\label{3.9}
\end{equation}
where $K_{\xi}$ is the off-shell Noether potential given as 
\begin{equation}
K_{\xi}=\frac{1}{2}\tilde{g}_{rs}i_{\xi}a^{r}\cdot a^{s}-\tilde{g}_{\omega s}\chi_{\xi}\cdot a^{s}.\label{3.10}
\end{equation}

As we discussed in Sec.~\ref{S1.4}, Chern-Simons like gravity theories
are not Lorentz covariant, so the surface term \eqref{1.48} is not
a Lorentz covariant quantity, and one has 
\begin{equation}
\delta_{\xi}\Theta(a,\delta a)=\mathfrak{L}_{\xi}\Theta(a,\delta a)+\Pi_{\xi}(a,\delta a),\label{3.11}
\end{equation}
with 
\begin{equation}
\Pi_{\xi}(a,\delta a)=\frac{1}{2}\tilde{g}_{\omega r}d\chi_{\xi}\cdot\delta a^{r},\label{3.12}
\end{equation}
using the fact that the difference of two dual spin-connections is
a Lorentz vector-valued 1-form, that is $\delta_{\xi}\delta\omega^{a}=\mathfrak{L}_{\xi}\delta\omega^{a}$.
To find the linearized off-shell conserved current, one needs the
variation of \eqref{3.3} as 
\begin{equation}
\begin{split}d\delta K_{\xi}= & \delta\Theta(a,\delta_{\xi}a)-i_{\xi}\delta L-\delta\psi_{\xi}+i_{\xi}\delta a^{r}\cdot E_{r}+i_{\xi}a^{r}\cdot\delta E_{r}-\chi_{\xi}\cdot\delta E_{\omega}\\
 & -i_{\delta\xi}L+i_{\delta\xi}a^{r}\cdot E_{r}-\chi_{\delta\xi}\cdot E_{\omega},
\end{split}
\label{C.1}
\end{equation}
where we used $\delta\chi_{\xi}=\chi_{\delta\xi}$ as $\chi_{\xi}$
is linear in $\xi$. Then, having $\xi\rightarrow\delta\xi$ in \eqref{3.3}
yields 
\begin{equation}
-i_{\delta\xi}L+i_{\delta\xi}a^{r}\cdot E_{r}-\chi_{\delta\xi}\cdot E_{\omega}=dK_{\delta\xi}-\Theta(a,\delta_{\delta\xi}a)+\psi_{\delta\xi}.\label{C.2}
\end{equation}
In addition, taking the interior product of \eqref{1.46} in $\xi$
yields 
\begin{equation}
i_{\xi}\delta L=i_{\xi}\delta a^{r}\cdot E_{r}-\delta a^{r}\cdot i_{\xi}E_{r}+i_{\xi}d\Theta(a,\delta a).\label{C.3}
\end{equation}
Furthermore, the LL-derivative of $\Theta(a,\delta a)$ is equal to
its ordinary Lie derivative as it has no free Lorentz index, and one
has 
\begin{equation}
i_{\xi}d\Theta(a,\delta a)=\delta_{\xi}\Theta(a,\delta a)-di_{\xi}\Theta(a,\delta a)-\Pi_{\xi}(a,\delta a),\label{C.4}
\end{equation}
from \eqref{3.11}. Using \eqref{C.3} and \eqref{C.4}, \eqref{C.1}
takes the form 
\begin{equation}
\begin{split}d\left[\delta K_{\xi}-K_{\delta\xi}-i_{\xi}\Theta(a,\delta a)\right]= & \delta\Theta(a,\delta_{\xi}a)-\delta_{\xi}\Theta(a,\delta a)-\Theta(a,\delta_{\delta\xi}a)\\
 & +\delta a^{r}\cdot i_{\xi}E_{r}+i_{\xi}a^{r}\cdot\delta E_{r}-\chi_{\xi}\cdot\delta E_{\omega},
\end{split}
\label{3.13}
\end{equation}
where we also used the explicit forms of $\psi_{\xi}$ and $\Pi_{\xi}$
given in \eqref{1.50} and \eqref{3.12}, respectively. Now, one can
define the extended off-shell ADT current as \cite{76} 
\begin{equation}
\begin{split}\mathcal{J}_{ADT}(a,\delta a,\delta_{\xi}a)= & \delta a^{r}\cdot i_{\xi}E_{r}+i_{\xi}a^{r}\cdot\delta E_{r}-\chi_{\xi}\cdot\delta E_{\omega}\\
 & +\delta\Theta(a,\delta_{\xi}a)-\delta_{\xi}\Theta(a,\delta a)-\Theta(a,\delta_{\delta\xi}a).
\end{split}
\label{3.14}
\end{equation}
Here, note that from the explicit form of $\Theta(a,\delta_{\xi}a)$
given in \eqref{1.48}, the last line of $\mathcal{J}_{ADT}$ becomes
\begin{equation}
\delta\Theta(a,\delta_{\xi}a)-\delta_{\xi}\Theta(a,\delta a)-\Theta(a,\delta_{\delta\xi}a)=\tilde{g}_{rs}\delta_{\xi}a^{r}\cdot\delta a^{s}.\label{3.15}
\end{equation}
Then, assuming that $\xi$ is a Killing vector field for which $\delta_{\xi}a^{r}=0$,
and if the field equations are satisfied, that is $E_{r}=0$, the
extended off-shell ADT current reduces to $\mathcal{J}_{\text{ADT}}=i_{\xi}a^{r}\cdot\delta E_{r}$
which is the on-shell ADT current given in \cite{Abbott,Deser_Tekin-PRL,28,Deser_Tekin-PRD,Senturk}.

Now, let us study some special cases which shed light on $\mathcal{J}_{ADT}$
definition. Consider the case where the $\xi$ is a Killing vector
field, then one can define the off-shell ADT current 
\begin{equation}
\tilde{\mathcal{J}}_{ADT}(a,\delta a,\delta_{\xi}a)=\delta a^{r}\cdot i_{\xi}E_{r}+i_{\xi}a^{r}\cdot\delta E_{r}-\chi_{\xi}\cdot\delta E_{\omega},\label{3.19}
\end{equation}
whose analog in metric formalism was given in \cite{Kim,32}. Taking
the exterior derivative of $\tilde{\mathcal{J}}_{ADT}$ yields 
\begin{equation}
d\tilde{\mathcal{J}}_{ADT}=\delta_{\xi}a^{r}\cdot\delta E_{r}-\delta a^{r}\cdot\delta_{\xi}E_{r},\label{3.20}
\end{equation}
which implies $\tilde{\mathcal{J}}_{ADT}$ is conserved, and one can
have the exact from 
\begin{equation}
\delta_{\xi}a^{r}\cdot\delta E_{r}-\delta a^{r}\cdot\delta_{\xi}E_{r}=-d\left(\tilde{g}_{rs}\delta_{\xi}a^{r}\cdot\delta a^{s}\right)=-d\mathcal{J}_{\Delta}.\label{3.21}
\end{equation}
With this result, one can write $\mathcal{J}_{ADT}$ as 
\begin{equation}
\mathcal{J}_{ADT}=\tilde{\mathcal{J}}_{ADT}+\mathcal{J}_{\Delta}.\label{3.23}
\end{equation}
For another special case, if the field equations and their linearizations
are assumed to be satisfied, then $\mathcal{J}_{ADT}$ becomes the
symplectic current \cite{9,33,34,Wald,37} 
\begin{equation}
\Omega_{\text{LW}}(a,\delta a,\delta_{\xi}a)=\delta\Theta(a,\delta_{\xi}a)-\delta_{\xi}\Theta(a,\delta a)-\Theta(a,\delta_{\delta\xi}a).\label{3.16}
\end{equation}
implying that $\mathcal{J}_{ADT}$ is the appropriate off-shell extension
of the on-shell ADT current so that $\xi$ is an asymptotic symmetry.
In the metric formalism, the analog of $\mathcal{J}_{ADT}$ was given
in \eqref{3.14}.

It is also possible to define an associated extended off-shell ADT
charge as 
\begin{equation}
\mathcal{Q}_{\text{ADT}}(a,\delta a,\delta_{\xi}a)=\delta K_{\xi}-K_{\delta\xi}-i_{\xi}\Theta(a,\delta a),\label{3.17}
\end{equation}
which defines quasi-local conserved charge for the (asymptotic) Killing
vector field $\xi$. Note that \eqref{3.13} can be written as 
\begin{equation}
\mathcal{J}_{\text{ADT}}(a,\delta a;\xi)=d\mathcal{Q}_{\text{ADT}}(a,\delta a;\xi),\label{3.18}
\end{equation}
in terms of $\mathcal{Q}_{\text{ADT}}$ definition.

Lastly, let us clarify that the variations $\delta$ and $\delta_{\xi}$
do not commute in general by considering the action of $(\delta\delta_{\xi}-\delta_{\xi}\delta)$
on $a^{r}$ as 
\begin{equation}
\begin{split}(\delta\delta_{\xi}-\delta_{\xi}\delta)a^{r} & =(\mathfrak{L}_{\xi}\delta a^{r}+\mathfrak{L}_{\delta\xi}a^{r}-\delta_{\omega}^{r}d\chi_{\delta\xi})-(\mathfrak{L}_{\xi}\delta a^{r})\\
 & =\delta_{\delta_{\xi}}a^{r},
\end{split}
\label{3.24}
\end{equation}
where again we used $\delta_{\xi}\delta\omega^{a}=\mathfrak{L}_{\xi}\delta\omega^{a}$.
Note that the commutator of the two variations vanishes when $\xi$
does not depend on the dynamical fields.

\section{Off-shell extension of the covariant phase space method}

\label{S.4.0} In this section, we define the Lee-Wald symplectic
current \cite{9,14,33,34,Wald,37} in the first order formalism. For
diffeomorphism covariant theories, the symplectic current was introduced
in \cite{33} while the generalization to noncovariant theories was
carried out in \cite{14}. As we discussed above, since the total
variation is covariant, the covariant phase space method can be used
to obtain conserved charges for the gravity theories formulated in
the first order formalism.

To derive Lee-Wald symplectic current, let us consider the commutator
$\delta_{[1,2]}$ of two arbitrary variations $\delta_{1}$ and $\delta_{2}$
acting on the Lagrangian. The second variations of the Lagrangian
simply become 
\begin{align}
\delta_{1}\delta_{2}L & =\delta_{1}\delta_{2}a^{r}\cdot E_{r}+\delta_{2}a^{r}\cdot\delta_{1}E_{r}+d\delta_{1}\Theta(a,\delta_{2}a),\label{4.3}\\
\delta_{2}\delta_{1}L & =\delta_{2}\delta_{1}a^{r}\cdot E_{r}+\delta_{1}a^{r}\cdot\delta_{2}E_{r}+d\delta_{2}\Theta(a,\delta_{1}a),\label{4.4}
\end{align}
by using \eqref{1.46}, that is $\delta L=\delta a^{r}\cdot E_{r}+d\Theta(a,\delta a)$;
and then, the commutator of $\delta_{1}$ and $\delta_{2}$ takes
the form 
\begin{equation}
\begin{split}\delta_{[1,2]}L= & \delta_{[1,2]}a^{r}\cdot E_{r}+\delta_{2}a^{r}\cdot\delta_{1}E_{r}-\delta_{1}a^{r}\cdot\delta_{2}E_{r}\\
 & +d\left[\delta_{1}\Theta(a,\delta_{2}a)-\delta_{2}\Theta(a,\delta_{1}a)\right].
\end{split}
\label{4.5}
\end{equation}
In addition, using \eqref{1.46}, one can also have 
\begin{equation}
\delta_{[1,2]}L=\delta_{[1,2]}a^{r}\cdot E_{r}+d\Theta(a,\delta_{[1,2]}a).\label{4.6}
\end{equation}
Using the two forms of $\delta_{[1,2]}L$, one can have 
\begin{equation}
d\Omega_{\text{LW}}(a;\delta_{1}a,\delta_{2}a)=\delta_{1}a^{r}\cdot\delta_{2}E_{r}-\delta_{2}a^{r}\cdot\delta_{1}E_{r},\label{4.7}
\end{equation}
where the Lee-Wald symplectic current $\Omega_{\text{LW}}(a;\delta_{1}a,\delta_{2}a)$
is defined as 
\begin{equation}
\Omega_{\text{LW}}(a;\delta_{1}a,\delta_{2}a)=\delta_{1}\Theta(a,\delta_{2}a)-\delta_{2}\Theta(a,\delta_{1}a)-\Theta(a,\delta_{[1,2]}a).\label{4.8}
\end{equation}
Note that the symplectic current is conserved when $a^{r}$ and $\delta a^{r}$
satisfy the field equations and the linearized field equations, respectively.
For Chern-Simons like theories of gravity, the symplectic current
can be written in the form 
\begin{equation}
\Omega_{\text{LW}}(a;\delta_{1}a,\delta_{2}a)=\tilde{g}_{rs}\delta_{2}a^{r}\cdot\delta_{1}a^{s},\label{4.9}
\end{equation}
which is closed, skew-symmetric, and nondegenerate.

Now, let us set $\delta_{1}=\delta$ and $\delta_{2}=\delta_{\xi}$,
so the commutator becomes $\delta_{[1,2]}=\delta_{\delta\xi}$ and
it vanishes when $\xi$ does not depend on the dynamical fields. For
this case, \eqref{4.8} and \eqref{4.9} becomes equal to \eqref{3.16}
and \eqref{3.15}, respectively. With this identification of the two
variations, $d\Omega_{\text{LW}}$ is 
\begin{equation}
d\Omega_{\text{LW}}(a;\delta a,\delta_{\xi}a)=\delta a^{r}\cdot\delta_{\xi}E_{r}-\delta_{\xi}a^{r}\cdot\delta E_{r},\label{4.10}
\end{equation}
so one has $d\Omega_{\text{LW}}=d\mathcal{J}_{\Delta}$ which yields
\begin{equation}
\mathcal{J}_{\Delta}=\Omega_{\text{LW}}(a;\delta a,\delta_{\xi}a)+dZ_{\xi}(a,\delta a),\label{4.11}
\end{equation}
where $Z_{\xi}(a,\delta a)$ is an arbitrary 1-form. One can use this
result in $\mathcal{J}_{ADT}=\tilde{\mathcal{J}}_{ADT}+\mathcal{J}_{\Delta}$
and compare the final form with the extended off-shell ADT current
$\mathcal{J}_{ADT}$ given in \eqref{4.11} to find that $Z_{\xi}(a,\delta a)$
is simply zero. Then, $\mathcal{J}_{ADT}$ has the final form in terms
of the Lee-Wald symplectic current as 
\begin{equation}
\mathcal{J}_{ADT}=\tilde{\mathcal{J}}_{ADT}+\Omega_{\text{LW}}(a;\delta a,\delta_{\xi}a).\label{4.12}
\end{equation}
Note that $\mathcal{J}_{ADT}$ reduces to the off-shell ADT current
when $\xi$ is a Killing vector field. Furthermore, if the field equations
and the linearized field equations are satisfied, then $\mathcal{J}_{ADT}$
reduces to $\Omega_{\text{LW}}$ as expected.

\section{Quasi-local conserved charge}

\label{S.5.0} In this section, we define quasi-local conserved charges
with the basic assumption that the spacetime is globally hyperbolic.
The quasi-local charge perturbation for a diffeomorphism generator
$\xi$ can be defined as 
\begin{equation}
\delta Q(\xi)=-\frac{1}{8\pi G}\int_{\mathcal{V}}\mathcal{J}_{ADT}(a,\delta a;\xi),\label{5.1}
\end{equation}
where $\mathcal{V}\subseteq\mathcal{C}$ for a Cauchy surface $\mathcal{C}$.
Using \eqref{3.18} and Stokes' theorem, one can have 
\begin{equation}
\delta Q(\xi)=-\frac{1}{8\pi G}\int_{\Sigma}\mathcal{Q}_{ADT}(a,\delta a;\xi),\label{5.2}
\end{equation}
or, using the definition of $\mathcal{Q}_{ADT}$, one explicitly has
\begin{equation}
\delta Q(\xi)=-\frac{1}{8\pi G}\int_{\Sigma}\left[\delta K_{\xi}-K_{\delta\xi}-i_{\xi}\Theta(a,\delta a)\right],\label{5.3}
\end{equation}
where $\Sigma$ is the boundary of $\mathcal{V}$. Note that the first
term is the charge perturbation given by Komar \cite{101a}. On the
other hand, the second term is due to the fact that $\xi$ depends
on the dynamical fields, and the third term is the contribution of
surface term.

To find quasi-local conserved charges, \eqref{5.2} should be integrated
over a one-parameter path in the solution space. Assume that $a^{r}(s\mathcal{N})$
are the collection of fields solving the field equations of the Chern-Simons
like theory. Here, $\mathcal{N}$ is a free parameter in the solution
space of field equations, and the parameter $s$ is in the domain
$0\leq s\leq1$. Expanding $a^{r}(s\mathcal{N})$ in $s$ yields $a^{r}(s\mathcal{N})=a^{r}(0)+\frac{\partial}{\partial s}a^{r}(0)s+\cdots$.
Then, using $a^{r}=a^{r}(s\mathcal{N})$ and $\delta a^{r}=\frac{\partial}{\partial s}a^{r}(0)$
in \eqref{5.3}, the quasi-local conserved charge can be defined for
the Killing vector field $\xi$ as 
\begin{equation}
Q(\xi)=-\frac{1}{8\pi G}\int_{0}^{1}ds\int_{\Sigma}\mathcal{Q}_{ADT}(a|s;\xi).\label{5.4}
\end{equation}
As suggested by \eqref{3.14}, the quasi-local conserved charge is
conserved for both the Killing vectors and the asymptotic Killing
vectors. The quasi-local conserved charge for a given solution (represented
by $s=1$) is finite as the contribution due to background (represented
by $s=0$) is subtracted. Note that $Q$ is independent of the $\mathcal{C}$
choice since $\mathcal{J}_{ADT}$ is conserved off-shell for any diffeomorphism
generator $\xi$. In addition, it is independent of the $\mathcal{V}$
choice due to the integration over one-parameter path on the solution
space.

Using the Noether potential \eqref{3.10} and the surface term \eqref{1.48},
the quasi-local conserved charge perturbation for a Chern-Simons like
theory becomes 
\begin{equation}
\delta Q(\xi)=-\frac{1}{8\pi G}\int_{\Sigma}\left(\tilde{g}_{rs}i_{\xi}a^{r}-\tilde{g}_{\omega s}\chi_{\xi}\right)\cdot\delta a^{s}.\label{5.5}
\end{equation}
The algebra of conserved charges is \cite{39,40}\textcolor{red}{}\footnote{In the paper \cite{28}, Eq.~\eqref{5.6} was proven under the assumption
of linearity and then was generalized in \cite{46} to the nonlinear
cases such as\eqref{5.4}.} 
\begin{equation}
\left\{ Q(\xi_{1}),Q(\xi_{2})\right\} _{\text{D.B.}}=Q\left(\left[\xi_{1},\xi_{2}\right]\right)+\mathcal{C}\left(\xi_{1},\xi_{2}\right),\label{5.6}
\end{equation}
where $\mathcal{C}\left(\xi_{1},\xi_{2}\right)$ is central extension
term, and the Dirac bracket is defined as 
\begin{equation}
\left\{ Q(\xi_{1}),Q(\xi_{2})\right\} _{\text{D.B.}}=\frac{1}{2}\left(\delta_{\xi_{2}}Q(\xi_{1})-\delta_{\xi_{1}}Q(\xi_{2})\right).\label{5.7}
\end{equation}
Then, the central extension term becomes 
\begin{equation}
\mathcal{C}\left(\xi_{1},\xi_{2}\right)=\frac{1}{2}\left(\delta_{\xi_{2}}Q(\xi_{1})-\delta_{\xi_{1}}Q(\xi_{2})\right)-Q\left(\left[\xi_{1},\xi_{2}\right]\right).\label{5.8}
\end{equation}
Note that in \eqref{5.6}, $\left[\xi_{1},\xi_{2}\right]$ is a modified
version of the Lie brackets defined as \cite{41} 
\begin{equation}
\left[\xi_{1},\xi_{2}\right]=\pounds_{\xi_{1}}\xi_{2}-\delta_{\xi_{1}}^{(g)}\xi_{2}+\delta_{\xi_{2}}^{(g)}\xi_{1},\label{5.9}
\end{equation}
where $\delta_{\xi_{1}}^{(g)}\xi_{2}$ denotes the change induced
in $\xi_{2}$ due to the variation of metric $\delta_{_{\xi_{1}}}g_{\mu\nu}=\pounds_{\xi_{1}}g_{\mu\nu}$.
In addition, \eqref{5.6} reduces to the ordinary Lie brackets $\left[\xi_{1},\xi_{2}\right]_{\text{Lie}}=\pounds_{\xi_{1}}\xi_{2}$
when $\xi$ does not depend on dynamical fields, that is $\delta_{\xi_{1}}^{(g)}\xi_{2}=\delta_{\xi_{2}}^{(g)}\xi_{1}=0$.

\section{Black hole entropy in {Chern-Simons-like theories of gravity}}

\label{S.6.0} Wald prescription will be utilized to find an expression
for black hole entropy in the Chern-Simons like theories. Wald's suggestion
is to identify the black hole entropy as the conserved charge associated
with the horizon-generating Killing vector field $\zeta$ which vanishes
on the bifurcation surface $\mathcal{B}$. Now, take $\Sigma$ in
\eqref{5.5} to be the bifurcation surface $\mathcal{B}$ then one
has 
\begin{equation}
Q(\zeta)=\frac{1}{8\pi G}\tilde{g}_{\omega r}\int_{\mathcal{B}}\chi_{\zeta}\cdot a^{r}.\label{6.1}
\end{equation}
Up to now, $\lambda_{\xi}^{ab}$ and so $\chi_{\xi}^{a}$ have been
considered to arbitrary functions of the spacetime coordinates and
of the diffeomorphism generator $\xi$. To further obtain an explicit
expression for $\lambda_{\xi}^{ab}$, in \cite{8} it was suggested
that the LL-derivative of $e^{a}$ vanishes when $\xi$ is a Killing
vector field. We follow \cite{8} to find an expression for $\chi_{\xi}^{a}$.
The LL-derivative of $e^{a}$ is given as 
\begin{equation}
\mathfrak{L}_{\xi}e_{\hspace{1.5mm}\mu}^{a}=\pounds_{\xi}e_{\hspace{1.5mm}\mu}^{a}+\lambda_{\xi\hspace{1mm}b}^{\hspace{1mm}a}e_{\hspace{1.5mm}\mu}^{b}.\label{D.1}
\end{equation}
Now, consider following contraction 
\begin{equation}
e_{a\mu}\mathfrak{L}_{\xi}e_{\hspace{1.5mm}\nu}^{a}=e_{a(\mu}\mathfrak{L}_{\xi}e_{\hspace{1.5mm}\nu)}^{a}+e_{a[\mu}\mathfrak{L}_{\xi}e_{\hspace{1.5mm}\nu]}^{a}.\label{D.2}
\end{equation}
The symmetric part of $e_{a\mu}\mathfrak{L}_{\xi}e_{\hspace{1.5mm}\nu}^{a}$
can be written as 
\begin{equation}
e_{a(\mu}\mathfrak{L}_{\xi}e_{\hspace{1.5mm}\nu)}^{a}=e_{a(\mu}\pounds_{\xi}e_{\hspace{1.5mm}\nu)}^{a}+\lambda_{\xi\hspace{0.5mm}ab}e_{\hspace{1.5mm}(\mu}^{a}e_{\hspace{1.5mm}\nu)}^{b}=e_{a(\mu}\pounds_{\xi}e_{\hspace{1.5mm}\nu)}^{a}\label{D.3}
\end{equation}
where anti-symmetric property of $\lambda_{\xi}^{ab}$ in the Latin
indices was used. From the Lie derivative of the metric, it follows
that 
\begin{equation}
\pounds_{\xi}g_{\mu\nu}=\pounds_{\xi}\left(e_{a\mu}e_{\hspace{1.5mm}\nu}^{a}\right)=e_{a\mu}\pounds_{\xi}e_{\hspace{1.5mm}\nu}^{a}+e_{a\nu}\pounds_{\xi}e_{\hspace{1.5mm}\mu}^{a}=2e_{a(\mu}\pounds_{\xi}e_{\hspace{1.5mm}\nu)}^{a}\label{D.4}
\end{equation}
By substituting the last expression in the previous one, one finds
that the symmetric part of $e_{a\mu}\mathfrak{L}_{\xi}e_{\hspace{1.5mm}\nu}^{a}$
reduces to 
\begin{equation}
e_{a(\mu}\mathfrak{L}_{\xi}e_{\hspace{1.5mm}\nu)}^{a}=\frac{1}{2}\pounds_{\xi}g_{\mu\nu}.\label{D.5}
\end{equation}
The anti-symmetric part of $e_{a\mu}\mathfrak{L}_{\xi}e_{\hspace{1.5mm}\nu}^{a}$
becomes 
\begin{equation}
\begin{split}e_{a[\mu}\mathfrak{L}_{\xi}e_{\hspace{1.5mm}\nu]}^{a}= & e_{a[\mu}\pounds_{\xi}e_{\hspace{1.5mm}\nu]}^{a}+\lambda_{\xi\hspace{0.5mm}ab}e_{\hspace{1.5mm}[\mu}^{a}e_{\hspace{1.5mm}\nu]}^{b}\\
= & e_{a[\mu}\pounds_{\xi}e_{\hspace{1.5mm}\nu]}^{a}+\lambda_{\xi\hspace{0.5mm}ab}e_{\hspace{1.5mm}\mu}^{a}e_{\hspace{1.5mm}\nu}^{b}.
\end{split}
\label{D.6}
\end{equation}
Using the last two equations in \eqref{D.2} yields 
\begin{equation}
e_{a\mu}\mathfrak{L}_{\xi}e_{\hspace{1.5mm}\nu}^{a}=\frac{1}{2}\pounds_{\xi}g_{\mu\nu}+e_{a[\mu}\pounds_{\xi}e_{\hspace{1.5mm}\nu]}^{a}+\lambda_{\xi\hspace{0.5mm}ab}e_{\hspace{1.5mm}\mu}^{a}e_{\hspace{1.5mm}\nu}^{b}.\label{D.7}
\end{equation}
Assuming $\pounds_{\xi}g_{\mu\nu}=0$ and demanding that the LL-derivative
vanishes for a Killing vector field, i.e. $\mathfrak{L}_{\xi}e_{\hspace{1.5mm}\mu}^{a}=0$,
we find from the last equation an expression for $\lambda_{\xi}^{ab}$:
\begin{equation}
\lambda_{\xi}^{ab}=e^{\sigma[a}\pounds_{\xi}e_{\hspace{1.5mm}\sigma}^{b]}.\label{D.8}
\end{equation}
By substituting this into \eqref{D.1}, one gets 
\begin{equation}
\mathfrak{L}_{\xi}e_{\hspace{1.5mm}\mu}^{a}=\frac{1}{2}e^{a\nu}\pounds_{\xi}g_{\mu\nu}.\label{D.9}
\end{equation}
It is clear that that the LL-derivative of the dreibein will vanish
when $\xi$ is a Killing vector field. Consider the usual Lie derivative
of the dreibein 
\begin{equation}
\begin{split}\pounds_{\xi}e_{\hspace{1.5mm}\mu}^{a}= & \xi^{\sigma}\partial_{\sigma}e_{\hspace{1.5mm}\mu}^{a}+e_{\hspace{1.5mm}\sigma}^{a}\partial_{\mu}\xi^{\sigma}\\
= & \xi^{\sigma}\nabla_{\sigma}e_{\hspace{1.5mm}\mu}^{a}+e_{\hspace{1.5mm}\sigma}^{a}\nabla_{\mu}\xi^{\sigma}+e_{\hspace{1.5mm}\alpha}^{a}\xi^{\sigma}\left(\Gamma_{\sigma\mu}^{\alpha}-\Gamma_{\mu\sigma}^{\alpha}\right)\\
= & \xi^{\sigma}\nabla_{\sigma}e_{\hspace{1.5mm}\mu}^{a}+e_{\hspace{1.5mm}\sigma}^{a}\nabla_{\mu}\xi^{\sigma}+\left(i_{\xi}T^{a}\right)_{\mu}
\end{split}
\label{D.10}
\end{equation}
where we used $T^{a}=e_{\hspace{1.5mm}\alpha}^{a}\Gamma_{\mu\nu}^{\alpha}dx^{\mu}\wedge dx^{\nu}$.
Contracting with $e^{b\mu}$ and using \eqref{1.12}, one arrives
at 
\begin{equation}
\lambda_{\xi}^{ab}=e^{\sigma[a}\pounds_{\xi}e_{\hspace{1.5mm}\sigma}^{b]}=i_{\xi}\omega^{ab}+e_{\hspace{1.5mm}\mu}^{[a}e_{\hspace{1.5mm}\nu}^{b]}\nabla^{\mu}\xi^{\nu}+e^{\sigma[a}\left(i_{\xi}T^{b]}\right)_{\sigma}\label{D.11}
\end{equation}
Since $\chi_{\xi}^{a}$ is related to $\lambda_{\xi}^{ab}$ as in
\eqref{1.45}, one has 
\begin{equation}
\chi_{\xi}^{a}=i_{\xi}\omega^{a}+\frac{1}{2}\varepsilon_{\hspace{1.5mm}bc}^{a}e^{\nu b}(i_{\xi}T^{c})_{\nu}+\frac{1}{2}\varepsilon_{\hspace{1.5mm}bc}^{a}e^{b\mu}e^{c\nu}\nabla_{\mu}\xi_{\nu}.\label{D.12}
\end{equation}
where \eqref{1.29} was used. One can write \eqref{1.54} as 
\begin{equation}
T(\omega)=\kappa_{\hspace{1.5mm}b}^{a}\wedge e^{b}\label{D.13}
\end{equation}
where an antisymmetric tensor is introduced: 
\begin{equation}
\kappa_{\hspace{1.5mm}b}^{a}=-\varepsilon_{\hspace{1.5mm}bc}^{a}k^{c}.\label{D.14}
\end{equation}
Equivalently, one has $\kappa_{\mu\nu\lambda}=e_{\hspace{1.5mm}\mu}^{a}e_{\hspace{1.5mm}\nu}^{b}\kappa_{ab\lambda}$
which is anti-symmetric in $\mu$ and $\nu$. One can write the Cartan
torsion tensor in terms of $\kappa_{\mu\nu\lambda}$, 
\begin{equation}
T_{\hspace{1.5mm}\mu\nu}^{\alpha}=-\frac{1}{2}\left(\kappa_{\hspace{1.5mm}\mu\nu}^{\alpha}-\kappa_{\hspace{1.5mm}\nu\mu}^{\alpha}\right).\label{D.15}
\end{equation}
By substituting \eqref{D.15} into \eqref{1.9} and using anti-symmetric
property of $\kappa_{\mu\nu\lambda}$, one can show that 
\begin{equation}
C_{\hspace{1.5mm}\mu\nu}^{\alpha}=2T_{\hspace{1.5mm}\mu\nu}^{\alpha}+\kappa_{\hspace{1.5mm}\mu\nu}^{\alpha}.\label{D.16}
\end{equation}
Now, we consider the covariant derivative of a vector field $\xi$
\begin{equation}
\nabla_{\mu}\xi^{\nu}=\partial_{\mu}\xi^{\nu}+\Gamma_{\mu\sigma}^{\nu}\xi^{\sigma}=\hat{\nabla}_{\mu}\xi^{\nu}+C_{\hspace{1.5mm}\mu\sigma}^{\nu}\xi^{\sigma}=\hat{\nabla}_{\mu}\xi^{\nu}-2\left(i_{\xi}T^{\nu}\right)_{\mu}+\kappa_{\hspace{1.5mm}\mu\sigma}^{\nu}\xi^{\sigma}\label{D.17}
\end{equation}
where \eqref{1.8} and \eqref{D.16} were used in second and in third
equalities. Playing with the last expression, one has 
\begin{equation}
\begin{split}\frac{1}{2}\varepsilon_{\hspace{1.5mm}bc}^{a}e^{b\mu}e^{c\nu}\nabla_{\mu}\xi_{\nu}= & \frac{1}{2}\varepsilon_{\hspace{1.5mm}bc}^{a}e^{b\mu}e^{c\nu}\hat{\nabla}_{\mu}\xi_{\nu}-\frac{1}{2}\varepsilon_{\hspace{1.5mm}bc}^{a}e^{\nu b}(i_{\xi}T^{c})_{\nu}-\frac{1}{2}\varepsilon_{\hspace{1.5mm}bc}^{a}i_{\xi}\kappa^{bc}\\
= & \frac{1}{2}\varepsilon_{\hspace{1.5mm}bc}^{a}e^{b\mu}e^{c\nu}\hat{\nabla}_{\mu}\xi_{\nu}-\frac{1}{2}\varepsilon_{\hspace{1.5mm}bc}^{a}e^{\nu b}(i_{\xi}T^{c})_{\nu}-i_{\xi}k^{a}
\end{split}
\label{D.18}
\end{equation}
where we made use of \eqref{D.14}. By substituting \eqref{1.30}
and \eqref{D.18} into \eqref{D.12}, one finds 
\begin{equation}
\chi_{\xi}^{a}=i_{\xi}\Omega^{a}+\frac{1}{2}e_{\hspace{1.5mm}\sigma}^{a}\epsilon^{\sigma\mu\nu}\hat{\nabla}_{\mu}\xi_{\nu}.\label{6.2}
\end{equation}
This equation is valid for both torsion-free theories and theories
with torsion. This choice for $\chi_{\xi}$ yields 
\begin{equation}
\mathfrak{L}_{\xi}e_{\hspace{1.5mm}\mu}^{a}=\pounds_{\xi}e_{\hspace{1.5mm}\mu}^{a}+\lambda_{\xi\hspace{1mm}b}^{\hspace{1mm}a}e_{\hspace{1.5mm}\mu}^{b}=\pounds_{\xi}e_{\hspace{1.5mm}\mu}^{a}+\left(\chi_{\xi}\times e_{\mu}\right)^{a}=\frac{1}{2}e^{a\nu}\pounds_{\xi}g_{\mu\nu}.\label{6.3}
\end{equation}
On the bifurcation surface, $\zeta|_{\mathcal{B}}=0$ and 
\begin{equation}
\hat{\nabla}_{\mu}\zeta_{\nu}|_{\mathcal{B}}=\kappa_{H}n_{\mu\nu}\label{6.4}
\end{equation}
where $\kappa_{H}$ and $n_{\mu\nu}$ are the surface gravity of the
black hole and the bi-normal vector to the bifurcation surface. Thus,
from \eqref{6.2}, one has 
\begin{equation}
\chi_{\zeta}^{a}|_{\mathcal{B}}=\frac{1}{2}\kappa_{H}e_{\hspace{1.5mm}\sigma}^{a}\epsilon^{\sigma\mu\nu}n_{\mu\nu}\label{6.5}
\end{equation}
By defining the dual bi-normal to the bifurcation surface 
\begin{equation}
N^{a}=\frac{1}{2}e_{\hspace{1.5mm}\sigma}^{a}\epsilon^{\sigma\mu\nu}n_{\mu\nu},\label{6.6}
\end{equation}
we can write \eqref{6.5} as 
\begin{equation}
\chi_{\zeta}^{a}|_{\mathcal{B}}=\kappa_{H}N^{a}.\label{6.7}
\end{equation}
The bi-normal to the bifurcation surface is normalized to $-2$, and
$N^{a}$ is normalized to $+1$. Finally, using \eqref{6.1} the black
hole entropy in the Chern-Simons like theory can be defined as \cite{75}
\begin{equation}
S=\frac{2\pi}{\kappa_{H}}Q(\zeta)=-\frac{1}{4G}\tilde{g}_{\omega r}\int_{\mathcal{B}}N\cdot a^{r}\label{6.9}
\end{equation}
For a stationary black hole solution, the horizon is a circle of radius
$r=r_{H}$, hence the non-zero components of bi-normal to that horizon
are $n_{tr}=-n_{rt}$. The only non-zero component of $N^{a}$ is
$N^{\phi}=(g_{\phi\phi})^{-1/2}$. Thus, \eqref{6.9} reduces to 
\begin{equation}
S=-\frac{1}{4G}\tilde{g}_{\omega r}\int_{r=r_{h}}(g_{\phi\phi})^{-1/2}a_{\hspace{1.5mm}\phi\phi}^{r}d\phi,\label{6.11}
\end{equation}
which is a generic expression for the entropy of black holes of the
Chern-Simons like theories.

\section{Asymptotically AdS$_{3}$ spacetimes}

\label{S.7.0} The Brown-Henneaux boundary conditions \cite{39} are
appropriate for both cosmological Einstein's gravity and general massive
gravity models.

\subsection{Brown-Henneaux boundary conditions}

\label{S.7.1} Now, we summarize the Brown-Henneaux (BH) boundary
conditions \cite{39}. Let $r$ and $x^{\pm}=\frac{t}{l}\pm\phi$
be the radial and the null coordinates. The BH boundary conditions
for asymptotically AdS$_{3}$ spacetimes are defined as 
\begin{equation}
\begin{split}g_{\pm\pm}= & f_{\pm\pm}+\mathcal{O}\left(\frac{1}{r}\right),\\
g_{+-}= & -\frac{r^{2}}{2}+f_{+-}+\mathcal{O}\left(\frac{1}{r}\right),\\
g_{rr}= & \frac{l^{2}}{r^{2}}+\frac{f_{rr}}{r^{4}}+\mathcal{O}\left(\frac{1}{r^{5}}\right),\\
g_{r\pm}= & \mathcal{O}\left(\frac{1}{r^{3}}\right),
\end{split}
\label{7.2}
\end{equation}
where $f_{\pm\pm}$, $f_{+-}$, $f_{rr}$ are arbitrary functions
of the null coordinates $x^{\pm}$. As usual, under transformation
generated by a vector field $\xi$, one has 
\begin{equation}
\delta_{\chi}g_{\mu\nu}=\pounds_{\xi}g_{\mu\nu}.\label{7.3}
\end{equation}
The boundary conditions are preserved by the following Killing vector
field $\xi$ (a.k.a as asymptotic symmetries) 
\begin{equation}
\begin{split}\xi^{\pm}= & T^{\pm}+\frac{l^{2}}{2r^{2}}\partial_{\mp}^{2}T^{\mp}+\mathcal{O}\left(\frac{1}{r^{4}}\right),\\
\xi^{r}= & -\frac{r}{2}\left(\partial_{+}T^{+}+\partial_{-}T^{-}\right)+\mathcal{O}\left(\frac{1}{r}\right),
\end{split}
\label{7.4}
\end{equation}
where $T^{\pm}=T^{\pm}(x^{\pm})$ are arbitrary functions. Under the
action of an asymptotic symmetry generator $\xi$, the dynamical fields
transform as 
\begin{equation}
\begin{split}\delta_{\xi}f_{rr}= & \partial_{+}(T^{+}f_{rr})+\partial_{-}(T^{-}f_{rr}),\\
\delta_{\xi}f_{+-}= & \partial_{+}(T^{+}f_{+-})+\partial_{-}(T^{-}f_{+-}),\\
\delta_{\xi}f_{\pm\pm}= & 2f_{\pm\pm}\partial_{\pm}T^{\pm}+T^{\pm}\partial_{\pm}f_{\pm\pm}-\frac{l^{2}}{2}\left(\partial_{\pm}T^{\pm}+\partial_{\pm}^{3}T^{\pm}\right).
\end{split}
\label{7.5}
\end{equation}
One can consider the Fourier modes, $T^{\pm}(x^{\pm})=e^{imx^{\pm}}$,
then the asymptotic Killing vectors are 
\begin{equation}
\xi_{m}^{\pm}=\frac{1}{2}e^{imx^{\pm}}\left[l\left(1-\frac{l^{2}m^{2}}{2r^{2}}\right)\partial_{t}-imr\partial_{r}\pm\left(1+\frac{l^{2}m^{2}}{2r^{2}}\right)\partial_{\phi}\right],\label{7.6}
\end{equation}
and they satisfy the Witt algebra 
\begin{equation}
i[\xi_{m}^{\pm},\xi_{n}^{\pm}]=(m-n)\xi_{m+n}^{\pm},\label{7.7}
\end{equation}
which can be obtained by substituting \eqref{7.6} into \eqref{5.9}.
The dreibeins of the global AdS$_{3}$ spacetime are 
\begin{equation}
\bar{e}^{0}=\frac{r}{l}dt,\hspace{0.5cm}\bar{e}^{1}=\frac{l}{r}dr,\hspace{0.5cm}\bar{e}^{2}=rd\phi,\label{7.8}
\end{equation}
Using \eqref{6.2}, the interior product of the spin-connection yields
\begin{equation}
i_{\xi_{n}^{\pm}}\Omega^{a}-\chi_{\xi_{n}^{\pm}}^{a}=\pm\frac{1}{l}(\xi_{n}^{\pm})^{a}.\label{7.9}
\end{equation}
Since the AdS$_{3}$ spacetime solves \eqref{2.2} and \eqref{2.3},
one finds 
\begin{equation}
\begin{split} & \delta_{\xi_{n}^{\pm}}\Omega_{\hspace{1.5mm}\phi}^{0}\pm\frac{1}{l}\delta_{\xi_{n}^{\pm}}e_{\hspace{1.5mm}\phi}^{0}=-\frac{iln^{3}}{2r}e^{inx^{\pm}},\\
 & \delta_{\xi_{n}^{\pm}}\Omega_{\hspace{1.5mm}\phi}^{1}\pm\frac{1}{l}\delta_{\xi_{n}^{\pm}}e_{\hspace{1.5mm}\phi}^{1}=0,\\
 & \delta_{\xi_{n}^{\pm}}\Omega_{\hspace{1.5mm}\phi}^{2}\pm\frac{1}{l}\delta_{\xi_{n}^{\pm}}e_{\hspace{1.5mm}\phi}^{2}=\pm\frac{iln^{3}}{2r}e^{inx^{\pm}}.
\end{split}
\label{7.10}
\end{equation}
where equations \eqref{1.58} and \eqref{1.59} were used.

The Lagrangian of Einstein's gravity with a negative cosmological
constant (EGN) is given by \eqref{1.33} and the non-zero components
of the flavor metric are $\tilde{g}_{e\omega}=\tilde{g}_{\omega e}=-1$.
Then, the conserved charge \eqref{5.4} for this case becomes 
\begin{equation}
Q_{E}(\xi_{m}^{\pm})=\frac{1}{8\pi G}\lim_{r\rightarrow\infty}\int_{0}^{2\pi}(\xi_{m}^{\pm})\cdot\left(\Delta\Omega_{\phi}\pm\frac{1}{l}\Delta e_{\phi}\right)d\phi,\label{7.11}
\end{equation}
where \eqref{7.9} was used. Also, $\Delta a^{r}=a_{(s=1)}^{r}-a_{(s=0)}^{r}$.
The connections corresponding to the two $SO(2,1)$ gauge groups are
defined as \eqref{1.42}, then \eqref{7.11} can be written as 
\begin{equation}
Q_{E}(\xi_{m}^{\pm})=\frac{1}{8\pi G}\lim_{r\rightarrow\infty}\int_{0}^{2\pi}(\xi_{m}^{\pm})\cdot\Delta A_{\phi}^{\pm}d\phi.\label{7.12}
\end{equation}
For BTZ black hole spacetime \eqref{2.4.2} at spatial infinity 
\begin{equation}
\Delta e_{\hspace{1.5mm}\phi}^{a}=0,\hspace{0.7cm}\Delta\Omega_{\hspace{1.5mm}\phi}^{0}=-\frac{r_{+}^{2}+r_{-}^{2}}{2lr},\hspace{0.7cm}\Delta\Omega_{\hspace{1.5mm}\phi}^{1}=0,\hspace{0.7cm}\Delta\Omega_{\hspace{1.5mm}\phi}^{2}=-\frac{r_{+}r_{-}}{lr},\label{7.13}
\end{equation}
and all these yield 
\begin{equation}
Q_{E}(\xi_{m}^{\pm})=\frac{l}{16G}\left(\frac{r_{+}\mp r_{-}}{l}\right)^{2}\delta_{m,0}.\label{7.14}
\end{equation}
Similarly\eqref{5.3} reduces to 
\begin{equation}
\delta_{\xi_{n}^{\pm}}Q_{E}(\xi_{m}^{\pm})=\frac{1}{8\pi G}\lim_{r\rightarrow\infty}\int_{0}^{2\pi}(\xi_{m}^{\pm})\cdot\delta_{\xi_{n}^{\pm}}A_{\hspace{1.5mm}\phi}^{\pm}d\phi.\label{7.15}
\end{equation}
where we set $\delta=\delta_{\xi}$. By using \eqref{7.10} and \eqref{1.42},
one can show that \eqref{7.15} yields 
\begin{equation}
\delta_{\xi_{n}^{\pm}}Q_{E}(\xi_{m}^{\pm})=\frac{iln^{3}}{8G}\delta_{m+n,0}.\label{7.16}
\end{equation}
Plugging \eqref{7.14} and \eqref{7.16} into \eqref{5.8} one finds
the central extension 
\begin{equation}
C_{E}(\xi_{m}^{\pm},\xi_{n}^{\pm})=i\frac{l}{8G}\left[n^{3}-\left(\frac{r_{+}\mp r_{-}}{l}\right)^{2}n\right]\delta_{m+n,0}.\label{7.17}
\end{equation}
This result agrees with the previous results (for instance, see \cite{42})
but to obtain the usual $m$ dependence, one should make a constant
shift on $Q$ \cite{43}. Now, we set $Q(\xi_{n}^{\pm})\equiv\hat{L}_{n}^{\pm}$
and replace the brackets with commutators, namely $\{Q(\xi_{m}^{\pm}),Q(\xi_{n}^{\pm})\}_{\text{D.B.}}\equiv i[\hat{L}_{m}^{\pm},\hat{L}_{n}^{\pm}]$,
then \eqref{5.7} becomes the usual Virasoro algebra 
\begin{equation}
[\hat{L}_{m}^{\pm},\hat{L}_{n}^{\pm}]=(m-n)\hat{L}_{m+n}^{\pm}+\frac{c_{\pm}}{12}m(m^{2}-1)\delta_{m+n,0},\label{7.18}
\end{equation}
where $c_{\pm}=\frac{3l}{2G}$ are the central charges, and $\hat{L}_{n}^{\pm}$
are generators. S The conclusion is that the classical algebra of
the conserved charges is isomorphic to two copies of the Virasoro
algebra.

\subsection{General massive gravity}

\label{S.7.3} We introduced the general massive gravity (GMG) in
the subsection \ref{S.2.4}. For the global AdS$_{3}$ the background
curvature 2-form is given by \eqref{2.4}. Therefore, the Schouten
tensor and the Cotton tensor are 
\begin{equation}
S^{a}=-\frac{1}{2l^{2}}e^{a},\hspace{1cm}C^{a}=0,\label{7.19}
\end{equation}
and so the auxiliary fields \eqref{2.26} become 
\begin{equation}
h^{a}=\frac{1}{2\mu l^{2}}e^{a},\hspace{1.5cm}f^{a}=\frac{1}{2l^{2}}e^{a}.\label{7.20}
\end{equation}
A similar construction along the lines of the previous section yields
\cite{77} 
\begin{equation}
\begin{split}Q_{GMG}(\xi_{n}^{\pm})= & \left(\sigma\mp\frac{1}{\mu l}+\frac{1}{2m^{2}l^{2}}\right)Q_{E}(\xi_{n}^{\pm})\\
 & -\frac{1}{8\pi G}\lim_{r\rightarrow\infty}\int_{0}^{2\pi}d\phi(\Delta h_{\hspace{1.5mm}\phi}^{\mu}\pm\frac{1}{m^{2}l}\Delta f_{\hspace{1.5mm}\phi}^{\mu})(\xi_{n}^{\pm})_{\mu},
\end{split}
\label{7.21}
\end{equation}
where we used \eqref{2.25}. The relevant variations are 
\begin{equation}
\delta S_{\mu\nu}=\delta\mathcal{R}_{\mu\nu}-\frac{1}{4}g_{\mu\nu}\delta\mathcal{R}+\frac{3}{2l^{2}}\delta g_{\mu\nu},\label{7.22}
\end{equation}
where 
\begin{equation}
\begin{split} & \delta\mathcal{R}_{\mu\nu}=\frac{1}{2}\left(-\Box\delta g_{\mu\nu}-\nabla_{\mu}\nabla_{\nu}(g^{\alpha\beta}\delta g_{\alpha\beta})+\nabla^{\lambda}\nabla_{\mu}\delta g_{\lambda\nu}+\nabla^{\lambda}\nabla_{\nu}\delta g_{\lambda\mu}\right),\\
 & \delta\mathcal{R}=-\Box(g^{\alpha\beta}\delta g_{\alpha\beta})+\nabla^{\mu}\nabla^{\nu}\delta g_{\mu\nu}+\frac{2}{l^{2}}(g^{\alpha\beta}\delta g_{\alpha\beta}).
\end{split}
\label{7.23}
\end{equation}
The variation of the Cotton tensor is $\delta C_{\hspace{1.5mm}\nu}^{\mu}=\epsilon_{\nu}^{\hspace{1.5mm}\alpha\beta}\nabla_{\alpha}\delta S_{\hspace{1.5mm}\beta}^{\mu}$.
For the BTZ black hole solution at spatial infinity we have 
\begin{equation}
\Delta g_{tt}=\frac{r_{+}^{2}+r_{-}^{2}}{l^{2}},\hspace{1cm}\Delta g_{t\phi}=-\frac{r_{+}r_{-}}{l},\hspace{1cm}\Delta g_{rr}=\frac{l^{2}(r_{+}^{2}+r_{-}^{2})}{r^{4}},\label{7.24}
\end{equation}
then, $\Delta C_{\hspace{1.5mm}\phi}^{\mu}=\Delta S_{\hspace{1.5mm}\phi}^{\mu}=0$.
Therefore, \eqref{7.22} reduces to 
\begin{equation}
Q_{GMG}(\xi_{n}^{\pm})=\left(\sigma\mp\frac{1}{\mu l}+\frac{1}{2m^{2}l^{2}}\right)Q_{E}(\xi_{n}^{\pm})\label{7.25}
\end{equation}
which shows that the conserved charges of the BTZ black hole in generalized
massive gravity are equal to the conserved charge of the same solution
in Einstein's gravity multiplied by a constant. In a similar way,
one can show that \eqref{5.5} can be simplified as 
\begin{equation}
\begin{split}\delta_{\xi_{m}^{\pm}}Q_{GMG}(\xi_{n}^{\pm})= & \left(\sigma\mp\frac{1}{\mu l}+\frac{1}{2m^{2}l^{2}}\right)\delta_{\xi_{m}^{\pm}}Q_{E}(\xi_{n}^{\pm})\\
 & -\frac{1}{8\pi G}\lim_{r\rightarrow\infty}\int_{0}^{2\pi}d\phi(\delta_{\xi_{m}^{\pm}}h_{\hspace{1.5mm}\phi}^{\mu}\pm\frac{1}{m^{2}l}\delta_{\xi_{m}^{\pm}}f_{\hspace{1.5mm}\phi}^{\mu})(\xi_{n}^{\pm})_{\mu}.
\end{split}
\label{7.26}
\end{equation}
It can be shown that $\delta_{\xi_{m}^{\pm}}h_{\hspace{1.5mm}\phi}^{\mu}=\delta_{\xi_{m}^{\pm}}f_{\hspace{1.5mm}\phi}^{\mu}=0$,
then one has 
\begin{equation}
\delta_{\xi_{m}^{\pm}}Q_{GMG}(\xi_{n}^{\pm})=\left(\sigma\mp\frac{1}{\mu l}+\frac{1}{2m^{2}l^{2}}\right)\delta_{\xi_{m}^{\pm}}Q_{E}(\xi_{n}^{\pm}).\label{7.27}
\end{equation}
Using \eqref{7.24} and \eqref{7.27} in \eqref{5.8}, one finds 
\begin{equation}
C_{GMG}(\xi_{m}^{\pm},\xi_{n}^{\pm})=\left(\sigma\mp\frac{1}{\mu l}+\frac{1}{2m^{2}l^{2}}\right)C_{E}(\xi_{m}^{\pm},\xi_{n}^{\pm}),\label{7.28}
\end{equation}
from which one can read off the central charges of the general massive
gravity as 
\begin{equation}
c_{\pm}=\frac{3l}{2G}\left(\sigma\mp\frac{1}{\mu l}+\frac{1}{2m^{2}l^{2}}\right),\label{7.29}
\end{equation}
which agrees with \cite{44}. The eigenvalues of the Virasoro generators
$\hat{L}_{n}^{\pm}$ can also be found as 
\begin{equation}
l_{n}^{\pm}=\frac{l}{16G}\left(\sigma\mp\frac{1}{\mu l}+\frac{1}{2m^{2}l^{2}}\right)\left(\frac{r_{+}\mp r_{-}}{l}\right)^{2}\delta_{n,0},\label{7.30}
\end{equation}
The eigenvalues of the Virasoro generators $\hat{L}_{n}^{\pm}$ are
related to the energy $E$ and the angular momentum $J$ via 
\begin{equation}
E=l^{-1}(l_{0}^{+}+l_{0}^{-})=\frac{1}{8G}\left[\left(\sigma+\frac{1}{2m^{2}l^{2}}\right)\frac{r_{+}^{2}+r_{-}^{2}}{l^{2}}+\frac{2r_{+}r_{-}}{\mu l^{3}}\right],\label{7.31}
\end{equation}
\begin{equation}
J=l^{-1}(l_{0}^{+}-l_{0}^{-})=\frac{1}{8G}\left[\left(\sigma+\frac{1}{2m^{2}l^{2}}\right)\frac{2r_{+}r_{-}}{l}+\frac{r_{+}^{2}+r_{-}^{2}}{\mu l^{2}}\right].\label{7.32}
\end{equation}
and using the Cardy's formula \cite{45,46} (see also \cite{47}),
one can find the entropy of the black hole as 
\begin{equation}
S=2\pi\sqrt{\frac{c_{+}l_{0}^{+}}{6}}+2\pi\sqrt{\frac{c_{-}l_{0}^{-}}{6}},\label{7.33}
\end{equation}
or more explicitly 
\begin{equation}
S=\frac{\pi}{2G}\left[\left(\sigma+\frac{1}{2m^{2}l^{2}}\right)r_{+}+\frac{r_{-}}{\mu l}\right].\label{7.34}
\end{equation}
One can find the energy, the angular momentum and the entropy of a
given black hole from \eqref{5.4} by the corresponding vector fields.
Thus, for the GMG \eqref{2.25} becomes 
\begin{equation}
\delta Q(\xi)=\frac{1}{8\pi G}\int_{0}^{2\pi}d\phi\left[\left(\sigma+\frac{1}{2m^{2}l^{2}}\right)\xi_{a}+\frac{1}{\mu}\Xi_{a}\right]\cdot\delta\omega_{\hspace{1.5mm}\phi}^{a},\label{7.35}
\end{equation}
where the integration runs over a circle of arbitrary radius. Here
\eqref{7.35}, $\xi^{a}$ and $\Xi^{a}$ are given as 
\begin{equation}
\xi^{a}=e_{\hspace{1.5mm}\mu}^{a}\xi^{\mu},\hspace{1.5cm}\Xi^{a}=-\frac{1}{2}e_{\hspace{1.5mm}\lambda}^{a}\epsilon^{\lambda\mu\nu}\nabla_{\mu}\xi_{\nu}.\label{7.36}
\end{equation}
The energy, angular momentum and the entropy of the BTZ black hole
correspond to the following Killing vectors, respectively 
\begin{equation}
\xi_{(E)}=\partial_{t},\hspace{1cm}\xi_{(J)}=-\partial_{\phi},\hspace{1cm}\xi_{(S)}=\frac{4\pi}{\kappa}(\partial_{t}+\Omega_{H}\partial_{\phi}).\label{7.37}
\end{equation}
Here $\Omega_{H}=\frac{r_{-}}{lr_{+}}$ is the angular velocity of
the black hole horizon and $\kappa=\frac{r_{+}^{2}-r_{-}^{2}}{l^{2}r_{+}}$
is the surface gravity. By substituting these Killing vectors in \eqref{7.35}
and carrying the integration over a one-parameter path on the solution
space, we find \eqref{7.31}, \eqref{7.32} and \eqref{7.34} exactly.
It is easy to show that these results satisfy the first law of black
hole mechanics.

\section{Asymptotically spacelike warped anti-de Sitter spacetimes in {General
Minimal Massive Gravity}}

\label{S.8.0} In this section, we consider asymptotically spacelike
warped anti-de Sitter spacetimes in the context of GMMG model. We
find conserved charges and algebra among them in the given model.

\subsection{Asymptotically spacelike warped AdS$_{3}$ spacetimes}

\label{S.8.1}

Boundary conditions for asymptotically warped AdS$_{3}$ spacetimes
in TMG were first introduced in \cite{Compere:2009zj}. A few years
later, in order to switch on the local degree of freedom (the massive
graviton), a set of asymptotic conditions was presented \cite{48}.
Here we follow \cite{48} to introduce the appropriate boundary conditions
\cite{81}: 
\begin{equation}
\begin{split} & g_{tt}=l^{2}+\mathcal{O}(r^{-3}),\hspace{0.7cm}g_{tr}=\mathcal{O}(r^{-3}),\hspace{0.7cm}g_{r\phi}=\mathcal{O}(r^{-2}),\\
 & g_{t\phi}=\frac{1}{2}l^{2}\left|\zeta\right|\left[r+A_{t\phi}(\phi)+\frac{1}{r}B_{t\phi}(\phi)\right]+\mathcal{O}(r^{-2}),\\
 & g_{rr}=\frac{l^{2}}{\zeta^{2}\nu^{2}}\left[\frac{1}{r^{2}}+\frac{1}{r^{3}}A_{rr}(\phi)+\frac{1}{r^{4}}B_{rr}(\phi)\right]+\mathcal{O}(r^{-5}),\\
 & g_{\phi\phi}=\frac{1}{4}l^{2}\zeta^{2}\left[\left(1-\nu^{2}\right)r^{2}+rA_{\phi\phi}(\phi)+B_{\phi\phi}(\phi)\right]+\mathcal{O}(r^{-1}),
\end{split}
\label{8.1}
\end{equation}
which are consistent with the metric \eqref{2.44}. The corresponding
components of the dreibeins are 
\begin{equation}
\begin{split}e_{\hspace{1.5mm}t}^{0}= & \frac{l\nu}{\sqrt{1-\nu^{2}}}-\frac{l\left[2\left(\nu^{2}-1\right)A_{t\phi}+A_{\phi\phi}\right]}{2r\nu\left(1-\nu^{2}\right)^{\frac{3}{2}}}\\
 & +\frac{l}{8r^{2}\nu^{3}\left(1-\nu^{2}\right)^{\frac{5}{2}}}\biggl[4A_{t\phi}^{2}\left(\nu^{2}-1\right)^{3}+A_{\phi\phi}^{2}\left(4\nu^{2}-1\right)\\
 & +4A_{t\phi}A_{\phi\phi}\left(\nu^{2}-1\right)\left(2\nu^{2}-1\right)+8B_{t\phi}\nu^{2}\left(\nu^{2}-1\right)^{2}\\
 & +4B_{\phi\phi}\nu^{2}\left(\nu^{2}-1\right)\biggr]+\mathcal{O}(r^{-3})
\end{split}
\label{8.2}
\end{equation}
\begin{equation}
e_{\hspace{1.5mm}r}^{1}=\frac{l}{\zeta\nu r}+\frac{lA_{rr}}{2\zeta\nu r^{2}}-\frac{l}{8\zeta\nu r^{3}}\left[A_{rr}^{2}-4B_{rr}\right]+\mathcal{O}(r^{-4})\label{8.3}
\end{equation}
\begin{equation}
\begin{split}e_{\hspace{1.5mm}t}^{2}= & \frac{l}{\sqrt{1-\nu^{2}}}-\frac{l\left[2\left(\nu^{2}-1\right)A_{t\phi}+A_{\phi\phi}\right]}{2r\left(1-\nu^{2}\right)^{\frac{3}{2}}}\\
 & +\frac{l}{8r^{2}\left(1-\nu^{2}\right)^{\frac{5}{2}}}\biggl[8B_{t\phi}\left(\nu^{2}-1\right)^{2}+4A_{t\phi}A_{\phi\phi}\left(\nu^{2}-1\right)\\
 & +4B_{\phi\phi}\left(\nu^{2}-1\right)+3A_{\phi\phi}^{2}\biggr]+\mathcal{O}(r^{-3})
\end{split}
\label{8.4}
\end{equation}
\begin{equation}
\begin{split}e_{\hspace{1.5mm}\phi}^{2}= & \frac{1}{2}rl\left|\zeta\right|\sqrt{1-\nu^{2}}+\frac{l\left|\zeta\right|A_{\phi\phi}}{4\sqrt{1-\nu^{2}}}\\
 & -\frac{l\left|\zeta\right|}{16r\left(1-\nu^{2}\right)^{\frac{3}{2}}}\left[4B_{\phi\phi}\left(\nu^{2}-1\right)+A_{\phi\phi}^{2}\right]+\mathcal{O}(r^{-2}),
\end{split}
\label{8.5}
\end{equation}
and the rest of them are $\mathcal{O}(r^{-4})$.The metric, under
transformation generated by vector field $\xi$, transforms as $\delta_{\xi}g_{\mu\nu}=\pounds_{\xi}g_{\mu\nu}$.
The variation generated by the following Killing vector field preserves
the boundary conditions \eqref{8.1} 
\begin{equation}
\begin{split}\xi^{t}(T,Y)= & T(\phi)-\frac{2\partial_{\phi}^{2}Y(\phi)}{\left|\zeta\right|^{3}\nu^{4}r}+\mathcal{O}(r^{-2}),\\
\xi^{r}(T,Y)= & -r\partial_{\phi}Y(\phi)+\mathcal{O}(r^{-2}),\\
\xi^{\phi}(T,Y)= & Y(\phi)+\frac{2\partial_{\phi}^{2}Y(\phi)}{\zeta^{4}\nu^{4}r^{2}}+\mathcal{O}(r^{-3}),
\end{split}
\label{8.6}
\end{equation}
where $T(\phi)$ and $Y(\phi)$ are two arbitrary periodic functions.
The asymptotic Killing vectors \eqref{8.6} are closed under the Lie
bracket 
\begin{equation}
\left[\xi(T_{1},Y_{1}),\xi(T_{2},Y_{2})\right]=\xi(T_{12},Y_{12}),\label{8.7}
\end{equation}
where 
\begin{equation}
\begin{split}T_{12}(\phi)= & Y_{1}(\phi)\partial_{\phi}T_{2}(\phi)-Y_{2}(\phi)\partial_{\phi}T_{1}(\phi),\\
Y_{12}(\phi)= & Y_{1}(\phi)\partial_{\phi}Y_{2}(\phi)-Y_{2}(\phi)\partial_{\phi}Y_{1}(\phi).
\end{split}
\label{8.8}
\end{equation}
Again, introducing Fourier modes $u_{m}=\xi(e^{im\phi},0)$ and $v_{m}=\xi(0,e^{im\phi})$,
one finds that 
\begin{equation}
\begin{split} & \left[u_{m},u_{n}\right]=0,\\
 & \left[v_{m},u_{n}\right]=-nu_{m+n},\\
 & \left[v_{m},v_{n}\right]=(m-n)v_{m+n},
\end{split}
\label{8.9}
\end{equation}
which is a semi direct product of the Witt algebra with the $U(1)$
current algebra. Under the action of a generic asymptotic symmetry
generator $\xi$ \eqref{8.6}, the dynamical fields transform as 
\begin{equation}
\begin{split}\delta_{\xi}A_{t\phi}= & \partial_{\phi}\left[Y(\phi)A_{t\phi}(\phi)\right]+\frac{2}{\left|\zeta\right|}\partial_{\phi}T(\phi),\\
\delta_{\xi}A_{rr}= & \partial_{\phi}\left[Y(\phi)A_{rr}(\phi)\right],\\
\delta_{\xi}A_{\phi\phi}= & \partial_{\phi}\left[Y(\phi)A_{\phi\phi}(\phi)\right]+\frac{4}{\left|\zeta\right|}\partial_{\phi}T(\phi),
\end{split}
\label{8.10}
\end{equation}
\begin{equation}
\begin{split}\delta_{\xi}B_{t\phi}= & Y(\phi)\partial_{\phi}B_{t\phi}(\phi)+2B_{t\phi}(\phi)\partial_{\phi}Y(\phi)-\frac{2}{\zeta^{4}\nu^{4}}\partial_{\phi}^{3}Y(\phi),\\
\delta_{\xi}B_{rr}= & Y(\phi)\partial_{\phi}B_{rr}(\phi)+2B_{rr}(\phi)\partial_{\phi}Y(\phi),\\
\delta_{\xi}B_{\phi\phi}= & Y(\phi)\partial_{\phi}B_{\phi\phi}(\phi)+2B_{\phi\phi}(\phi)\partial_{\phi}Y(\phi)+\frac{4}{\left|\zeta\right|}A_{t\phi}(\phi)\partial_{\phi}T(\phi)\\
 & -\frac{4\left(1+\nu^{2}\right)}{\zeta^{4}\nu^{4}}\partial_{\phi}^{3}Y(\phi).
\end{split}
\label{8.11}
\end{equation}
We are interested in solutions which are asymptotically spacelike
warped AdS$_{3}$. Thus, we demand that equations \eqref{2.48} and
\eqref{2.49} hold asymptotically: 
\begin{equation}
\nabla_{\mu}J_{\nu}-\frac{\left|\zeta\right|}{2l}\epsilon_{\mu\nu\lambda}J^{\lambda}=\mathcal{O}(r^{-1}),\label{8.12}
\end{equation}
and the Ricci tensor reads 
\begin{equation}
\mathcal{R}_{\mu\nu}-\frac{\zeta^{2}}{2l^{2}}\left(1-2\nu^{2}\right)g_{\mu\nu}+\frac{\zeta^{2}}{l^{4}}\left(1-\nu^{2}\right)J_{\mu}J_{\nu}=\mathcal{O}(r^{-1}).\label{8.13}
\end{equation}
By substituting \eqref{8.1} into the last two equations \eqref{8.12}
one gets 
\begin{equation}
\begin{split} & A_{\phi\phi}(\phi)=\nu^{2}A_{rr}(\phi)+2A_{t\phi}(\phi),\\
 & B_{\phi\phi}(\phi)=\nu^{2}\left[B_{rr}(\phi)+2B_{t\phi}(\phi)-A_{rr}(\phi)^{2}\right]+A_{t\phi}(\phi)^{2}+2B_{t\phi}(\phi).
\end{split}
\label{8.14}
\end{equation}
Hence, the metric \eqref{8.1} solves equations of motion of general
minimal massive gravity asymptotically when equations \eqref{2.53}-\eqref{2.58}
and \eqref{8.14} are satisfied. It is important to note that, in
the light of the preceding discussion, \eqref{2.48}, \eqref{2.49}
and \eqref{2.52} hold only asymptotically.

\subsection{Conserved charges of asymptotically spacelike warped AdS$_{3}$ spacetimes
in {General Minimal Massive Gravity}}

\label{S.8.2} We first want to simplify the expression for conserved
charge perturbation \eqref{5.5} for asymptotically spacelike warped
AdS$_{3}$ spacetimes \eqref{8.1} in the context of general minimal
massive gravity. So, we use \eqref{2.52}-\eqref{2.58}, \eqref{2.27},
\eqref{2.9} and \eqref{2.48}. After some calculations we find the
following expression for the variation of the charges 
\begin{equation}
\begin{split}\delta Q(\xi)=-\frac{1}{8\pi G}\int_{\Sigma}\biggl\{ & -\left(\sigma+\frac{\alpha H_{1}}{\mu}+\frac{F_{1}}{m^{2}}\right)\left[i_{\xi}e\cdot\delta\Omega+\left(i_{\xi}\Omega-\chi_{\xi}\right)\cdot\delta e\right]\\
 & +\frac{1}{\mu}\left(i_{\xi}\Omega-\chi_{\xi}\right)\cdot\delta\Omega+\alpha H_{2}\left(\frac{\alpha H_{2}}{\mu}+\frac{2F_{2}}{m^{2}}\right)i_{\xi}\mathfrak{J}\cdot\delta\mathfrak{J}\\
 & +\left[-\frac{\zeta^{2}}{\mu l^{2}}\left(\frac{3}{4}-\nu^{2}\right)+l\left|\zeta\right|\left(\frac{\alpha H_{2}}{\mu}+\frac{F_{2}}{m^{2}}\right)\right]i_{\xi}e\cdot\delta e\\
 & -\left(\frac{\alpha H_{2}}{\mu}+\frac{F_{2}}{m^{2}}\right)\left[i_{\xi}\mathfrak{J}\cdot\delta\Omega+\left(i_{\xi}\Omega-\chi_{\xi}\right)\cdot\delta\mathfrak{J}\right]\\
 & +\left[\frac{\zeta^{2}}{\mu l^{4}}\left(1-\nu^{2}\right)-\frac{3\left|\zeta\right|}{2l}\left(\frac{\alpha H_{2}}{\mu}+\frac{F_{2}}{m^{2}}\right)\right]\left(i_{\xi}\mathfrak{J}\cdot\delta e+i_{\xi}e\cdot\delta\mathfrak{J}\right)\biggr\}.
\end{split}
\label{8.15}
\end{equation}
where $\mathfrak{J}_{\hspace{1.5mm}\mu}^{a}=J^{a}J_{\mu}$.

Now we take the spacelike warped AdS$_{3}$ spacetime as the background
which can be described by the following dreibeins 
\begin{equation}
\begin{split}\bar{e}^{0}= & \frac{l\nu}{\sqrt{1-\nu^{2}}}dt,\\
\bar{e}^{1}= & \frac{l}{\zeta\nu r}dr,\\
\bar{e}^{2}= & \frac{l}{\sqrt{1-\nu^{2}}}dt+\frac{1}{2}rl\left|\zeta\right|\sqrt{1-\nu^{2}}d\phi.
\end{split}
\label{8.16}
\end{equation}
As we mentioned in the section \ref{S.5.0}, one can take an integration
of \eqref{8.15} over a one-parameter path on the space of solutions
to find the conserved charge corresponds to the Killing vector field
$\xi$: 
\begin{equation}
\begin{split}Q(\xi)=-\frac{1}{8\pi G}\int_{\Sigma}\biggl\{ & -\left(\sigma+\frac{\alpha H_{1}}{\mu}+\frac{F_{1}}{m^{2}}\right)\left[i_{\xi}\bar{e}\cdot\Delta\Omega+\left(i_{\xi}\bar{\Omega}-\bar{\chi}_{\xi}\right)\cdot\Delta e\right]\\
 & +\frac{1}{\mu}\left(i_{\xi}\bar{\Omega}-\bar{\chi}_{\xi}\right)\cdot\Delta\Omega+\alpha H_{2}\left(\frac{\alpha H_{2}}{\mu}+\frac{2F_{2}}{m^{2}}\right)i_{\xi}\bar{\mathfrak{J}}\cdot\Delta\mathfrak{J}\\
 & +\left[-\frac{\zeta^{2}}{\mu l^{2}}\left(\frac{3}{4}-\nu^{2}\right)+l\left|\zeta\right|\left(\frac{\alpha H_{2}}{\mu}+\frac{F_{2}}{m^{2}}\right)\right]i_{\xi}\bar{e}\cdot\Delta e\\
 & -\left(\frac{\alpha H_{2}}{\mu}+\frac{F_{2}}{m^{2}}\right)\left[i_{\xi}\bar{\mathfrak{J}}\cdot\Delta\Omega+\left(i_{\xi}\bar{\Omega}-\bar{\chi}_{\xi}\right)\cdot\Delta\mathfrak{J}\right]\\
 & +\left[\frac{\zeta^{2}}{\mu l^{4}}\left(1-\nu^{2}\right)-\frac{3\left|\zeta\right|}{2l}\left(\frac{\alpha H_{2}}{\mu}+\frac{F_{2}}{m^{2}}\right)\right]\left(i_{\xi}\bar{\mathfrak{J}}\cdot\Delta e+i_{\xi}\bar{e}\cdot\Delta\mathfrak{J}\right)\biggr\},
\end{split}
\label{8.17}
\end{equation}
with $\Delta\Phi=\Phi_{(s=1)}-\Phi_{(s=0)}$, where $\Phi_{(s=1)}$
and $\Phi_{(s=0)}$ are calculated on the spacetime solution and on
the background spacetime, respectively\footnote{For instance, $\Delta e=e-\bar{e}$, where $e$ and $\bar{e}$ are
given by \eqref{8.2}-\eqref{8.5} and \eqref{8.16}, respectively.}.

\subsection{The algebra of conserved charges}

\label{S.8.3} As we laid out in \ref{S.8.1}, the asymptotic Killing
vector field is given by \eqref{8.6}. Now we shall find the conserved
charge corresponding to the asymptotic Killing vector field given
in \eqref{8.6}. We need to use $J=J^{\mu}\partial_{\mu}=\partial_{t}$,
\eqref{2.53}-\eqref{2.58}, \eqref{8.2}-\eqref{8.6}, \eqref{8.14},
\eqref{8.15} and \eqref{8.16}. After some cumbersome calculations
one eventually arrives at 
\begin{equation}
Q(T,Y)=P(T)+L(Y),\label{8.18}
\end{equation}
with 
\begin{equation}
P(T)=\frac{\left|\zeta\right|}{96\pi}c_{U}\int_{0}^{2\pi}T(\phi)\left[A_{rr}(\phi)+2A_{t\phi}(\phi)\right]d\phi,\label{8.19}
\end{equation}
\begin{equation}
L(Y)=-\frac{\zeta^{4}\nu^{4}}{768\pi}c_{V}\int_{0}^{2\pi}Y(\phi)\left[-3A_{rr}(\phi)^{2}+4B_{rr}(\phi)+16B_{t\phi}(\phi)\right]d\phi,\label{8.20}
\end{equation}
where 
\begin{equation}
c_{U}=\frac{3l\left|\zeta\right|\nu^{2}}{G}\left\{ \sigma+\frac{\alpha}{\mu}\left(H_{1}+l^{2}H_{2}\right)+\frac{1}{m^{2}}\left(F_{1}+l^{2}F_{2}\right)-\frac{\left|\zeta\right|}{2\mu l}\right\} ,\label{8.21}
\end{equation}
\begin{equation}
c_{V}=\frac{3l}{\left|\zeta\right|\nu^{2}G}\left\{ \sigma+\frac{\alpha}{\mu}\left(H_{1}+l^{2}H_{2}\right)+\frac{1}{m^{2}}\left(F_{1}+l^{2}F_{2}\right)-\frac{\left|\zeta\right|}{2\mu l}\left(1-2\nu^{2}\right)\right\} .\label{8.22}
\end{equation}
Note that $c_{U}$ and $c_{V}$ are related via 
\begin{equation}
\zeta^{2}\nu^{4}c_{V}-c_{U}=\frac{3\zeta^{2}\nu^{4}}{\mu G}.\label{8.23}
\end{equation}
Plugging \eqref{8.18} in \eqref{5.6} and making use of \eqref{8.10},
\eqref{5.7} and \eqref{8.11}, we find 
\begin{equation}
\begin{split}\left\{ Q(T_{1},Y_{1}),Q(T_{2},Y_{2})\right\} _{\text{D.B}}= & Q(T_{12},Y_{12})\\
 & -\frac{\left|\zeta\right|}{192\pi}c_{U}\int_{0}^{2\pi}T_{12}(\phi)\left[A_{rr}(\phi)+2A_{t\phi}(\phi)\right]d\phi\\
 & +\frac{1}{48\pi}c_{U}\int_{0}^{2\pi}\left(T_{1}\partial_{\phi}T_{2}-T_{2}\partial_{\phi}T_{1}\right)d\phi\\
 & +\frac{1}{48\pi}c_{V}\int_{0}^{2\pi}\left(Y_{1}\partial_{\phi}^{3}Y_{2}-Y_{2}\partial_{\phi}^{3}Y_{1}\right)d\phi.
\end{split}
\label{8.24}
\end{equation}
Introducing the Fourier modes as 
\begin{equation}
\begin{split}P_{m}= & Q(e^{im\phi},0)=P(e^{im\phi}),\\
L_{m}= & Q(0,e^{im\phi})=L(e^{im\phi}),
\end{split}
\label{8.25}
\end{equation}
one can read off the algebra of conserved charges as follows: 
\begin{equation}
\begin{split}i\left\{ P_{m},P_{n}\right\} _{\text{D.B.}}= & -\frac{c_{U}}{12}n\delta_{m+n,0}\\
i\left\{ L_{m},P_{n}\right\} _{\text{D.B.}}= & -nP_{m+n}+\frac{\left|\zeta\right|c_{U}}{192\pi}n\int_{0}^{2\pi}e^{i(m+n)\phi}\left[A_{rr}(\phi)+2A_{t\phi}(\phi)\right]d\phi\\
i\left\{ L_{m},L_{n}\right\} _{\text{D.B.}}= & (m-n)L_{m+n}+\frac{c_{V}}{12}n^{3}\delta_{m+n,0}.
\end{split}
\label{8.26}
\end{equation}
Now we consider the warped black hole solution as an example. For
this case \eqref{2.44} with\eqref{2.45}, one has 
\begin{equation}
\begin{split} & A_{rr}=r_{+}+r_{-},\hspace{0.7cm}A_{t\phi}=\nu\sqrt{r_{+}r_{-}},\\
 & B_{rr}=r_{+}^{2}+r_{-}^{2}+r_{+}r_{-},\hspace{0.7cm}B_{t\phi}=0
\end{split}
\label{8.27}
\end{equation}
and \eqref{8.26} reduce to 
\begin{equation}
\begin{split}i\left\{ P_{m},P_{n}\right\} _{\text{D.B.}}= & -\frac{c_{U}}{12}n\delta_{m+n,0}\\
i\left\{ L_{m},P_{n}\right\} _{\text{D.B.}}= & -nP_{m+n}+\frac{\left|\zeta\right|c_{U}}{96}n\left(r_{+}+r_{-}+2\nu\sqrt{r_{+}r_{-}}\right)\delta_{m+n,0}\\
i\left\{ L_{m},L_{n}\right\} _{\text{D.B.}}= & (m-n)L_{m+n}+\frac{c_{V}}{12}n^{3}\delta_{m+n,0}.
\end{split}
\label{8.28}
\end{equation}
Setting $\hat{P}_{m}\equiv P_{m}$ and $\hat{L}_{m}\equiv L_{m}$,
and replacing the brackets with commutators $i\{,\}\rightarrow[,]$,
Eq.~\eqref{8.28} become\textcolor{red}{}\footnote{It is worth mentioning that the algebra \eqref{8.29} was first obtained
in \cite{Compere:2008cv}.} 
\begin{equation}
\begin{split}\left[\hat{P}_{m},\hat{P}_{n}\right]= & -\frac{c_{U}}{12}n\delta_{m+n,0}\\
\left[\hat{L}_{m},\hat{P}_{n}\right]= & -n\hat{P}_{m+n}+\frac{n}{2}p_{0}\delta_{m+n,0}\\
\left[\hat{L}_{m},\hat{L}_{n}\right]= & (m-n)\hat{L}_{m+n}+\frac{c_{V}}{12}n^{3}\delta_{m+n,0}.
\end{split}
\label{8.29}
\end{equation}
Here $p_{0}$ is the zero mode eigenvalue of $\hat{P}_{m}$. From
\eqref{8.25} and using \eqref{8.19}, \eqref{8.20} and \eqref{8.27},
one can read off the eigenvalues of $\hat{P}_{m}$ and $\hat{L}_{m}$
as 
\begin{equation}
p_{m}=\frac{\left|\zeta\right|c_{U}}{48}\left(r_{+}+r_{-}+2\nu\sqrt{r_{+}r_{-}}\right)\delta_{m,0},\label{8.30}
\end{equation}
\begin{equation}
l_{m}=-\frac{\zeta^{4}\nu^{4}c_{V}}{384}\left(r_{+}-r_{-}\right)^{2}\delta_{m,0},\label{8.31}
\end{equation}
respectively. So, it is apparent that the algebra of asymptotic conserved
charges is given as the semi direct product of the Virasoro algebra
with $U(1)$ current algebra, with central charges $c_{V}$ and $c_{U}$.

\subsection{Mass, angular momentum and entropy of warped black hole solution
of {general minimal massive Gravity}}

\label{S.8.4} Even though the algebra of conserved charges \eqref{8.29}
does not describe the conformal symmetry \cite{Anninos}, one can
use a particular Sugawara construction \cite{50} to construct the
conformal algebra as was done in the case of topologically massive
gravity \cite{51} and on the spacelike warped AdS$_{3}$ black hole
solutions of NMG \cite{52}. For this purpose, two new operators can
be introduced as 
\begin{equation}
\hat{L}_{m}^{+}=\frac{im}{\left|\zeta\right|\nu^{2}}\hat{P}_{-m}-\frac{6}{c_{U}}\hat{K}_{-m},\label{8.32}
\end{equation}
\begin{equation}
\hat{L}_{m}^{-}=\hat{L}_{m}+\frac{6p_{0}}{c_{U}}\hat{P}_{m}-\frac{6}{c_{U}}\hat{K}_{m},\label{8.33}
\end{equation}
where 
\begin{equation}
\hat{K}_{m}=\sum_{q\in\mathbb{Z}}\hat{P}_{m+q}\hat{P}_{-q}.\label{8.34}
\end{equation}
These operators satisfy the following algebra 
\begin{equation}
\begin{split}\left[\hat{L}_{m}^{+},\hat{L}_{n}^{+}\right]= & (m-n)\hat{L}_{m+n}^{+}+\frac{c_{+}}{12}n^{3}\delta_{m+n,0}\\
\left[\hat{L}_{m}^{-},\hat{L}_{n}^{-}\right]= & (m-n)\hat{L}_{m+n}^{-}+\frac{c_{-}}{12}n^{3}\delta_{m+n,0}+\frac{3p_{0}^{2}}{c_{U}}n\delta_{m+n,0}\\
\left[\hat{L}_{m}^{+},\hat{L}_{n}^{-}\right]= & 0
\end{split}
\label{8.35}
\end{equation}
where 
\begin{equation}
c_{+}=\frac{c_{U}}{\zeta^{2}\nu^{4}},\hspace{0.7cm}c_{-}=c_{V},\label{8.36}
\end{equation}
In this way, \eqref{8.23} can be rewritten as 
\begin{equation}
c_{-}-c_{+}=\frac{3}{\mu G}.\label{8.37}
\end{equation}
From \eqref{8.32} and \eqref{8.33}, one can read the eigenvalues
of $\hat{L}_{0}^{\pm}$ as 
\begin{equation}
l_{0}^{+}=-\frac{6p_{0}^{2}}{c_{U}},\hspace{0.7cm}l_{0}^{-}=l_{0}.\label{8.38}
\end{equation}
Finally, Cardy's formula \eqref{7.33} leads to the following expression
for the CFT entropy 
\begin{equation}
\begin{split}S_{\text{CFT}}= & \frac{\pi l\left|\zeta\right|}{4G}\biggl\{\left[\sigma+\frac{\alpha}{\mu}\left(H_{1}+l^{2}H_{2}\right)+\frac{1}{m^{2}}\left(F_{1}+l^{2}F_{2}\right)-\frac{\left|\zeta\right|}{2\mu l}\right]\left(r_{+}+\nu\sqrt{r_{+}r_{-}}\right)\\
 & +\frac{\left|\zeta\right|\nu^{2}}{2\mu l}\left(r_{+}-r_{-}\right)\biggr\}.
\end{split}
\label{8.39}
\end{equation}
where CFT stands for emphasizing that the entropy \eqref{8.39} comes
from Conformal field theory considerations. Mass and angular momentum
can also be obtained by $M=p_{0}$ and $J=\left(l_{0}^{+}-l_{0}^{-}\right)$,
respectively \cite{51} yielding 
\begin{equation}
M=\frac{\left|\zeta\right|c_{U}}{48}\left(r_{+}+r_{-}+2\nu\sqrt{r_{+}r_{-}}\right),\label{8.40}
\end{equation}
\begin{equation}
J=-\frac{\zeta^{2}}{384}\left\{ c_{U}\left(r_{+}+r_{-}+2\nu\sqrt{r_{+}r_{-}}\right)^{2}+\zeta^{2}\nu^{4}c_{V}\left(r_{+}-r_{-}\right)^{2}\right\} .\label{8.41}
\end{equation}
The angular velocity of the black hole is 
\begin{equation}
\Omega_{H}=-N^{\phi}(r_{+})=-\frac{2}{\left|\zeta\right|\left(r_{+}+\nu\sqrt{r_{+}r_{-}}\right)},\label{8.42}
\end{equation}
and the surface gravity of the event horizon is 
\begin{equation}
\kappa_{H}=\left[-\frac{1}{2}\nabla^{\mu}\zeta^{\nu}\nabla_{\mu}\zeta_{\nu}\right]_{r=r_{+}}^{\frac{1}{2}}=\frac{\left|\zeta\right|\nu^{2}\left(r_{+}-r_{-}\right)}{2\left(r_{+}+\nu\sqrt{r_{+}r_{-}}\right)},\label{8.43}
\end{equation}
where $\zeta=\partial_{t}+\Omega_{H}\partial_{\phi}$ is the horizon-generating
Killing vector field. One can readily show that these conserved quantities
for the black hole satisfy the first law of black hole mechanics,
\begin{equation}
\delta M=T_{H}\delta S+\Omega_{H}\delta J,\label{8.44}
\end{equation}
where $T_{H}=\kappa_{H}/2\pi$ is the Hawking temperature.

We now end this section with a discussion of the symmetries. The $SL(2,\mathbb{R})\times SL(2,\mathbb{R})$
isometry group of AdS$_{3}$ space reduces to the $SL(2,\mathbb{R})\times U(1)$
for the warped AdS$_{3}$ .Therefore, the asymptotic symmetry of the
warped AdS$_{3}$ also differs from the full conformal symmetry. So
the dual theory of the warped AdS$_{3}$ is not a 2-dimensional CFT,
but a, so called, warped CFT (WCFT), which should exhibit partial
conformal symmetry. Let us define $\tilde{P}_{m}=\left|\zeta\right|^{-1}\nu^{-2}\hat{P}_{m}$
after which the algebra \eqref{8.29} becomes 
\begin{equation}
\begin{split}\left[\tilde{P}_{m},\tilde{P}_{n}\right]= & -\frac{\tilde{c}_{U}}{12}n\delta_{m+n,0}\\
\left[\hat{L}_{m},\tilde{P}_{n}\right]= & -n\tilde{P}_{m+n}+\frac{n}{2}\tilde{p}_{0}\delta_{m+n,0}\\
\left[\hat{L}_{m},\hat{L}_{n}\right]= & (m-n)\hat{L}_{m+n}+\frac{c_{V}}{12}n^{3}\delta_{m+n,0}
\end{split}
\label{8.45}
\end{equation}
where $\tilde{c}_{U}=\zeta^{-2}\nu^{-4}c_{U}$ and $\tilde{p}_{m}=\left|\zeta\right|^{-1}\nu^{-2}p_{m}$.
This is the semi direct product of the Virasoro algebra with the $U(1)$
algebra, with central charges $c_{V}$ and $\tilde{c}_{U}$ which
is the symmetry of warped CFT. Therefore, the dual theory of the warped
black hole solution of GMMG is a WCFT. A warped version of the Cardy
formula was introduced by Detournay et al. \cite{53} which leads
to 
\begin{equation}
S_{\text{WCFT}}=\frac{24\pi}{\tilde{c}_{U}}\tilde{p}_{0}^{(vac)}\tilde{p}_{0}+4\pi\sqrt{-l_{0}^{(vac)}l_{0}},\label{8.46}
\end{equation}
where $\tilde{p}_{0}^{(vac)}$ and $l_{0}^{(vac)}$ correspond to
the minimum values of $\tilde{p}_{0}$ and $l_{0}$, i.e. the value
of the vacuum geometry. Since the vacuum corresponds to $r_{\pm}=0$,
from \eqref{8.45}, one finds 
\begin{equation}
\tilde{p}_{0}^{(vac)}=-\frac{\tilde{c}_{U}}{24},\hspace{0.7cm}l_{0}^{(vac)}=-\frac{c_{V}}{24}.\label{8.47}
\end{equation}
With all these one arrives at $S_{\text{WCFT}}=S_{\text{CFT}}$.

\section{Conserved charges of the {Rotating Oliva-Tempo-Troncoso black hole}}

\label{S.10.0} The rotating OTT black hole spacetime \eqref{2.22},
is a solution of the NMG in the special choice of the parameters that
leads to a unique vacuum as discussed. To find the conserved charges
of this black hole, we take the AdS$_{3}$ spacetime \eqref{7.8}
as the background, i.e. the AdS$_{3}$ spacetime corresponds to $s=0$
\cite{82}. The non-zero components of the flavor metric are given
in \eqref{2.14}, with \eqref{2.21}, one can show that the extended
off-shell ADT charge \eqref{3.17} 
\begin{equation}
\mathcal{Q}_{\text{ADT}}(\xi)=\left(\tilde{g}_{rs}i_{\xi}a^{r}-\tilde{g}_{\omega s}\chi_{\xi}\right)\cdot\delta a^{s},\label{10.1}
\end{equation}
reduces to 
\begin{equation}
\mathcal{Q}_{\text{ADT}}(\xi)=\left\{ -2i_{\xi}\bar{e}\cdot\delta\Omega_{\phi}+2l^{2}(i_{\xi}\bar{\Omega}-\bar{\chi}_{\xi})\cdot\delta S_{\phi}\right\} d\phi,\label{10.2}
\end{equation}
on $\Sigma$. The energy of the metric corresponds to the Killing
vector $\xi_{E}=\partial_{t}$. For this Killing vector on the background,
\eqref{6.2} becomes 
\begin{equation}
i_{\xi_{(E)}}\bar{\Omega}^{a}-\bar{\chi}_{\xi_{(E)}}^{a}=\frac{1}{l^{2}}e_{\hspace{1.5mm}\phi}^{a}.\label{10.3}
\end{equation}
Making use of \eqref{7.8} and \eqref{10.3} in\eqref{10.2}, one
finds 
\begin{equation}
\mathcal{Q}_{\text{ADT}}(\xi_{(E)})=2r\{\frac{1}{l}\delta\Omega_{\hspace{1.5mm}\phi}^{0}+\delta S_{\hspace{1.5mm}\phi}^{2}\}d\phi\label{10.4}
\end{equation}
and again, integrating over a path on the solution space, yields 
\begin{equation}
\begin{split}\int_{0}^{1}Q_{\text{ADT}}(\xi_{(E)})ds= & 2r\{\frac{1}{l}\left[\Omega_{\hspace{1.5mm}\phi(s=1)}^{0}-\Omega_{\hspace{1.5mm}\phi(s=0)}^{0}\right]\\
 & +\left[S_{\hspace{1.5mm}\phi(s=1)}^{2}-S_{\hspace{1.5mm}\phi(s=0)}^{2}\right]\}d\phi
\end{split}
\label{10.5}
\end{equation}
where again $s=1$ for the rotating black hole. Expansion of $\Omega_{\hspace{1.5mm}\phi(s=1)}^{0}$
and $S_{\hspace{1.5mm}\phi(s=1)}^{2}$ at infinity yield 
\begin{equation}
\begin{split} & \Omega_{\hspace{1.5mm}\phi(s=1)}^{0}=\Omega_{\hspace{1.5mm}\phi(s=0)}^{0}+\frac{bl}{4}(1+\eta)-\frac{l}{16r}\left[b^{2}l^{2}(1+\eta^{2})+8\mu\right]+\mathcal{O}(r^{-2}),\\
 & S_{\hspace{1.5mm}\phi(s=1)}^{2}=S_{\hspace{1.5mm}\phi(s=0)}^{2}-\frac{b}{4}(1+\eta)-\frac{b^{2}l^{2}}{16r}(1-\eta^{2})+\mathcal{O}(r^{-2}),
\end{split}
\label{10.6}
\end{equation}
which reduces \eqref{10.5} to 
\begin{equation}
\int_{0}^{1}Q_{\text{ADT}}(\xi_{(E)})ds=\left\{ \mu+\frac{1}{4}b^{2}l^{2}+\mathcal{O}(r^{-1})\right\} d\phi.\label{10.7}
\end{equation}
Plugging \eqref{10.7} into \eqref{5.4} and by taking the limit $r\rightarrow\infty$,
one arrives at 
\begin{equation}
E=\frac{1}{4}\left(\mu+\frac{1}{4}b^{2}l^{2}\right).\label{10.8}
\end{equation}
For the rotation Killing vector $\xi_{(J)}=\partial_{\phi}$, 
\begin{equation}
i_{\xi_{(J)}}\bar{\Omega}^{a}-\bar{\chi}_{\xi_{(J)}}^{a}=e_{\hspace{1.5mm}t}^{a},\label{10.9}
\end{equation}
which reduces \eqref{10.2} to 
\begin{equation}
\mathcal{Q}_{\text{ADT}}(\xi_{(J)})=-2r\left\{ \delta\omega_{\hspace{1.5mm}\phi}^{2}+l\delta S_{\hspace{1.5mm}\phi}^{0}\right\} d\phi,\label{10.10}
\end{equation}
A similar expansion as above yield 
\begin{equation}
\begin{split} & \omega_{\hspace{1.5mm}\phi(s=1)}^{2}=\omega_{\hspace{1.5mm}\phi(s=0)}^{2}-\frac{bl}{4}\sqrt{1-\eta^{2}}+\frac{l}{16r}\left[b^{2}l^{2}(1-\eta)+8\mu\right]\sqrt{1-\eta^{2}}+\mathcal{O}(r^{-2}),\\
 & S_{\hspace{1.5mm}\phi(s=1)}^{0}=S_{\hspace{1.5mm}\phi(s=0)}^{0}+\frac{b}{4}\sqrt{1-\eta^{2}}+\frac{b^{2}l^{2}}{16r}(1+\eta)\sqrt{1-\eta^{2}}+\mathcal{O}(r^{-2}),
\end{split}
\label{10.11}
\end{equation}
which then leads to the angular momentum of the black hole 
\begin{equation}
J=l\sqrt{1-\eta^{2}}E.\label{10.12}
\end{equation}

\section{Explicit examples of the black hole entropy}

\label{S.9.0} In this section we shall use the black hole entropy
formula \eqref{6.11} to compute entropy of some black holes for various
models. In fact, based on papers \cite{75,81,82,104}, we simplify
entropy formula in the context of MMG, GMG, GMMG and NMG.

\subsection{{Banados-Teitelboim-Zanelli} black hole entropy}

\label{S.9.2} In this subsection, we calculate the entropy of BTZ
black hole, as a solution of the EGN, in the context of some models.

\subsubsection{{Minimal Massive Gravity}}

\label{S.9.2.1}For MMG, the non-zero components of the flavor metric
are are given by \eqref{2.5}. The MMG is not a torsion-free theory
so that the spin-connection $\omega$ is given by \eqref{2.9}, where
$\Omega$ is the torsion-free dual spin-connection given by \eqref{1.52}.
Also, there is one auxiliary field which is given by \eqref{2.12}.
For this model \eqref{6.11} can be simply rewritten as 
\begin{equation}
S=-\frac{1}{4G}\int_{r=r_{h}}\frac{d\phi}{\sqrt{g_{\phi\phi}}}\left(-\sigma g_{\phi\phi}+\frac{1}{\mu}\Omega_{\phi\phi}-\frac{\alpha}{\mu}h_{\phi\phi}\right).\label{9.1}
\end{equation}
For the BTZ black hole solution \eqref{2.4.2}, the relevant quantities
are 
\begin{equation}
g_{\phi\phi}=r^{2},\hspace{0.7cm}\Omega_{\phi\phi}=-\frac{r_{+}r_{-}}{l},\hspace{0.7cm}S_{\phi\phi}=-\frac{r^{2}}{2l^{2}},\hspace{0.7cm}C_{\phi\phi}=0,\label{9.2}
\end{equation}
\begin{equation}
h_{\phi\phi}=\frac{(1-\alpha\Lambda_{0}l^{2})r^{2}}{2\mu l^{2}(1+\alpha\sigma)^{2}}.\label{9.3}
\end{equation}
Making use of these in \eqref{9.1}, one arrives at the entropy of
the BTZ black hole in MMG: 
\begin{equation}
S=\frac{\pi}{2G}\left[\left(\sigma+\frac{\alpha(1-\alpha\Lambda_{0}l^{2})}{2\mu^{2}l^{2}(1+\alpha\sigma)^{2}}\right)r_{+}+\frac{r_{-}}{\mu l}\right].\label{9.4}
\end{equation}

\subsubsection{{General Massive Gravity}}

\label{S.9.2.2} Let us study the generalized massive gravity as another
example. In this model, there are four flavors of one-form, $a^{s}=\{e,\omega,h,f\}$
and the non-zero components of the flavor metric are given by \eqref{2.25}.
The model is torsion-free. The auxiliary fields $f$ and $h$ are
given by \eqref{2.26}, then one has 
\begin{equation}
S=-\frac{1}{4G}\int_{r=r_{h}}\frac{d\phi}{\sqrt{g_{\phi\phi}}}\left(-\sigma g_{\phi\phi}+\frac{1}{\mu}\Omega_{\phi\phi}+\frac{1}{m^{2}}S_{\phi\phi}\right).\label{9.5}
\end{equation}
Substituting \eqref{9.2} into \eqref{9.5}, one finds the entropy
of BTZ black hole in GMG: 
\begin{equation}
S=\frac{\pi}{2G}\left[\left(\sigma+\frac{1}{2m^{2}l^{2}}\right)r_{+}+\frac{r_{-}}{\mu l}\right].\label{9.6}
\end{equation}

\subsubsection{{General Minimal Massive Gravity}}

\label{S.9.2.3} We discussed that the BTZ black hole is a solution
of GMMG provided that the equations \eqref{2.36}-\eqref{2.39} are
satisfied. Using equations \eqref{2.27}, \eqref{2.36}-\eqref{2.39}
and \eqref{2.9}, one can show that \eqref{6.11} for this theory
becomes 
\begin{equation}
S=\frac{1}{4G}\int_{r=r_{h}}\frac{d\phi}{\sqrt{g_{\phi\phi}}}\left[\left(\sigma+\frac{\alpha H}{\mu}+\frac{F}{m^{2}}\right)g_{\phi\phi}-\frac{1}{\mu}\Omega_{\phi\phi}\right].\label{9.14}
\end{equation}
Again, making use of \eqref{9.2} in \eqref{9.14}, we find that the
entropy of the BTZ black hole in this theory becomes 
\begin{equation}
S=\frac{\pi}{2G}\left[\left(\sigma+\frac{\alpha H}{\mu}+\frac{F}{m^{2}}\right)r_{+}+\frac{r_{-}}{\mu l}\right],\label{9.15}
\end{equation}
where $F$ and $H$ must satisfy the equations \eqref{2.37}-\eqref{2.39}.
The entropy of the BTZ black hole in MMG, GMG and GMMG were first
obtained in \cite{104,Setare:2011jt,Setare:2015vea} respectively
and here we reproduced them by using new entropy formula.

\subsection{Warped black hole entropy from the gravity side}

\label{S.9.1} We would like to simplify \eqref{6.11} for GMMG. Along
the way, we would like to compute the entropy of the Warped black
hole (WBH) \eqref{2.44}, with \eqref{2.45}, as a solution of GMMG
model. One can use \eqref{2.27}, \eqref{2.9} and \eqref{2.52} to
show that, for this particular case, the gravitational black hole
entropy reduces to 
\begin{equation}
\begin{split}S_{\text{WBH}}=-\frac{1}{4G}\int_{\text{Horizon}}\frac{d\phi}{\sqrt{g_{\phi\phi}}}\biggl\{ & -\left(\sigma+\frac{\alpha H_{1}}{\mu}+\frac{F_{1}}{m^{2}}\right)g_{\phi\phi}+\frac{1}{\mu}\Omega_{\phi\phi}\\
 & -\left(\frac{\alpha H_{2}}{\mu}+\frac{F_{2}}{m^{2}}\right)J_{\phi}J_{\phi}\biggr\}.
\end{split}
\label{9.7}
\end{equation}
For warped black hole solution \eqref{2.44} with \eqref{2.45}, we
have 
\begin{equation}
\begin{split}g_{\phi\phi}\mid_{r=r_{+}}= & \frac{1}{4}l^{2}\zeta^{2}\left(r_{+}+\nu\sqrt{r_{+}r_{-}}\right)^{2},\\
\Omega_{\phi\phi}\mid_{r=r_{+}}= & -\frac{1}{4}\zeta^{2}\nu^{2}\sqrt{g_{\phi\phi}}\mid_{r=r_{+}}\left(r_{+}-r_{-}\right)+\frac{\left|\zeta\right|}{2l}g_{\phi\phi}\mid_{r=r_{+}},\\
\left(J_{\phi}J_{\phi}\right)\mid_{r=r_{+}}= & l^{2}g_{\phi\phi}\mid_{r=r_{+}}.
\end{split}
\label{9.8}
\end{equation}
By substituting \eqref{9.8} into \eqref{9.7}, we find the gravitational
entropy of the warped black hole as 
\begin{equation}
\begin{split}S_{\text{WBH}}= & \frac{\pi l\left|\zeta\right|}{4G}\biggl\{\left[\sigma+\frac{\alpha}{\mu}\left(H_{1}+l^{2}H_{2}\right)+\frac{1}{m^{2}}\left(F_{1}+l^{2}F_{2}\right)-\frac{\left|\zeta\right|}{2\mu l}\right]\left(r_{+}+\nu\sqrt{r_{+}r_{-}}\right)\\
 & +\frac{\left|\zeta\right|\nu^{2}}{2\mu l}\left(r_{+}-r_{-}\right)\biggr\}.
\end{split}
\label{9.9}
\end{equation}
This result matches the one obtained in the subsection \ref{S.8.4},
namely \eqref{8.39}, via the Cardy's formula.

\subsection{Entropy of the rotating Oliva-Tempo-Troncoso black hole}

\label{S.9.3} Now, we will compute the entropy of the rotating OTT
black hole in the context of the NMG (See subsection \ref{S.2.3}).
In the NMG model the non-zero components of the flavor metric are
given in the \eqref{2.14}, with \eqref{2.21}, with these one has
\begin{equation}
S=\frac{1}{4G}\int_{0(r=r_{h})}^{2\pi}\frac{d\phi}{\sqrt{g_{\phi\phi}}}\left(g_{\phi\phi}-\frac{1}{m^{2}}S_{\phi\phi}\right),\label{9.10}
\end{equation}
where $r_{h}$ is the radius of the Killing horizon located at $r_{h}=r_{+}$
and $r_{+}$ is given \cite{54} by 
\begin{equation}
r_{+}=l\sqrt{\frac{1+\eta}{2}}\left(-\frac{bl}{2}\sqrt{\eta}+\sqrt{\mu+\frac{1}{4}b^{2}l^{2}}\right).\label{9.11}
\end{equation}
Following the similar lines as the previous calculations, one arrives
at 
\begin{equation}
S_{\phi\phi}=-\frac{1}{2}\left(\frac{H(r)^{2}}{l^{2}}+\frac{b}{2}H(r)(1+\eta)+\frac{\mu}{2}(1-\eta)-\frac{b^{2}l^{2}}{16}(1+\eta)^{2}+\frac{b^{2}l^{2}}{4}\right).\label{9.12}
\end{equation}
and 
\begin{equation}
S=\frac{2\pi l}{G}\sqrt{\frac{(1+\eta)E}{2}},\label{9.13}
\end{equation}
which matches with the results of \cite{54}.

\section{Near horizon symmetries of the non-extremal black hole solutions
of the {General Minimal Massive Gravity}}

\label{S.11.0} An arbitrary variation of the Lagrangian the GMMG
\eqref{2.28} is given by 
\begin{equation}
\delta L_{\text{GMMG}}=\delta e\cdot E_{e}+\delta\omega\cdot E_{\omega}+\delta f\cdot E_{f}+\delta h\cdot E_{h}+d\tilde{\Theta}(a,\delta a),\label{11.1}
\end{equation}
where $E_{e}=E_{\omega}=E_{f}=E_{h}=0$ are the field equations \eqref{2.29}-\eqref{2.32}
and $\tilde{\Theta}(a,\delta a)$ is a surface term which reads explicitly
as 
\begin{equation}
\tilde{\Theta}(a,\delta a)=-\sigma\delta\omega\cdot e+\frac{1}{2\mu}\delta\omega\cdot\omega-\frac{1}{m^{2}}\delta\omega\cdot f+\delta e\cdot h.\label{11.2}
\end{equation}
Now, assume that the variation is due to a diffeomorphism which is
generated by the vector field $\xi$, then the variation of the Lagrangian
\eqref{2.28} is 
\begin{equation}
\delta_{\xi}L=\delta_{\xi}e\cdot E_{e}+\delta_{\xi}\omega\cdot E_{\omega}+\delta_{\xi}f\cdot E_{f}+\delta_{\xi}h\cdot E_{h}+d\tilde{\Theta}(a,\delta_{\xi}a).\label{11.3}
\end{equation}
The Lorentz Chern-Simons term in the Lagrangian \eqref{2.28} makes
the theory to be a non-covariant under the Lorentz gauge transformations.
So, the total variation due to diffeomorphism generator $\xi$ can
be written as 
\begin{equation}
\delta_{\xi}L=\mathfrak{L}_{\xi}L+d\psi_{\xi},\label{11.4}
\end{equation}
with the surface part 
\begin{equation}
\psi_{\xi}=\frac{1}{2\mu}d\chi_{\xi}\cdot\omega.\label{11.5}
\end{equation}
Using equations \eqref{1.55}-\eqref{1.57} and \eqref{11.4}, one
can write \eqref{11.3} as 
\begin{equation}
\begin{split}dJ_{\text{N}}(\xi)= & (i_{\xi}\omega-\chi_{\xi})\cdot\left[D(\omega)E_{\omega}+e\times E_{e}+f\times E_{f}+h\times E_{h}\right]\\
 & +i_{\xi}e\cdot D(\omega)E_{e}+i_{\xi}f\cdot D(\omega)E_{f}+i_{\xi}h\cdot D(\omega)E_{h}\\
 & -i_{\xi}T(\omega)\cdot E_{e}-i_{\xi}R(\omega)\cdot E_{\omega}-i_{\xi}D(\omega)f\cdot E_{f}-i_{\xi}D(\omega)h\cdot E_{h},
\end{split}
\label{11.6}
\end{equation}
where 
\begin{equation}
\begin{split}J_{\text{N}}(\xi)= & \tilde{\Theta}(a,\delta_{\xi}a)-i_{\xi}L_{\text{GMMG}}-\psi_{\xi}+(i_{\xi}\omega-\chi_{\xi})\cdot E_{\omega}\\
 & +i_{\xi}e\cdot E_{e}+i_{\xi}f\cdot E_{f}+i_{\xi}h\cdot E_{h}.
\end{split}
\label{11.7}
\end{equation}
By substituting the explicit forms of $E_{e}$, $E_{\omega}$, $E_{f}$
and $E_{h}$ (i.e. the equations \eqref{2.29}-\eqref{2.32} \textit{without}
imposing that they vanish, one arrives at 
\begin{equation}
\begin{split}dJ_{\text{N}}(\xi)= & (i_{\xi}\omega-\chi_{\xi})\cdot\left[-\sigma\left(D(\omega)T(\omega)-R(\omega)\times e\right)+\frac{1}{\mu}D(\omega)R(\omega)\right]\\
 & -\sigma i_{\xi}e\cdot D(\omega)R(\omega)-\frac{1}{m^{2}}i_{\xi}f\cdot D(\omega)R\\
 & +i_{\xi}h\cdot\left[D(\omega)T(\omega)-R(\omega)\times e\right].
\end{split}
\label{11.8}
\end{equation}
The right hand side of the \eqref{11.8} vanishes due to the Bianchi
identities \eqref{1.51}. Thus, $J_{\text{N}}(\xi)$ is the off-shell
conserved Noether current associated to $\xi$.Locally, by the virtue
of the {Poincare} lemma, 
\begin{equation}
J_{\text{N}}(\xi)=dK_{\text{N}}(\xi),\label{11.9}
\end{equation}
where $K_{\text{N}}(\xi)$ is a potential given as 
\begin{equation}
K_{\text{N}}(\xi)=(i_{\xi}\omega-\chi_{\xi})\cdot\left(-\sigma e+\frac{1}{2\mu}\omega-\frac{1}{m^{2}}f\right)+i_{\xi}e\cdot h.\label{11.10}
\end{equation}
We can define the off-shell charge as \cite{83} 
\begin{equation}
Q_{\text{N}}(\xi)=-\frac{1}{8\pi G}\int_{\Sigma}K_{\text{N}}(\xi),\label{11.11}
\end{equation}
where $G$ denotes Newton's gravitational constant and $\Sigma$ is
a space-like codimension two surface. Now, we consider all of the
solutions of cosmological Einstein's gravity that solve GMMG provided
that the equations \eqref{2.36}-\eqref{2.39} hold. For this class
of solutions $K_{\text{N}}(\xi)$ becomes 
\begin{equation}
\begin{split}K_{\text{N}}(\xi)= & -\left(\sigma+\frac{\alpha H}{2\mu}+\frac{F}{m^{2}}\right)(i_{\xi}\Omega-\chi_{\xi})\cdot e+\frac{1}{2\mu}(i_{\xi}\Omega-\chi_{\xi})\cdot\Omega\\
 & -\frac{\alpha H}{2\mu}i_{\xi}e\cdot\Omega+\frac{1}{2\mu l^{2}}i_{\xi}e\cdot e,
\end{split}
\label{11.12}
\end{equation}
where \eqref{2.9} was used. This last formula expresses potential
in terms of the ordinary torsion-free spin-connection.

\subsection{Near horizon symmetries of non-extremal black holes}

\label{S.11.1} Here we summarize some results of of \cite{55}. The
near horizon geometry of a 3 dimensional black hole in the Gaussian
null coordinates (in the dreiben form) can be given as 
\begin{equation}
\begin{split}e^{0}= & \sqrt{-A+\frac{C^{2}}{R^{2}}}dv-\frac{B}{\sqrt{-A+\frac{C^{2}}{R^{2}}}}d\rho\\
e^{1}= & \frac{B}{\sqrt{-A+\frac{C^{2}}{R^{2}}}}d\rho\\
e^{2}= & \frac{C}{R}dv+Rd\phi
\end{split}
\label{11.13}
\end{equation}
where $v$, $\rho$ and $\phi$ are the retarded time, the radial
distance to the horizon and the angular coordinate, respectively.
The horizon of black hole is located at $\rho=0$. The metric reads
\begin{equation}
ds^{2}=Adv^{2}+2Bdvd\rho+2Cdvd\phi+R^{2}d\phi^{2}.\label{11.14}
\end{equation}
We consider the case in which $A$, $B$, $C$ and $R$ obey the following
fall-off conditions near the horizon \cite{55} 
\begin{equation}
\begin{split} & A=-2\kappa_{H}\rho+\mathcal{O}(\rho^{2}),\hspace{0.7cm}B=1+\mathcal{O}(\rho^{2})\\
 & C=\theta(\phi)\rho+\mathcal{O}(\rho^{2}),\hspace{0.7cm}R^{2}=\gamma(\phi)^{2}+\beta(\phi)\rho+\mathcal{O}(\rho^{2}),
\end{split}
\label{11.15}
\end{equation}
where $\kappa_{H}$ is the surface gravity. The near horizon geometry
of a non-extremal black hole is invariant under the following scaling
\begin{equation}
v\rightarrow v/a,\hspace{0.7cm}\rho\rightarrow a\rho,\hspace{0.7cm}\kappa_{H}\rightarrow a\kappa_{H},\label{11.16}
\end{equation}
where $a$ is a scale factor. The fall-off conditions \eqref{11.15}
obey the scaling property. These fall-off conditions yield finite
charges. We also demand that $g_{\rho\rho}=\mathcal{O}(\rho^{2})$
and $g_{\rho\phi}=\mathcal{O}(\rho^{2})$. We should mention that
the boundary conditions \eqref{11.15} break the {Poincare} symmetry.
The near horizon Killing vectors preserving the fall-off conditions
are \cite{55} 
\begin{equation}
\begin{split}\xi^{v}= & T(\phi)+\mathcal{O}(\rho^{3})\\
\xi^{\rho}= & \frac{\theta(\phi)T^{\prime}(\phi)}{2\gamma(\phi)^{2}}\rho^{2}+\mathcal{O}(\rho^{3})\\
\xi^{\phi}= & Y(\phi)-\frac{T^{\prime}(\phi)}{\gamma(\phi)^{2}}\rho+\frac{\beta(\phi)T^{\prime}(\phi)}{2\gamma(\phi)^{4}}\rho^{2}+\mathcal{O}(\rho^{3}),
\end{split}
\label{11.17}
\end{equation}
where $T(\phi)$ and $Y(\phi)$ are arbitrary functions of their arguments,
and the prime denotes differentiation with respect to $\phi$. Under
a transformation generated by the Killing vector fields \eqref{11.17},
the arbitrary functions $\theta(\phi)$, $\gamma(\phi)$ and $\beta(\phi)$,
that appear in the metric transform as 
\begin{equation}
\begin{split} & \delta_{\xi}\theta=\left(\theta Y\right)^{\prime}-2\kappa_{H}T^{\prime},\hspace{0.7cm}\delta_{\xi}\gamma=\left(\gamma Y\right)^{\prime},\\
 & \delta_{\xi}\beta=2Y^{\prime}\beta+2T^{\prime}\theta+Y\beta^{\prime}-2T^{\prime\prime}+\frac{2\gamma^{\prime}T^{\prime}}{\gamma}.
\end{split}
\label{11.18}
\end{equation}
Using the modified version of the Lie brackets \eqref{5.9}, we have
\begin{equation}
\left[\xi(T_{1},Y_{1}),\xi(T_{2},Y_{2})\right]=\xi(T_{12},Y_{12}),\label{11.19}
\end{equation}
where 
\begin{equation}
T_{12}=Y_{1}T_{2}^{\prime}-Y_{2}T_{1}^{\prime},\hspace{0.7cm}Y_{12}=Y_{1}Y_{2}^{\prime}-Y_{2}Y_{1}^{\prime}.\label{11.20}
\end{equation}
Again introducing the Fourier modes $T_{n}=\xi\left(e^{in\phi},0\right)$
and $Y_{n}=\xi\left(0,e^{in\phi}\right)$, one can find that $T_{n}$
and $Y_{n}$ satisfy the algebra 
\begin{equation}
\begin{split}i\left[T_{m},T_{n}\right]= & 0,\\
i\left[Y_{m},Y_{n}\right]= & (m-n)Y_{m+n},\\
i\left[Y_{m},T_{n}\right]= & -nT_{m+n},
\end{split}
\label{11.21}
\end{equation}
where $T_{n}$ and $Y_{n}$ are generators of the supertranslations
and superrotations respectively.

\subsection{Near horizon conserved charges and their algebra for {General Minimal
Massive Gravity}}

\label{S.11.2} Let us take the space-like codimension two surface
$\Sigma$ to be a circle with radius $\rho\rightarrow0$. Then \eqref{11.11}
can be written as 
\begin{equation}
Q_{\text{N}}(\xi)=-\frac{1}{8\pi G}\lim_{\rho\rightarrow0}\int_{0}^{2\pi}K_{\text{N}\phi}d\phi,\label{11.22}
\end{equation}
where $K_{\text{N}}(\xi)$ is given by \eqref{11.12}. Plugging \eqref{11.12},
\eqref{11.13} and \eqref{11.17} in \eqref{11.22} the conserved
charge for the Killing vector \eqref{11.17} reads 
\begin{equation}
\begin{split}Q_{\text{N}}(\xi)=\frac{1}{16\pi G}\int_{0}^{2\pi}d\phi\biggl\{ & \left(\sigma+\frac{\alpha H}{2\mu}+\frac{F}{m^{2}}\right)\gamma(\phi)\left[2\kappa_{H}T(\phi)-\theta(\phi)Y(\phi)\right]\\
 & +\frac{1}{4\mu}\theta(\phi)\left[2\kappa_{H}T(\phi)-\theta(\phi)Y(\phi)\right]\\
 & -\frac{\alpha H}{2\mu}\gamma(\phi)\theta(\phi)Y(\phi)-\frac{1}{\mu l^{2}}\gamma(\phi)^{2}Y(\phi)\biggr\}
\end{split}
\label{11.23}
\end{equation}
where we also used\eqref{6.2}. In the limit $\mu\rightarrow\infty$
and $m\rightarrow\infty$ and choosing $\sigma=1$, at which the GMMG
model reduces to the the Einstein's theory, the last expression reduces
to the result of \cite{55}.

The algebra of the charges is given by \eqref{5.6}, then for \eqref{11.23}
one finds 
\begin{multline}
\left\{ Q(\xi_{1}),Q(\xi_{2})\right\} =Q\left(\left[\xi_{1},\xi_{2}\right]\right)\\
-\frac{\kappa_{H}}{64\pi\mu G}\int_{0}^{2\pi}d\phi\left\{ 2\kappa_{H}\left(T_{1}T_{2}^{\prime}-T_{2}T_{1}^{\prime}\right)-\left(\theta(\phi)+2\alpha H\gamma(\phi)\right)T_{12}\right\} .\label{11.24}
\end{multline}
For the Fourier modes $\mathcal{T}_{n}=Q(e^{in\phi},0)$ and $\mathcal{Y}_{n}=Q(0,e^{in\phi})$,
we have 
\begin{equation}
\begin{split}i\left[\mathcal{T}_{m},\mathcal{T}_{n}\right]= & \frac{\kappa_{H}^{2}n}{8\mu G}\delta_{m+n,0},\\
i\left[\mathcal{Y}_{m},\mathcal{Y}_{n}\right]= & (m-n)\mathcal{Y}_{m+n},\\
i\left[\mathcal{Y}_{m},\mathcal{T}_{n}\right]= & -n\mathcal{T}_{m+n}-\frac{\kappa_{H}n}{64\pi\mu G}\int_{0}^{2\pi}e^{i(m+n)\phi}\left\{ \theta(\phi)+2\alpha H\gamma(\phi)\right\} d\phi.
\end{split}
\label{11.25}
\end{equation}
In the limit $\mu\rightarrow\infty$, this algebra reduces to the
result given in \cite{55}. In other words, the central extension
term comes from just the Lorentz Chern-Simons term. In the framework
of the cosmological Einstein's gravity, the algebra spanned by $\mathcal{T}_{m}$
and $\mathcal{Y}_{n}$ is isomorphic to \eqref{11.21}, with no central
extension \cite{55}. By looking at the expression of conserved charge
\eqref{11.23} and the algebra \eqref{11.25}, one can find the algebra
of \cite{55} by turning-off the topological term, i.e $\mu\rightarrow\infty$.
The presence of the term proportional to $\frac{1}{m^{2}}$ (the term
of the NMG), does not lead to a centrally extended algebra. If one
introduces the generator 
\begin{equation}
\mathcal{P}_{n}=\sum_{k\in Z}\mathcal{T}_{k}\mathcal{T}_{n-k},\label{11.26}
\end{equation}
one can show that the algebra spanned by $\mathcal{P}_{n}$ and $\mathcal{Y}_{n}$
is $BMS_{3}$ \cite{55,56}. So according to the above discussion
we conclude that that the near horizon geometry of a non-extremal
black hole solution of NMG, has a $BMS_{3}$ symmetry which can be
recovered by means of the mentioned Sugawara construction.

One can easily read off the eigenvalues of $\mathcal{T}_{n}$ and
$\mathcal{Y}_{n}$ from \eqref{11.23} 
\begin{equation}
\mathcal{T}_{n}=\frac{\kappa_{H}}{8\pi G}\int_{0}^{2\pi}e^{in\phi}\left\{ \left(\sigma+\frac{\alpha H}{2\mu}+\frac{F}{m^{2}}\right)\gamma(\phi)+\frac{1}{4\mu}\theta(\phi)\right\} d\phi,\label{11.27}
\end{equation}
\begin{equation}
\mathcal{Y}_{n}=-\frac{1}{16\pi G}\int_{0}^{2\pi}e^{in\phi}\left\{ \left(\sigma+\frac{\alpha H}{\mu}+\frac{F}{m^{2}}\right)\gamma(\phi)\theta(\phi)+\frac{\theta(\phi)^{2}}{4\mu}+\frac{\gamma(\phi)^{2}}{\mu l^{2}}\right\} d\phi.\label{11.28}
\end{equation}
For the BTZ black hole, we have \cite{55} 
\begin{equation}
\gamma=r_{+},\hspace{0.7cm}\theta=\frac{2r_{-}}{l},\hspace{0.7cm}\kappa_{H}=\frac{r_{+}^{2}-r_{-}^{2}}{l^{2}r_{+}},\label{11.29}
\end{equation}
Thus, the algebra spanned by $\mathcal{T}_{n}$ and $\mathcal{Y}_{n}$
reduce to 
\begin{equation}
\begin{split}i\left[\mathcal{T}_{m},\mathcal{T}_{n}\right]= & \frac{\kappa_{H}^{2}n}{8\mu G}\delta_{m+n,0},\\
i\left[\mathcal{Y}_{m},\mathcal{Y}_{n}\right]= & (m-n)\mathcal{Y}_{m+n},\\
i\left[\mathcal{Y}_{m},\mathcal{T}_{n}\right]= & -n\mathcal{T}_{m+n}-\frac{\kappa_{H}n}{8\mu G}\left(\frac{r_{-}}{l}+\alpha Hr_{+}\right)\delta_{m+n,0}.
\end{split}
\label{11.30}
\end{equation}
where we made a shift on spectrum of $\mathcal{T}_{n}$ by a constant
$+\frac{\kappa_{H}}{8\mu G}\left(\frac{r_{-}}{l}+\alpha Hr_{+}\right)$
which is suitable for the following discussion. In this case, \eqref{11.27}
and \eqref{11.28} reduce to 
\begin{equation}
\mathcal{T}_{n}=+\frac{\kappa_{H}}{4G}\left\{ \left(\sigma+\frac{\alpha H}{\mu}+\frac{F}{m^{2}}\right)r_{+}+\frac{r_{-}}{\mu l}\right\} \delta_{n,0},\label{11.32}
\end{equation}
\begin{equation}
\mathcal{Y}_{n}=-\frac{1}{8G}\left\{ \left(\sigma+\frac{\alpha H}{\mu}+\frac{F}{m^{2}}\right)\frac{2r_{+}r_{-}}{l}+\frac{(r_{+}^{2}+r_{-}^{2})}{\mu l^{2}}\right\} \delta_{n,0}.\label{11.33}
\end{equation}
Using the results \cite{104}, we find that the zero mode charge $\mathcal{T}_{0}$
is proportional to the entropy of the BTZ black hole solution of GMMG,
i.e. $\mathcal{T}_{0}=\frac{\kappa_{H}}{2\pi}S$, where $S$ is the
entropy of the BTZ black hole which is given by \eqref{9.15}. Also,
$\mathcal{Y}_{0}$ gives the angular momentum, i.e. $J=-\mathcal{Y}_{0}$

\section{Extended near horizon geometry}

\label{S.12.0} In \cite{57}, the authors proposed the following
metric with new fall-off conditions for the near horizon of a non-extremal
black hole 
\begin{equation}
\begin{split}ds^{2}= & \left[l\rho\left(f_{+}\zeta^{+}+f_{-}\zeta^{-}\right)+\frac{l^{2}}{4}\left(\zeta^{+}-\zeta^{-}\right)^{2}\right]dv^{2}+2ldvd\rho\\
 & +l\left(\frac{\mathcal{J}^{+}}{\zeta^{+}}-\frac{\mathcal{J}^{-}}{\zeta^{-}}\right)d\rho d\phi+l\rho\left(\frac{\mathcal{J}^{+}}{\zeta^{+}}-\frac{\mathcal{J}^{-}}{\zeta^{-}}\right)\left(f_{+}\zeta^{+}+f_{-}\zeta^{-}\right)dvd\phi\\
 & +\left[\frac{l^{2}}{4}\left(\mathcal{J}^{+}+\mathcal{J}^{-}\right)^{2}-\frac{l\rho}{\zeta^{+}\zeta^{-}}\left(f_{+}\zeta^{+}+f_{-}\zeta^{-}\right)\mathcal{J}^{+}\mathcal{J}^{-}\right]d\phi^{2},
\end{split}
\label{12.1}
\end{equation}
where $\zeta^{\pm}$ are constant parameters, $\mathcal{J}^{\pm}=\mathcal{J}^{\pm}(\phi)$
are arbitrary functions of $\phi$ and $f_{\pm}=f_{\pm}(\rho)$ are
given as 
\begin{equation}
f_{\pm}(\rho)=1-\frac{\rho}{2l\zeta^{\pm}}.\label{12.2}
\end{equation}
This metric is written in the ingoing Eddington-Finkelstein coordinates:
$v$, $\rho$ and $\phi$ are the advanced time, the radial coordinate
and the angular coordinate, respectively. In the particular case of
$\zeta^{\pm}=-a$, where the constant $a$ is the Rindler acceleration,
the metric reduces to 
\begin{equation}
\begin{split}ds^{2}= & -2al\rho f(\rho)dv^{2}+2ldvd\rho-2a^{-1}\theta(\phi)d\phi d\rho+4\rho\theta(\phi)f(\rho)dvd\phi\\
 & +\left[\gamma(\phi)^{2}+\frac{2\rho}{al}f(\rho)\left(\gamma(\phi)^{2}-\theta(\phi)^{2}\right)\right]d\phi^{2},
\end{split}
\label{12.3}
\end{equation}
where $l\mathcal{J}^{\pm}=\gamma\pm\theta$ and $f(\rho)=1+\frac{\rho}{2la}$.
It describes a spacetime which possesses an event horizon located
at $\rho=0$ and solves the cosmological Einstein's gravity \eqref{2.4}.\\
 The following Killing vector 
\begin{equation}
\begin{split} & \xi^{v}=\frac{1}{2}\left\{ -\left(\frac{1}{\zeta^{+}}-\frac{1}{\zeta^{-}}\right)\left(\frac{\mathcal{J}^{+}}{\zeta^{+}}-\frac{\mathcal{J}^{-}}{\zeta^{-}}\right)\left(\frac{\mathcal{J}^{+}}{\zeta^{+}}+\frac{\mathcal{J}^{-}}{\zeta^{-}}\right)^{-1}+\left(\frac{1}{\zeta^{+}}+\frac{1}{\zeta^{-}}\right)\right\} \Xi(\phi)\\
 & \xi^{\rho}=0\\
 & \xi^{\phi}=\left(\frac{1}{\zeta^{+}}-\frac{1}{\zeta^{-}}\right)\left(\frac{\mathcal{J}^{+}}{\zeta^{+}}+\frac{\mathcal{J}^{-}}{\zeta^{-}}\right)^{-1}\Xi(\phi)
\end{split}
\label{12.4}
\end{equation}
preserves the fall-off conditions \eqref{12.1}, up to terms that
involve powers of $\delta\mathcal{J}$, i.e. we ignore the terms of
order $\mathcal{O}(\delta\mathcal{J}^{2})$. $\Xi(\phi)$ is an arbitrary
function of $\phi$. Under the transformations generated by the Killing
vector field, the arbitrary functions $\mathcal{J}^{\pm}(\phi)$ transform
as 
\begin{equation}
\delta_{\xi}\mathcal{J}^{\pm}=\pm\Xi^{\prime},\label{12.5}
\end{equation}
We can use the modified version of the Lie brackets \eqref{5.9} to
show that 
\begin{equation}
\left[\xi_{1},\xi_{2}\right]=0.\label{12.6}
\end{equation}
Therefore, the Killing vectors $\xi_{1}=\xi(\Xi_{1})$ and $\xi_{2}=\xi(\Xi_{2})$
commute. The dreibeins for \eqref{12.1} are 
\begin{equation}
\begin{split}e^{0}= & -\frac{1}{2}\left[2-\frac{l\rho}{2}\left(f_{+}\zeta^{+}+f_{-}\zeta^{-}\right)\right]dv+\frac{l}{2}d\rho\\
 & +\frac{1}{2}\left[-\left(\frac{\mathcal{J}^{+}}{\zeta^{+}}-\frac{\mathcal{J}^{-}}{\zeta^{-}}\right)+\frac{l\rho}{2}\left(f_{+}\mathcal{J}^{+}-f_{-}\mathcal{J}^{-}\right)\right]d\phi\\
e^{1}= & \frac{l}{2}\left(\zeta^{+}-\zeta^{-}\right)dv+\frac{l}{2}\left[\left(\mathcal{J}^{+}+\mathcal{J}^{-}\right)-\frac{\rho}{l}\left(\frac{\mathcal{J}^{+}}{\zeta^{+}}+\frac{\mathcal{J}^{-}}{\zeta^{-}}\right)\right]d\phi\\
e^{2}= & -\frac{1}{2}\left[2+\frac{l\rho}{2}\left(f_{+}\zeta^{+}+f_{-}\zeta^{-}\right)\right]dv-\frac{l}{2}d\rho\\
 & -\frac{1}{2}\left[\left(\frac{\mathcal{J}^{+}}{\zeta^{+}}-\frac{\mathcal{J}^{-}}{\zeta^{-}}\right)+\frac{l\rho}{2}\left(f_{+}\mathcal{J}^{+}-f_{-}\mathcal{J}^{-}\right)\right]d\phi.
\end{split}
\label{12.7}
\end{equation}

\subsection{Application to the general minimal massive gravity}

\label{S.12.1} Now we want to simplify \eqref{5.5} in the GMMG model
for for Einstein spaces, for which we have \eqref{2.4}. To this end,
we use the equations \eqref{2.36}-\eqref{2.39} and \eqref{2.9},
then we find 
\begin{equation}
\begin{split}\delta Q(\xi)=-\frac{1}{8\pi G}\int_{\Sigma}\biggl\{ & -\left(\sigma+\frac{\alpha H}{\mu}+\frac{F}{m^{2}}\right)\left[(i_{\xi}\Omega-\chi_{\xi})\cdot\delta e+i_{\xi}e\cdot\delta\Omega\right]\\
 & +\frac{1}{\mu}\left[(i_{\xi}\Omega-\chi_{\xi})\cdot\delta\Omega+\frac{1}{l^{2}}i_{\xi}e\cdot\delta e\right]\biggr\}.
\end{split}
\label{12.8}
\end{equation}
The torsion free spin-connection is given by \eqref{1.52}, then by
substituting \eqref{12.7} into \eqref{1.52} one gets 
\begin{equation}
\begin{split}\Omega^{0}= & -\frac{1}{4}\left(\zeta^{+}-\zeta^{-}\right)\rho dv+\frac{1}{2l}\left[\left(\frac{\mathcal{J}^{+}}{\zeta^{+}}+\frac{\mathcal{J}^{-}}{\zeta^{-}}\right)-\frac{l\rho}{2}\left(f_{+}\mathcal{J}^{+}+f_{-}\mathcal{J}^{-}\right)\right]d\phi\\
\Omega^{1}= & -\frac{1}{2}\left[\left(\zeta^{+}+\zeta^{-}\right)-\frac{2\rho}{l}\right]dv-\frac{1}{2}\left[\left(1-\frac{\rho}{l\zeta^{+}}\right)\mathcal{J}^{+}-\left(1-\frac{\rho}{l\zeta^{-}}\right)\mathcal{J}^{-}\right]d\phi\\
\Omega^{2}= & \frac{1}{4}\left(\zeta^{+}-\zeta^{-}\right)\rho dv+\frac{1}{2l}\left[\left(\frac{\mathcal{J}^{+}}{\zeta^{+}}+\frac{\mathcal{J}^{-}}{\zeta^{-}}\right)+\frac{l\rho}{2}\left(f_{+}\mathcal{J}^{+}+f_{-}\mathcal{J}^{-}\right)\right]d\phi.
\end{split}
\label{12.9}
\end{equation}
On the other hand, $\chi_{\xi}^{a}$ is given by \eqref{6.2}. Therefore,
using equations \eqref{12.4}, \eqref{12.7}, \eqref{12.9}, one obtains
\begin{equation}
\begin{split}(i_{\xi}\Omega-\chi_{\xi})\cdot\delta e+i_{\xi}e\cdot\delta\Omega= & -\frac{l}{2}\left(\Xi\delta\mathcal{J}^{+}+\Xi\delta\mathcal{J}^{-}\right)d\phi+\mathcal{O}(\delta\mathcal{J}^{2}),\\
(i_{\xi}\Omega-\chi_{\xi})\cdot\delta\Omega+\frac{1}{l^{2}}i_{\xi}e\cdot\delta e= & \frac{1}{2}\left(\Xi\delta\mathcal{J}^{+}-\Xi\delta\mathcal{J}^{-}\right)d\phi+\mathcal{O}(\hat{\delta}\mathcal{J}^{2}).
\end{split}
\label{12.10}
\end{equation}
By substituting \eqref{12.10} into \eqref{12.8} and carrying an
integration over the one-parameter path on the solution space, one
arrives at 
\begin{equation}
Q(\xi)=Q(\tau^{+})+Q(\tau^{-})\label{12.11}
\end{equation}
where $\tau^{\pm}=\pm\Xi(\phi)$ and $Q(\tau^{\pm})$ reads 
\begin{equation}
Q(\tau^{\pm})=\mp\frac{k}{4\pi}\left(\sigma\mp\frac{1}{\mu l}+\frac{\alpha H}{\mu}+\frac{F}{m^{2}}\right)\int_{0}^{2\pi}\tau^{\pm}(\phi)\mathcal{J}^{\pm}(\phi)d\phi.\label{12.12}
\end{equation}
In the equation \eqref{12.12}, we set $k=\frac{l}{4G}$. The algebra
of conserved charges is \eqref{5.6}. Due to \eqref{5.8} and \eqref{12.6},
one can deduce that $\delta_{\xi_{2}}Q(\xi_{1})=\mathcal{C}\left(\xi_{1},\xi_{2}\right)$.
By varying \eqref{12.11} with respect to the dynamical fields so
that the variation is generated by a Killing vector, one has 
\begin{equation}
\begin{split}\delta_{\tau_{2}^{\pm}}Q(\tau_{1}^{\pm})= & \mp\frac{k}{8\pi}\left(\sigma\mp\frac{1}{\mu l}+\frac{\alpha H}{\mu}+\frac{F}{m^{2}}\right)\int_{0}^{2\pi}\Xi_{12}(\phi)d\phi,\\
\hat{\delta}_{\tau_{2}^{\pm}}Q(\tau_{1}^{\mp})= & 0,
\end{split}
\label{12.13}
\end{equation}
where 
\begin{equation}
\Xi_{12}=\Xi_{1}\Xi_{2}^{\prime}-\Xi_{2}\Xi_{1}^{\prime}.\label{12.14}
\end{equation}
By setting $\tau^{\pm}=\pm\Xi(\phi)=\pm e^{in\phi}$, one can expand
$Q(\tau^{\pm})$ in the Fourier modes 
\begin{equation}
J_{n}^{\pm}=-\frac{k}{4\pi}\left(\sigma\mp\frac{1}{\mu l}+\frac{\alpha H}{\mu}+\frac{F}{m^{2}}\right)\int_{0}^{2\pi}e^{in\phi}\mathcal{J}^{\pm}(\phi)d\phi.\label{12.15}
\end{equation}
Also, by substituting $\Xi_{1}=e^{in\phi}$, $\Xi_{2}=e^{im\phi}$
into \eqref{12.13} and replacing the Dirac brackets with commutators,
we get 
\begin{equation}
\begin{split}\left[\hat{J}_{n}^{\pm},\hat{J}_{m}^{\pm}\right]= & \pm\frac{k}{2}\left(\sigma\mp\frac{1}{\mu l}+\frac{\alpha H}{\mu}+\frac{F}{m^{2}}\right)n\delta_{m+n,0},\\
\left[\hat{J}_{n}^{\pm},\hat{J}_{m}^{\mp}\right]= & 0.
\end{split}
\label{12.16}
\end{equation}
Similar to the near horizon symmetry algebra of the black flower solutions
of the Einstein's gravity, the above algebra consists of two $U(1)$
algebras, but with levels $\pm\frac{k}{2}$. The level of the algebra
is given by $\pm\frac{k}{2}\left(\sigma\mp\frac{1}{\mu l}+\frac{\alpha H}{\mu}+\frac{F}{m^{2}}\right)$.

One can change the basis as 
\begin{equation}
\begin{split}\hat{X}_{n}= & \frac{1}{\sqrt{2u_{+}}}\hat{J}_{n}^{+}-\frac{i}{\sqrt{2u_{-}}}\hat{J}_{n}^{-}\hspace{0.5cm}\text{for}\hspace{0.3cm}n\in\mathbb{Z}\\
\hat{P}_{n}= & \frac{i}{n\sqrt{2u_{+}}}\hat{J}_{-n}^{+}-\frac{1}{n\sqrt{2u_{-}}}\hat{J}_{-n}^{-}\hspace{0.5cm}\text{for}\hspace{0.3cm}n\neq0\\
\hat{P}_{0}= & \hat{J}_{0}^{+}+\hat{J}_{0}^{-}\hspace{0.5cm}\text{for}\hspace{0.3cm}n=0,
\end{split}
\label{12.17}
\end{equation}
where 
\begin{equation}
u_{\pm}=\pm\frac{k}{2}\left(\sigma\mp\frac{1}{\mu l}+\frac{\alpha H}{\mu}+\frac{F}{m^{2}}\right).\label{12.18}
\end{equation}
Then the algebra \eqref{12.16}, takes the following form 
\begin{equation}
\left[\hat{X}_{n},\hat{X}_{m}\right]=\left[\hat{P}_{n},\hat{P}_{m}\right]=\left[\hat{X}_{0},\hat{P}_{n}\right]=\left[\hat{P}_{0},\hat{X}_{n}\right]=0\label{12.19}
\end{equation}
\begin{equation}
\left[\hat{X}_{n},\hat{P}_{m}\right]=i\delta_{nm}\hspace{0.5cm}\text{for}\hspace{0.3cm}n,m\neq0,\label{12.20}
\end{equation}
which is the Heisenberg algebra and $\hat{X}_{0}$ and $\hat{P}_{0}$
are the two Casimirs. It is interesting that, for the GMMG, the Heisenberg
algebra appears as the near horizon symmetry algebra of the black
flower solutions. Comparing the definition of $\hat{P}_{0}$ and \eqref{12.11},
one notes that $\hat{P}_{0}$ is just the Hamiltonian, i.e. $\hat{H}\equiv\hat{P}_{0}$.

Setting $\sigma=1$, $\mu\rightarrow\infty$ and $m^{2}\rightarrow\infty$,
the above results \eqref{12.12}, \eqref{12.15} and \eqref{12.16},
which we obtained for the Chern-Simons-like theories of gravity, reduce
to the Cosmological Einstein's theory obtained in \cite{57} with
different methods.

\subsection{Soft hair and the soft hairy black hole entropy}

\label{S.12.2} We know that the Hamiltonian $H=P_{0}$ gives the
dynamics of the system near the horizon. Let us consider all the descendants
of the vacuum \cite{57} 
\begin{equation}
|\psi(q)\rangle=N(q)\prod_{i=1}^{N^{+}}\left(J_{-n_{i}^{+}}^{+}\right)^{m_{i}^{+}}\prod_{i=1}^{N^{-}}\left(J_{-n_{i}^{-}}^{-}\right)^{m_{i}^{-}}|0\rangle\label{12.21}
\end{equation}
where $q$ is a set of arbitrary non-negative integer quantum numbers
$N^{\pm}$, $n_{i}^{\pm}$ and $m_{i}^{\pm}$. Also, $N(q)$ is a
normalization constant such that $\langle\psi(q)|\psi(q)\rangle=1$.
The Hamiltonian $\hat{H}=\hat{J}_{0}^{+}+\hat{J}_{0}^{-}$ commutes
with all the generators $\hat{J}_{n}^{\pm}$, so the energy of all
states are the same. The energy of the vacuum state is given by 
\begin{equation}
H|0\rangle=E_{\text{vac}}|0\rangle.\label{12.22}
\end{equation}
Also, for all descendants, we have 
\begin{equation}
E_{\psi}=\langle\psi(q)|H|\psi(q)\rangle.\label{12.23}
\end{equation}
Due to the mentioned property of the Hamiltonian, we find that all
descendants of the vacuum have the same energy as the vacuum, 
\begin{equation}
E_{\psi}=E_{\text{vac}},\label{12.24}
\end{equation}
in other words, they are \textit{soft hairs } in the sense of having
zero-energy \cite{57,HPS}.

For the case of the BTZ black hole, we have 
\begin{equation}
\mathcal{J}^{\pm}=-\frac{1}{l}\left(r_{+}\mp r_{-}\right),\hspace{0.7cm}\zeta^{\pm}=-\kappa_{H}=-\frac{r_{+}^{2}-r_{-}^{2}}{l^{2}r_{+}},\label{12.25}
\end{equation}
By substituting \eqref{12.25} into \eqref{12.15}, we find the eigenvalues
of $\hat{J}_{n}^{\pm}$ as 
\begin{equation}
j_{n}^{\pm}=\frac{1}{8G}\left(\sigma\mp\frac{1}{\mu l}+\frac{\alpha H}{\mu}+\frac{F}{m^{2}}\right)\left(r_{+}\mp r_{-}\right)\delta_{n,0}.\label{12.26}
\end{equation}
The entropy of a soft hairy black hole is related to the zero mode
charges $j_{0}^{\pm}$ by the following {formula \cite{57,HPS,59,60,61}}
\begin{equation}
S=2\pi\left(j_{0}^{+}+j_{0}^{-}\right).\label{12.27}
\end{equation}
Let us discuss about relation between this result and Iyer-Wald one
which states that the entropy is a Noether charge (See \ref{S.6.0}).
To do this, we restrict ourselves to the case in which $\zeta^{\pm}=-a$.
In this case, asymptotic Killing vector \eqref{12.4} reduces to $\xi=-a^{-1}\Xi(\phi)\partial_{v}$
which means that $\Xi(\phi)$ generator is a supertranslation associated
with the symmetry 
\begin{equation}
v\rightarrow v-a^{-1}\Xi(\phi).
\end{equation}
Therefore, the charge $P_{0}$ conjugate to time translations $\xi_{0}=-a^{-1}\partial_{v}$
is proportional to the black hole entropy. Such a discussion makes
it clear that Eq.~\eqref{12.27} is consistent with the Iyer-Wald
statement but Eq.~\eqref{12.27} naturally motivates performing a
microstate counting in the spirit of \cite{Carlip:1994gy,Strominger:1997eq}.
Hence, by substituting \eqref{12.26} into \eqref{12.27}, we find
the entropy of the BTZ black hole solution of GMMG as 
\begin{equation}
S=\frac{\pi}{2G}\left\{ \left(\sigma+\frac{\alpha H}{\mu}+\frac{F}{m^{2}}\right)r_{+}+\frac{r_{-}}{\mu l}\right\} \label{12.28}
\end{equation}
which is equivalent to \eqref{9.15}. Since, $\hat{J}_{0}^{+}+\hat{J}_{0}^{-}=\hat{P}_{0}$
is one of two Casimirs of algebra, i.e. $\hat{P}_{0}$ is a constant
of motion, we expect that the zero mode eigenvalue of $\hat{P}_{0}$
is a conserved charge. We have shown that the entropy is intended
conserved charge in the general minimal massive gravity and cosmological
Einstein's gravity.

\section{Horizon Fluffs In Generalized Minimal Massive Gravity}

\label{S.13.0} Banados geometries have the metric \cite{62} 
\begin{equation}
ds^{2}=l^{2}\frac{dr^{2}}{r^{2}}-\left(rdx^{+}-\frac{l^{2}\mathcal{L}_{-}}{r}dx^{-}\right)\left(rdx^{-}-\frac{l^{2}\mathcal{L}_{+}}{r}dx^{+}\right),\label{13.1}
\end{equation}
where $x^{\pm}=t/l\pm\phi$. $r$, $t$ and $\phi\sim\phi+2\pi$ are
respectively radial, time and angular coordinates and $\mathcal{L}_{\pm}=\mathcal{L}_{\pm}(x^{\pm})$
are two arbitrary periodic functions. This metric solves cosmological
Einstein's gravity, then we can use the expression \eqref{12.8}.
The metric under transformations generated by $\xi$ transforms as
$\delta_{_{\xi}}g_{\mu\nu}=\pounds_{\xi}g_{\mu\nu}$. The variation
generated by the following Killing vector field preserves the form
of the metric \cite{63} 
\begin{equation}
\begin{split}\xi^{r}= & -\frac{r}{2}\left(\partial_{+}T^{+}+\partial_{-}T^{-}\right),\\
\xi^{\pm}= & T^{\pm}+\frac{l^{2}r^{2}\partial_{\mp}^{2}T^{\mp}+l^{4}\mathcal{L}_{\mp}\partial_{\pm}^{2}T^{\pm}}{2\left(r^{4}-l^{4}\mathcal{L}_{+}\mathcal{L}_{-}\right)},
\end{split}
\label{13.2}
\end{equation}
where $T^{\pm}=T^{\pm}(x^{\pm})$ are two arbitrary periodic functions.
In other words, under transformation generated by the Killing vector
field \eqref{13.2}, the metric \eqref{13.1} transforms as \cite{63}
\begin{equation}
g_{\mu\nu}[\mathcal{L}_{+},\mathcal{L}_{-}]\rightarrow g_{\mu\nu}[\mathcal{L}_{+}+\delta_{\xi}\mathcal{L}_{+},\mathcal{L}_{-}+\delta_{\xi}\mathcal{L}_{-}],\label{13.3}
\end{equation}
with 
\begin{equation}
\delta_{\xi}\mathcal{L}_{\pm}=2\mathcal{L}_{\pm}\partial_{\pm}T^{\pm}+T^{\pm}\partial_{\pm}\mathcal{L}_{\pm}-\frac{1}{2}\partial_{\pm}^{3}T^{\pm},\label{13.4}
\end{equation}
Banados geometries obey the standard Brown-Henneaux boundary conditions
at spatial infinity (see \eqref{7.2}). So, by expanding the Killing
vector field \eqref{13.2} at spatial infinity we will find the asymptotic
Killing vectors \eqref{7.4}. If we set $\delta_{\xi}\mathcal{L}_{\pm}=0$,
we can get the exact symmetries of the {Banados} geometries. In
this case $T^{\pm}$ are not arbitrary functions and this was considered
in \cite{64,65}. Here, we consider the general case in which $\delta_{\xi}\mathcal{L}_{\pm}\neq0$.
Since $\xi$ depends on the dynamical fields, we need to use the modified
version of the Lie brackets \eqref{5.9}. By substituting \eqref{13.2}
into \eqref{5.9}, we find a closed algebra 
\begin{equation}
\left[\xi(T_{1}^{+},T_{1}^{-}),\xi(T_{2}^{+},T_{2}^{-})\right]=\xi(T_{12}^{+},T_{12}^{-}),\label{13.5}
\end{equation}
where $T_{12}^{\pm}=T_{1}^{\pm}\partial_{\pm}T_{2}^{\pm}-T_{2}^{\pm}\partial_{\pm}T_{1}^{\pm}$.

\subsection{\textquotedbl Asymptotic\textquotedbl{} and \textquotedbl near horizon\textquotedbl{}
conserved charges and their algebras}

\label{S.13.1}

\subsubsection{Asymptotic conserved charges}

\label{S.13.1.1} Now, we are going to obtain the conserved charges
that correspond to the asymptotic symmetries of generic black holes
in the class of {Banados} geometries for GMMG \cite{79}. Then we
will obtain the algebra satisfied by these charges. The dreibein that
correspond to the line-element \eqref{13.1} are 
\begin{equation}
\begin{split}e^{0}= & \frac{1}{2}\left(r-\frac{l^{2}\mathcal{L}_{+}}{r}\right)dx^{+}+\frac{1}{2}\left(r-\frac{l^{2}\mathcal{L}_{-}}{r}\right)dx^{-},\\
e^{1}= & \frac{1}{2}\left(r+\frac{l^{2}\mathcal{L}_{+}}{r}\right)dx^{+}-\frac{1}{2}\left(r+\frac{l^{2}\mathcal{L}_{-}}{r}\right)dx^{-},\\
e^{2}= & \frac{l}{r}dr.
\end{split}
\label{13.7}
\end{equation}
The torsion-free spin-connection is 
\begin{equation}
\begin{split}\Omega^{0}= & \frac{1}{2l}\left(r-\frac{l^{2}\mathcal{L}_{+}}{r}\right)dx^{+}-\frac{1}{2l}\left(r-\frac{l^{2}\mathcal{L}_{-}}{r}\right)dx^{-},\\
\Omega^{1}= & \frac{1}{2l}\left(r+\frac{l^{2}\mathcal{L}_{+}}{r}\right)dx^{+}+\frac{1}{2l}\left(r+\frac{l^{2}\mathcal{L}_{-}}{r}\right)dx^{-},\\
\Omega^{2}= & 0,
\end{split}
\label{13.8}
\end{equation}
where \eqref{1.52} was used. One can use \eqref{6.2}, \eqref{13.2},
\eqref{13.7} and \eqref{13.8} to find 
\begin{equation}
\begin{split}(i_{\xi}\Omega-\chi_{\xi})\cdot\hat{\delta}e+i_{\xi}e\cdot\hat{\delta}\Omega= & l\left(T^{+}\hat{\delta}\mathcal{L}_{+}dx^{+}-T^{-}\hat{\delta}\mathcal{L}_{-}dx^{-}\right),\\
(i_{\xi}\Omega-\chi_{\xi})\cdot\hat{\delta}\Omega+\frac{1}{l^{2}}i_{\xi}e\cdot\hat{\delta}e= & T^{+}\hat{\delta}\mathcal{L}_{+}dx^{+}+T^{-}\hat{\delta}\mathcal{L}_{-}dx^{-}.
\end{split}
\label{13.9}
\end{equation}
By substituting \eqref{13.9} into \eqref{12.8} and taking an integration
over a one-parameter path on the solution space, we obtain the conserved
charge corresponding to the Killing vector field \eqref{13.2} as
\begin{equation}
Q(\xi)=Q^{+}(T^{+})+Q^{-}(T^{-}),\label{13.10}
\end{equation}
where 
\begin{equation}
Q^{\pm}(T^{\pm})=\pm\frac{l}{8\pi G}\left(\sigma+\frac{\alpha H}{\mu}+\frac{F}{m^{2}}\mp\frac{1}{\mu l}\right)\int_{\Sigma}T^{\pm}\mathcal{L}_{\pm}dx^{\pm}.\label{13.11}
\end{equation}
The algebra of the conserved charges can be obtained by \eqref{5.6}.
Hence, by substituting \eqref{13.4}, \eqref{13.5} and \eqref{13.10}
into \eqref{5.6}, we find 
\begin{equation}
\begin{split}\left\{ Q^{\pm}(T_{1}^{\pm}),Q^{\pm}(T_{2}^{\pm})\right\} _{\text{D.B.}}= & Q^{\pm}(T_{12}^{\pm})\\
-\frac{l}{4\pi G}\left(\sigma+\frac{\alpha H}{\mu}+\frac{F}{m^{2}}\mp\frac{1}{\mu l}\right)\biggl\{ & \int_{\Sigma}T_{12}^{\pm}\mathcal{L}_{\pm}dx^{\pm}\\
 & -\frac{1}{8}\int_{\Sigma}\left(T_{1}^{\pm}\partial_{\pm}^{3}T_{2}^{\pm}-T_{2}^{\pm}\partial_{\pm}^{3}T_{1}^{\pm}\right)dx^{\pm}\biggr\},
\end{split}
\label{13.12}
\end{equation}
\begin{equation}
\left\{ Q^{\pm}(T_{1}^{\pm}),Q^{\mp}(T_{2}^{\mp})\right\} _{\text{D.B.}}=0.\label{13.13}
\end{equation}
By introducing the Fourier modes $Q_{m}^{\pm}=Q^{\pm}(e^{imx^{\pm}})$,
one arrives at 
\begin{equation}
\begin{split}i\left\{ Q_{m}^{\pm},Q_{n}^{\pm}\right\} _{\text{D.B.}}= & (m-n)Q_{m+n}^{\pm}\\
-\frac{l}{8G} & \left(\sigma+\frac{\alpha H}{\mu}+\frac{F}{m^{2}}\mp\frac{1}{\mu l}\right)\left\{ n^{3}\delta_{m+n,0}+\frac{2}{\pi}(m-n)\tilde{\mathcal{L}}_{\pm(m+n)}\right\} ,
\end{split}
\label{13.14}
\end{equation}
\begin{equation}
i\left\{ Q_{m}^{\pm},Q_{n}^{\mp}\right\} _{\text{D.B.}}=0,\label{13.15}
\end{equation}
where 
\begin{equation}
\tilde{\mathcal{L}}_{\pm m}=\int_{\Sigma}e^{imx^{\pm}}\mathcal{L}_{\pm}dx^{\pm}.\label{13.16}
\end{equation}
Now we set $\hat{L}_{m}^{\pm}\equiv Q_{m}^{\pm}$ and replace the
Dirac brackets by commutators and after making a constant shift on
the spectrum of $\hat{L}_{m}^{\pm}$, we have 
\begin{equation}
\left[\hat{L}_{m}^{\pm},\hat{L}_{n}^{\pm}\right]=(m-n)\hat{L}_{m+n}^{\pm}+\frac{c_{\pm}}{12}m^{2}(m-1)\delta_{m+n,0},\label{13.17}
\end{equation}
\begin{equation}
\left[\hat{L}_{m}^{+},\hat{L}_{n}^{-}\right]=0,\label{13.18}
\end{equation}
where $c_{\pm}$ are the central charges given by 
\begin{equation}
c_{\pm}=\frac{3l}{2G}\left(\sigma+\frac{\alpha H}{\mu}+\frac{F}{m^{2}}\mp\frac{1}{\mu l}\right).\label{13.19}
\end{equation}
It is clear that $\hat{L}_{m}^{\pm}$ are the generators of the Virasoro
algebra and the algebra of the asymptotic conserved charges is isomorphic
to two copies of the Virasoro algebra.

\subsubsection{Near horizon conserved charges}

\label{S.13.1.2} In the section \ref{S.12.0}, we found the near
horizon conserved charges of non-extremal black holes in the GMMG.
Also, we showed that the obtained near horizon conserved charges obey
the following algebra (see \eqref{12.16} and compare with \eqref{13.19})
\begin{equation}
\left[\hat{J}_{m}^{\pm},\hat{J}_{n}^{\pm}\right]=\pm\frac{c_{\pm}}{12}m\delta_{m+n,0}\hspace{2mm},\label{13.20}
\end{equation}
\begin{equation}
\left[\hat{J}_{m}^{+},\hat{J}_{n}^{-}\right]=0.\label{13.21}
\end{equation}
Similar to the near horizon symmetry algebra in the Einstein's gravity,
the algebra \eqref{13.20} and \eqref{13.21} consists of two $U(1)$
current algebras, but with levels $\pm\frac{l}{8G}$, here the level
of algebra is given as $\pm\frac{c_{\pm}}{12}$.

\subsection{Relation between the asymptotic and near horizon algebras}

\label{S.13.2} To relate the asymptotic Virasoro algebra \eqref{13.17}
and the near horizon algebra \eqref{13.20}, one needs a twisted Sugawara
construction \cite{66}: 
\begin{equation}
\hat{\mathfrak{L}}_{m}^{\pm}=\frac{im}{\sqrt{\pm2}}\hat{J}_{m}^{\pm}\pm\frac{6}{c_{\pm}}\sum_{p\in\mathbb{Z}}\hat{J}_{m-p}^{\pm}\hat{J}_{p}^{\pm}.\label{13.22}
\end{equation}
It is straightforward to show that they satisfy 
\begin{equation}
\left[\hat{\mathfrak{L}}_{m}^{\pm},\hat{\mathfrak{L}}_{n}^{\pm}\right]=(m-n)\hat{\mathfrak{L}}_{m+n}^{\pm}+\frac{c_{\pm}}{12}m^{3}\delta_{m+n,0},\label{13.23}
\end{equation}
Now, we can get to the asymptotic Virasoro algebra by making a shift
on the spectrum of $\hat{\mathfrak{L}}_{m}^{\pm}$ by a constant,
\begin{equation}
\hat{L}_{m}^{\pm}=\hat{\mathfrak{L}}_{m}^{\pm}+\frac{c_{\pm}}{24}\delta_{m+n,0}\hspace{2mm}.\label{13.24}
\end{equation}
In this way, we can relate the near horizon symmetry algebra to the
asymptotic one in the context of GMMG. This relation shows that the
main part of the horizon fluffs proposed by \cite{Afshar:2016uax,64}
appear for generic black holes in the class of Banados geometries
in GMMG model.

\section{Asymptotically flat spacetimes in {General Minimal Massive Gravity}}

\label{S.14.0} We apply the fall of conditions presented in \cite{67}
for the asymptotically flat spacetime solutions of GMMG model.

\subsection{Asymptotically 2+1 Dimensional Flat Spacetimes}

\label{S.14.1} In this subsection, we consider the following fall
of conditions for asymptotically flat spacetimes 
\begin{equation}
\begin{split}g_{uu}= & \mathcal{M}(\phi)+\mathcal{O}(r^{-2})\\
g_{ur}= & -e^{\mathcal{A}(\phi)}+\mathcal{O}(r^{-2})\\
g_{u\phi}= & \mathcal{N}(u,\phi)+\mathcal{O}(r^{-1})\\
g_{rr}= & \mathcal{O}(r^{-2})\\
g_{r\phi}= & -e^{\mathcal{A}(\phi)}\mathcal{E}(u,\phi)+\mathcal{O}(r^{-1})\\
g_{\phi\phi}= & e^{2\mathcal{A}(\phi)}r^{2}+\mathcal{E}(u,\phi)\left[2\mathcal{N}(u,\phi)-\mathcal{M}(\phi)\mathcal{E}(u,\phi)\right]+\mathcal{O}(r^{-1})
\end{split}
\label{14.1}
\end{equation}
with 
\begin{equation}
\mathcal{N}(u,\phi)=\mathcal{L}(\phi)+\frac{1}{2}u\partial_{\phi}\mathcal{M}(\phi),\hspace{0.7cm}\mathcal{E}(u,\phi)=\mathcal{B}(\phi)+u\partial_{\phi}\mathcal{A}(\phi)\label{14.2}
\end{equation}
as given in \cite{67}. In the this metric $\mathcal{M}(\phi)$, $\mathcal{A}(\phi)$,
$\mathcal{B}(\phi)$ and $\mathcal{L}(\phi)$ are arbitrary functions.
The variation generated by the following Killing vector field preserves
the boundary conditions 
\begin{equation}
\begin{split}\xi^{u}= & \alpha(u,\phi)-\frac{1}{r}e^{-\mathcal{A}(\phi)}\mathcal{E}(u,\phi)\beta(u,\phi)+\mathcal{O}(r^{-2}),\\
\xi^{r}= & rX(\phi)+e^{-\mathcal{A}(\phi)}\left[\mathcal{E}(u,\phi)\partial_{\phi}X(\phi)-\partial_{\phi}\beta(u,\phi)\right],\\
 & +\frac{1}{r}e^{-2\mathcal{A}(\phi)}\beta(u,\phi)\left[\mathcal{N}(u,\phi)-\mathcal{M}(\phi)\mathcal{E}(u,\phi)\right]+\mathcal{O}(r^{-2}),\\
\xi^{\phi}= & Y(\phi)+\frac{1}{r}e^{-\mathcal{A}(\phi)}\beta(u,\phi)+\mathcal{O}(r^{-2}),
\end{split}
\label{14.3}
\end{equation}
with 
\begin{equation}
\alpha(u,\phi)=T(\phi)+u\partial_{\phi}Y(\phi),\hspace{0.7cm}\beta(u,\phi)=Z(\phi)+u\partial_{\phi}X(\phi),\label{14.4}
\end{equation}
where $T(\phi)$, $X(\phi)$, $Y(\phi)$ and $Z(\phi)$ are arbitrary
functions of $\phi$. Since $\xi$ depends on the dynamical fields,
we have to use the modified version of the Lie brackets \eqref{5.9}.
By substituting \eqref{14.3} into \eqref{5.9}, one finds 
\begin{equation}
\left[\xi_{1},\xi_{2}\right]=\xi_{12},\label{14.5}
\end{equation}
where $\xi_{12}=\xi(T_{12},X_{12},Y_{12},Z_{12})$, with 
\begin{equation}
\begin{split}T_{12}= & Y_{1}\partial_{\phi}T_{2}-Y_{2}\partial_{\phi}T_{1}+T_{1}\partial_{\phi}Y_{2}-T_{2}\partial_{\phi}Y_{1},\\
X_{12}= & Y_{1}\partial_{\phi}X_{2}-Y_{2}\partial_{\phi}X_{1},\\
Y_{12}= & Y_{1}\partial_{\phi}Y_{2}-Y_{2}\partial_{\phi}Y_{1},\\
Z_{12}= & Y_{1}\partial_{\phi}Z_{2}-Y_{2}\partial_{\phi}Z_{1}+T_{1}\partial_{\phi}X_{2}-T_{2}\partial_{\phi}X_{1}.
\end{split}
\label{14.6}
\end{equation}
Under a transformation generated by the Killing vector fields \eqref{14.3},
the arbitrary functions $\mathcal{M}(\phi)$, $\mathcal{A}(\phi)$,
$\mathcal{B}(\phi)$ and $\mathcal{L}(\phi)$ transform as 
\begin{equation}
\begin{split}\delta_{\xi}\mathcal{M}(\phi)= & -2\partial_{\phi}X(\phi)\partial_{\phi}\mathcal{A}(\phi)+2\partial_{\phi}Y(\phi)\mathcal{M}(\phi)+Y(\phi)\partial_{\phi}\mathcal{M}(\phi)\\
 & +2\partial_{\phi}^{2}X(\phi),
\end{split}
\label{14.7}
\end{equation}
\begin{equation}
\delta_{\xi}\mathcal{A}(\phi)=Y(\phi)\partial_{\phi}\mathcal{A}(\phi)+\partial_{\phi}Y(\phi)+X(\phi),\label{14.8}
\end{equation}
\begin{equation}
\delta_{\xi}\mathcal{B}(\phi)=T(\phi)\partial_{\phi}\mathcal{A}(\phi)+Y(\phi)\partial_{\phi}\mathcal{B}(\phi)+Z(\phi)+\partial_{\phi}T(\phi),\label{14.9}
\end{equation}
\begin{equation}
\begin{split}\delta_{\xi}\mathcal{L}(\phi)= & \partial_{\phi}T(\phi)\mathcal{M}(\phi)+Y(\phi)\partial_{\phi}\mathcal{L}(\phi)+2\partial_{\phi}Y(\phi)\mathcal{L}(\phi)+\frac{1}{2}T(\phi)\partial_{\phi}\mathcal{M}(\phi)\\
 & -\partial_{\phi}Z(\phi)\partial_{\phi}\mathcal{A}(\phi)-\partial_{\phi}X(\phi)\partial_{\phi}\mathcal{B}(\phi)+\partial_{\phi}^{2}Z(\phi).
\end{split}
\label{14.10}
\end{equation}
By introducing the Fourier modes 
\begin{equation}
\begin{split}\xi_{m}^{(T)}= & \xi(e^{im\phi},0,0,0),\\
\xi_{m}^{(X)}= & \xi(0,e^{im\phi},0,0),\\
\xi_{m}^{(Y)}= & \xi(0,0,e^{im\phi},0),\\
\xi_{m}^{(Z)}= & \xi(0,0,0,e^{im\phi}),
\end{split}
\label{14.11}
\end{equation}
one has 
\begin{equation}
\begin{split} & i\left[\xi_{m}^{(T)},\xi_{n}^{(T)}\right]=0,\hspace{0.5cm}i\left[\xi_{m}^{(T)},\xi_{n}^{(Z)}\right]=0,\hspace{0.5cm}i\left[\xi_{m}^{(X)},\xi_{n}^{(X)}\right]=0,\\
 & i\left[\xi_{m}^{(X)},\xi_{n}^{(Z)}\right]=0,\hspace{0.5cm}i\left[\xi_{m}^{(Z)},\xi_{n}^{(Z)}\right]=0,\hspace{0.5cm}i\left[\xi_{m}^{(T)},\xi_{n}^{(X)}\right]=-n\xi_{m+n}^{(Z)},\\
 & i\left[\xi_{m}^{(X)},\xi_{n}^{(Y)}\right]=m\xi_{m+n}^{(X)},\hspace{0.5cm}i\left[\xi_{m}^{(Y)},\xi_{n}^{(Z)}\right]=-n\xi_{m+n}^{(Z)},\\
 & i\left[\xi_{m}^{(T)},\xi_{n}^{(Y)}\right]=(m-n)\xi_{m+n}^{(T)},\hspace{0.5cm}i\left[\xi_{m}^{(Y)},\xi_{n}^{(Y)}\right]=(m-n)\xi_{m+n}^{(Y)}.
\end{split}
\label{14.12}
\end{equation}
Now we introduce the following dreibeins 
\begin{equation}
\begin{split} & e_{\hspace{1.5mm}u}^{0}=r-\frac{1}{4r}\mathcal{M}(\phi)+\mathcal{O}(r^{-2}),\hspace{0.5cm}e_{\hspace{1.5mm}r}^{0}=\frac{1}{2r}e^{\mathcal{A}(\phi)}+\mathcal{O}(r^{-2}),\\
 & e_{\hspace{1.5mm}\phi}^{0}=r\mathcal{E}(u,\phi)-\frac{1}{4r}\left[2\mathcal{N}(u,\phi)-\mathcal{M}(\phi)\mathcal{E}(u,\phi)\right]+\mathcal{O}(r^{-2}),\\
 & e_{\hspace{1.5mm}u}^{1}=\mathcal{O}(r^{-2}),\hspace{0.5cm}e_{\hspace{1.5mm}r}^{1}=\mathcal{O}(r^{-2}),\hspace{0.5cm}e_{\hspace{1.5mm}\phi}^{1}=re^{\mathcal{A}(\phi)}+\mathcal{O}(r^{-2}),\\
 & e_{\hspace{1.5mm}u}^{2}=-r-\frac{1}{4r}\mathcal{M}(\phi)+\mathcal{O}(r^{-2}),\hspace{0.5cm}e_{\hspace{1.5mm}r}^{2}=\frac{1}{2r}e^{\mathcal{A}(\phi)}+\mathcal{O}(r^{-2}),\\
 & e_{\hspace{1.5mm}\phi}^{2}=-r\mathcal{E}(u,\phi)-\frac{1}{4r}\left[2\mathcal{N}(u,\phi)-\mathcal{M}(\phi)\mathcal{E}(u,\phi)\right]+\mathcal{O}(r^{-2}).
\end{split}
\label{14.13}
\end{equation}
The torsion-free dual curvature 2-form is 
\begin{equation}
R^{a}(\Omega)=\frac{1}{2}e_{\hspace{1.5mm}\lambda}^{a}\epsilon^{\lambda\alpha\beta}\mathcal{R}_{\alpha\beta\mu\nu}dx^{\mu}\wedge dx^{\nu},\label{14.14}
\end{equation}
therefore, for this spacetime, we have 
\begin{equation}
R(\Omega)=\mathcal{O}(r^{-2}).\label{14.15}
\end{equation}
Now, in the context of GMMG, we consider the following ansatz 
\begin{equation}
f=Fe,\hspace{0.7cm}h=He,\label{14.16}
\end{equation}
where $F$ and $H$ are just two constant parameters. By substituting
\eqref{14.15} and \eqref{14.16} into the field equations of GMMG
\eqref{2.33}- \eqref{2.35}, we find 
\begin{equation}
\begin{split} & \Lambda_{0}=\alpha\left(1+\alpha\sigma\right)H^{2}+\frac{F^{2}}{m^{2}},\\
 & F=\mu\left(1+\alpha\sigma\right)H+\frac{\mu\alpha}{m^{2}}HF,\\
 & F+\frac{1}{2}\alpha^{2}H^{2}=0.
\end{split}
\label{14.17}
\end{equation}
Thus, the metric \eqref{14.1} solves equations of motion of GMMG
asymptotically provided that $\Lambda_{0}$, $F$ and $H$ satisfy
equations \eqref{14.17} which admit the following trivial solution
\begin{equation}
\Lambda_{0}=F=H=0.\label{14.18}
\end{equation}
Now, we consider the case in which $\alpha\neq0$. In that case we
have two non-trivial solution 
\begin{equation}
\begin{split}H_{\pm}= & \frac{m^{2}}{2\mu\alpha}\pm\left[\frac{m^{4}}{4\mu^{2}\alpha^{2}}+\frac{m^{2}}{\alpha^{3}}\left(1+\alpha\sigma\right)\right]^{\frac{1}{2}},\\
F_{\pm}= & -\frac{1}{2}\alpha^{2}H_{\pm}^{2},\\
\Lambda_{0\pm}= & \alpha H_{\pm}^{2}\left[\left(1+\alpha\sigma\right)+\frac{\alpha^{3}}{4m^{2}}H_{\pm}^{2}\right].
\end{split}
\label{14.19}
\end{equation}
If one considers the case in which $\alpha=0$ then one again gets
the trivial solution \eqref{14.18}. Thus, for asymptotically flat
spacetimes in GMMG, in contrast to Einstein gravity, cosmological
parameter could be non-zero.

\subsection{Conserved charges of asymptotically flat spacetimes in {general
minimal massive gravity}}

\label{S.14.2} One can use equations \eqref{2.27}, \eqref{2.9},
\eqref{14.18} and \eqref{14.19} to simplify the expression \eqref{5.5}
So for asymptotically flat spacetimes \cite{80} 
\begin{equation}
\begin{split}\delta Q(\xi)=\frac{1}{8\pi G}\lim_{r\rightarrow\infty}\int_{0}^{2\pi}\biggl\{ & \left(\sigma+\frac{\alpha H}{\mu}+\frac{F}{m^{2}}\right)\left[i_{\xi}e\cdot\delta\Omega_{\phi}+\left(i_{\xi}\Omega-\chi_{\xi}\right)\cdot\delta e_{\phi}\right]\\
 & -\frac{1}{\mu}\left(i_{\xi}\Omega-\chi_{\xi}\right)\cdot\delta\Omega_{\phi}\biggr\} d\phi.
\end{split}
\label{14.20}
\end{equation}
By substituting \eqref{14.13} into \eqref{14.20}, using \eqref{1.52}
and \eqref{6.2}, and taking an integration over the one-parameter
path on the solution space to find the conserved charge corresponding
to the Killing vector field \eqref{14.3}, we obtain 
\begin{equation}
Q(T,X,Y,Z)=M(T)+J(X)+L(Y)+P(Z),\label{14.21}
\end{equation}
with 
\begin{equation}
M(T)=\frac{1}{16\pi G}\left(\sigma+\frac{\alpha H}{\mu}+\frac{F}{m^{2}}\right)\int_{0}^{2\pi}T(\phi)\mathcal{M}(\phi)d\phi,\label{14.22}
\end{equation}
\begin{equation}
J(X)=-\frac{1}{8\pi G}\int_{0}^{2\pi}X(\phi)\left[\left(\sigma+\frac{\alpha H}{\mu}+\frac{F}{m^{2}}\right)\partial_{\phi}\mathcal{B}(\phi)-\frac{1}{2\mu}\partial_{\phi}\mathcal{A}(\phi)\right]d\phi,\label{14.23}
\end{equation}
\begin{equation}
\begin{split}L(Y)=\frac{1}{8\pi G}\int_{0}^{2\pi}Y(\phi)\biggl\{ & \left(\sigma+\frac{\alpha H}{\mu}+\frac{F}{m^{2}}\right)\mathcal{L}(\phi)\\
 & -\frac{1}{4\mu}\left[2\mathcal{M}(\phi)+\left(\partial_{\phi}\mathcal{A}(\phi)\right)^{2}-2\partial_{\phi}^{2}\mathcal{A}(\phi)\right]\biggr\} d\phi,
\end{split}
\label{14.24}
\end{equation}
\begin{equation}
P(Z)=-\frac{1}{8\pi G}\left(\sigma+\frac{\alpha H}{\mu}+\frac{F}{m^{2}}\right)\int_{0}^{2\pi}Z(\phi)\partial_{\phi}\mathcal{A}(\phi)d\phi.\label{14.25}
\end{equation}
The algebra of conserved charges is given by \eqref{5.6}. The left
hand side of the equation \eqref{5.6} is defined by \eqref{5.7}.
Therefore one can find the central extension by using \eqref{5.8}.
By substituting \eqref{14.21} and the equations \eqref{14.5}-\eqref{14.10}
into \eqref{5.8}, we obtain the central extension 
\begin{equation}
\begin{split}\mathcal{C}\left(\xi_{1},\xi_{2}\right)= & \frac{1}{8\pi G}\left(\sigma+\frac{\alpha H}{\mu}+\frac{F}{m^{2}}\right)\int_{0}^{2\pi}\biggl\{\left(T_{1}\partial_{\phi}^{2}X_{2}-T_{2}\partial_{\phi}^{2}X_{1}\right)\\
 & \hspace{2cm}+\left(Y_{1}\partial_{\phi}^{2}Z_{2}-Y_{2}\partial_{\phi}^{2}Z_{1}\right)-\left(X_{1}\partial_{\phi}Z_{2}-X_{2}\partial_{\phi}Z_{1}\right)\biggr\} d\phi\\
 & -\frac{1}{16\pi G\mu}\int_{0}^{2\pi}\left\{ \left(Y_{1}\partial_{\phi}^{2}X_{2}-Y_{2}\partial_{\phi}^{2}X_{1}\right)-X_{1}\partial_{\phi}X_{2}-Y_{1}\partial_{\phi}^{3}Y_{2}\right\} d\phi.
\end{split}
\label{14.26}
\end{equation}
By introducing the Fourier modes 
\begin{equation}
\begin{split}M_{m}= & Q(e^{im\phi},0,0,0)=M(e^{im\phi}),\\
J_{m}= & Q(0,e^{im\phi},0,0)=J(e^{im\phi}),\\
L_{m}= & Q(0,0,e^{im\phi},0)=L(e^{im\phi}),\\
P_{m}= & Q(0,0,0,e^{im\phi})=P(e^{im\phi}),
\end{split}
\label{14.27}
\end{equation}
we find 
\begin{equation}
\begin{split} & i\{M_{m},M_{n}\}=0,\hspace{0.7cm}i\{M_{m},P_{n}\}=0,\hspace{0.7cm}i\{P_{m},P_{n}\}=0,\\
 & i\{J_{m},J_{n}\}=k_{J}n\delta_{m+n,0},\hspace{0.7cm}i\{J_{m},P_{n}\}=k_{P}n\delta_{m+n,0},\\
 & i\{M_{m},J_{n}\}=-nP_{m+n}-ik_{P}n^{2}\delta_{m+n,0},\\
 & i\{J_{m},L_{n}\}=mJ_{m+n}+ik_{J}m^{2}\delta_{m+n,0},\\
 & i\{L_{m},P_{n}\}=-nP_{m+n}-ik_{P}n^{2}\delta_{m+n,0},\\
 & i\{M_{m},L_{n}\}=(m-n)M_{m+n},\\
 & i\{L_{m},L_{n}\}=(m-n)L_{m+n}+k_{J}m^{3}\delta_{m+n,0},
\end{split}
\label{14.28}
\end{equation}
where $k_{P}$ and $k_{J}$ are given as 
\begin{equation}
k_{P}=\frac{1}{4G}\left(\sigma+\frac{\alpha H}{\mu}+\frac{F}{m^{2}}\right),\hspace{0.7cm}k_{J}=-\frac{1}{8G\mu}.\label{14.29}
\end{equation}
Now we set $\hat{M}_{m}\equiv M_{m}$, $\hat{J}_{m}\equiv J_{m}$,
$\hat{L}_{m}\equiv L_{m}$ and $\hat{P}_{m}\equiv P_{m}$, also we
replace Dirac brackets by commutators $i\{,\}_{\text{D.B.}}\rightarrow[,]$,
therefore we can rewrite equations \eqref{14.28} as 
\begin{equation}
\begin{split} & [\tilde{L}_{m},\tilde{L}_{n}]=(m-n)\tilde{L}_{m+n}+\frac{c_{L}}{12}m^{3}\delta_{m+n,0}\\
 & [\tilde{M}_{m},\tilde{L}_{n}]=(m-n)\tilde{M}_{m+n}+\frac{c_{M}}{12}m^{3}\delta_{m+n,0},\hspace{0.7cm}[\tilde{M}_{m},\tilde{M}_{n}]=0,
\end{split}
\label{14.30}
\end{equation}
\begin{equation}
\begin{split} & [\tilde{M}_{m},\hat{J}_{n}]=-n\hat{P}_{m+n},\hspace{0.7cm}[\tilde{M}_{m},\hat{P}_{n}]=0,\\
 & [\tilde{L}_{m},\hat{J}_{n}]=-n\hat{J}_{m+n},\hspace{0.7cm}[\tilde{L}_{m},\hat{P}_{n}]=-n\hat{P}_{m+n},\\
 & [\hat{P}_{m},\hat{P}_{n}]=0,\hspace{0.5cm}[\hat{J}_{m},\hat{J}_{n}]=k_{J}n\delta_{m+n,0},\hspace{0.5cm}[\hat{J}_{m},\hat{P}_{n}]=k_{p}n\delta_{m+n,0},
\end{split}
\label{14.31}
\end{equation}
with 
\begin{equation}
c_{L}=24k_{J}=-\frac{3}{G\mu},\hspace{0.7cm}c_{M}=12k_{P}=\frac{3}{G}\left(\sigma+\frac{\alpha H}{\mu}+\frac{F}{m^{2}}\right),\label{14.32}
\end{equation}
where we have performed the shift: 
\begin{equation}
\tilde{M}_{m}=\hat{M}_{m}+im\hat{P}_{m},\hspace{0.7cm}\tilde{L}_{m}=\hat{L}_{m}+im\hat{J}_{m}.\label{14.33}
\end{equation}
The resulting asymptotic symmetry algebra \eqref{14.30} and \eqref{14.31}
is a semidirect product of a $\mathfrak{bms}_{3}$ algebra, with central
charges $c_{L}$ and $c_{M}$, and two $\mathfrak{u}(1)$ current
algebras \cite{67}. If we set $\sigma=+1$, $\alpha=0$ and $m^{2}\rightarrow\infty$
the algebra \eqref{14.30} and \eqref{14.31} reduce to the one presented
in \cite{67} for topologically massive gravity.

The the algebra of the asymptotic conserved charges of asymptotically
AdS$_{3}$ spacetimes in GMMG is isomorphic to two copies of the Virasoro
algebra (see \eqref{13.17}-\eqref{13.19}). The BMS algebra \eqref{14.30}
can be obtained by a contraction of the AdS$_{3}$ asymptotic symmetry
algebra 
\begin{equation}
\tilde{L}_{m}=\hat{L}_{m}^{+}-\hat{L}_{-m}^{-},\hspace{0.7cm}\tilde{M}_{m}=\frac{1}{l}\left(\hat{L}_{m}^{+}+\hat{L}_{-m}^{-}\right),\label{14.34}
\end{equation}
when the AdS$_{3}$ radius tends to infinity in the flat-space limit
\cite{68,69}. Then the corresponding BMS central charges in the algebra
\eqref{14.30} become 
\begin{equation}
c_{M}=\lim_{l\rightarrow\infty}\frac{1}{l}\left(c_{+}+c_{-}\right),\hspace{0.7cm}c_{L}=\lim_{l\rightarrow\infty}\left(c_{+}-c_{-}\right).\label{14.35}
\end{equation}

\subsection{Thermodynamics}

\label{S.14.3} We know that the energy and the angular momentum are
conserved charges corresponding to two asymptotic Killing vector fields
$\partial_{u}$ and $-\partial_{\phi}$, respectively. It can be seen
that $\partial_{u}$ and $-\partial_{\phi}$ are asymptotic Killing
vector fields admitted by spacetimes which behave asymptotically as
\eqref{14.1} when we have $\mathcal{M}(\phi)=\mathcal{M}$, $\mathcal{A}(\phi)=\mathcal{A}$,
$\mathcal{L}(\phi)=\mathcal{L}$ and $\mathcal{B}(\phi)=\mathcal{B}$,
where $\mathcal{M}$, $\mathcal{A}$, $\mathcal{L}$ and $\mathcal{B}$
are constants. Hence, with this assumption, one can use \eqref{14.20}
to find the energy and the angular momentum: 
\begin{equation}
E=Q(\partial_{u})=\frac{1}{8G}\left(\sigma+\frac{\alpha H}{\mu}+\frac{F}{m^{2}}\right)\mathcal{M},\label{14.36}
\end{equation}
\begin{equation}
J=Q(-\partial_{\phi})=-\frac{1}{4G}\left[\left(\sigma+\frac{\alpha H}{\mu}+\frac{F}{m^{2}}\right)\mathcal{L}-\frac{1}{2\mu}\mathcal{M}\right],\label{14.37}
\end{equation}
respectively.The cosmological horizon is located at the solution to
the following equation 
\begin{equation}
g_{uu}g_{\phi\phi}-\left(g_{u\phi}\right)^{2}=0,\label{14.38}
\end{equation}
which is 
\begin{equation}
r_{H}=\frac{e^{-\mathcal{A}}}{\sqrt{\mathcal{M}}}\left|\mathcal{L}-\mathcal{M}\mathcal{B}\right|.\label{14.39}
\end{equation}
One can associate an angular velocity to the cosmological horizon
as 
\begin{equation}
\Omega_{H}=-\frac{g_{u\phi}}{g_{\phi\phi}}\biggr|_{r=r_{H}}=-\frac{\mathcal{M}}{\mathcal{L}}.\label{14.40}
\end{equation}
Since the norm of Killing vector $\zeta=\partial_{u}+\Omega_{H}\partial_{\phi}$
vanishes on the cosmological horizon, one can associate a temperature
to the cosmological horizon as 
\begin{equation}
T_{H}=\frac{\kappa_{H}}{2\pi}\label{14.41}
\end{equation}
where 
\begin{equation}
\kappa_{H}=\left[-\frac{1}{2}\nabla_{\mu}\zeta_{\nu}\nabla^{\mu}\zeta^{\nu}\right]_{r=r_{H}}^{\frac{1}{2}},\label{14.42}
\end{equation}
therefore we have 
\begin{equation}
T_{H}=\frac{\mathcal{M}^{\frac{3}{2}}}{2\pi\mathcal{L}}.\label{14.43}
\end{equation}
One can obtain entropy in the context of GMMG using \eqref{9.14}.
On the cosmological horizon we have 
\begin{equation}
g_{\phi\phi}\bigr|_{r=r_{H}}=\frac{\mathcal{L}^{2}}{\mathcal{M}},\hspace{0.7cm}\Omega_{\phi\phi}\bigr|_{r=r_{H}}=\mathcal{L},\label{14.44}
\end{equation}
then \eqref{9.14} becomes 
\begin{equation}
S=\frac{\pi}{2G}\left[\left(\sigma+\frac{\alpha H}{\mu}+\frac{F}{m^{2}}\right)\frac{\mathcal{L}}{\sqrt{\mathcal{M}}}-\frac{1}{\mu}\sqrt{\mathcal{M}}\right].\label{14.45}
\end{equation}
One can easily check that \eqref{14.36}, \eqref{14.37}, \eqref{14.40},
\eqref{14.43} and \eqref{14.45} satisfy the first law of flat space
cosmologies \cite{19} which is given as 
\begin{equation}
\delta E=-T_{H}\delta S+\Omega_{H}\delta J.\label{14.46}
\end{equation}
It is easy to see that the obtained results \eqref{14.36}, \eqref{14.37}
and \eqref{14.45} will reduce to the corresponding results in topologically
massive gravity \cite{67} for $\sigma=+1$, $\alpha=0$ and $m^{2}\rightarrow\infty$.

\section{Conserved Charges of Minimal Massive Gravity Coupled to Scalar Field}

\label{S.15.0}

\subsection{Minimal massive gravity coupled to a scalar field}

\label{S.15.1} In \cite{70}, it was demonstrated that from the Lorentz-Chern-Simons
action, a topologically massive gravity (TMG) non-minimally coupled
to a scalar field can be constructed. Given 
\begin{equation}
L_{\text{CS}}(\omega)=\omega_{\hspace{1.5mm}b}^{a}\wedge d\omega_{\hspace{1.5mm}a}^{b}+\frac{2}{3}\omega_{\hspace{1.5mm}b}^{a}\wedge\omega_{\hspace{1.5mm}c}^{b}\wedge\omega_{\hspace{1.5mm}a}^{c},\label{15.1}
\end{equation}
where $\omega^{ab}$ are the components of spin-connection 1-form,
we can decompose the spin-connection in two independent parts \eqref{1.13},
where $\Omega^{ab}$ is the torsion-free part which is known as Riemannian
spin-connection and $C^{ab}$ is contorsion 1-form. As discussed before,
we denote contorsion 1-form by $\kappa^{ab}$ (see \eqref{D.13}).
The field equations for the Lorentz-Chern-Simons Lagrangian are 
\begin{equation}
R^{ab}(\omega)=d\omega^{ab}+\omega_{\hspace{1.5mm}c}^{a}\wedge\omega^{cb}=0,\label{15.2}
\end{equation}
where $R^{ab}(\omega)$ is curvature 2-form. Using the Bianchi identities
\eqref{1.18}, we find that $D(\omega)T^{a}(\omega)=0$. This equation
has the following solution in three dimensions 
\begin{equation}
T^{a}(\omega)=\varphi_{0}\varepsilon_{\hspace{1.5mm}bc}^{a}e^{b}\wedge e^{c},\label{15.3}
\end{equation}
where $\varphi_{0}$ is a constant. By comparing \eqref{D.13} and
\eqref{15.3}, we find 
\begin{equation}
\kappa^{ab}=-\varphi_{0}\varepsilon_{\hspace{2.5mm}c}^{ab}e^{c}.\label{15.4}
\end{equation}
In \cite{70}, the authors upgrade $\varphi_{0}$ to be a local dynamical
field $\varphi=\varphi(x)$. By substituting $\omega^{ab}=\Omega^{ab}+\kappa^{ab}$
with $\kappa^{ab}=-\varphi\varepsilon_{\hspace{2.5mm}c}^{ab}e^{c}$
into the Lagrangian \eqref{15.1}, we have 
\begin{equation}
\begin{split}L_{CS}(\omega)= & \frac{1}{2}L_{CS}(\Omega)+\varphi\varepsilon_{abc}e^{a}\wedge R^{bc}(\Omega)+\frac{1}{3}\varphi^{3}\varepsilon_{abc}e^{a}\wedge e^{b}\wedge e^{c}\\
 & +\varphi^{2}e_{a}\wedge T^{a}(\Omega)+\frac{1}{2}d\left(\varphi\varepsilon_{abc}\Omega^{ab}\wedge e^{c}\right).
\end{split}
\label{15.5}
\end{equation}
Eventually, in \cite{70} the following Lagrangian is arrived at 
\begin{equation}
\begin{split}L_{[\lambda,m]}= & \varphi\varepsilon_{abc}e^{a}\wedge R^{bc}(\Omega)+\frac{1}{3!}\lambda\varphi^{3}\varepsilon_{abc}e^{a}\wedge e^{b}\wedge e^{c}+\frac{1}{2m}L_{CS}(\Omega)\\
 & +\varphi^{2}e_{a}\wedge T^{a}(\Omega)+\frac{1}{2}d\left(\varphi\varepsilon_{abc}\Omega^{ab}\wedge e^{c}\right)+\frac{1}{2m}\zeta_{a}\wedge T^{a}(\Omega),
\end{split}
\label{15.6}
\end{equation}
where $\lambda$ and $m$ are two parameters and they are introduced
to denote the cosmological constant and the mass parameter of the
TMG, respectively. The last term in the Lagrangian \eqref{15.6} makes
this theory to be torsion free. This theory describes the topologically
massive gravity coupled non-minimally to a scalar field.

We discussed before that, in thee dimensions, it is convenient to
define the dual spin connection 1-form and the dual curvature 2-form
as \eqref{1.29}. By using a 3D-vector algebra notation for Lorentz
vectors, the dual curvature and torsion 2-forms can be written as
\eqref{1.31} and \eqref{1.32} in terms of the dual spin-connection,
respectively. So far we have reviewed the main idea of the paper \cite{70}.
Now, we are going to generalize the Lagrangian \eqref{15.6} such
that it describes the minimal massive gravity non-minimally coupled
to a scalar field \cite{78}. In order to generalize the Lagrangian
\eqref{15.6}, we consider the spin connection $\Omega^{ab}$ and
the dreibein $e^{a}$ as two independent dynamical fields on equal
footing. First of all, we consider the following redefinitions in
the Lagrangian \eqref{15.6} 
\begin{equation}
\begin{split} & \lambda\rightarrow2\lambda,\hspace{1cm}m\rightarrow\mu,\hspace{1cm}L_{[\lambda,m]}\rightarrow2L,\\
 & \hspace{1cm}\frac{1}{4m}\zeta\rightarrow h,\hspace{1cm}\Omega\rightarrow\omega,
\end{split}
\label{15.7}
\end{equation}
and then, we add the following term to it 
\begin{equation}
\frac{1}{2}\alpha e\cdot h\times h,\label{15.8}
\end{equation}
where $\alpha$ is just a parameter with dimension length. Thus, we
have 
\begin{equation}
\begin{split}L= & \varphi e\cdot R(\omega)+\frac{1}{3!}\lambda\varphi^{3}e\cdot e\times e+\frac{1}{2\mu}(\omega\cdot d\omega+\frac{1}{3}\omega\cdot\omega\times\omega)\\
 & +\frac{1}{2}\varphi^{2}e\cdot T(\omega)+h\cdot T(\omega)+\frac{1}{2}d\left(\varphi\omega\cdot e\right)+\frac{1}{2}\alpha e\cdot h\times h.
\end{split}
\label{15.9}
\end{equation}
It is easy to see that the theory described by the above Lagrangian
is not a torsion free one (for 3D gravities with torsion see \cite{71,72,73}).
If one eliminates the last two terms in the Lagrangian \eqref{15.9},
one may interpret the obtained Lagrangian as a Lagrangian of the Mielke-Baekler
model \cite{74} non-minimally coupled to a scalar field. But we should
note that the Mielke-Baekler Lagrangian is given by \cite{74} 
\begin{equation}
\begin{split}L_{MB}= & \theta_{C}e\cdot R(\omega)+\theta_{\Lambda}e\cdot e\times e+\theta_{L}(\omega\cdot d\omega+\frac{1}{3}\omega\cdot\omega\times\omega)\\
 & +\theta_{T}e\cdot T(\omega)+h\cdot T(\omega).
\end{split}
\label{15.10}
\end{equation}
where $\theta_{C}$, $\theta_{\Lambda}$, $\theta_{L}$ and $\theta_{T}$
are constants. By comparing \eqref{15.9} and \eqref{15.10}, one
may guess that, by changing the frame as $e\rightarrow\varphi e$,
the Mielke-Baekler Lagrangian \eqref{15.10} turns to the Lagrangian
\eqref{15.9} without the last two terms. But this is not correct,
because the torsion 2-form will change as $T(\omega)\rightarrow T(\omega)+d\varphi e$
under the change of frame. Thus, it seems that the considered model,
which is described by the Lagrangian \eqref{15.9}, cannot be simply
seen as a change of frame in the Mielke-Baekler theory.

\subsection{Field equations}

\label{S.15.2} To find the field equations, consider the variation
of \eqref{15.9} 
\begin{equation}
\delta L=\delta\varphi E_{\varphi}+\delta e\cdot E_{e}+\delta\omega\cdot E_{\omega}+\delta h\cdot E_{h}+d\Theta(\Phi,\delta\Phi)\label{15.11}
\end{equation}
where $\Phi$ is a collection of all the fields, i.e. $\Phi=\{\varphi,e,\omega,h\}$.
In the above equation, we have the following definitions 
\begin{equation}
E_{\varphi}=e\cdot R(\omega)+\frac{\lambda}{2}\varphi^{2}e\cdot e\times e+\varphi e\cdot T(\omega),\label{15.12}
\end{equation}
\begin{equation}
E_{e}=\varphi R(\omega)+\frac{\lambda}{2}\varphi^{3}e\times e+\frac{1}{2}\varphi^{2}T(\omega)+\frac{1}{2}\alpha h\times h+\frac{1}{2}D(\omega)\left(\varphi^{2}e\right)+D(\omega)h,\label{15.13}
\end{equation}
\begin{equation}
E_{\omega}=\frac{1}{\mu}R(\omega)+\frac{1}{2}\varphi^{2}e\times e+e\times h+D(\omega)\left(\varphi e\right),\label{15.14}
\end{equation}
\begin{equation}
E_{h}=T(\omega)+\alpha e\times h,\label{15.15}
\end{equation}
\begin{equation}
\Theta(\Phi,\delta\Phi)=\varphi\delta\omega\cdot e+\frac{1}{2\mu}\delta\omega\cdot\omega+\frac{1}{2}\varphi^{2}\delta e\cdot e+\delta e\cdot h+\frac{1}{2}\delta(\varphi\omega\cdot e).\label{15.16}
\end{equation}
So the field equations are 
\begin{equation}
E_{\varphi}=E_{e}=E_{\omega}=E_{h}=0,\label{15.17}
\end{equation}
and $\Theta(\Phi,\delta\Phi)$ is a surface term. We can write \eqref{15.15},
namely $E_{h}=0$, as 
\begin{equation}
T(\omega)=de+(\omega+\alpha h)\times e=0,\label{15.18}
\end{equation}
It is clear that one can use \eqref{2.9} to rewrite the field equations
of in terms of the torsion free dual spin-connection as 
\begin{equation}
e\cdot R(\Omega)+\frac{\lambda}{2}\varphi^{2}e\cdot e\times e-\alpha\varphi e\cdot e\times h+\frac{1}{2}\alpha^{2}e\cdot h\times h-\alpha e\cdot D(\Omega)h=0,\label{15.19}
\end{equation}
\begin{equation}
\varphi R(\Omega)+\frac{\lambda}{2}\varphi^{3}e\times e-\alpha\varphi^{2}e\times h-\frac{1}{2}\alpha(1-\alpha\varphi)h\times h+(1-\alpha\varphi)D(\Omega)h+\varphi d\varphi\hspace{0.5mm}e=0,\label{15.20}
\end{equation}
\begin{equation}
R(\Omega)+\frac{1}{2}\mu\varphi^{2}e\times e+\mu(1-\alpha\varphi)e\times h-\alpha D(\Omega)h+\frac{1}{2}\alpha^{2}h\times h+\mu d\varphi\hspace{0.5mm}e=0,\label{15.21}
\end{equation}
\begin{equation}
T(\Omega)=0.\label{15.22}
\end{equation}
To obtain the above equations we have used 
\begin{equation}
D(\omega)f=D(\Omega)f-\alpha h\times f,\label{15.23}
\end{equation}
where $f$ is an arbitrary Lorentz vector valued 1-form. By combining
the equations \eqref{15.20} and \eqref{15.21}, we have 
\begin{equation}
\begin{split} & R(\Omega)+\frac{1}{2}\left[\alpha\lambda\varphi+\mu(1-\alpha\varphi)\right]\varphi^{2}e\times e+\left[\mu(1-\alpha\varphi)^{2}-\alpha^{2}\varphi^{2}\right]e\times h\\
 & +\left[\alpha\varphi+\mu(1-\alpha\varphi)\right]d\varphi\hspace{0.5mm}e=0.
\end{split}
\label{15.24}
\end{equation}
We can solve this equation to find 
\begin{equation}
\begin{split}h_{\hspace{1.5mm}\mu}^{a}=-\frac{1}{\left[\mu(1-\alpha\varphi)^{2}-\alpha^{2}\varphi^{2}\right]} & \{S_{\hspace{1.5mm}\mu}^{a}+\frac{1}{2}\left[\alpha\lambda\varphi+\mu(1-\alpha\varphi)\right]\varphi^{2}e_{\hspace{1.5mm}\mu}^{a}\\
 & +\left[\alpha\varphi+\mu(1-\alpha\varphi)\right]\varepsilon_{\hspace{3mm}c}^{ab}e_{b}^{\hspace{1.5mm}\nu}e_{\hspace{1.5mm}\mu}^{c}\partial_{\nu}\varphi\}.
\end{split}
\label{15.25}
\end{equation}
In contrast to ordinary MMG, $h_{\mu\nu}$ is not a symmetric tensor,
i.e. in the given model, the condition $e\cdot h=0$ no longer holds.
In the equation \eqref{15.25}, $S_{\mu\nu}$ is 3D Schouten tensor
\eqref{2.13}.

For the BTZ black hole spacetime \eqref{2.4.2}, we have 
\begin{equation}
R(\Omega)=-\frac{1}{2l^{2}}e\times e,\hspace{1cm}S^{a}=-\frac{1}{2l^{2}}e^{a}.\label{15.26}
\end{equation}
Assuming that $\varphi$ is a constant, say $\varphi=\varphi_{0}$,
the BTZ black hole spacetime solves the field equations \eqref{15.19}-\eqref{15.22}.
So, by taking $\varphi=\varphi_{0}$, for BTZ black hole spacetime
the equation \eqref{15.25} reduces to 
\begin{equation}
h^{a}=\beta e^{a},\label{15.27}
\end{equation}
where 
\begin{equation}
\beta=\frac{1-\alpha\lambda l^{2}\varphi_{0}^{3}-\mu l^{2}(1-\alpha\varphi_{0})\varphi_{0}^{2}}{2l^{2}\left[\mu(1-\alpha\varphi_{0})^{2}-\alpha^{2}\varphi_{0}^{2}\right]}.\label{15.28}
\end{equation}
By substituting \eqref{15.26} and \eqref{15.27} into the equations
\eqref{15.19}-\eqref{15.22}, we have 
\begin{equation}
-\frac{1}{2l^{2}}+\frac{1}{2}\lambda\varphi_{0}^{2}-\alpha\beta\varphi_{0}+\frac{1}{2}\alpha^{2}\beta^{2}=0,\label{15.29}
\end{equation}
\begin{equation}
-\frac{\varphi_{0}}{2l^{2}}+\frac{1}{2}\lambda\varphi_{0}^{3}-\alpha\beta\varphi_{0}^{2}-\frac{1}{2}\alpha\beta^{2}(1-\alpha\varphi_{0})=0,\label{15.30}
\end{equation}
\begin{equation}
-\frac{1}{2l^{2}}+\frac{1}{2}\mu\varphi_{0}^{2}+\mu\beta(1-\alpha\varphi_{0})+\frac{1}{2}\alpha^{2}\beta^{2}=0.\label{15.31}
\end{equation}
It is obvious that by combining \eqref{15.30} and \eqref{15.31},
the equation \eqref{15.28} can be obtained. Thus, the BTZ black hole
spacetime together with $\varphi=\varphi_{0}$ is be a solution of
the considered model when the equation \eqref{15.29} is satisfied,
where $\beta$ is given by \eqref{15.28}. When we combine equations
\eqref{15.29} and \eqref{15.30} we find 
\begin{equation}
\alpha\beta=0,\hspace{1cm}\varphi_{0}=\pm\frac{1}{l\sqrt{\lambda}}.\label{15.32}
\end{equation}
By substituting \eqref{15.32} into \eqref{15.31} we obtain 
\begin{equation}
\beta=\frac{1}{2l^{2}}\left(\frac{1}{\mu}-\frac{1}{\lambda}\right).\label{15.33}
\end{equation}
Now we have two types of solutions, one of those is 
\begin{equation}
\alpha=0,\hspace{0.7cm}\varphi_{0}=\pm\frac{1}{l\sqrt{\lambda}},\hspace{0.7cm}\beta=\frac{1}{2l^{2}}\left(\frac{1}{\mu}-\frac{1}{\lambda}\right),\label{15.34}
\end{equation}
and the other is given by 
\begin{equation}
\alpha\neq0,\hspace{0.7cm}\varphi_{0}=\pm\frac{1}{l\sqrt{\lambda}},\hspace{0.7cm}\beta=0,\hspace{0.7cm}\mu=\lambda\label{15.35}
\end{equation}
In both cases we have $T(\omega)=0$, i.e. the BTZ black hole spacetime
together with $\varphi=\varphi_{0}$ will be a torsion-free solution
of the given model when one of set of equations \eqref{15.34} and
\eqref{15.35} are satisfied.

\subsection{Quasi-local conserved charges}

\label{S.15.3} In this subsection, we will find an expression to
conserved charges of the above theory, associated with the asymptotic
Killing vector field $\xi$ based on the quasi-local formalism for
conserved charges.\\
 Now, suppose that the variation of the Lagrangian \eqref{15.9},
i.e. \eqref{15.11}, is due to a diffeomorphism which is generated
by the vector field $\xi$, then 
\begin{equation}
\delta_{\xi}L=\delta_{\xi}\varphi E_{\varphi}+\delta_{\xi}e\cdot E_{e}+\delta_{\xi}\omega\cdot E_{\omega}+\delta_{\xi}h\cdot E_{h}+d\Theta(\Phi,\delta_{\xi}\Phi).\label{15.36}
\end{equation}
On the one hand, presence of topological Chern-Simons term in the
Lagrangian \eqref{15.9} makes this model to be Lorentz non-covariant.
So, the total variation of the Lagrangian \eqref{15.9} due to diffeomorphism
generator $\xi$ can be written as 
\begin{equation}
\delta_{\xi}L=\mathfrak{L}_{\xi}L+d\psi_{\xi},\label{15.37}
\end{equation}
Although the Lagrangian \eqref{15.9} is not invariant under general
Lorentz gauge transformation, it is invariant under the infinitesimal
Lorentz gauge transformation (see \eqref{15.37} and \eqref{15.49}),
and also general coordinate transformation. By substituting \eqref{15.37},
Eqs.\eqref{1.58}-\eqref{1.60} and 
\begin{equation}
\delta_{\xi}\varphi=i_{\xi}D(\omega)\varphi,\label{15.38}
\end{equation}
into \eqref{15.36}, we have 
\begin{equation}
\begin{split} & d\left[\Theta(\Phi,\delta_{\xi}\Phi)-i_{\xi}L-\psi_{\xi}+i_{\xi}e\cdot E_{e}+(i_{\xi}\omega-\chi_{\xi})\cdot E_{\omega}+i_{\xi}h\cdot E_{h}\right]=\\
 & (i_{\xi}\omega-\chi_{\xi})\cdot\left[D(\omega)E_{\omega}+e\times E_{e}+h\times E_{h}\right]+i_{\xi}e\cdot D(\omega)E_{e}+i_{\xi}h\cdot D(\omega)E_{h}\\
 & -i_{\xi}T(\omega)\cdot E_{e}-i_{\xi}R(\omega)\cdot E_{\omega}-i_{\xi}D(\omega)h\cdot E_{h}-E_{\varphi}i_{\xi}D(\omega)\varphi.
\end{split}
\label{15.39}
\end{equation}
The right hand side of above equation becomes zero by virtue of the
Bianchi identities \eqref{1.51}. Therefore, we find that 
\begin{equation}
dJ_{\xi}=0,\label{15.40}
\end{equation}
where 
\begin{equation}
J_{\xi}=\Theta(\Phi,\delta_{\xi}\Phi)-i_{\xi}L-\psi_{\xi}+i_{\xi}e\cdot E_{e}+(i_{\xi}\omega-\chi_{\xi})\cdot E_{\omega}+i_{\xi}h\cdot E_{h}.\label{15.41}
\end{equation}
Thus, the quantity $J_{\xi}$ defined above is conserved off-shell.
Again locally, $J_{\xi}=dK_{\xi}$. Since this model is not Lorentz
covariant we expect that the total variation of surface term differs
from its LL-derivative 
\begin{equation}
\delta_{\xi}\Theta(\Phi,\delta\Phi)=\mathfrak{L}_{\xi}\Theta(\Phi,\delta\Phi)+\Pi_{\xi}.\label{15.42}
\end{equation}
Now, we take an arbitrary variation from \eqref{15.41} and we find
\begin{equation}
\mathfrak{J}_{ADT}(\Phi,\delta\Phi;\xi)=d\left[\delta K_{\xi}-i_{\xi}\Theta(\Phi,\delta\Phi)\right]+\delta\psi_{\xi}-\Pi_{\xi},\label{15.43}
\end{equation}
where $\mathfrak{J}_{ADT}(\Phi,\delta\Phi;\xi)$ is defined as 
\begin{equation}
\begin{split}\mathfrak{J}_{ADT}(\Phi,\delta\Phi;\xi)= & \delta e\cdot i_{\xi}E_{e}+\delta\omega\cdot i_{\xi}E_{\omega}+\delta h\cdot i_{\xi}E_{h}-\delta\varphi i_{\xi}E_{\varphi}\\
 & +i_{\xi}e\cdot\delta E_{e}+(i_{\xi}\omega-\chi_{\xi})\cdot\delta E_{\omega}+i_{\xi}h\cdot\delta E_{h}\\
 & +\delta\Theta(\Phi,\delta_{\xi}\Phi)-\delta_{\xi}\Theta(\Phi,\delta\Phi),
\end{split}
\label{15.44}
\end{equation}
and we will refer to that as \textquotedbl extended off-shell ADT
current\textquotedbl{} in the given model. In this subsection, we consider
just the cases in which $\xi$ is independent of dynamical fields.
It seems that we can write 
\begin{equation}
\delta\psi_{\xi}-\Pi_{\xi}=dZ_{\xi},\label{15.45}
\end{equation}
so the equation \eqref{15.44} can be rewritten as 
\begin{equation}
\mathfrak{J}_{ADT}(\Phi,\delta\Phi;\xi)=d\mathfrak{Q}_{ADT}(\Phi,\delta\Phi;\xi),\label{15.46}
\end{equation}
where $\mathfrak{Q}_{ADT}(\Phi,\delta\Phi;\xi)$ is the extended off-shell
ADT conserved charge associated to asymptotically Killing vector field
$\xi$ which is given as 
\begin{equation}
\mathfrak{Q}_{ADT}(\Phi,\delta\Phi;\xi)=\delta K_{\xi}-i_{\xi}\Theta(\Phi,\delta\Phi)+Z_{\xi}.\label{15.47}
\end{equation}
The quasi-local conserved charge associated to the Killing vector
field can be define as $\xi$ as 
\begin{equation}
Q(\xi)=-\frac{1}{8\pi G}\int_{0}^{1}ds\int_{\Sigma}\mathfrak{Q}_{ADT}(\Phi|s),\label{15.48}
\end{equation}
The integration over $s$ is just integration over a one-parameter
path in the solution space and $s=0$ and $s=1$ correspond to the
background solution and the solution of interest, respectively.

It is straightforward to calculate $\psi_{\xi}$ in \eqref{15.37}
using the fact that exterior derivative and LL-derivative do not commute
(see \eqref{E.3}). Thus, we have 
\begin{equation}
\psi_{\xi}=d\chi_{\xi}\cdot\left[-\frac{1}{2}\varphi e+\frac{1}{2\mu}\omega\right].\label{15.49}
\end{equation}
In a similar way, we can obtain $\Pi_{\xi}$ in the equation \eqref{15.42}
as 
\begin{equation}
\Pi_{\xi}=d\chi_{\xi}\cdot\left[-\frac{1}{2}\delta\varphi e-\frac{1}{2}\varphi\delta e+\frac{1}{2\mu}\delta\omega\right].\label{15.50}
\end{equation}
It is easy to see form equations \eqref{15.45}, \eqref{15.49} and
\eqref{15.50} that $dZ_{\xi}=0$ then we can choose $Z_{\xi}$ to
be zero. As mentioned earlier we can write $J_{\xi}=dK_{\xi}$ by
Poincare lemma, so from \eqref{15.41}, we can find $K_{\xi}$ as
follows: 
\begin{equation}
\begin{split}K_{\xi}= & \varphi(i_{\xi}\omega-\chi_{\xi})\cdot e+\frac{1}{2\mu}i_{\xi}\omega\cdot\omega-\frac{1}{\mu}\chi_{\xi}\cdot\omega+\frac{1}{2}\varphi^{2}i_{\xi}e\cdot e\\
 & +i_{\xi}e\cdot h+\frac{1}{2}\varphi i_{\xi}\omega\cdot e-\frac{1}{2}\varphi i_{\xi}e\cdot\omega.
\end{split}
\label{15.51}
\end{equation}
Considering the above results, namely equations \eqref{15.49}-\eqref{15.51},
and by taking into account \eqref{2.9}, one can calculate the extended
ADT conserved charge \eqref{15.47} as 
\begin{equation}
\begin{split}\mathfrak{Q}_{ADT}(\Phi,\delta\Phi;\xi)= & \left[(i_{\xi}\Omega-\chi_{\xi})\cdot e-\alpha i_{\xi}h\cdot e+\varphi i_{\xi}e\cdot e\right]\delta\varphi\\
 & +\left[\varphi(i_{\xi}\Omega-\chi_{\xi})+\varphi^{2}i_{\xi}e+(1-\alpha\varphi)i_{\xi}h\right]\cdot\delta e\\
 & +\left[\varphi i_{\xi}e+\frac{1}{\mu}(i_{\xi}\Omega-\chi_{\xi})-\frac{\alpha}{\mu}i_{\xi}h\right]\cdot\delta\Omega\\
 & +\left[(1-\alpha\varphi)i_{\xi}e-\frac{\alpha}{\mu}(i_{\xi}\Omega-\chi_{\xi})+\frac{\alpha^{2}}{\mu}i_{\xi}h\right]\cdot\delta h.
\end{split}
\label{15.52}
\end{equation}
To calculate the conserved charges of the considered solutions by
using \eqref{15.52}, we can employ the expression \eqref{6.2} for
$\chi_{\xi}$. In the above procedure to find conserved charges, we
assumed that $\varphi$, like $e$ and $\omega$, is a dynamical field.
So it is clear that one can use \eqref{15.48}, \eqref{15.52} and
\eqref{6.2} to obtain the charges associated to the solutions of
our model, that may have non-constant scalar field.

\subsection{General formula for entropy of black holes in minimal massive gravity
coupled to a scalar field}

\label{S.15.4} Let us consider a stationary black hole solution of
the minimal massive gravity coupled to a scalar field. We know that
the entropy of a black hole is the conserved charge associated to
the Killing horizon generated by the Killing field $\zeta$. We take
the codimension two surface $\Sigma$ to be the bifurcate surface
$\mathcal{B}$. Assuming that $\zeta$ is the Killing vector field
which generates the Killing horizon, we must set $\zeta=0$ on $\mathcal{B}$.
Thus, the equation \eqref{15.52} reduces to 
\begin{equation}
\mathfrak{Q}_{ADT}(\Phi,\delta\Phi;\zeta)=-\delta\left[\chi_{\zeta}\cdot\left(\varphi e+\frac{1}{\mu}\Omega-\frac{\alpha}{\mu}h\right)\right]\label{15.53}
\end{equation}
on the bifurcate surface. $s=0$ and $s=1$ correspond to the considered
black hole spacetime and the perturbed one, respectively. Therefore,
by integrating over a one-parameter path in the solution space, we
have 
\begin{equation}
\int_{0}^{1}ds\mathfrak{Q}_{ADT}(\Phi,\delta\Phi;\zeta)=-\chi_{\zeta}\cdot\left[\varphi e+\frac{1}{\mu}\Omega-\frac{\alpha}{\mu}h\right].\label{15.54}
\end{equation}
On the bifurcate surface of a stationary black hole, we have \eqref{6.7}
with $N^{\mu}=\left(0,0,1/\sqrt{g_{\phi\phi}}\right)$. By substituting
\eqref{6.7} and \eqref{15.54} in \eqref{15.48}, one finds 
\begin{equation}
Q(\zeta)=\frac{\kappa}{8\pi G}\int_{0}^{2\pi}\frac{d\phi}{\sqrt{g_{\phi\phi}}}\left[\varphi g_{\phi\phi}+\frac{1}{\mu}\Omega_{\phi\phi}-\frac{\alpha}{\mu}h_{\phi\phi}\right].\label{15.55}
\end{equation}
which should be calculated on the horizon. Now, we can define entropy
of a stationary black hole as 
\begin{equation}
S=-\frac{2\pi}{\kappa}Q(\zeta),\label{15.56}
\end{equation}
therefore 
\begin{equation}
S=-\frac{1}{4G}\int_{\mathcal{B}}\frac{d\phi}{\sqrt{g_{\phi\phi}}}\left[\varphi g_{\phi\phi}+\frac{1}{\mu}\Omega_{\phi\phi}-\frac{\alpha}{\mu}h_{\phi\phi}\right].\label{15.57}
\end{equation}
In the above formula, $h_{\phi\phi}$ is given by 
\begin{equation}
h_{\phi\phi}=-\frac{1}{\left[\mu(1-\alpha\varphi)^{2}-\alpha^{2}\varphi^{2}\right]}\left\{ S_{\phi\phi}+\frac{1}{2}\left[\alpha\lambda\varphi+\mu(1-\alpha\varphi)\right]\varphi^{2}g_{\phi\phi}\right\} .\label{15.58}
\end{equation}
The formula \eqref{15.57} will be similar to the that of minimal
massive gravity for $\varphi=\varphi_{0}$.

\subsection{Application for the BTZ black hole with $\varphi=\varphi_{0}$}

\label{S.15.5} We now calculate the conserved charges and entropy
of the BTZ black hole solution \eqref{2.4.2} with $\varphi=\varphi_{0}$
in the context of the above model. We take the integration surface
$\Sigma$ to be a circle with a radius of infinity. Therefore, we
can consider the AdS$_{3}$ spacetime \eqref{7.8} to be background.
Thus, \eqref{15.52} reduces to 
\begin{equation}
\begin{split}\mathfrak{Q}_{ADT}(\Phi,\delta\Phi;\xi)= & \left[\left(\varphi_{0}-\frac{\alpha\beta}{\mu}\right)(i_{\xi}\bar{\Omega}-\bar{\chi}_{\xi})+\frac{1}{\mu l^{2}}i_{\xi}\bar{e}\right]\cdot\delta e\\
 & +\left[\left(\varphi_{0}-\frac{\alpha\beta}{\mu}\right)i_{\xi}\bar{e}+\frac{1}{\mu}(i_{\xi}\bar{\Omega}-\bar{\chi}_{\xi})\right]\cdot\delta\Omega
\end{split}
\label{15.59}
\end{equation}
where \eqref{15.32} was used. Then integration yields 
\begin{equation}
\begin{split}\int_{0}^{1}ds\mathfrak{Q}_{ADT}(\Phi,\delta\Phi;\xi)= & \left[\left(\varphi_{0}-\frac{\alpha\beta}{\mu}\right)(i_{\xi}\bar{\Omega}-\bar{\chi}_{\xi})+\frac{1}{\mu l^{2}}i_{\xi}\bar{e}\right]\cdot\Delta e\\
 & +\left[\left(\varphi_{0}-\frac{\alpha\beta}{\mu}\right)i_{\xi}\bar{e}+\frac{1}{\mu}(i_{\xi}\bar{\Omega}-\bar{\chi}_{\xi})\right]\cdot\Delta\Omega,
\end{split}
\label{15.60}
\end{equation}
where $\Delta\Phi=\Phi_{(s=1)}-\Phi_{(s=0)}$. By substituting \eqref{15.60}
into \eqref{15.48}, we find 
\begin{equation}
\begin{split}Q(\xi)=-\frac{1}{8\pi G}\lim_{r\rightarrow\infty}\int_{0}^{2\pi} & \biggl\{\left[\left(\varphi_{0}-\frac{\alpha\beta}{\mu}\right)(i_{\xi}\bar{\Omega}-\bar{\chi}_{\xi})+\frac{1}{\mu l^{2}}i_{\xi}\bar{e}\right]\cdot\Delta e_{\phi}\\
 & +\left[\left(\varphi_{0}-\frac{\alpha\beta}{\mu}\right)i_{\xi}\bar{e}+\frac{1}{\mu}(i_{\xi}\bar{\Omega}-\bar{\chi}_{\xi})\right]\cdot\Delta\Omega_{\phi}\biggr\} d\phi.
\end{split}
\label{15.61}
\end{equation}
For the BTZ black hole spacetime at spatial infinity, we have \eqref{7.13}.
Energy corresponds to the Killing vector $\xi_{(E)}=\partial_{t}$:
\begin{equation}
E=\frac{1}{8G}\left[\left(-\varphi_{0}+\frac{\alpha\beta}{\mu}\right)\left(\frac{r_{+}^{2}+r_{-}^{2}}{l^{2}}\right)+\frac{2r_{+}r_{-}}{\mu l^{3}}\right],\label{15.62}
\end{equation}
For BTZ, we have $\alpha\beta=0$ (see \eqref{15.32}), thus the contribution
from MMG vanishes and the expression for energy \eqref{15.62} becomes
\begin{equation}
E=\frac{1}{8G}\left[-\varphi_{0}\left(\frac{r_{+}^{2}+r_{-}^{2}}{l^{2}}\right)+\frac{2r_{+}r_{-}}{\mu l^{3}}\right],\label{15.63}
\end{equation}
Similarly, the angular momentum ( for the Killing vector $\xi_{(J)}=-\partial_{\phi}$)
reads 
\begin{equation}
J=\frac{1}{8G}\left[-\varphi_{0}\left(\frac{2r_{+}r_{-}}{l}\right)+\frac{r_{+}^{2}+r_{-}^{2}}{\mu l^{2}}\right].\label{15.65}
\end{equation}
Since on the horizon of the BTZ black hole we have \eqref{9.2}. Hence,
by substituting \eqref{9.2} into \eqref{15.57}, we find the entropy
of the BTZ black hole solution to be 
\begin{equation}
S=\frac{\pi}{2G}\left[-\varphi_{0}r_{+}+\frac{r_{-}}{\mu l}\right].\label{15.67}
\end{equation}
It is straightforward to check that these results satisfy the first
law of black hole mechanics.

\subsection{Virasoro algebra and the central extension }

\label{S.15.6} Now, we want to find the central extension term for
the our model and subsequently we can read off the central charges.
In this subsection, we take AdS$_{3}$ spacetime with $\varphi=\varphi_{0}$
as background and the integration surface $\Sigma$ to be a circle
with a radius of infinity. Two copies of the Witt algebra, are given
by \eqref{7.7}, where $\xi_{m}^{\pm}$ ($m\in\mathbb{Z}$) are the
asymptotic Killing vector fields \eqref{7.6}. Also, the square brackets
in \eqref{7.7} denote the Lie bracket.

It is clear from \eqref{15.48} that we can write at spatial infinity
\begin{equation}
\delta Q(\xi)=-\frac{1}{8\pi G}\int_{\infty}\mathfrak{Q}_{ADT}(\bar{\Phi},\delta\Phi;\xi).\label{15.69}
\end{equation}
Therefore, integration yields 
\begin{equation}
Q(\xi)=-\frac{1}{8\pi G}\int_{\infty}\mathfrak{Q}_{ADT}(\bar{\Phi},\Delta\Phi;\xi).\label{15.70}
\end{equation}
From \eqref{15.69}, we can easily deduce that 
\begin{equation}
\delta_{\xi_{n}^{\pm}}Q(\xi_{m}^{\pm})=-\frac{1}{8\pi G}\int_{\infty}\mathfrak{Q}_{ADT}(\bar{\Phi},\delta_{\xi_{n}^{\pm}}\Phi;\xi_{m}^{\pm}).\label{15.71}
\end{equation}
Thus, by substituting \eqref{15.70} and \eqref{15.71} into \eqref{5.8},
we find an expression for the central extension term and consequently
we can read off the central charges of the considered model. Since
we take the AdS$_{3}$ spacetime with $\varphi=\varphi_{0}$, the
equation \eqref{15.71} can be rewritten as 
\begin{equation}
\delta_{\xi_{n}^{\pm}}Q(\xi_{m}^{\pm})=\frac{1}{8\pi G}\left(-\varphi_{0}+\frac{\alpha\beta}{\mu}\mp\frac{1}{\mu l}\right)\lim_{r\rightarrow\infty}\int_{0}^{2\pi}i_{\xi_{m}^{\pm}}\bar{e}\cdot\delta_{\xi_{n}^{\pm}}A_{\phi}^{\pm}d\phi,\label{15.72}
\end{equation}
where $A^{\pm}$ are connections that correspond to the two $so(2,1)$
algebras and \eqref{7.9} was used. By substituting equations \eqref{7.10}
into \eqref{15.72}, we find 
\begin{equation}
\delta_{\xi_{n}^{\pm}}Q(\xi_{m}^{\pm})=\frac{iln^{3}}{8G}\left(-\varphi_{0}+\frac{\alpha\beta}{\mu}\mp\frac{1}{\mu l}\right)\delta_{m+n,0}.\label{15.73}
\end{equation}
Suppose that $\varphi=\varphi_{0}$ and $h=\beta e$, as they are
meaningful for the BTZ black hole, then \eqref{15.70} for $\xi=\xi_{m}^{\pm}$
becomes 
\begin{equation}
Q(\xi_{m}^{\pm})=\frac{1}{8\pi G}\left(-\varphi_{0}+\frac{\alpha\beta}{\mu}\mp\frac{1}{\mu l}\right)\lim_{r\rightarrow\infty}\int_{0}^{2\pi}i_{\xi_{m}^{\pm}}\bar{e}\cdot\Delta A_{\phi}^{\pm}d\phi.\label{15.74}
\end{equation}
By substituting equations \eqref{7.13} into \eqref{15.74}, we find
\begin{equation}
Q(\xi_{m}^{\pm})=\frac{l}{16G}\left(-\varphi_{0}+\frac{\alpha\beta}{\mu}\mp\frac{1}{\mu l}\right)\left(\frac{r_{+}\mp r_{-}}{l}\right)^{2}\delta_{m,0}.\label{15.75}
\end{equation}
Now, to find the central extension term we substitute \eqref{15.73}
and \eqref{15.75} into \eqref{5.8} and arrive at 
\begin{equation}
C(\xi_{m}^{\pm},\xi_{n}^{\pm})=\frac{il}{8G}\left(-\varphi_{0}+\frac{\alpha\beta}{\mu}\mp\frac{1}{\mu l}\right)\left[n^{3}-\left(\frac{r_{+}\mp r_{-}}{l}\right)^{2}n\right]\delta_{m+n,0}.\label{15.76}
\end{equation}
To obtain the usual $n$ dependence, that is $n(n^{2}-1)$, in the
the above expression, it is sufficient one make a shift on $Q(\xi_{m}^{\pm})$
by a constant. Thus, by the following substitution 
\begin{equation}
Q(\xi_{n}^{\pm})\equiv\hat{L}_{n}^{\pm},\hspace{1cm}\{Q(\xi_{m}^{\pm}),Q(\xi_{n}^{\pm})\}_{\text{D.B.}}\equiv i[\hat{L}_{m}^{\pm},\hat{L}_{n}^{\pm}],\label{15.77}
\end{equation}
the algebra among conserved charges becomes 
\begin{equation}
[\hat{L}_{m}^{\pm},\hat{L}_{n}^{\pm}]=(m-n)\hat{L}_{m+n}^{\pm}+\frac{c_{\pm}}{12}m(m^{2}-1)\delta_{m+n,0},\label{15.78}
\end{equation}
where 
\begin{equation}
c_{\pm}=\frac{3l}{2G}\left(-\varphi_{0}+\frac{\alpha\beta}{\mu}\mp\frac{1}{\mu l}\right),\label{15.79}
\end{equation}
are central charges and $\hat{L}_{n}^{\pm}$ are the generators of
the Virasoro algebra. We can read off the eigenvalues of the Virasoro
generators $\hat{L}_{n}^{\pm}$ from \eqref{15.75} as 
\begin{equation}
l_{n}^{\pm}=\frac{l}{16G}\left(-\varphi_{0}+\frac{\alpha\beta}{\mu}\mp\frac{1}{\mu l}\right)\left(\frac{r_{+}\mp r_{-}}{l}\right)^{2}\delta_{m,0}.\label{15.80}
\end{equation}
By virtue of \eqref{15.32}, the contribution from MMG for the central
charges and eigenvalues of the Virasoro generators vanish, thus we
have 
\begin{equation}
c_{\pm}=\frac{3l}{2G}\left(-\varphi_{0}\mp\frac{1}{\mu l}\right),\label{15.81}
\end{equation}
\begin{equation}
l_{n}^{\pm}=\frac{l}{16G}\left(-\varphi_{0}\mp\frac{1}{\mu l}\right)\left(\frac{r_{+}\mp r_{-}}{l}\right)^{2}\delta_{m,0}.\label{15.82}
\end{equation}
for the central charges and the eigenvalues of the Virasoro generators,
respectively. The eigenvalues of the Virasoro generators $\hat{L}_{n}^{\pm}$
are related to the energy $E$ and the angular momentum $J$ of the
BTZ black hole by the following equations respectively 
\begin{equation}
E=l^{-1}(l_{0}^{+}+l_{0}^{-})=\frac{1}{8G}\left[-\varphi_{0}\left(\frac{r_{+}^{2}+r_{-}^{2}}{l^{2}}\right)+\frac{2r_{+}r_{-}}{\mu l^{3}}\right],\label{15.83}
\end{equation}
\begin{equation}
J=l^{-1}(l_{0}^{+}-l_{0}^{-})=\frac{1}{8G}\left[-\varphi_{0}\left(\frac{2r_{+}r_{-}}{l}\right)+\frac{r_{+}^{2}+r_{-}^{2}}{\mu l^{2}}\right].\label{15.84}
\end{equation}
Also, to calculate the entropy of the considered black hole one can
use the Cardy's formula 
\begin{equation}
S=2\pi\sqrt{\frac{c_{+}l_{0}^{+}}{6}}+2\pi\sqrt{\frac{c_{-}l_{0}^{-}}{6}}=\frac{\pi}{2G}\left[-\varphi_{0}r_{+}+\frac{r_{-}}{\mu l}\right].\label{15.85}
\end{equation}
By comparing above results, equations \eqref{15.83}-\eqref{15.85},
with equations \eqref{15.63}, \eqref{15.65} and \eqref{15.67} we
can see that they they indeed match.

\section{Conclusions}

In the first part of this review, we gave detailed account of Killing
charge construction of global conserved quantities in generic gravity
theories following the works of Deser-Tekin which also relied on the
work of Abbott-Deser that was carried out for cosmological Einstein's
theory. All of these constructions of course extend the well-known
ADM mass for asymptotically flat spacetimes. We have discussed subtle
issues about the decay conditions, large gauge transformations, and
linearization instability and the apparent infinite degeneracy of
the vacuum solution in some critical extended gravity theories. We
applied the formalism to many explicitly known spacetimes and computed
their charges. We studied some known extended gravity theories, such
as Born-Infeld gravity in 2+1 dimensions. Using both the covariant
symplectic structure construction and the usual linearization construction,
we have also shown that the AD charges are valid not only for asymptotically
flat and anti-de Sitter spacetimes (which are maximally symmetric)
but they are also valid for any Einstein space with at least one Killing
vector field. We also studied the conformal properties of these charges
under the conformal transformation of the metric, a subject which
is quite relevant for the Jordan-String frame formulation versus the
Einstein frame formulation of scalar tensor theories.

In the second part of the review, we give a very exhaustive study
of the conserved charges for a large number of 2+1 dimensional gravity
theories that can be considered as a Chern-Simons-like theory of gravity.
Such theories play the role of theoretical laboratories for ideas
in quantum gravity. Our approach has been in the context of spin-connection
and dreibein formulation both for global construction and quasi-local
charge constructions. We also give the extended off-shell formulation
of the ADT charges. We also discussed the notions of entropy and thermodynamics
of black holes, computed the near horizon geometries and studied the
relevant Virasoro algebras for various asymptotic conditions.

\section{Acknowledgments}

Bayram Tekin would like to thank Stanley Deser for many useful discussions
since 2001 to this day on conserved charges of gravity theories, particularly
generic gravity theories. The work of Mohammad Reza Setare and Hamed
Adami has been financially supported by Research Institute for Astronomy
Astrophysics of Maragha (RIAAM).

\appendix

\section{EQCA and Linearization of the Field Equations}

In this Appendix, we linearize the field equations of $f\left(R_{\alpha\beta}^{\mu\nu}\right)$
theory and show that they are equal to the linearized field equations
of a quadratic curvature theory with specific parameters given in
(\ref{eq:a_b_g_definitions}), (\ref{eq:kappa_tilde}), and (\ref{eq:Lambda_0_tilde}).
To lay out the setting for linearization, first recall that the field
equations of a $f\left(R_{\alpha\beta}^{\mu\nu}\right)$ theory is
\begin{align}
\frac{1}{2}\left(g_{\nu\rho}\nabla^{\lambda}\nabla_{\sigma}-g_{\nu\sigma}\nabla^{\lambda}\nabla_{\rho}\right)\frac{\partial f}{\partial R_{\rho\sigma}^{\mu\lambda}}-\frac{1}{2}\left(g_{\mu\rho}\nabla^{\lambda}\nabla_{\sigma}-g_{\mu\sigma}\nabla^{\lambda}\nabla_{\rho}\right)\frac{\partial f}{\partial R_{\rho\sigma}^{\lambda\nu}}\nonumber \\
-\frac{1}{2}\left(\frac{\partial f}{\partial R_{\rho\sigma}^{\mu\lambda}}R_{\rho\sigma\phantom{\lambda}\nu}^{\phantom{\rho\sigma}\lambda}-\frac{\partial f}{\partial R_{\rho\sigma}^{\lambda\nu}}R_{\rho\sigma\phantom{\lambda}\mu}^{\phantom{\rho\sigma}\lambda}\right)-\frac{1}{2}g_{\mu\nu}f\left(R_{\alpha\beta}^{\mu\nu}\right) & =0.\label{field_eq}
\end{align}
Then, for maximally symmetric spacetimes, (\ref{field_eq}) reduces
to (\ref{eq:Vacuum_eq}), that is 
\begin{equation}
-\frac{2}{\kappa_{l}}\bar{R}_{\mu\nu}+\bar{g}_{\mu\nu}\bar{f}=0,\label{eq:AdS_back}
\end{equation}
where $\kappa_{l}$ and $\bar{f}$ have the definitions 
\begin{equation}
\left[\frac{\partial f}{\partial R_{\rho\sigma}^{\mu\nu}}\right]_{\bar{g}}=:\frac{1}{\kappa_{l}}\delta_{\mu}^{[\rho}\delta_{\nu}^{\sigma]},\qquad\qquad\bar{f}:=f\left(\bar{R}_{\rho\sigma}^{\alpha\beta}\right),\label{eq:kappa-l_and_fbar_again}
\end{equation}
respectively.

With this setting, we aim to linearize (\ref{field_eq}) in the metric
perturbation $h_{\mu\nu}:=g_{\mu\nu}-\bar{g}_{\mu\nu}$ where $\bar{g}_{\mu\nu}$
represents the (A)dS background that solves (\ref{eq:AdS_back}).
The linearization of the last term in (\ref{field_eq}) is relatively
simple and one has 
\begin{equation}
\left[g_{\mu\nu}f\left(R_{\alpha\beta}^{\mu\nu}\right)\right]_{\left(1\right)}=h_{\mu\nu}\bar{f}+\bar{g}_{\mu\nu}\left[\frac{\partial f}{\partial R_{\rho\sigma}^{\alpha\beta}}\right]_{\bar{g}}\left(R_{\rho\sigma}^{\alpha\beta}\right)_{\left(1\right)}=h_{\mu\nu}\bar{f}+\frac{1}{\kappa_{l}}\bar{g}_{\mu\nu}R_{\left(1\right)}.\label{gf}
\end{equation}
Now, let us move to the first term in the second line of (\ref{field_eq})
which can be linearized as 
\begin{equation}
\left(\frac{\partial f}{\partial R_{\rho\sigma}^{\mu\lambda}}R_{\rho\sigma\phantom{\lambda}\nu}^{\phantom{\rho\sigma}\lambda}\right)_{\left(1\right)}=\left[\frac{\partial^{2}f}{\partial R_{\alpha\tau}^{\eta\theta}\partial R_{\rho\sigma}^{\mu\lambda}}\right]_{\bar{g}}\left(R_{\alpha\tau}^{\eta\theta}\right)_{\left(1\right)}\bar{R}_{\rho\sigma\phantom{\lambda}\nu}^{\phantom{\rho\sigma}\lambda}+\left[\frac{\partial f}{\partial R_{\rho\sigma}^{\mu\lambda}}\right]_{\bar{g}}\left(R_{\rho\sigma\phantom{\lambda}\nu}^{\phantom{\rho\sigma}\lambda}\right)_{\left(1\right)}.\label{eq:Middle_structure}
\end{equation}
Using (\ref{AdS_background}) and (\ref{eq:d2f/dRiem2_gbar}), that
are 
\begin{equation}
\bar{R}_{\rho\sigma}^{\mu\nu}=\frac{2\Lambda}{\left(n-1\right)\left(n-2\right)}\left(\delta_{\rho}^{\mu}\delta_{\sigma}^{\nu}-\delta_{\sigma}^{\mu}\delta_{\rho}^{\nu}\right),
\end{equation}
and 
\begin{equation}
\left[\frac{\partial^{2}f}{\partial R_{\alpha\tau}^{\eta\theta}\partial R_{\rho\sigma}^{\mu\lambda}}\right]_{\bar{g}}=2\alpha\delta_{\eta}^{[\alpha}\delta_{\theta}^{\tau]}\delta_{\mu}^{[\rho}\delta_{\lambda}^{\sigma]}+\beta\left(\delta_{[\eta}^{\alpha}\delta_{\theta]}^{[\rho}\delta_{[\mu}^{\left|\tau\right|}\delta_{\lambda]}^{\sigma]}-\delta_{[\eta}^{\tau}\delta_{\theta]}^{[\rho}\delta_{[\mu}^{\left|\alpha\right|}\delta_{\lambda]}^{\sigma]}\right)+12\gamma\delta_{\eta}^{[\alpha}\delta_{\theta}^{\tau}\delta_{\mu}^{\rho}\delta_{\lambda}^{\sigma]},\label{eq:Back_val_of_d2fdR2_again}
\end{equation}
respectively, (\ref{eq:Middle_structure}) takes the final form 
\begin{align}
\left(\frac{\partial f}{\partial R_{\rho\sigma}^{\mu\lambda}}R_{\rho\sigma\phantom{\lambda}\nu}^{\phantom{\rho\sigma}\lambda}\right)_{\left(1\right)}= & -\left(\alpha\frac{4\Lambda}{\left(n-2\right)}+\beta\frac{n\Lambda}{\left(n-1\right)\left(n-2\right)}\right)R_{\left(1\right)}\bar{g}_{\mu\nu}\nonumber \\
 & +\left(\gamma\frac{8\Lambda\left(n-3\right)}{\left(n-1\right)\left(n-2\right)}-\beta\frac{2\Lambda}{n-1}\right)\left(R_{\mu\nu}^{\left(1\right)}-\frac{1}{2}\bar{g}_{\mu\nu}R_{\left(1\right)}-\frac{2\Lambda}{n-2}h_{\mu\nu}\right)\nonumber \\
 & -\frac{1}{\kappa_{l}}R_{\mu\nu}^{\left(1\right)}.\label{eq:Lin_dfdR-Riem}
\end{align}
Afterwards, let us turn to the first term in (\ref{field_eq}), that
is $g_{\nu\rho}\nabla^{\lambda}\nabla_{\sigma}\frac{\partial f}{\partial R_{\rho\sigma}^{\mu\lambda}}$,
and the first step in its linearization is 
\begin{equation}
\left(g_{\nu\rho}\nabla^{\lambda}\nabla_{\sigma}\frac{\partial f}{\partial R_{\rho\sigma}^{\mu\lambda}}\right)_{\left(1\right)}=\bar{g}_{\nu\rho}\bar{g}^{\lambda\beta}\left(\nabla_{\beta}\nabla_{\sigma}\frac{\partial f}{\partial R_{\rho\sigma}^{\mu\lambda}}\right)_{\left(1\right)}+\left(g_{\nu\rho}g^{\lambda\beta}\right)_{\left(1\right)}\left(\nabla_{\beta}\nabla_{\sigma}\frac{\partial f}{\partial R_{\rho\sigma}^{\mu\lambda}}\right)_{\bar{g}},\label{eq:Lin_of_nabla2dfdR_initial}
\end{equation}
where the second term is zero since $\left(\partial f/\partial R_{\rho\sigma}^{\mu\lambda}\right)_{\bar{g}}$
is a four tensor that is composed of two background metrics as $\bar{g}\bar{g}$
for the AdS background, and then the metric compatibility reduce $\left(\nabla_{\beta}\nabla_{\sigma}\partial f/\partial R_{\rho\sigma}^{\mu\lambda}\right)_{\bar{g}}$
to zero. To linearize the term $\left(\nabla_{\beta}\nabla_{\sigma}\partial f/\partial R_{\rho\sigma}^{\mu\lambda}\right)_{\left(1\right)}$,
remember the linearization of the covariant derivative of a vector
as 
\begin{equation}
\left(\nabla_{\mu}A^{\nu}\right)_{\left(1\right)}=\bar{\nabla}_{\mu}A_{\left(1\right)}^{\nu}+\left(\Gamma_{\mu\rho}^{\nu}\right)_{\left(1\right)}\bar{A}^{\nu},\label{eq:Linearization_of_nabla}
\end{equation}
where the barred quantities represent the background values as usual
and $\left(\Gamma_{\mu\rho}^{\nu}\right)_{\left(1\right)}$ is the
linearized Christoffel connection whose specific form is not important
for our purposes as it would be clear with the below calculations.
Using (\ref{eq:Linearization_of_nabla}), the first step in the linearization
of $\left(\nabla_{\beta}\nabla_{\sigma}\partial f/\partial R_{\rho\sigma}^{\mu\lambda}\right)_{\left(1\right)}$
is 
\begin{align}
\left(\nabla_{\beta}\nabla_{\sigma}\frac{\partial f}{\partial R_{\rho\sigma}^{\mu\lambda}}\right)_{\left(1\right)}= & \bar{\nabla}_{\beta}\left(\nabla_{\sigma}\frac{\partial f}{\partial R_{\rho\sigma}^{\mu\lambda}}\right)_{\left(1\right)}+\left(\Gamma_{\beta\alpha}^{\rho}\right)_{\left(1\right)}\left(\nabla_{\sigma}\frac{\partial f}{\partial R_{\alpha\sigma}^{\mu\lambda}}\right)_{\bar{g}}\nonumber \\
 & -\left(\Gamma_{\beta\mu}^{\alpha}\right)_{\left(1\right)}\left(\nabla_{\sigma}\frac{\partial f}{\partial R_{\rho\sigma}^{\alpha\lambda}}\right)_{\bar{g}}-\left(\Gamma_{\beta\lambda}^{\alpha}\right)_{\left(1\right)}\left(\nabla_{\sigma}\frac{\partial f}{\partial R_{\rho\sigma}^{\mu\alpha}}\right)_{\bar{g}},
\end{align}
where the last three terms are zero for AdS background due to metric
compatibility, and the first term $\nabla_{\sigma}\partial f/\partial R_{\rho\sigma}^{\mu\lambda}$
can be linearized as 
\begin{align}
\left(\nabla_{\sigma}\frac{\partial f}{\partial R_{\rho\sigma}^{\mu\lambda}}\right)_{\left(1\right)}= & \bar{\nabla}_{\sigma}\left(\frac{\partial f}{\partial R_{\rho\sigma}^{\mu\lambda}}\right)_{\left(1\right)}+\left(\Gamma_{\sigma\alpha}^{\rho}\right)_{\left(1\right)}\left(\frac{\partial f}{\partial R_{\alpha\sigma}^{\mu\lambda}}\right)_{\bar{g}}+\left(\Gamma_{\sigma\alpha}^{\sigma}\right)_{\left(1\right)}\left(\frac{\partial f}{\partial R_{\rho\alpha}^{\mu\lambda}}\right)_{\bar{g}}\nonumber \\
 & -\left(\Gamma_{\sigma\mu}^{\alpha}\right)_{\left(1\right)}\left(\frac{\partial f}{\partial R_{\rho\sigma}^{\alpha\lambda}}\right)_{\bar{g}}-\left(\Gamma_{\sigma\lambda}^{\alpha}\right)_{\left(1\right)}\left(\frac{\partial f}{\partial R_{\rho\sigma}^{\mu\alpha}}\right)_{\bar{g}}.\label{eq:Lin_of_nabla-dfdR}
\end{align}
Note that the second term drops out since a term symmetric in $\alpha$and
$\sigma$ is multiplied a term antisymmetric in the same indices.
The last three terms also reduce to zero altogether by using the definition
of $\kappa_{l}$ in (\ref{eq:kappa-l_and_fbar_again}) as 
\begin{align}
 & \left(\Gamma_{\sigma\alpha}^{\sigma}\right)_{\left(1\right)}\left(\frac{\partial f}{\partial R_{\rho\alpha}^{\mu\lambda}}\right)_{\bar{g}}-\left(\Gamma_{\sigma\mu}^{\alpha}\right)_{\left(1\right)}\left(\frac{\partial f}{\partial R_{\rho\sigma}^{\alpha\lambda}}\right)_{\bar{g}}-\left(\Gamma_{\sigma\lambda}^{\alpha}\right)_{\left(1\right)}\left(\frac{\partial f}{\partial R_{\rho\sigma}^{\mu\alpha}}\right)_{\bar{g}}\nonumber \\
= & \frac{1}{\kappa_{l}}\left[\left(\Gamma_{\sigma\alpha}^{\sigma}\right)_{\left(1\right)}\delta_{\mu}^{[\rho}\delta_{\lambda}^{\alpha]}-\left(\Gamma_{\sigma\mu}^{\alpha}\right)_{\left(1\right)}\delta_{\alpha}^{[\rho}\delta_{\lambda}^{\sigma]}-\left(\Gamma_{\sigma\lambda}^{\alpha}\right)_{\left(1\right)}\delta_{\mu}^{[\rho}\delta_{\alpha}^{\sigma]}\right]\nonumber \\
= & \frac{1}{\kappa_{l}}\left(\Gamma_{\sigma\beta}^{\alpha}\right)_{\left(1\right)}\left(\delta_{[\mu}^{\rho}\delta_{\lambda]}^{\sigma}\delta_{\alpha}^{\beta}+\delta_{[\lambda}^{\rho}\delta_{\alpha]}^{\sigma}\delta_{\mu}^{\beta}+\delta_{[\alpha}^{\rho}\delta_{\mu]}^{\sigma}\delta_{\lambda}^{\beta}\right)=\frac{1}{\kappa_{l}}\left(\Gamma_{\sigma\beta}^{\alpha}\right)_{\left(1\right)}\left(\delta_{\mu}^{[\rho}\delta_{\lambda}^{\sigma}\delta_{\alpha}^{\beta]}\right)\nonumber \\
= & 0.
\end{align}
Lastly, $\partial f/\partial R_{\rho\sigma}^{\mu\lambda}$ appearing
in (\ref{eq:Lin_of_nabla-dfdR}) can be linearized by using (\ref{eq:Back_val_of_d2fdR2_again})
as 
\begin{align}
\left(\frac{\partial f}{\partial R_{\rho\sigma}^{\mu\lambda}}\right)_{\left(1\right)}= & \left[\frac{\partial^{2}f}{\partial R_{\alpha\tau}^{\eta\theta}\partial R_{\rho\sigma}^{\mu\lambda}}\right]_{\bar{g}}\left(R_{\alpha\tau}^{\eta\theta}\right)_{\left(1\right)}\nonumber \\
= & 2\alpha\delta_{\mu}^{[\rho}\delta_{\lambda}^{\sigma]}R_{\left(1\right)}+2\beta\delta_{\theta}^{[\rho}\delta_{[\mu}^{\left|\tau\right|}\delta_{\lambda]}^{\sigma]}\left(R_{\tau}^{\theta}\right)_{\left(1\right)}\nonumber \\
 & +2\gamma\delta_{\mu}^{[\rho}\delta_{\lambda}^{\sigma]}R_{\left(1\right)}-8\gamma\delta_{\theta}^{[\rho}\delta_{[\mu}^{\left|\tau\right|}\delta_{\lambda]}^{\sigma]}\left(R_{\tau}^{\theta}\right)_{\left(1\right)}+2\gamma\delta_{\eta}^{[\rho}\delta_{\theta}^{\sigma]}\left(R_{\mu\lambda}^{\eta\theta}\right)_{\left(1\right)}.
\end{align}
Putting all these results step by step back into (\ref{eq:Lin_of_nabla2dfdR_initial})
yields 
\begin{align}
\left(g_{\nu\rho}\nabla^{\lambda}\nabla_{\sigma}\frac{\partial f}{\partial R_{\rho\sigma}^{\mu\lambda}}\right)_{L}= & \bar{g}_{\nu\rho}\bar{g}^{\lambda\beta}\bar{\nabla}_{\beta}\bar{\nabla}_{\sigma}\left[2\alpha\delta_{\mu}^{[\rho}\delta_{\lambda}^{\sigma]}R_{\left(1\right)}+2\beta\delta_{\theta}^{[\rho}\delta_{[\mu}^{\left|\tau\right|}\delta_{\lambda]}^{\sigma]}\left(R_{\tau}^{\theta}\right)_{\left(1\right)}\right]\nonumber \\
 & +\bar{g}_{\nu\rho}\bar{g}^{\lambda\beta}\bar{\nabla}_{\beta}\bar{\nabla}_{\sigma}\left[2\gamma\delta_{\mu}^{[\rho}\delta_{\lambda}^{\sigma]}R_{\left(1\right)}-8\gamma\delta_{\theta}^{[\rho}\delta_{[\mu}^{\left|\tau\right|}\delta_{\lambda]}^{\sigma]}\left(R_{\tau}^{\theta}\right)_{\left(1\right)}+2\gamma\delta_{\eta}^{[\rho}\delta_{\theta}^{\sigma]}\left(R_{\mu\lambda}^{\eta\theta}\right)_{\left(1\right)}\right],
\end{align}
which reduces to 
\begin{align}
\left(g_{\nu\rho}\nabla^{\lambda}\nabla_{\sigma}\frac{\partial f}{\partial R_{\rho\sigma}^{\mu\lambda}}\right)_{\left(1\right)}= & \alpha\left(\bar{g}_{\mu\nu}\bar{\nabla}^{\lambda}\bar{\nabla}_{\lambda}R_{\left(1\right)}-\bar{\nabla}_{\nu}\bar{\nabla}_{\mu}R_{\left(1\right)}\right)\nonumber \\
 & +\frac{\beta}{2}\Biggl[\bar{g}_{\nu\rho}\bar{\nabla}^{\lambda}\bar{\nabla}_{\lambda}\left(R_{\mu}^{\rho}\right)_{\left(1\right)}-\bar{g}_{\nu\rho}\bar{\nabla}^{\lambda}\bar{\nabla}_{\mu}\left(R_{\lambda}^{\rho}\right)_{\left(1\right)}\nonumber \\
 & \phantom{+\frac{\beta}{2}\Biggl[}-\bar{\nabla}_{\nu}\bar{\nabla}_{\sigma}\left(R_{\mu}^{\sigma}\right)_{\left(1\right)}+\bar{g}_{\mu\nu}\bar{\nabla}^{\lambda}\bar{\nabla}_{\sigma}\left(R_{\lambda}^{\sigma}\right)_{\left(1\right)}\Biggr]\nonumber \\
 & +\gamma\Biggl[\bar{g}_{\mu\nu}\bar{\nabla}^{\lambda}\bar{\nabla}_{\lambda}R_{\left(1\right)}-2\bar{g}_{\nu\rho}\bar{\nabla}^{\sigma}\bar{\nabla}_{\sigma}\left(R_{\mu}^{\rho}\right)_{\left(1\right)}\nonumber \\
 & \phantom{+\gamma\Biggl[}+2\bar{g}_{\nu\rho}\bar{\nabla}^{\lambda}\bar{\nabla}_{\mu}\left(R_{\lambda}^{\rho}\right)_{\left(1\right)}+2\bar{g}_{\nu\rho}\bar{\nabla}^{\lambda}\bar{\nabla}_{\sigma}\left(R_{\mu\lambda}^{\rho\sigma}\right)_{\left(1\right)}\Biggr]\nonumber \\
 & -\gamma\left[\bar{\nabla}_{\nu}\bar{\nabla}_{\mu}R_{\left(1\right)}-2\bar{\nabla}_{\nu}\bar{\nabla}_{\sigma}\left(R_{\mu}^{\sigma}\right)_{\left(1\right)}+2\bar{g}_{\mu\nu}\bar{\nabla}^{\lambda}\bar{\nabla}_{\sigma}\left(R_{\lambda}^{\sigma}\right)_{\left(1\right)}\right].\label{eq:Lin_of_nabla2dfdR_middle}
\end{align}
With this result, the linearization of $g_{\nu\rho}\nabla^{\lambda}\nabla_{\sigma}\partial f/\partial R_{\rho\sigma}^{\mu\lambda}$
is achieved.

However, this result cannot be readily used since several terms in
the result are not in a form that comply with the forms appearing
in the linearized field equation of the quadratic curvature gravity
in (\ref{Linearized_eom}), that is 
\begin{align}
\left(\frac{1}{\kappa}+\frac{4\Lambda n}{n-2}\alpha+\frac{4\Lambda}{n-1}\beta+\frac{4\Lambda\left(n-3\right)\left(n-4\right)}{\left(n-1\right)\left(n-2\right)}\gamma\right)\mathcal{G}_{\mu\nu}^{\left(1\right)}\nonumber \\
+\left(2\alpha+\beta\right)\left(\bar{g}_{\mu\nu}\bar{\square}-\bar{\nabla}_{\mu}\bar{\nabla}_{\nu}+\frac{2\Lambda}{n-2}\bar{g}_{\mu\nu}\right)R_{\left(1\right)}+\beta\left(\bar{\square}\mathcal{G}_{\mu\nu}^{\left(1\right)}-\frac{2\Lambda}{n-1}\bar{g}_{\mu\nu}R_{\left(1\right)}\right) & =T_{\mu\nu}.\label{eq:Linearized_eom_again}
\end{align}
To put the terms appearing in (\ref{eq:Lin_of_nabla2dfdR_middle})
into the desired forms, first note that $\left(R_{\mu}^{\rho}\right)_{\left(1\right)}$
can be written in terms of $R_{\mu\alpha}^{\left(1\right)}$ as 
\begin{equation}
\left(R_{\mu}^{\rho}\right)_{\left(1\right)}=\left(g^{\rho\alpha}R_{\mu\alpha}\right)_{\left(1\right)}=\bar{g}^{\rho\alpha}R_{\mu\alpha}^{\left(1\right)}-\frac{2\Lambda}{n-2}h_{\mu}^{\rho},
\end{equation}
and from the linearized Bianchi identity $\bar{\nabla}^{\mu}\mathcal{G}_{\mu\nu}^{L}=0$,
the term $\bar{\nabla}_{\mu}\left(R_{\nu}^{\mu}\right)_{\left(1\right)}$
becomes simply 
\begin{equation}
\bar{\nabla}_{\mu}\left(R_{\nu}^{\mu}\right)_{\left(1\right)}=\frac{1}{2}\bar{\nabla}_{\nu}R_{\left(1\right)}.
\end{equation}
Using these in (\ref{eq:Lin_of_nabla2dfdR_middle}) yields 
\begin{align}
\left(g_{\nu\rho}\nabla^{\lambda}\nabla_{\sigma}\frac{\partial f}{\partial R_{\rho\sigma}^{\mu\lambda}}\right)_{\left(1\right)}= & \alpha\left(\bar{g}_{\mu\nu}\bar{\square}R_{\left(1\right)}-\bar{\nabla}_{\mu}\bar{\nabla}_{\nu}R_{\left(1\right)}\right)\nonumber \\
 & +\frac{\beta}{2}\left[\bar{\square}\mathcal{G}_{\mu\nu}^{\left(1\right)}-\bar{g}_{\nu\rho}\bar{\nabla}^{\lambda}\bar{\nabla}_{\mu}\left(R_{\lambda}^{\rho}\right)_{\left(1\right)}-\frac{1}{2}\bar{\nabla}_{\nu}\bar{\nabla}_{\mu}R_{\left(1\right)}+\bar{g}_{\mu\nu}\bar{\square}R_{\left(1\right)}\right]\nonumber \\
 & +\gamma\left[-2\bar{g}_{\nu\rho}\bar{\square}\left(R_{\mu}^{\rho}\right)_{\left(1\right)}+2\bar{g}_{\nu\rho}\bar{\nabla}^{\lambda}\bar{\nabla}_{\mu}\left(R_{\lambda}^{\rho}\right)_{\left(1\right)}+2\bar{g}_{\nu\rho}\bar{\nabla}^{\lambda}\bar{\nabla}_{\sigma}\left(R_{\mu\lambda}^{\rho\sigma}\right)_{\left(1\right)}\right],\label{eq:Intermediate_eqn_in_lin}
\end{align}
where $\bar{g}_{\nu\rho}\bar{\nabla}^{\lambda}\bar{\nabla}_{\mu}\left(R_{\lambda}^{\rho}\right)_{\left(1\right)}$
and $\bar{g}_{\nu\rho}\bar{\nabla}^{\lambda}\bar{\nabla}_{\sigma}\left(R_{\mu\lambda}^{\rho\sigma}\right)_{\left(1\right)}$
are the remaining terms which must be rewritten in terms of $R_{\left(1\right)}$,
$\mathcal{G}_{\mu\nu}^{\left(1\right)}$, and their derivatives. Note
that the first two terms in the $\gamma$ parenthesis are kept in
the up-down indexed Ricci tensor form since in this form the last
term $\bar{g}_{\nu\rho}\bar{\nabla}^{\lambda}\bar{\nabla}_{\sigma}\left(R_{\mu\lambda}^{\rho\sigma}\right)_{\left(1\right)}$
will yield an immediate cancellation as shown below. For the term
$\bar{g}_{\nu\rho}\bar{\nabla}^{\lambda}\bar{\nabla}_{\mu}\left(R_{\lambda}^{\rho}\right)_{\left(1\right)}$
, changing the order of derivatives, using the linearized Bianchi
identity, and rearranging the terms yield 
\begin{equation}
\bar{g}_{\nu\rho}\bar{\nabla}^{\lambda}\bar{\nabla}_{\mu}\left(R_{\lambda}^{\rho}\right)_{\left(1\right)}=\frac{1}{2}\bar{\nabla}_{\mu}\bar{\nabla}_{\nu}R_{\left(1\right)}+\frac{2n\Lambda}{\left(n-1\right)\left(n-2\right)}\mathcal{G}_{\mu\nu}^{\left(1\right)}+\frac{\Lambda}{n-1}\bar{g}_{\mu\nu}R_{\left(1\right)}.
\end{equation}
To rewrite the term $\bar{g}_{\nu\rho}\bar{\nabla}^{\lambda}\bar{\nabla}_{\sigma}\left(R_{\mu\lambda}^{\rho\sigma}\right)_{\left(1\right)}$,
note that $\nabla^{\lambda}\nabla_{\sigma}R_{\mu\lambda}^{\rho\sigma}$
can be written in terms of the derivatives of the Ricci tensor by
taking the covariant derivative of the once-contracted Bianchi identity
\begin{equation}
\nabla^{\nu}R_{\mu\alpha\nu\beta}=\nabla_{\mu}R_{\alpha\beta}-\nabla_{\alpha}R_{\mu\beta},
\end{equation}
as 
\begin{equation}
\nabla^{\mu}\nabla_{\nu}R_{\mu\alpha}^{\nu\beta}=\square R_{\alpha}^{\beta}-\nabla^{\mu}\nabla_{\alpha}R_{\mu}^{\beta}.
\end{equation}
Then, linearization of this equation yields 
\begin{equation}
\bar{\nabla}^{\mu}\bar{\nabla}_{\nu}\left(R_{\mu\alpha}^{\nu\beta}\right)_{\left(1\right)}=\bar{\square}\left(R_{\alpha}^{\beta}\right)_{\left(1\right)}-\bar{\nabla}^{\mu}\bar{\nabla}_{\alpha}\left(R_{\mu}^{\beta}\right)_{\left(1\right)},
\end{equation}
where the right-hand-side terms matches with the first two terms of
the $\gamma$ parenthesis in (\ref{eq:Intermediate_eqn_in_lin}).
Now, we can write the final form of the linearization of $g_{\nu\rho}\nabla^{\lambda}\nabla_{\sigma}\partial f/\partial R_{\rho\sigma}^{\mu\lambda}$
which becomes 
\begin{align}
\left(g_{\nu\rho}\nabla^{\lambda}\nabla_{\sigma}\frac{\partial f}{\partial R_{\rho\sigma}^{\mu\lambda}}\right)_{\left(1\right)}= & \frac{\left(2\alpha+\beta\right)}{2}\left(\bar{g}_{\mu\nu}\bar{\square}R_{\left(1\right)}-\bar{\nabla}_{\mu}\bar{\nabla}_{\nu}R_{\left(1\right)}\right)+\frac{\beta}{2}\bar{\square}\mathcal{G}_{\mu\nu}^{\left(1\right)}\nonumber \\
 & -\frac{\beta}{2}\left(\frac{2n\Lambda}{\left(n-1\right)\left(n-2\right)}\mathcal{G}_{\mu\nu}^{\left(1\right)}+\frac{\Lambda}{n-1}\bar{g}_{\mu\nu}R_{\left(1\right)}\right),\label{eq:Lin_of_nabla2dfdR_final}
\end{align}
where all the terms on the right-hand side matches with the ones appearing
in (\ref{eq:Linearized_eom_again}) and notice that $\gamma$ term
drops out in the final form.

We finished the linearization of every key component appearing in
the field equations of the $f\left(R_{\alpha\beta}^{\mu\nu}\right)$
theory given in (\ref{field_eq}) and we can write the linearized
form of this field equation. To do this, observe that the first line
of (\ref{field_eq}) can be written as 
\begin{align}
 & \frac{1}{2}\left(g_{\nu\rho}\nabla^{\lambda}\nabla_{\sigma}-g_{\nu\sigma}\nabla^{\lambda}\nabla_{\rho}\right)\frac{\partial f}{\partial R_{\rho\sigma}^{\mu\lambda}}-\frac{1}{2}\left(g_{\mu\rho}\nabla^{\lambda}\nabla_{\sigma}-g_{\mu\sigma}\nabla^{\lambda}\nabla_{\rho}\right)\frac{\partial f}{\partial R_{\rho\sigma}^{\lambda\nu}}\nonumber \\
= & g_{\nu\rho}\nabla^{\lambda}\nabla_{\sigma}\frac{\partial f}{\partial R_{\rho\sigma}^{\mu\lambda}}+g_{\mu\sigma}\nabla^{\lambda}\nabla_{\rho}\frac{\partial f}{\partial R_{\rho\sigma}^{\lambda\nu}},
\end{align}
where on the left-hand side, only the antisymmetry of $\partial f/\partial R_{\rho\sigma}^{\lambda\nu}$
in $\rho$ and $\sigma$ is made explicit. In addition, the right-hand
side only represents the symmetry in $\mu$ and $\nu$, and note that
(\ref{eq:Lin_of_nabla2dfdR_final}) is already symmetric in $\mu$
and $\nu$. Thus, $\left(g_{\nu\rho}\nabla^{\lambda}\nabla_{\sigma}\partial f/\partial R_{\rho\sigma}^{\mu\lambda}\right)_{\left(1\right)}$
represents the contribution coming from the first two terms in the
first line of (\ref{field_eq}), and the last two terms in the first
line of (\ref{field_eq}) also yield the same $\left(\mu,\nu\right)$-symmetric
contribution as $\left(g_{\nu\rho}\nabla^{\lambda}\nabla_{\sigma}\partial f/\partial R_{\rho\sigma}^{\mu\lambda}\right)_{\left(1\right)}$.
Furthermore, observe that the first two terms in the second line of
(\ref{field_eq}) is only a representation of $\mu$ and $\nu$ symmetry,
and (\ref{eq:Lin_dfdR-Riem}) is already symmetric in $\mu$ and $\nu$.
Thus, the first two terms in the second line of (\ref{field_eq})
have the same linearized form given in (\ref{eq:Lin_dfdR-Riem}).
With these observations, the linearized field equation of the $f\left(R_{\alpha\beta}^{\mu\nu}\right)$
theory can be found as 
\begin{align}
\left(\frac{1}{\kappa_{l}}-\beta\frac{2\Lambda}{\left(n-1\right)\left(n-2\right)}-\gamma\frac{4\Lambda\left(n-3\right)}{\left(n-1\right)\left(n-2\right)}\right)\mathcal{G}_{\mu\nu}^{\left(1\right)}\nonumber \\
+\left(2\alpha+\beta\right)\left(\bar{g}_{\mu\nu}\bar{\square}-\bar{\nabla}_{\mu}\bar{\nabla}_{\nu}+\frac{2\Lambda}{n-2}\bar{g}_{\mu\nu}\right)R_{\left(1\right)}+\beta\left(\bar{\square}\mathcal{G}_{\mu\nu}^{\left(1\right)}-\frac{2\Lambda}{n-1}\bar{g}_{\mu\nu}R_{\left(1\right)}\right) & =0,\label{eq:Lin_field_eqn_of_fRiem}
\end{align}
where we also used the background field equation (\ref{eq:AdS_back})
to remove a term involving $h_{\mu\nu}$ as the tensor structure.

Notice that (\ref{eq:Lin_field_eqn_of_fRiem}) matches with the linearized
field equation of the quadratic curvature theory restated in (\ref{eq:Linearized_eom_again})
except the coefficient of $\mathcal{G}_{\mu\nu}^{\left(1\right)}$.
Thus, the linearized field equation of the $f\left(R_{\alpha\beta}^{\mu\nu}\right)$
theory is the same as the linearized field equation of the quadratic
curvature gravity defined with the Lagrangian density 
\begin{equation}
\mathcal{L}=\frac{1}{\tilde{\kappa}}\left(R-2\tilde{\Lambda}_{0}\right)+\alpha R^{2}+\beta R_{\sigma}^{\lambda}R_{\lambda}^{\sigma}+\gamma\chi_{\text{GB}},\label{eq:Equiv_quad}
\end{equation}
where $\tilde{\kappa}$ must satisfy (\ref{eq:kappa_tilde}), that
is 
\begin{equation}
\frac{1}{\tilde{\kappa}}=\frac{1}{\kappa_{l}}-\frac{4\Lambda}{n-2}\left(n\alpha+\beta+\gamma\frac{\left(n-2\right)\left(n-3\right)}{\left(n-1\right)}\right),
\end{equation}
to yield the same coefficient appearing in (\ref{eq:Lin_field_eqn_of_fRiem}).

Finally, let us discuss the fine point related to the equivalence
of the linearized field equations of the $f\left(R_{\alpha\beta}^{\mu\nu}\right)$
theory and the quadratic curvature gravity (\ref{eq:Equiv_quad}):
the effective cosmological constant $\Lambda$ appearing in both linearized
field equations must satisfy the same field equation, that is both
theories must have the same vacua. To achieve this, $\tilde{\Lambda}_{0}$
needs to be defined as (\ref{eq:Lambda_0_tilde}), that is 
\begin{equation}
\frac{\tilde{\Lambda}_{0}}{\tilde{\kappa}}=-\frac{1}{2}\bar{f}+\left(\frac{n\Lambda}{n-2}\right)\frac{1}{\kappa_{l}}-\frac{2\Lambda^{2}n}{\left(n-2\right)^{2}}\left(n\alpha+\beta+\gamma\frac{\left(n-2\right)\left(n-3\right)}{\left(n-1\right)}\right),
\end{equation}
for which the vacuum field equation of (\ref{eq:Equiv_quad}) given
as 
\begin{equation}
\frac{\Lambda-\tilde{\Lambda}_{0}}{2\tilde{\kappa}}+\Lambda^{2}\left(\left(n\alpha+\beta\right)\frac{\left(n-4\right)}{\left(n-2\right)^{2}}+\gamma\frac{\left(n-3\right)\left(n-4\right)}{\left(n-1\right)\left(n-2\right)}\right)=0,
\end{equation}
reduces to (\ref{eq:AdS_back}).

\section{Procedure of obtaining equations \eqref{2.53}-\eqref{2.58}}

\label{A} Here we present the procedure of arriving at equations
\eqref{2.53}-\eqref{2.58}. Let's start by introducing the following
identity 
\begin{equation}
R(\Omega)=e\times S,\label{A.1}
\end{equation}
where $S_{\mu\nu}$ is the 3D Schouten tensor \eqref{2.13}. By taking
the Hodge star \eqref{A.1}, one can easily check the identity \eqref{A.1}.
By substituting \eqref{2.49} and \eqref{2.50} into \eqref{A.1},
one can write the dual curvature 2-form as 
\begin{equation}
R^{a}(\Omega)=\frac{\zeta^{2}}{2l^{2}}\left(\frac{3}{4}-\nu^{2}\right)\varepsilon_{\hspace{1.5mm}bc}^{a}e^{b}\wedge e^{c}-\frac{\zeta^{2}}{l^{4}}\left(1-\nu^{2}\right)\varepsilon_{\hspace{1.5mm}bc}^{a}e_{\hspace{1.5mm}\nu}^{b}J^{c}J_{\nu}dx^{\mu}\wedge dx^{\nu}.\label{A.2}
\end{equation}
The action of the Hodge star $\star$ on a 2-form $C=C_{\mu\nu}dx^{\mu}\wedge dx^{\nu}$
is given by $\star C=\epsilon_{\lambda\mu\nu}C^{\mu\nu}dx^{\lambda}$,
therefore 
\begin{equation}
\star R^{a}(\Omega)=\frac{\zeta^{2}}{4l^{2}}e^{a}-\frac{\zeta^{2}}{l^{4}}\left(1-\nu^{2}\right)J^{a}J_{\mu}dx^{\mu}.\label{A.3}
\end{equation}
Consider the following two Lorentz vector valued 1-forms 
\begin{equation}
\begin{split}A_{\hspace{1.5mm\mu}}^{a}= & a_{1}e_{\hspace{1.5mm\mu}}^{a}+a_{2}J^{a}J_{\mu},\\
B_{\hspace{1.5mm\mu}}^{a}= & b_{1}e_{\hspace{1.5mm\mu}}^{a}+b_{2}J^{a}J_{\mu}.
\end{split}
\label{A.4}
\end{equation}
The Hodge dual of $(A\times B)^{a}$ becomes 
\begin{equation}
\star(A\times B)^{a}=-\left[2a_{1}b_{1}+l^{2}\left(a_{1}b_{2}+a_{2}b_{1}\right)\right]e^{a}+\left(a_{1}b_{2}+a_{2}b_{1}\right)J^{a}J_{\mu}dx^{\mu}.\label{A.5}
\end{equation}
One can use \eqref{1.14} to show 
\begin{equation}
D(\Omega)A^{a}=e^{a\lambda}\hat{\nabla}_{\mu}A_{\nu\lambda}dx^{\mu}\wedge dx^{\nu},\label{A.6}
\end{equation}
then one has 
\begin{equation}
D(\Omega)A^{a}=\frac{1}{l}\left|\zeta\right|a_{2}e^{a\lambda}J^{\alpha}\epsilon_{\alpha\mu(\nu}J_{\lambda)}dx^{\mu}\wedge dx^{\nu}\label{A.7}
\end{equation}
where we have used \eqref{2.48}. The Hodge dual of \eqref{A.7} is
\begin{equation}
\star D(\Omega)A^{a}=\frac{l}{2}\left|\zeta\right|a_{2}e^{a}-\frac{3}{2l}\left|\zeta\right|a_{2}J^{a}J_{\mu}dx^{\mu}.\label{A.8}
\end{equation}
By taking the Hodge dual field equations of GMMG \eqref{2.33}-\eqref{2.35}
and using equations \eqref{A.3},\eqref{A.5} and \eqref{A.8}, one
can show that equations \eqref{2.33}-\eqref{2.35} become three equations
of the form of 
\begin{equation}
(\cdots)e^{a}+(\cdots)J^{a}J_{\mu}dx^{\mu}=0.\label{A.9}
\end{equation}
By setting the coefficients of $e^{a}$ and $J^{a}J_{\mu}dx^{\mu}$
to zero, we will arrive at the equations \eqref{2.53}-\eqref{2.58}.

\section{Technical proof of equation \eqref{3.2}}

\label{B} By substituting \eqref{1.44} and \eqref{1.49} into \eqref{3.1},
one has 
\begin{equation}
dJ_{\xi}=i_{\xi}a^{r}\cdot dE_{r}-i_{\xi}da^{r}\cdot E_{r}-\chi_{\xi}\cdot\left(dE_{\omega}+a^{r}\times E_{r}\right).\label{B.1}
\end{equation}
where $J_{\xi}$ is given by \eqref{3.3}. Since the exterior covariant
derivative of a Lorentz vector valued 1-form $A^{a}$ in terms of
the dual spin-connection can be written as 
\begin{equation}
D(\omega)A=dA+\omega\times A,\label{B.2}
\end{equation}
then 
\begin{equation}
D(\omega)a^{r}=da^{r}+\omega\times a^{r},\label{B.3}
\end{equation}
\begin{equation}
D(\omega)E_{r}=dE_{r}+\omega\times E_{r}.\label{B.4}
\end{equation}
By substituting \eqref{B.3} and \eqref{B.4} into \eqref{B.1}, we
have 
\begin{equation}
\begin{split}dJ_{\xi}= & i_{\xi}a^{r}\cdot D(\omega)E_{r}-i_{\xi}D(\omega)a^{r}\cdot E_{r}+i_{\xi}\omega\cdot a^{r}\times E_{r}\\
 & -\chi_{\xi}\cdot\left(D(\omega)E_{\omega}+a^{r}\times E_{r}-\omega\times E_{\omega}\right)
\end{split}
\label{B.5}
\end{equation}
This equation can be rewritten as 
\begin{equation}
\begin{split}dJ_{\xi}= & i_{\xi}e\cdot D(\omega)E_{e}+i_{\xi}\omega\cdot D(\omega)E_{\omega}+i_{\xi}a^{r^{\prime}}\cdot D(\omega)E_{r^{\prime}}\\
 & -i_{\xi}D(\omega)e\cdot E_{e}-i_{\xi}D(\omega)\omega\cdot E_{\omega}-i_{\xi}D(\omega)a^{r^{\prime}}\cdot E_{r^{\prime}}\\
 & +i_{\xi}\omega\cdot e\times E_{e}+i_{\xi}\omega\cdot\omega\times E_{\omega}+i_{\xi}\omega\cdot a^{r^{\prime}}\times E_{r^{\prime}}\\
 & -\chi_{\xi}\cdot\left(D(\omega)E_{\omega}+a^{r^{\prime}}\times E_{r^{\prime}}\right)-\chi_{\xi}\cdot e\times E_{e}
\end{split}
\label{B.6}
\end{equation}
where $r^{\prime}$ runs over all the flavor indices except $e$ and
$\omega$. Equations \eqref{1.31} and \eqref{1.33} imply the following
\begin{equation}
T(\omega)=D(\omega)e,\hspace{1cm}R(\omega)=D(\omega)\omega-\frac{1}{2}\omega\times\omega.\label{B.7}
\end{equation}
Using \eqref{B.7}, one can rearrange \eqref{B.6} as \eqref{3.2}.


\begin{thebibliography}{100}
\bibitem{Noether} E.~Noether, \emph{``Invariante Variationsprobleme,''}Nachr.~d.~König.~Gesellsch.~d.~Wiss.~zu
Göttingen, Math-phys. Klasse, 235--257 (1918).

\bibitem{Witten_dS} E.~Witten, \emph{``Quantum gravity in de Sitter
space,''} hep-th/0106109.

\bibitem{adm} R.~Arnowitt, S.~Deser and C.~Misner, \emph{``The
Dynamics of General Relativity,''} Phys. \ Rev.\ \textbf{116},
1322 (1959); \textbf{117}, 1595 (1960); in \textit{{Gravitation:
An Introduction to Current Research}}, ed L.~Witten (Wiley, New
York, 1962).

\bibitem{Abbott} L.~F.~Abbott and S.~Deser, \emph{``Stability
Of Gravity With A Cosmological Constant,''} Nucl.\ Phys.\ B \textbf{195},
76 (1982).

\bibitem{Deser_Tekin-PRL} S.~Deser and B.~Tekin, \emph{``Gravitational
Energy in Quadratic Curvature Gravities,''} Phys.\ Rev.\ Lett.\ \textbf{89},
101101 (2002).

\bibitem{Deser_Tekin-PRD} S.~Deser and B.~Tekin, \emph{``Energy
in generic higher curvature gravity theories,''} Phys.\ Rev.\ D
\textbf{67}, 084009 (2003).

\bibitem{Bartnik} R.~Bartnik \emph{``The mass of an asymptotically
flat manifold,''} Comm.\ Pure Appl.\ Math.\ \textbf{39}: 661-693
(1986).

\bibitem{Ashtekar} A.~Ashtekar and V.~Petkov, \emph{Springer Handbook
of Spacetime} (Springer, Berlin, Germany, 2014).

\bibitem{Regge:1974zd}  T.~Regge and C.~Teitelboim, \emph{``Role
of Surface Integrals in the Hamiltonian Formulation of General Relativity,''}
Annals Phys.\ \textbf{88} 286 (1974).

\bibitem{hen84}  M.~Henneaux and C.~Teitelboim, \emph{``Hamiltonian
Treatment Of Asymptotically Anti-de Sitter Spaces,''} Phys.\ Lett.\ \textbf{142B},
355 (1984).

\bibitem{jamsin}  E.~Jamsin, \emph{``A Note on conserved charges
of asymptotically flat and anti-de Sitter spaces in arbitrary dimensions,''}
Gen.\ Rel.\ Grav.\ \textbf{40}, 2569 (2008).

\bibitem{Magnon} A.~Ashtekar and A.~Magnon, \emph{``Asymptotically
anti-de Sitter space-times,''} Class.\ Quant.\ Grav.\ \textbf{1},
L39 (1984).

\bibitem{HawkingHorowitz}  S.~W.~Hawking and G.~T.~Horowitz,
\emph{``The Gravitational Hamiltonian, action, entropy and surface
terms,''} \textit{\emph{Class.\ Quant.\ Grav.\ }}\textbf{13},
1487 (1996).

\bibitem{97} V.~Balasubramanian and P.~Kraus, \emph{``A stress
tensor for anti-de Sitter gravity,''} \textit{\emph{Commun.\ Math.\ Phys.\ }}\textbf{208},
413 (1999).

\bibitem{Aros} R.~Aros, M.~Contreras, R.~Olea, R.~Troncoso and
J.~Zanelli, \emph{``Conserved charges for gravity with locally AdS
asymptotics,''} Phys.\ Rev.\ Lett.\ \textbf{84}, 1647 (2000).

\bibitem{40} G.~Barnich and F.~Brandt, \emph{``Covariant theory
of asymptotic symmetries, conservation laws and central charges,''}
Nucl.\ Phys.\ B\textbf{ 633}, 3 (2002).

\bibitem{Hollands} S.~Hollands, A.~Ishibashi and D.~Marolf, \emph{``Comparison
between various notions of conserved charges in asymptotically AdS-spacetimes,''}
Class.\ Quant.\ Grav.\ \textbf{22}, 2881 (2005).

\bibitem{Tekin_et_al}  A.~Petrov, S.~Kopeikin, R.~Lompay, and
B.~Tekin, \emph{Metric Theories of Gravity. Perturbations and Conservation
Laws}. Berlin, Boston: De Gruyter, (2017).

\bibitem{Banados_rev} M.~Banados and I.~A.~Reyes, \emph{``A short
review on Noether's theorems, gauge symmetries and boundary terms,''}
Int.\ J.\ Mod.\ Phys.\ D \textbf{25}, no. 10, 1630021 (2016).

\bibitem{Compere} G.~Compère and A.~Fiorucci, \emph{``Advanced
Lectures on General Relativity,''} Lect.\ Notes Phys.\ \textbf{952},
(2019).

\bibitem{Besse} A.~L.~Besse, Einstein manifolds, (Springer-Verlag,
Berlin, Germany, 1987).

\bibitem{Henneaux85} M.~Henneaux and C.~Teitelboim, \emph{``Asymptotically
anti-De Sitter Spaces,''} Commun.\ Math.\ Phys.\ \textbf{98}, 391
(1985).

\bibitem{Bouchareb} A.~Bouchareb and G.~Clement, ``\emph{Black
hole mass and angular momentum in topologically massive gravity,}''
Class.\ Quant.\ Grav.\ \textbf{24}, 5581 (2007).

\bibitem{Nazaroglu} C.~Nazaroglu, Y.~Nutku and B.~Tekin, ``\emph{Covariant
Symplectic Structure and Conserved Charges of Topologically Massive
Gravity,}'' Phys.\ Rev.\ D \textbf{83}, 124039 (2011).

\bibitem{Devecioglu} D.~O.~Devecioglu and O.~Sarioglu, ``\emph{Conserved
Killing charges of quadratic curvature gravity theories in arbitrary
backgrounds},'' Phys.\ Rev.\ D \textbf{83}, 021503 (2011).

\bibitem{39}  J.~D.~Brown and M.~Henneaux, \emph{``Central charges
in the canonical realization of asymptotic symmetries: an example
from three-dimensional gravity,''} Commun.\ Math.\ Phys.\ \textbf{104},
207 (1986).

\bibitem{ST}  O.~Sarioglu and B.~Tekin, \emph{``Another proof
of the positive energy theorem in gravity,''} arXiv:0709.0407 {[}gr-qc{]}.

\bibitem{DenisovSolovev}  V.I. Denisov and V.O. Solovev, \emph{``The
energy determined in general relativity on the basis of the traditional
Hamiltonian approach does not have physical meaning,''} Theor. and
Math. Phys. \textbf{56}, 832 (1983), English translation.

\bibitem{BrayChrusciel} H.~L.~Bray and P.~T.~Chrusciel, \emph{``The
Penrose inequality,''} The Einstein equations and the large scale
behavior of gravitational fields, {Birkh$\ddot{\text{a}}$user},
Basel, 2004. 39-70.

\bibitem{Schoen} R.~Schoen and S.~T.~Yau, \emph{``On the Proof
of the positive mass conjecture in general relativity,''} Commun.\ Math.\ Phys.\textbf{\ 65},
45 (1979); \emph{``Proof Of The Positive Mass Theorem. 2,''} Commun.\ Math.\ Phys.\textbf{\ 79},
231 (1981).

\bibitem{Witten} E.~Witten, \emph{``A Simple Proof Of The Positive
Energy Theorem,''} Commun. Math. Phys. \textbf{80}, 381 (1981).

\bibitem{Deser} S.~Deser and C.~Teitelboim, \emph{``Supergravity
Has Positive Energy,''} Phys. Rev. Lett. \textbf{39}, 249 (1977).

\bibitem{Grisaru} M.~T.~Grisaru, \emph{``Positivity Of The Energy
In Einstein Theory,''} Phys. Lett. B \textbf{73}, 207 (1978).

\bibitem{Parker} T.~Parker and C.~H.~Taubes, \emph{``On Witten's
Proof Of The Positive Energy Theorem,''} Commun. Math. Phys. \textbf{84},
223 (1982).

\bibitem{Nester} J.~A.~Nester, \emph{``A New Gravitational Energy
Expression With A Simple Positivity Proof,''} Phys. Lett. A \textbf{83},
241 (1981).

\bibitem{Gibbons} G.~W.~Gibbons, H.~Lu, D.~N.~Page and C.~N.~Pope,
\emph{``The general Kerr-de Sitter metrics in all dimensions,''}
J.\ Geom.\ Phys.\ \textbf{53}, 49 (2005).

\bibitem{KerrSchild} R.~P.~Kerr and A.~Schild, \emph{``Some algebraically
degenerate solutions of Einstein's gravitattional field equations,\textquotedbl}
Proc.\ Symp.\ Appl.\ Math.\ \textbf{17}, 199 (1965).

\bibitem{GursesGursey} M.~Gurses and F.~Gursey, \emph{``Lorentz
invariant treatment of Kerr-Schild geometry,\textquotedbl} J.\ Math.\ Phys.\ \textbf{16},
2385 (1975).

\bibitem{MyersPerry} R.~C.~Myers and M.~J.~Perry, \emph{``Black
Holes in Higher Dimensional Space-Times,''} Annals Phys\ \textbf{172},
304 (1986).

\bibitem{Kanik} S.~Deser, I.~Kanik and B.~Tekin, \emph{``Conserved
charges of higher D Kerr-AdS spacetimes,''} Class.\ Quant.\ Grav.\ \textbf{22},
3383 (2005).

\bibitem{Pope} G.~W.~Gibbons, M.~J.~Perry and C.~N.~Pope, \emph{``The
first law of thermodynamics for Kerr-{}-anti-de Sitter black holes,''}
arXiv:hep-th/0408217. A relevant discussion of first law in $D=4$
is given in M.~M.~Caldarelli, G.~Cognola and D.~Klemm, \emph{``Thermodynamics
of Kerr-Newman-AdS black holes and conformal field theories,''} Class.\ Quant.\ Grav.\ \textbf{17},
399 (2000).

\bibitem{Deruelle} N.~Deruelle and J.~Katz, \emph{``On the mass
of a Kerr-{}-anti-de Sitter spacetime in D dimensions,''} Class.\ Quant.\ Grav.\ \textbf{22},
421 (2005).

\bibitem{Cebeci} H.~Cebeci, O.~Sarioglu and B.~Tekin, \emph{``Negative
mass solitons in gravity,''} Phys.\ Rev.\ D \textbf{73}, 064020
(2006).

\bibitem{Horowitz} G.~T.~Horowitz and R.~C.~Myers, \emph{``The
AdS/CFT correspondence and a new positive energy conjecture for general
relativity,''} \textit{\emph{Phys.\ Rev.\ D}} \textbf{59}, 026005
(1999).

\bibitem{clarkson} R.~Clarkson and R.~B.~Mann, \textit{``Soliton
solutions to the Einstein equations in five dimensions,'' }\textit{\emph{Phys.\ Rev.\ Lett.}}\textbf{\ 96}
, 051104 (2006); ``Eguchi-Hanson solitons in odd dimensions,'' Class.\ Quant.\ Grav.\ \textbf{23},
1507 (2006).

\bibitem{eguchi} T.~Eguchi and A.~J.~Hanson, \emph{``Asymptotically
Flat Selfdual Solutions To Euclidean Gravity,''} \textit{\emph{Phys.\ Lett.}}\ B
\textbf{74}, 249 (1978).

\bibitem{henningson} M.~Henningson and K.~Skenderis, \emph{``The
holographic Weyl anomaly,''} JHEP \textbf{9807}, 023 (1998).

\bibitem{stelea1} R.~B.~Mann and C.~Stelea, \emph{``New Taub-NUT-Reissner-{Nordstr$\ddot{\text{o}}$m}
spaces in higher dimensions,''} \textit{\emph{Phys.\ Lett.\ B}}
\textbf{632}, 537 (2006).

\bibitem{Zwiebach} B.~Zwiebach, \emph{``Curvature Squared Terms
and String Theories,''} Phys.\ Lett.\ \textbf{156B}, 315 (1985).

\bibitem{BoulwareDeser} D.~G.~Boulware and S.~Deser, \emph{``String
Generated Gravity Models,''} Phys.\ Rev.\ Lett.\ \textbf{55},
2656 (1985).

\bibitem{17} J.~Oliva, D.~Tempo and R.~Troncoso, \emph{``Three-dimensional
black holes, gravitational solitons, kinks and wormholes for BHT massive
gravity,''} JHEP \textbf{0907}, 011 (2009).

\bibitem{Dias} O.~J.~C.~Dias, H.~S.~Reall and J.~E.~Santos,
\emph{``Kerr-CFT and gravitational perturbations,''} JHEP \textbf{0908},
101 (2009).

\bibitem{Critical4D} H.~Lu and C.~N.~Pope, \emph{``Critical Gravity
in Four Dimensions,''} Phys.\ Rev.\ Lett.\ \textbf{106}, 181302
(2011).

\bibitem{Critical} S.~Deser, H.~Liu, H.~Lu, C.~N.~Pope, T.~C.~Sisman
and B.~Tekin, \emph{``Critical Points of D-Dimensional Extended
Gravities,''} Phys.\ Rev.\ D \textbf{83}, 061502 (2011).

\bibitem{Ghodsi1} A.~Ghodsi and M.~Moghadassi, \emph{``Charged
Black Holes in New Massive Gravity,''} Phys.\ Lett.\ B \textbf{695},
359 (2011).

\bibitem{Ghodsi2} A.~Ghodsi and D.~M.~Yekta, \emph{``Black Holes
in Born-Infeld Extended New Massive Gravity,''} Phys.\ Rev.\ D
\textbf{83}, 104004 (2011).

\bibitem{Lu} H.~Lu, Y.~Pang, C.~N.~Pope and J.~F.~Vazquez-Poritz,
\emph{``AdS and Lifshitz Black Holes in Conformal and Einstein-Weyl
Gravities,''} Phys.\ Rev.\ D \textbf{86}, 044011 (2012).

\bibitem{Ghodsi3} A.~Ghodsi and D.~M.~Yekta, \emph{``On Asymptotically
AdS-Like Solutions of Three Dimensional Massive Gravity,''} JHEP
\textbf{1206}, 131 (2012).

\bibitem{Kim} W.~Kim, S.~Kulkarni and S.~H.~Yi, \emph{``Quasilocal
Conserved Charges in a Covariant Theory of Gravity,''} Phys.\ Rev.\ Lett.\ \textbf{111},
no. 8, 081101 (2013) Erratum: {[}Phys.\ Rev.\ Lett.\ \textbf{112},
no. 7, 079902 (2014){]}.

\bibitem{Goya} G.~Giribet and A.~Goya, \emph{``The Brown-York
mass of black holes in Warped Anti-de Sitter space,''} JHEP \textbf{1303},
130 (2013).

\bibitem{Ghodsi4} A.~Ghodsi and D.~M.~Yekta, \emph{``Stability
of vacua in New Massive Gravity in different gauges,''} JHEP \textbf{1308},
095 (2013).

\bibitem{Gim} Y.~Gim, W.~Kim and S.~H.~Yi, \emph{``The first
law of thermodynamics in Lifshitz black holes revisited,''} JHEP
\textbf{1407}, 002 (2014).

\bibitem{GiribetVasquez} G.~Giribet and Y.~Vasquez, \emph{``Minimal
Log Gravity,''} Phys.\ Rev.\ D \textbf{91}, no. 2, 024026 (2015).

\bibitem{Ayon-Beato} E.~Ayon-Beato, M.~Bravo-Gaete, F.~Correa,
M.~Hassaine, M.~M.~Juarez-Aubry and J.~Oliva, \emph{``First law
and anisotropic Cardy formula for three-dimensional Lifshitz black
holes,''} Phys.\ Rev.\ D \textbf{91}, no. 6, 064006 (2015) Addendum:
{[}Phys.\ Rev.\ D \textbf{96}, no. 4, 049903 (2017){]}.

\bibitem{Bravo-Gaete} M.~Bravo-Gaete and M.~Hassaine, \emph{``Thermodynamics
of charged Lifshitz black holes with quadratic corrections,''} Phys.\ Rev.\ D
\textbf{91}, no. 6, 064038 (2015).

\bibitem{Vasquez} G.~Giribet and Y.~Vasquez, \emph{``Evanescent
gravitons in warped anti--de Sitter space,''} Phys.\ Rev.\ D \textbf{93},
no. 2, 024001 (2016).

\bibitem{Guajardo} M.~Bravo Gaete, L.~Guajardo and M.~Hassaine,
\emph{``A Cardy-like formula for rotating black holes with planar
horizon,''} JHEP \textbf{1704}, 092 (2017).

\bibitem{Cisterna} A.~Cisterna, L.~Guajardo, M.~Hassaine and J.~Oliva,
\emph{``Quintic quasi-topological gravity,''} JHEP \textbf{1704},
066 (2017).

\bibitem{Ghodsi5} A.~Ghodsi and F.~Najafi, \emph{``Ricci cubic
gravity in d dimensions, gravitons and SAdS/Lifshitz black holes,''}
Eur.\ Phys.\ J.\ C \textbf{77}, no. 8, 559 (2017).

\bibitem{Hindawi} A.~Hindawi, B.~A.~Ovrut and D.~Waldram, \emph{``Nontrivial
vacua in higher derivative gravitation,'' }Phys.\ Rev.\ D \textbf{53},
5597 (1996).

\bibitem{Gullu-UniBI} I.~Gullu, T.~C.~Sisman and B.~Tekin, \emph{``Unitarity
analysis of general Born-Infeld gravity theories,''} Phys.\ Rev.\ D
\textbf{82}, 124023

\bibitem{Gullu-AllUni3D} I.~Gullu, T.~C.~Sisman and B.~Tekin,
\emph{``All Bulk and Boundary Unitary Cubic Curvature Theories in
Three Dimensions,''} Phys.\ Rev.\ D \textbf{83}, 024033 (2011).

\bibitem{Sisman-AllUni} T.~C.~Sisman, I.~Gullu and B.~Tekin,
\emph{``All unitary cubic curvature gravities in D dimensions,''}
Class.\ Quant.\ Grav.\ \textbf{28}, 195004 (2011).

\bibitem{Senturk} C.~Senturk, T.~C.~Sisman and B.~Tekin, \emph{``Energy
and Angular Momentum in Generic F(Riemann) Theories,''} Phys.\ Rev.\ D
\textbf{86}, 124030 (2012).

\bibitem{UniBI4D} {I.~Gullu,} T.~C.~Sisman and B.~Tekin, \emph{``Born-Infeld
Gravity with a Massless Graviton in Four Dimensions,''} Phys.\ Rev.\ D
\textbf{91}, no. 4, 044007 (2015).

\bibitem{UniBIanyD} {I.~Gullu,} T.~C.~Sisman and B.~Tekin,
\emph{``Born-Infeld Gravity with a Unique Vacuum and a Massless Graviton,''}
Phys.\ Rev.\ D \textbf{92}, no. 10, 104014 (2015).

\bibitem{Bueno} P.~Bueno and P.~A.~Cano, \emph{``Einsteinian
cubic gravity,''} Phys.\ Rev.\ D \textbf{94}, no. 10, 104005 (2016).

\bibitem{Cano} P.~Bueno, P.~A.~Cano, V.~S.~Min and M.~R.~Visser,
\emph{``Aspects of general higher-order gravities,''} Phys.\ Rev.\ D\textbf{
95}, no. 4, 044010 (2017).

\bibitem{Tekin_rap} B.~Tekin, \emph{``Particle Content of Quadratic
and $f(R_{\mu\nu\sigma\rho})$ Theories in $(A)dS$,''} Phys.\ Rev.\ D
\textbf{93}, no. 10, 101502 (2016).

\bibitem{Azeyanagi} T.~Azeyanagi, G.~Compere, N.~Ogawa, Y.~Tachikawa
and S.~Terashima, \emph{``Higher-Derivative Corrections to the Asymptotic
Virasoro Symmetry of 4d Extremal Black Holes,''} Prog.\ Theor.\ Phys.\ \textbf{122},
355 (2009).

\bibitem{Gullu-BINMG} I.~Gullu, T.~C.~Sisman and B.~Tekin, \emph{``Born-Infeld
extension of new massive gravity,''} Class.\ Quant.\ Grav.\ \textbf{27},
162001 (2010).

\bibitem{NMG} E.~A.~Bergshoeff, O.~Hohm and P.~K.~Townsend,
\emph{``Massive Gravity in Three Dimensions,''} Phys.\ Rev.\ Lett.\ \textbf{102},
201301 (2009).

\bibitem{BTZ}  M.~Banados, C.~Teitelboim and J.~Zanelli, \emph{``The
Black hole in three-dimensional space-time,''} Phys.\ Rev.\ Lett.\ \textbf{69},
1849 (1992).

\bibitem{Nam-Extended} S.~Nam, J.~-D.~Park and S.~-H.~Yi, \emph{``AdS
Black Hole Solutions in the Extended New Massive Gravity,''} JHEP\textbf{
1007}, 058 (2010).

\bibitem{Gullu-cfunc} I.~Gullu, T.~C.~Sisman and B.~Tekin, \emph{``c-functions
in the Born-Infeld extended New Massive Gravity,''} Phys.\ Rev.\ D
\textbf{82}, 024032 (2010).

\bibitem{Wald} R.~M.~Wald, \emph{``Black hole entropy is the Noether
charge,''} Phys.\ Rev.\ D \textbf{48}, no. 8, R3427 (1993).%

\bibitem{DJT-PRL} S.~Deser, R.~Jackiw and S.~Templeton, \emph{``Three-Dimensional
Massive Gauge Theories,''} Phys.\ Rev.\ Lett.\ \textbf{48}, 975
(1982).

\bibitem{DJT} S.~Deser, R.~Jackiw and S.~Templeton, \emph{``Topologically
Massive Gauge Theories,'' }Annals Phys. \textbf{140}, 372 (1982).

\bibitem{DT-TMG} S.~Deser and B.~Tekin, \emph{``Energy in topologically
massive gravity,''} Class.\ Quant.\ Grav.\ \textbf{20}, L259 (2003).

\bibitem{Alkac} G.~Alkac and D.~O.~Devecioglu, \emph{``Covariant
Symplectic Structure and Conserved Charges of New Massive Gravity,''}
Phys.\ Rev.\ D \textbf{85}, 064048 (2012).

\bibitem{Witten-Symp} E.~Witten Nuc. Phys. B \textbf{276} 291 (1986);
C.~Crnkovic and E.~Witten in \textit{Three hundred years of gravitation}
S. W. Hawking and W. Israel, eds. Cambridge University Press (1987).

\bibitem{Olmez} S.~Olmez, O.~Sarioglu and B.~Tekin, \emph{``Mass
and angular momentum of asymptotically ads or flat solutions in the
topologically massive gravity,''} Class.\ Quant.\ Grav.\ \textbf{22},
4355 (2005).

\bibitem{ChiralGrav} W.~Li, W.~Song and A.~Strominger, \emph{``Chiral
Gravity in Three Dimensions,''} JHEP \textbf{0804}, 082 (2008).

\bibitem{Giribet} A.~Garbarz, G.~Giribet, Y.~Vasquez, \emph{``Asymptotically
AdS\_3 Solutions to Topologically Massive Gravity at Special Values
of the Coupling Constants,''} Phys.\ Rev.\ \textbf{D79}, 044036
(2009).

\bibitem{MiskovicOlea} O.~Miskovic and R.~Olea, \emph{``Background-independent
charges in Topologically Massive Gravity,''} JHEP \textbf{0912},
046 (2009).

\bibitem{Cvetkovic} M.~Blagojevic and B.~Cvetkovic, \emph{``Conserved
charges in 3D gravity,''} Phys.\ Rev.\ D \textbf{81}, 124024 (2010).

\bibitem{Nester-Charge} J.M.~Nester, \emph{``A covariant Hamiltonian
for gravity theories,''} Mod.\ Phys.\ Lett.\ A \textbf{6}, 2655
(1991).

\bibitem{Anninos} D.~Anninos, W.~Li, M.~Padi, W.~Song, A.~Strominger,
\emph{``Warped AdS(3) Black Holes,''} JHEP \textbf{0903}, 130 (2009).

\bibitem{Nutku} Y.~Nutku, ``Exact solutions of topologically massive
gravity with a cosmological constant,'' Class.\ Quant.\ Grav.\ \textbf{10}
2657 (1993).

\bibitem{Gurses} M.~Gurses, \emph{``Perfect Fluid Sources in 2+1
Dimensions,''} Class.\ Quant.\ Grav.\ \textbf{11}, no.~10 2585
(1994).

\bibitem{DeserTekin-Conformal} S.~Deser and B.~Tekin, \emph{``Conformal
Properties of Charges in Scalar-Tensor Gravities,''} Class.\ Quant.\ Grav.\ \textbf{23},
7479 (2006).

\bibitem{JacobsonKang} T.~Jacobson and G.~Kang, \emph{``Conformal
invariance of black hole temperature,''} Class.\ Quant.\ Grav.\ \textbf{10},
L201 (1993).

\bibitem{Murata} G.~Compere, K.~Murata and T.~Nishioka, \emph{``Central
Charges in Extreme Black Hole/CFT Correspondence,''} JHEP \textbf{0905},
077 (2009).

\bibitem{Witten-3DCS} E.~Witten, \emph{``(2+1)-Dimensional Gravity
as an Exactly Soluble System,''} Nucl.\ Phys.\ B \textbf{311},
46 (1988).

\bibitem{Horne-Witten} J.~H.~Horne and E.~Witten, \emph{``Conformal
Gravity in Three-dimensions as a Gauge Theory,''} Phys.\ Rev.\ Lett.\ \textbf{62},
501 (1989).

\bibitem{Brill_Deser} D.~R.~Brill and S.~Deser, \emph{``Instability
of Closed Spaces in General Relativity,''} Commun.\ Math.\ Phys.\ \textbf{32},
291 (1973).

\bibitem{Marsden_Fischer} A.~E.~Fischer and J.~E.~Marsden, \emph{``Linearization
stability of the Einstein equations,''} Bull.\ Amer.\ Math.\ Soc.,
\textbf{79}, 997-1003 (1973).

\bibitem{Moncrief} V.~Moncrief, \emph{``Spacetime symmetries and
linearization stability of the {Einstein} equations. {I},''}
J.\ Math.\ Phys.\textbf{\ 16} , 493-498 (1975).

\bibitem{Marsden_Arms} J.~M.~Arms and J.~E.~Marsden ``\emph{The
absence of Killing fields is necessary for linearization stability
of Einstein's equations,''} Indiana University Math.\ J., \textbf{28},
119-125 (1979).

\bibitem{Marsden} A.~E.~Fischer, J.~E.~Marsden and V.~Moncrief,
\emph{``The structure of the space of solutions of Einstein's equations.
I. One Killing field,''} Annales de l'I.H.P.\ Physique {theorique}
\textbf{33.2}, 147-194 (1980).

\bibitem{Marsden_lectures} J.~E.~Marsden, \emph{``Lectures on
Geometric Methods in Mathematical Physics,''} CBMS- NSF Regional
Conf.\ Ser.\ in Appl.\ Math., \textbf{37}, SIAM, Philadelphia,
Pa. (1981).

\bibitem{SaraykarRai} R.~V.~Saraykar and J.~H.~Rai, \emph{``Linearization
Stability of Einstein Field Equations is a Generic Property,''} Electron.\ J.\ Theor.\ Phys.\ \textbf{13},
no. 36, 229 (2016).

\bibitem{Altas} E.~Altas and B.~Tekin, \emph{``Linearization Instability
for Generic Gravity in AdS,''} arXiv:1705.10234 {[}hep-th{]}.

\bibitem{Choquet-Bruhat} Y.~Choquet-Bruhat, \emph{``General Relativity
and the Einstein Equations,''} Oxford Science Publications, (2009).

\bibitem{Girbau} J.~Girbau and L.~Bruna, \emph{``Stability by
linearization of Einstein's field equation,\textquotedbl} Springer,
(2010).

\bibitem{Deser_Choquet-Bruhat} S.~Deser and Y.~Choquet-Bruhat,
\emph{``On the Stability of Flat Space,''} Annals Phys.\ \textbf{81},
165 (1973).

\bibitem{99} J.~D.~Bekenstein, \emph{``Black holes and entropy,''}
Phys. Rev. D\ \textbf{7}, 2333 (1973).

\bibitem{Hajian} K.~Hajian, M.~M.~Sheikh-Jabbari, \emph{``Solution
Phase Space and Conserved Charges: A General Formulation for Charges
Associated with Exact Symmetries,''} Phys. Rev. D\ \textbf{93},
044074 (2016).

\bibitem{95} L.~B.~Szabados, \emph{``Quasi-local Energy-Momentum
and Angular Momentum in GR: A Review Article,''} Living Rev. Rel.\ \textbf{7},
4 (2004).

\bibitem{96} J.~D.~Brown and J.~W.~York, \emph{``Quasilocal
energy and conserved charges derived from the gravitational action,''}
Phys.Rev. D\ \textbf{47}, 1407 (1993).

\bibitem{Nojiri} S.~Nojiri, S.~D.~Odintsov, S.~Ogushi, \emph{``Cosmological
and black hole brane-world Universes in higher derivative gravity,''}
Phys. Rev. D\ \textbf{65}, 023521 (2002).

\bibitem{Cvetic} M. Cvetic, S. Nojiri, S.D. Odintsov, \emph{``Black
Hole Thermodynamics and Negative Entropy in deSitter and Anti-deSitter
Einstein-Gauss-Bonnet gravity,''} Nucl. Phys. B\ \textbf{628} 295
(2002).

\bibitem{101a} A.~Komar, \emph{``Covariant conservation laws in
general relativity,''} Phys. Rev.\ \textbf{113} 934 (1959).

\bibitem{9} V.~Iyer and R.~M.~Wald, \emph{``Some properties of
the Noether charge and a proposal for dynamical black hole entropy,''}
Phys. Rev. D\ \textbf{50}, 846 (1994).

\bibitem{Barnich:1995ap} G.~Barnich, F.~Brandt and M.~Henneaux,
\emph{``Local BRST cohomology in Einstein Yang-Mills theory,''}
Nucl.\ Phys.\ B \textbf{455} (1995) 357. 

\bibitem{Barnich:2003xg} G.~Barnich, \emph{``Boundary charges in
gauge theories: Using Stokes theorem in the bulk,''} Class.\ Quant.\ Grav.\ \textbf{20}
(2003) 3685.

\bibitem{33} J.~Lee and R.~M.~Wald, \emph{``Local symmetries
and constraints,''} J. Math. Phys.\ \textbf{31}, 725 (1990).

\bibitem{34} R.~M.~Wald and A.~Zoupas, \emph{``General definition
of conserved quantities in general relativity and other theories of
gravity,''} Phys. Rev. D\ \textbf{61}, 084027 (2000).

\bibitem{14}  Y.~Tachikawa, \emph{``Black Hole Entropy in the presence
of Chern-Simons Terms,''} Class. Quant. Grav.\ \textbf{24}, 737
(2007).

\bibitem{102} G.~Barnich and G.~Compere, \emph{``Surface charge
algebra in gauge theories and thermodynamic integrability,''} J.
Math. Phys.\ \textbf{49}, 042901 (2008).

\bibitem{103} G.~Barnich and G.~Compere, \emph{``Generalized Smarr
relation for Kerr AdS black holes from improved surface integrals,''}
Phys. Rev. D\ \textbf{71}, 044016 (2005).

\bibitem{32} W.~Kim, S.~Kulkarni, S.~H.~Yi, \emph{``Quasilocal
conserved charges in the presence of a gravitational Chern-Simons
term,''} Phys. Rev. D\ \textbf{88}, 124004 (2013).

\bibitem{36} S.~Hyun, J.~Jeong, S.~A.~Park, S.~H.~Yi, \emph{``Quasilocal
conserved charges and holography,''} Phys. Rev. D\ \textbf{90},
104016 (2014).

\bibitem{2} E.~A.~Bergshoeff, O.~Hohm and P.~K.~Townsend, \emph{``More
on massive 3D gravity,''} Phys. Rev. D\ \textbf{79}, 124042 (2009).

\bibitem{3} E.~Bergshoeff, O.~Hohm, W.~Merbis, A.~J.~Routh and
P.~K.~Townsend, \emph{``Minimal Massive 3D Gravity,''} Class.
Quantum Gravity\ \textbf{31}, 145008 (2014).

\bibitem{4} E.~A.~Bergshoeff, S.~de Haan, O.~Hohm, W.~Merbis
and P.~K.~Townsend, \emph{``Zwei-dreibein gravity: a two-frame-field
model of 3D massive gravity,''} Phys. Rev. Lett.\ \textbf{111},
111102 (2013).

\bibitem{5} M.~R.~Setare, \emph{``On the Generalized Minimal Massive
Gravity,''} Nucl. Phys. B\ \textbf{898}, 259 (2015).

\bibitem{11} O.~Hohm, A.~Routh, P.~K.~Townsend and B.~Zhang,
\emph{``On the Hamiltonian form of 3D massive gravity,''} Phys.
Rev. D\ \textbf{86}, 084035 (2012);\\
 E.~A.~Bergshoeff, O.~Hohm, W.~Merbis, A.~J.~Routh and P.~K.~Townsend,
\emph{``Chern-Simons-Like Gravity Theories,''} Lect. Notes Phys.\ \textbf{892},
181 (2015).

\bibitem{6} M.~Nakahara, \emph{``Geometry, topology, and physics,''}
Taylor and Francis (2003).

\bibitem{7} N.~J.~Poplawski, \emph{``Spacetime and fields,''}
arXiv:0911.0334 {[}gr-qc{]}.

\bibitem{8} T.~Jacobson and A.~Mohd, \emph{``Black hole entropy
and Lorentz-diffeomorphism Noether charge,''} Phys. Rev. D\ \textbf{92},
124010 (2015).

\bibitem{10} R.~M.~Wald, \emph{``General Relativity,''} University
of Chicago Press, Chicago (1980).

\bibitem{12} A.~Achucarro and P.~K.~Townsend, \emph{``A Chern-Simons
action for three-dimensional anti-de Sitter supergravity theories,''}
Phys. Lett. B\ \textbf{180}, 89 (1986).

\bibitem{15} W.~Merbis, \emph{``Chern-Simons-like Theories of Gravity,''}
arXiv: 1411.6888 {[}gr-qc{]}.

\bibitem{Yekta1} D.~Mahdavian Yekta, \emph{``Hamiltonian formalism
of Minimal Massive Gravity,''} Phys.\ Rev.\ D \textbf{92}, no.
6, 064044 (2015).

\bibitem{Yekta2} D.~Mahdavian Yekta, \emph{``Canonical structure
of BHT massive gravity in warped AdS}\textsubscript{\emph{3}}\emph{
sector,''} Phys.\ Lett.\ B \textbf{759}, 115 (2016).

\bibitem{18} G.~Giribet, J.~Oliva, D.~Tempo and R.~Troncoso,
\emph{``Microscopic entropy of the three-dimensional rotating black
hole of Bergshoeff-Hohm-Townsend massive gravity,''} Phys. Rev. D\ \textbf{80},
124046 (2009).

\bibitem{19} G.~Giribet and M.~Leston, \emph{``Boundary stress
tensor and counterterms for weakened AdS$_{3}$ asymptotic in New
Massive Gravity,''} JHEP\ \textbf{09}, 070 (2010).

\bibitem{21} L.~Bonora, M.~Cvitan, P.~D.~Prester, S.~Pallua,
I.~Smolic, \emph{``Gravitational Chern-Simons Lagrangians and black
hole entropy,''} JHEP\ \textbf{1107}, 085 (2011).

\bibitem{22} E.~Tonni, \emph{``Warped black holes in 3D general
massive gravity,''} JHEP\ \textbf{1008}, 070 (2010).

\bibitem{23} K.~A.~Moussa, G.~Clement and C.~Leygnac, \emph{``The
black holes of topologically massive gravity,''} Class. Quant. Grav.\ \textbf{20},
L277 (2003).

\bibitem{25} K.~A.~Moussa, G.~Clement, H.~Guennoune and C.~Leygnac,
\emph{``Three-dimensional Chern-Simons black holes,''} Phys. Rev.
D\ \textbf{78}, 064065 (2008).

\bibitem{26} G.~Clement, \emph{``Warped AdS(3) black holes in new
massive gravity,''} Class. Quant. Grav.\ \textbf{26}, 105015 (2009).

\bibitem{27} S.~Detournay, L.~A.~Douxchamps, G.~S.~Ng and C.~Zwikel,
\emph{``Warped AdS$_{3}$ black holes in higher derivative gravity
theories,''} JHEP\ \textbf{1006}, 014 (2016).

\bibitem{Regge:1974otg} T.~Regge and C.~Teitelboim, \emph{``Improved
Hamiltonian for general relativity,''} Phys.\ Lett.\ \textbf{53B}
(1974) 101.

\bibitem{28} L.~F.~Abbott, S.~Deser, \emph{``Charge Definition
in Nonabelian Gauge Theories,''} Phys. Lett. B\ \textbf{116}, 259
(1982).

\bibitem{76} M.~R.~Setare and H.~Adami, \emph{``The Heisenberg
algebra as near horizon symmetry of the black flower solutions of
Chern-Simons-like theories of gravity,''} Nucl. Phys. B\ \textbf{914},
220 (2017).

\bibitem{37} K.~Prabhu, \emph{``The First Law of Black Hole Mechanics
for Fields with Internal Gauge Freedom,''} Class. Quant. Grav.\ \textbf{34},
3, 035011 (2017).

\bibitem{41} G.~Barnich, C.~Troessaert, \emph{``Aspects of the
BMS/CFT correspondence,''} J.High Energy Phys.\ \textbf{1005}, 062
(2010).

\bibitem{46} H.~W.~J.~Blote, J.~A.~Cardy and M. P. Nightingale,
\emph{``Conformal invariance, the central charge, and universal finite-size
amplitudes at criticality,''} Phys.\ Rev.\ Lett.\textbf{\ 56},
742 (1986).

\bibitem{75} M.~R.~Setare and H.~Adami, \emph{``Black hole entropy
in the Chern-Simons-like theories of gravity and Lorentz-diffeomorphism
Noether charge,''} Nucl. Phys. B\ \textbf{902}, 115 (2016).

\bibitem{42} S.~Silva, \emph{``Black hole entropy and thermodynamics
from symmetries,''} Class. Quant. Grav.\ \textbf{19}, 3947 (2002).

\bibitem{43} S.~Carlip, \emph{``Black Hole Entropy from Conformal
Field Theory in Any Dimension,''} Phys. Rev. Lett.\ \textbf{82},
2828 (1999).

\bibitem{77}  M.~R.~Setare and H.~Adami, \emph{``Lorentz-diffeomorphism
quasi-local conserved charges and Virasoro algebra in Chern-Simons-like
theories of gravity,''} Nucl. Phys. B\ \textbf{909}, 345 (2016).

\bibitem{44} Y.~Liu and Y.~W.~Sun, \emph{``Generalized massive
gravity in AdS$_{3}$ spacetime,''} Phys. Rev. D\ \textbf{79} 126001
(2009).

\bibitem{45} J.~A.~Cardy, \emph{``Operator content of two-dimensional
conformally invariant theories,''} Nucl. Phys. B\ \textbf{270} 186
(1986).

\bibitem{47} K.~Hotta, Y.~Hyakutake, T.~Kubota and H.~Tanida,
\emph{``Brown-Henneaux's Canonical Approach to Topologically Massive
Gravity,''} JHEP\ \textbf{0807}, 066 (2008).

\bibitem{Compere:2009zj} G.~Compere and S.~Detournay, \emph{``Boundary
conditions for spacelike and timelike warped AdS$_{3}$ spaces in
topologically massive gravity,''} JHEP \textbf{0908} (2009) 092.

\bibitem{48} M.~Henneaux, C.~Martinez, and R.~Troncoso, \emph{``Asymptotically
warped anti-de Sitter spacetimes in topologically massive gravity,''}
Phys. Rev. D\ \textbf{84}, 124016 (2011).

\bibitem{81} M.~R.~Setare and H.~Adami, \emph{``Asymptotically
spacelike warped anti-de Sitter spacetimes in generalized minimal
massive gravity,''} Class. Quant. Grav.\ \textbf{34}, no.12, 125008
(2017).

\bibitem{Compere:2008cv} G.~Compere and S.~Detournay, \emph{``Semi-classical
central charge in topologically massive gravity,''} Class.\ Quant.\ Grav.\ \textbf{26}
(2009) 012001 Erratum: {[}Class.\ Quant.\ Grav.\ \textbf{26} (2009)
139801{]}.

\bibitem{50} H.~Sugawara, \emph{``A Field Theory of Currents,''}
Phys. Rev.\ \textbf{170}, 1659 (1968).

\bibitem{51} M.~Blagojevic, B.~Cvetkovic, \emph{``Asymptotic structure
of topologically massive gravity in spacelike stretched AdS sector,''}
JHEP\ \textbf{0909}, 006 (2009).

\bibitem{52} L.~Donnay and G.~Giribet, \emph{``Holographic entropy
of Warped-AdS$_{3}$ black holes,''} JHEP\ \textbf{1506}, 099 (2015).

\bibitem{53} S.~Detournay, T.~Hartman and D.~M.~Hofman, \emph{``Warped
conformal field theory,''} Phys. Rev. D\ \textbf{86}, 124018 (2012).

\bibitem{82} M.~R.~Setare and H. Adami, \emph{``Quasi-local conserved
charges in Lorenz-diffeomorphism covariant theory of gravity,''}
Eur. Phys. J. C\ \textbf{76}, 187 (2016).

\bibitem{104} M.~R.~Setare and H.~Adami, \emph{``The entropy
formula of black holes in Minimal Massive Gravity and its application
for BTZ black holes,''} Phys. Rev. D\ \textbf{91}, 104039 (2015).

\bibitem{Setare:2011jt} M.~R.~Setare and V.~Kamali, \emph{``Generalized
Massive Gravity and Galilean Conformal Algebra in two dimensions,''}
EPL \textbf{98} (2012) no.3, 31001.

\bibitem{Setare:2015vea} M.~R.~Setare and H.~Adami, \emph{``Black
hole conserved charges in Generalized Minimal Massive Gravity,''}
Phys.\ Lett.\ B \textbf{744} (2015) 280.

\bibitem{54} M.~Blagojevic and B.~Cvetkovic, \emph{``Conformally
flat black holes in Poincare gauge theory,''} Phys. Rev. D\ \textbf{93},
044018 (2016).

\bibitem{83} M.~R.~Setare and H.~Adami, \emph{``Near Horizon
Symmetries of the Non-Extremal Black Hole Solutions of Generalized
Minimal Massive Gravity,''} Phys. Lett. B\ \textbf{760}, 411 (2016).

\bibitem{55} L.~Donnay, G.~Giribet, H.~A.~Gonzalez, M.~Pino,
\emph{``Supertranslations and Superrotations at the Black Hole Horizon,''}
Phys. Rev. Lett.\ \textbf{116}, 091101 (2016).

\bibitem{56}G.~Barnich and G.~Compere, \emph{``Classical central
extension for asymptotic symmetries at null infinity in three spacetime
dimensions,''} Class. Quant. Grav.\ \textbf{24}, F15 (2007).

\bibitem{57} H.~Afshar, S.~Detournay, D.~Grumiller, W.~Merbis,
A.~Perez, D.~Tempo and R.~Troncoso, \emph{``Soft Heisenberg hair
on black holes in three dimensions,''} Phys. Rev. D\ \textbf{93}
(2016) 101503.

\bibitem{HPS}  S.~W.~Hawking, M.~J.~Perry and A.~Strominger,
\emph{``Soft Hair on Black Holes,''} Phys.\ Rev.\ Lett.\ \textbf{116},
no. 23, 231301 (2016).

\bibitem{59} J.~de~Boer, J.~I.~Jottar, \emph{``Thermodynamics
of Higher Spin Black Holes in AdS$_{3}$,''} JHEP\ \textbf{1401},
023 (2014).

\bibitem{60} C.~Bunster, M.~Henneaux, A.~Perez, D.~Tempo and
R.~Troncoso, \emph{``Generalized Black Holes in Three-dimensional
Spacetime,''} JHEP \textbf{1405}, 031 (2014).

\bibitem{61} A.~Perez, D.~Tempo and R.~Troncoso, \emph{``Higher
spin gravity in 3D: black holes, global charges and thermodynamics,''}
Phys. Lett. B\ \textbf{726}, 444 (2013).

\bibitem{Carlip:1994gy} S.~Carlip, \emph{``The Statistical mechanics
of the (2+1)-dimensional black hole,''} Phys.\ Rev.\ D \textbf{51}
(1995) 632.

\bibitem{Strominger:1997eq} A.~Strominger, \emph{``Black hole entropy
from near horizon microstates,''} JHEP \textbf{9802} (1998) 009.

\bibitem{62} M.~Banados, \emph{``Three-dimensional quantum geometry
and black holes,''} AIP Conf. Proc.\ \textbf{484}, 147 (1999), arXiv:hep-th/9901148v3
{[}hep-th{]}.

\bibitem{63} G.~Compere, P.~-J.~ Mao, A.~Seraj and M.~M.~Sheikh-Jabbari,
\emph{``Symplectic and Killing symmetries of AdS$_{3}$ gravity:
holographic vs boundary gravitons,''} JHEP\ \textbf{01}, 080 (2016).

\bibitem{64} M.~M.~Sheikh-Jabbari and H.~Yavartanoo, \emph{``On
Quantization of AdS3 Gravity I: Semi-Classical Analysis,''} JHEP\ \textbf{1407},
104 (2014).

\bibitem{65} M.~M.~Sheikh-Jabbari and H.~Yavartanoo, \emph{``Horizon
Fluffs: Near Horizon Soft Hairs as Microstates of Generic AdS3 Black
Holes,''} Phys. Rev. D\ \textbf{95}, 044007 (2017).

\bibitem{79} M.~R.~Setare and H. Adami, \emph{``Horizon Fluffs:
In the Context of Generalized Minimal Massive Gravity,''} arXiv:
1611.04259 {[}hep-th{]}.

\bibitem{66} P.~Di~Francesco, P.~Mathieu and D.~Senechal, \emph{``Conformal
Field Theory''} (Springer, 1997).

\bibitem{Afshar:2016uax} H.~Afshar, D.~Grumiller and M.~M.~Sheikh-Jabbari,
``Near horizon soft hair as microstates of three dimensional black
holes,'' Phys.\ Rev.\ D \textbf{96} (2017) no.8, 084032.

\bibitem{67} S.~Detournay and M.~Riegler, \emph{``Enhanced Asymptotic
Symmetry Algebra of 2+1 Dimensional Flat Space,''} Phys. Rev. D\ \textbf{95},
046008 (2017).

\bibitem{80} M.~R.~Setare and H.~Adami, \emph{``Enhanced asymptotic
BMS$_{3}$ algebra of the flat spacetime solutions of generalized
minimal massive gravity,''} arXiv:1703.00936v1 {[}hep-th{]}.

\bibitem{68} G.~Barnich, \emph{``Entropy of three-dimensional asymptotically
flat cosmological solutions,''} JHEP\ \textbf{10}, 095 (2012).

\bibitem{69} A.~Bagchi, S.~Detournay and D. Grumiller, \emph{``Flat-Space
Chiral Gravity,''} Phys. Rev. Lett.\ \textbf{109}, 151301 (2012).

\bibitem{70} S.~del~Pino, G.~Giribet,~A. Toloza and J.~Zanelli,
\emph{``From Lorentz-Chern-Simons to Massive Gravity in 2+1 Dimensions,''}
JHEP\ \textbf{06}, 113 (2015).

\bibitem{78} M.~R.~Setare and H.~Adami, \emph{``Conserved Charges
of Minimal Massive Gravity Coupled to Scalar Field,''} arXiv:1604.07837v1
{[}hep-th{]}.

\bibitem{71} M.~Blagojevic and B.~Cvetkovic, \emph{``Black hole
entropy in 3D gravity with torsion,''} Class. Quant .Grav.\
\textbf{23}, 4781 (2006).

\bibitem{72} M.~Blagojevic and B.~Cvetkovic, \emph{``Black hole
entropy from the boundary conformal structure in 3D gravity with torsion,''}
JHEP\ \textbf{0610}, 005 (2006).

\bibitem{73}M.~Blagojevic and B.~Cvetkovic, \emph{``Electric field
in 3D gravity with torsion,''} Phys. Rev. D\ \textbf{78}, 044036
(2008).

\bibitem{74} E.~W.~Mielke and P.~Baekler, \emph{``Topological
gauge model of gravity with torsion,''} Phys. Lett. A\ \textbf{156},
399 (1991).
\end{thebibliography}
\end{document}